\documentclass[aip,jap,amsmath,amssymb,reprint,groupedaddress]{revtex4-1}%
\usepackage{graphicx}
\usepackage{dcolumn}
\usepackage{natmove}
\usepackage[T1]{fontenc}
\usepackage{amsmath}
\usepackage{graphicx}
\usepackage[exponent-product=\cdot,range-phrase=-]{siunitx}
\usepackage{xcolor}
\usepackage{braket}
\usepackage{longtable}%
\usepackage{amsfonts}%
\usepackage{amssymb}
\usepackage{hyperref}

\DeclareRobustCommand{\orcid}[1]{\href{http://orcid.org/#1}{%
    \includegraphics[height=\fontcharht\font`\B]{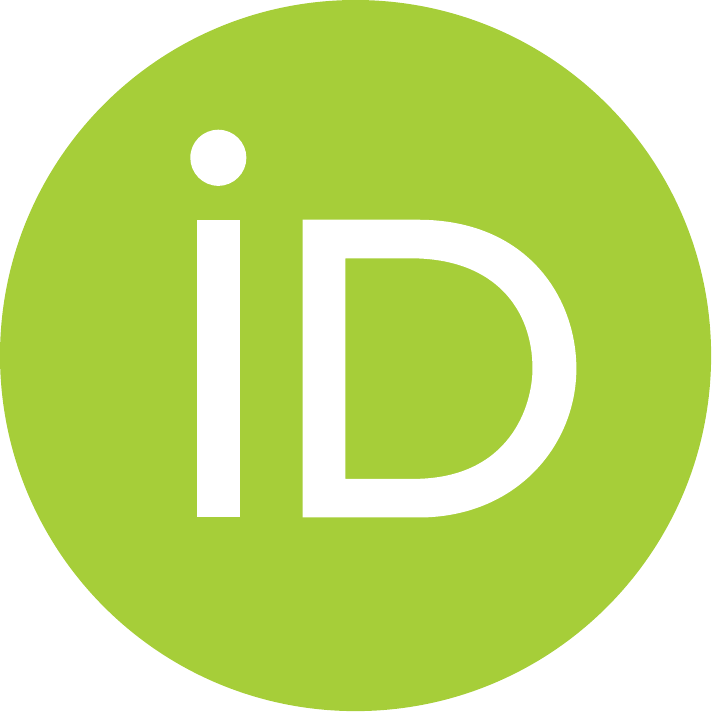}}
}

\begin{document}

\title{Optoelectronic Device Simulations based on Macroscopic Maxwell-Bloch Equations}
\author{Christian Jirauschek\orcid{0000-0003-0785-5530}}
\email{jirauschek@tum.de}
\author{Michael Riesch\orcid{0000-0002-4030-2818}}
\author{Petar Tzenov}
\affiliation{Department of Electrical and Computer Engineering, Technical University of Munich, Arcisstr. 21, 80333 Munich, Germany}
\date{9 June 2020, published as Adv. Theory Simul. 2, 1900018 (2019)}

\begin{abstract}
Due to their intuitiveness, flexibility and relative numerical efficiency, the
macroscopic Maxwell-Bloch (MB) equations are a widely used semiclassical and
semi-phenomenological model to describe optical propagation and coherent
light-matter interaction in media consisting of discrete-level quantum
systems. This review focuses on the application of this model to advanced
optoelectronic devices, such as quantum cascade and quantum dot lasers. The
Bloch equations are here treated as a density matrix model for driven quantum
systems with two or multiple discrete energy levels, where dissipation is
included by Lindblad terms. Furthermore, the one-dimensional MB equations for
semiconductor waveguide structures and optical fibers are rigorously derived.
Special analytical solutions and suitable numerical methods are presented. Due
to the importance of the MB equations in computational electrodynamics, an
emphasis is placed on the comparison of different numerical schemes, both with
and without the rotating wave approximation. The implementation of additional
effects which can become relevant in semiconductor structures, such as spatial
hole burning, inhomogeneous broadening and local-field corrections, is
discussed. Finally, links to microscopic models and suitable extensions of the
Lindblad formalism are briefly addressed.
This is the peer reviewed version of the following article: C. Jirauschek, M. Riesch, and P. Tzenov, ``Optoelectronic device simulations based on macroscopic Maxwell-Bloch equations'', Adv. Theory Simul. 2, 1900018 (2019), which has been published in final form at \url{https://doi.org/10.1002/adts.201900018}. This article may be used for non-commercial purposes in accordance with Wiley Terms and Conditions for Use of Self-Archived Versions.
\end{abstract}

\maketitle

\tableofcontents

\section*{List of Symbols and Acronyms}%

\begin{longtable}
[c]{ll} Symbol & Description\\\hline$\mathbf{A}$ & Magnetic vector potential\\
$A_{\mathrm{cv}}$ & Spontaneous emission coefficient\\ $A_{\mathrm{eff}}$ &
Effective mode area\\ $A_{\mathrm{q}}$ & Quantum active region cross section\\
$a$ & Power loss coefficient\\ $c$ & Vacuum speed of light\\ $D$ & Diffusion
coefficient\\ $\mathbf{D}$ & Displacement field\\ $\mathcal{D}$ & Dissipation
superoperator\\ $\mathcal{D}_{ijmn}$ & Tensor element of $\mathcal{D}$\\
$\mathbf{\hat{d}}$ & Dipole operator\\ $\mathbf{d}_{ij}$ & Dipole matrix
element vector\\ $\mathbf{E}$ & Electric field\\ $\underline{\mathbf{E}}$ &
Slowly varying amplitude of $\mathbf{E}$\\ $\underline{E}$ & $\underline
{E}=\underline{\mathbf{E}}\mathbf{e}$\\ $\underline{E}_{\pm}$ & $\underline
{E}$ for forward/backward propagating field\\ $E_{\mathrm{g}}$ & Bandgap
energy\\ $E_{i}$ & Eigenenergy of level $i$\\ $\underline{E}_{p}^{\omega}$ &
Fourier amplitude of waveguide E-field\\ $\underline{\mathbf{E}}^{\mathrm{t}}$
& Transverse dependency of $\underline{\mathbf{E}}$\\ EIT &
Electromagnetically induced transparency\\ $e$ & Elementary charge\\
$\mathbf{e}$ & Polarization direction of electric field\\ $F$ & Modal field
distribution\\ $F_{i}$ & Wavefunction of semiconductor state $i$,
$F_{i}=\varphi_{i}u_{v_{i} }$\\ FDTD & Finite difference time-domain\\
$f^{\pm}$ & Forward/backward normalized polarization amplitude\\ $g$ & Power
gain coefficient\\ $g\left(                \omega\right)                  $ &
Distribution function of resonance frequencies\\ $\mathbf{H}$ & Magnetic
field\\ $\underline{\mathbf{H}}$ & Slowly varying amplitude of $\mathbf{H}$\\
$\hat{H}$ & Hamiltonian\\ $\hat{H}_{0}$ & Unperturbed system Hamiltonian\\
$\hat{H}_{1}$ & Interaction Hamiltonian\\ $\underline{H}_{p}^{\omega}$ &
Fourier amplitude of waveguide H-field\\ $\underline{\mathbf{H}}^{\mathrm{t}}$
& Transverse dependency of $\underline{\mathbf{H}}$\\ $\hbar$ & Reduced Planck
constant\\ $I$ & Optical intensity\\ $I_{\mathrm{s}}$ & Saturation intensity\\
$\mathbf{J}_{\mathrm{f} }$ & Current density due to free carriers\\
$\mathbf{J}_{\mathrm{q}}$ & Total current density due to quantum systems\\
$\mathbf{k}$ & 3D bulk/2D in-plane wavevector\\ $k_{0}$ & $\omega/c$\\
$k_{\mathrm{c}}$ & Carrier wavenumber with $\left|   k_{\mathrm{c}
}\right|                 =n_{0}\omega_{\mathrm{c}}/c$\\ $\mathcal{L}$ &
Liouville superoperator\\ $\hat{L}$ & Lindblad operator\\ $\hat{L}
_{\alpha\rightarrow\beta}$ & $\hat{L}$ for incoherent transition\\ $\ell$ &
$v_{\mathrm{g}}a/2$\\ MB & Maxwell-Bloch\\ $m$ & Mass\\ $m^{\ast}$ & Effective
mass\\ $m_{\mathrm{e}}$ & Electron mass\\ $N$ & Number of system levels\\
$\underline{n}$ & Complex refractive index\\ $n_{0}$ & Background refractive
index\\ $n_{\mathrm{2D}}$ & QD sheet density\\ $n_{\mathrm{3D}}$ & Carrier
number density\\ $n_{\mathrm{cv}}$ & Number density of electron-hole pairs\\
$n_{\mathrm{eff}}$ & Effective waveguide index $\beta/k_{0}$\\ $\underline
{n}_{\mathrm{eff}}$ & Complex effective waveguide index $\underline{\beta
}/k_{0}$\\ $P$ & Optical power\\ $\mathbf{P}$ & Macroscopic polarization\\
$\underline{\mathbf{P}}$ & Slowly varying amplitude of $\mathbf{P}
_{\mathrm{q}}$\\ $\underline{P}_{p}^{\omega}$ & Fourier amplitude of quantum
system polarization\\ $\mathbf{P}_{\mathrm{q}}$ & Contribution of $\mathbf{P}$
due to quantum systems\\ $P_{\mathrm{s}}$ & Saturation power\\ $\mathbf
{\hat{p}}$ & Momentum operator\\ $p_{s}$ & Carrier fraction for inhomogeneous
broadening\\ QCL & Quantum cascade laser\\ QD & Quantum dot\\ $q$ & Carrier
charge\\ RNFD & Risken-Nummedal finite differences\\ RWA & Rotating wave
approximation\\ $\mathbf{r}$ & Microscopic position vector\\ $\mathbf{\hat{r}
}$ & Position operator\\ $r_{i}$ & Outscattering rate from level $i$\\
$r_{i\rightarrow j}$ & Transition rate from level $i$ to $j$\\ $S$ & In-plane
area of quantum well\\ $\mathbf{S}$ & Coherence vector, Bloch vector
$\left[               u,v,w\right]                 ^{\mathrm{T}}$\\ $S_{z}$ &
$z$ component of Poynting vector\\ SHB & Spatial hole burning\\ SVAA & Slowly
varying amplitude approximation\\ $s$ & Complex frequency variable\\
$\mathbf{s}$ & Bloch vector in RWA\\ $T_{1}$ & Energy relaxation time\\
$T_{2}$ & Phase relaxation rate\\ TE & Transverse electric\\ TM & Transverse
magnetic\\ $t$ & Time variable\\ $u$ & Bloch vector component\\ $u_{v}$ &
Periodic Bloch function of band $v$\\ $V$ & Potential energy\\ $V_{\mathrm{p}
}$ & Probe volume\\ $v$ & Bloch vector component\\ $\mathbf{\hat{v}}$ &
Velocity operator\\ $v_{\mathrm{g}}$ & Group velocity\\ $W_{\alpha
\beta,\mathbf{k} }^{\gamma}$ & Boltzmann rates\\ $w$ & Population inversion\\
$x$ & Coordinate, e.g., in growth direction\\ $\mathbf{x}$ & Macroscopic
position vector\\ $y$ & In-plane coordinate\\ $z$ & Propagation coordinate\\
$\beta$ & $\Re\left\{                  \underline{\beta}\right\}
$\\ $\underline{\beta}$ & Complex propagation constant\\ $\beta_{n}$ &
$\left[                  \mathrm{d}_{\omega}^{n}\beta\right]   _{\omega
=\omega_{\mathrm{c} }} $\\ $\Gamma$ & Overlap factor\\ $\Gamma_{ij}$ & Pumping
rate from level $i$ to $j$\\ $\gamma$ & Self-phase modulation coefficient\\
$\gamma_{1}$ & Energy relaxation rate\\ $\gamma_{2}$ & Phase relaxation rate\\
$\gamma_{ij}$ & Dephasing rate between levels $i$ and $j$\\ $\gamma
_{ij}^{\prime}$ & Pure dephasing rate between levels $i$ and $j$\\ $\Delta$ &
Frequency detuning $\Delta_{21}$ in two-level system\\ $\Delta_{ij}$ &
Frequency detuning $\mathrm{sgn}\left(        \omega_{ij} \right)   \left(
\omega_{\mathrm{c} }-\left|        \omega_{ij} \right|      \right)    $\\
$\Delta_{n}$ & Transverse refractive index profile\\ $\Delta_{t}$ & Time
step\\ $\Delta_{z}$ & Spatial increment\\ $\Delta_{\omega}$ & $\omega
-\omega_{\mathrm{c}}$\\ $\epsilon_{0}$ & Vacuum permittivity\\ $\epsilon
_{\mathrm{r}}$ & Dielectric constant\\ $\underline{\epsilon} _{\mathrm{r}}$ &
Complex dielectric constant\\ $\eta_{ij}$ & Slowly varying envelope function
of $\rho_{ij}$\\ $\eta_{ij}^{\pm}$ & $\eta_{ij}$ for forward/backward
propagating field\\ $\mu_{0}$ & Vacuum permeability\\ $\hat{\rho}$ & Density
operator\\ $\rho_{ii}^{\pm}$ & Population grating amplitude, $\rho_{ii}
^{+}=\left(                 \rho_{ii} ^{-}\right)      ^{\ast} $\\ $\rho
_{ii}^{0}$ & Center value of population grating\\ $\rho_{ij}$ & Density matrix
element\\ $\sigma$ & Conductivity\\ $\varphi_{i}$ & Envelope wavefunction of
level $i$\\ $\Psi$ & Time dependent wavefunction\\ $\psi_{i}$ & 1D
wavefunction of level $i$ in quantum wells\\ $\Omega$ & Instantaneous Rabi
frequency\\ $\underline{\Omega}$ & Slowly varying amplitude of $\Omega$\\
$\Omega_{\mathrm{g}}$ & Generalized Rabi frequency\\ $\omega$ & Frequency
variable\\ $\omega_{\mathrm{c}}$ & Carrier frequency\\ $\omega_{ij}$ &
Transition frequency between levels $i$ and $j$
\end{longtable}
\setcounter{table}{0}

\section{\label{sec:intro}Introduction}

\begin{figure}[ptb]
\includegraphics{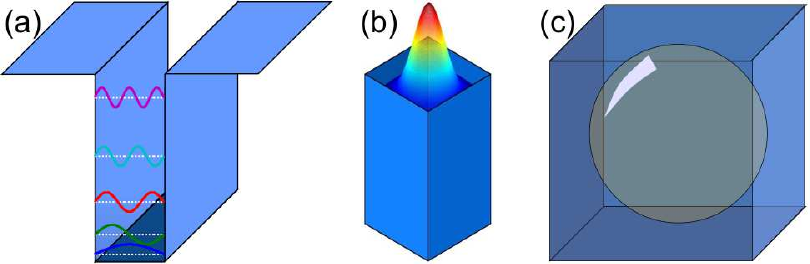}
\caption{Schematic illustration of quantum confinement structures: (a) Quantum
well, (b) quantum wire, (c) quantum dot.}%
\label{fig:conf}%
\end{figure}

Due to advancements in nanotechnology, structuring in the nanometer range is
meanwhile routinely exploited in electronics and photonics. For example, in
optoelectronic devices such as semiconductor optical amplifiers and lasers,
quantum confinement is widely used to concentrate the carriers in certain
energy states, yielding improved wall-plug efficiencies and higher output
powers. As a further effect, the wavelength can be tuned by changing the size
of the confinement structure. On a commercial basis, mostly one-dimensional
confinement is used in form of quantum well structures, which are fabricated
based on deposition of nanometer-thin semiconductor layers of different
compositions. In such structures, a quantum well is formed by a layer
consisting of a lower bandgap material than the adjacent layers, which
restricts the free electron motion in that layer to the in-plane directions
and gives rise to quantized energy states in growth direction. As a
consequence of the further restriction of the energy spectrum and the even
stronger carrier localization, additional improvement can be expected from
two- or three-dimensional confinement, resulting in quantum wire/dash and
quantum dot (QD) structures, respectively. Indeed, QD
\cite{kirstaedter1994low,ledentsov1998quantum,huffaker19981} and quantum dash
\cite{reithmaier2007inas} lasers and laser amplifiers have been shown to
exhibit excellent characteristics. In Fig.\thinspace\ref{fig:conf}, the
formation of quantized states in quantum wells, wires and dots is
schematically illustrated. The term quantum dash refers to an elongated
nanostructure, i.e., some kind of short quantum wire. By contrast, the term
nanowire does not necessarily indicate strong quantum confinement. For
example, in nanowire lasers the nanowire geometry typically serves as a
single-mode optical waveguide resonator, while the active region is based on a
heterostructure or quantum well, as in a conventional laser diode
\cite{duan2003single,mayer2017long}.

Semiconductor optoelectronic devices usually rely on electron-hole
recombination, i.e., optical transitions between conduction and valence band
states. The associated resonance wavelength is largely determined by the
semiconductor bandgap, which establishes a lower bound on the transition
energy. Thus, the coverage of a certain spectral region depends on the
existence of suitable semiconductor materials, which for example restricts the
availability of practical optoelectronic sources and detectors in the
mid-infrared and terahertz regions. An alternative concept is based on
intersubband devices, which employ so-called intersubband transitions between
quantized energy states in the conduction (or, in some cases, valence) band of
a nanostructure, and thus allow quantum engineering of the transition
wavelength independent of the bandgap. Quantum well devices based on this
concept include quantum cascade lasers \cite{1994Sci...264..553F}, quantum
cascade detectors \cite{hofstetter2002quantum,gendron2004quantum} and quantum
well infrared photodetectors \cite{levine1993quantum}. Furthermore,
intersubband transitions are used for QD infrared photodetectors
\cite{phillips1998far,lee1999bound,liu2001quantum}.

Along with quantum confinement, also quantum coherent effects are found to be
increasingly relevant for modern optoelectronic devices. Such effects result
from the coherent light-matter interaction, which requires that the states
involved in the optical transition maintain a well-defined phase relationship
over a significant time. The coherent interaction manifests itself in
so-called Rabi flopping \cite{rabi1937space}, i.e., carrier population
oscillations between the states, which are driven by the optical field. The
resulting carrier dynamics couples back to the optical field via the
polarization, thus also affecting the propagating optical waveform. Besides
being an essential prerequisite for the emerging field of quantum information
technology \cite{zrenner2002coherent}, quantum coherence plays an increasingly
important role for modern optoelectronic devices in general. Due to the strong
interaction with the semiconductor environment, e.g., in the form of phonon
scattering and carrier-carrier interactions, this phase relationship tends to
be quickly destroyed, which is commonly referred to as dephasing. However,
under favorable conditions, signatures of Rabi oscillations have been observed
in nanostructured optoelectronic systems and devices. These include quantum
well structures \cite{cundiff1994rabi,schulzgen1999direct}, nanowire lasers
\cite{mayer2017long}, quantum cascade lasers \cite{choi2010ultrafast}, and
single QDs \cite{stievater2001rabi,kamada2001exciton} at cryogenic
temperatures, as well as QD \cite{kolarczik2013quantum,karni2013rabi} and
quantum dash \cite{capua2014} amplifiers at room temperature. Closely related
is self-induced transparency \cite{mccall1967self,mccall1969self}, where Rabi
flopping enables a special optical pulse form to propagate without being
attenuated or disturbed. This phenomenon has meanwhile also been observed in
semiconductor structures such as QD waveguides
\cite{schneider2003self,karni2013rabi}, with potential applications such as
the generation of ultrashort optical pulses in QD and quantum cascade lasers
\cite{kozlov1997self,kalosha1999theory,kozlov2011obtaining,2009PhRvL.102b3903M,arkhipov2016self2}%
. Another effect that relies on quantum coherence is slow light propagation or
even complete halting of
light~\cite{hau1999light,liu2001observation,phillips2001storage}, with
possible applications such as optical buffers~\cite{khurgin2005optical},
imaging~\cite{camacho2007,firstenberg2009elimination} and quantum
memory~\cite{lukin2000entanglement}. This effect has meanwhile also been
demonstrated in solid-state media, namely in doped crystals
\cite{turukhin2001observation,bigelow2003observation}. The use of suitably
engineered semiconductor structures would be especially attractive from a
practical point of view
\cite{ginzburg2006slow,borges2012tunneling,tzenov2017slow}. Furthermore,
quantum interference, e.g., in QD or intersubband quantum well systems, is an
interesting candidate to realize all-optical
switching~\cite{wu2005ultrafast,gea2006optical}. Due to the discrete energy
level structure of QDs, semiconductor devices based thereupon are especially
likely to be, at least in part, governed by coherence effects
\cite{zrenner2002coherent}, although for QD ensembles the dephasing tends to
be strong \cite{kolarczik2013quantum}. The same applies to intersubband
quantum well devices, where the levels close to the band edge have parallel
dispersion relations, and thus the quantum dynamics resembles that of
discrete-level systems \cite{choi2010ultrafast}. Especially in such devices,
coherence effects can significantly influence the dynamic operation even for
considerable dephasing.

Suitable theoretical models are required for an in-depth understanding of the
often quite complex interplay of effects determining the dynamic device
characteristics, as well as for quantitative simulation and systematic device
optimization. For the optoelectronic devices and structures discussed above,
an adequate theoretical description must include the coherent carrier
dynamics, incoherent processes such as scattering or spontaneous emission, as
well as the interaction with the optical field. Our focus is here on the very
widely used Maxwell-Bloch (MB) equations. The Bloch equations provide a
compact model for the discrete-level carrier dynamics, which is described by
the density matrix formalism. The coherent single-carrier dynamics is here
modeled by the Hamiltonian of the quantum system, such as a QD, and also
includes the interaction with a classical optical field. Effects beyond the
single-electron quantum evolution are regarded as interaction with the
environment in form of the semiconductor host, which gives rise to incoherent
effects such as scattering with other carriers and phonons. The resulting
dissipation in the quantum system is in the Bloch equations phenomenologically
modeled by relaxation rate terms, which introduce dephasing and incoherent
carrier transitions. The Bloch equations were first devised to describe the
evolution of the nuclear magnetic moment in a magnetic field
\cite{bloch1946nuclear}, and later on extended to a pair of levels in
resonance with a classical optical field
\cite{feynman1957geometrical,abella1966photon}. The model is closed by
coupling the Bloch equations to Maxwell's equations
\cite{feynman1957geometrical,allen1987optical,mccall1967self}, which describe
the evolution of the classical optical field. This review paper is concerned
with the resulting MB equations, where we go beyond the often applied
two-level approximation
\cite{feynman1957geometrical,allen1987optical,abella1966photon,mccall1967self}
by considering multiple, albeit discrete,\ energy levels. Furthermore, we root
the phenomenological dissipation terms in the Lindblad formalism, which
ensures physical behavior of the quantum system and allows for the
construction of more general dissipation terms
\cite{lindblad1976generators,gorini1976completely}.

The MB equations offer a generic description of semiclassical light-matter
interaction, which can be applied to different media such as semiconductor
structures or gases. The focus of this review lies on semiconductor
structures, which is reflected in the treatment of some specific issues, such
as the concrete embodiment of Maxwell's equations, or the inclusion of spatial
hole burning in linear resonators. Independent of the modeled system, the main
attractiveness of the MB equations lies in the relatively compact description
of the carrier dynamics, which is helpful for providing intuitive insight into
the device behavior and even allows for closed analytical solutions in some
special cases \cite{allen1987optical,mccall1967self}. From a computational
point of view, the Bloch equations are widely used in combination with
electromagnetic simulations, e.g., based on~the finite-difference time-domain
method, as a quantum model of the medium \cite{taflove2005}, replacing simpler
classical descriptions such as the Lorentz model. Due to the relative
compactness of the Bloch model, also computationally demanding two- or
three-dimensional simulations can be carried out
\cite{slavcheva2002coupled,klaedtke2006ultrafast,sukharev2011,pusch2012coherent,lopata2009nonlinear,takeda2011self,dridi2013model,cartar2017}%
. Likewise, the MB equations enable systematic device optimizations over a
large parameter range, as well as long-term simulations, e.g., to investigate
the steady-state laser dynamics
\cite{kozlov2011obtaining,riesch2018dynamic,tzenov2016time}. Another important
advantage of the MB equations is that they can easily be adapted to specific
problems by adding further effects, such as inhomogeneous broadening
\cite{allen1987optical,mccall1967self} or local-field corrections
\cite{bowden1993near,slepyan2002quantum}.

Clearly, the Bloch equations constitute a compromise between accuracy and
compactness of the model. A full microscopic treatment of light-matter
interaction in a semiconductor, accounting for carrier-phonon and many-body
Coulomb interactions as well as for free carrier motion in the unconfined
directions, results in the so-called semiconductor MB equations
\cite{chow2012semiconductor,haug2009quantum}. These do not require
phenomenological input parameters, but the significantly increased model
complexity usually restricts the modeling to one spatial dimension and
short-term simulations. While the semiconductor MB equations are beyond the
scope of this review, they can be used as a basis to derive macroscopic
discrete-level MB equations with Lindblad dissipation and additional
correction terms for specific semiconductor structures
\cite{ning1997effective,yao1995semiconductor,balle1995effective}.

In detail, our paper is organized as follows: In Section \ref{sec:lindbl}, the
density matrix formalism and Lindblad model are introduced, serving as a basis
for the Bloch equations. These are treated in Section \ref{sec:Bloch}, which
also includes a discussion of the widely used rotating wave approximation
(RWA). In Section \ref{sec:maxwellbloch}, the MB equations are introduced in
full-wave treatment and invoking the RWA, along with the slowly varying
amplitude approximation (SVAA) for the field propagation. Section \ref{sec:1D}
treats the reduction of the MB equations for semiconductor waveguide
structures and optical fibers to a spatially one-dimensional model, which is a
widely used simplification. Section \ref{sec:Ana} deals with available
analytical solutions for the Bloch and MB equations, while in Section
\ref{sec:Num}, numerical methods for the MB equations are covered. Section
\ref{sec:eff} is dedicated to the inclusion of further effects, such as
local-field corrections, inhomogeneous broadening and noise. Section
\ref{sec:Appl} deals with the application of the MB model to concrete
optoelectronic devices, including bulk as well as inter- and intraband quantum
well and QD devices. The paper is concluded in Section \ref{sec:concl}, where
dissipation models beyond the Lindblad formalism are discussed.

\section{\label{sec:lindbl}Lindblad equation}

In the following, we consider discrete quantum systems with $N$ states
$\left|  i\right\rangle $, where $i=1..N$. We restrict ourselves to a
single-particle description, valid for carrier densities which are
sufficiently low to neglect Pauli blocking, but sufficiently high to neglect
electron-hole Coulomb correlation \cite{rosati2014scattering}. It has been
pointed out that these requirements are often fulfilled in state-of-the-art
semiconductor quantum devices which are the main scope of this paper, and that
the Lindblad approach introduced below is then well justified
\cite{rosati2014scattering}. Furthermore, we do not explicitly consider spin
dependent effects, even though the Lindblad formalism can be extended
accordingly \cite{stepanenko2006enhancement}.

The time evolution of an ideal quantum system is famously described by the
time dependent Schr\"{o}dinger equation
\begin{equation}
\mathrm{i}\hbar\partial_{t}\left|  \Psi\right\rangle =\hat{H}\left|
\Psi\right\rangle \label{eq:schr}%
\end{equation}
with the reduced Planck constant $\hbar$, where the system state vector
$\left|  \Psi\right\rangle $, and generally also the system Hamiltonian
$\hat{H}$, depend on time $t$. $\left|  \Psi\right\rangle $ is a pure state,
i.e., a coherent superposition of the basis states $\left|  i\right\rangle $
with $\left|  \Psi\left(  t\right)  \right\rangle =\sum_{i}c_{i}\left(
t\right)  \left|  i\right\rangle $, where $c_{i}$ are complex coefficients. In
reality, however, no quantum system is perfectly isolated, but rather
interacts with its environment. This induces decoherence, i.e., loss of
quantum coherence in the system, which must be included into any realistic
description. The resulting statistical state of the system is generally a
mixed state which cannot be represented by the system state vector $\left|
\Psi\right\rangle $, but rather requires an extended description in terms of
the density operator $\hat{\rho}$. The corresponding density matrix with
respect to the chosen basis states $\left|  i\right\rangle $ has the elements
$\rho_{ij}=\left\langle i\right|  \hat{\rho}\left|  j\right\rangle $, where
the diagonal elements $\rho_{ii}$ give the occupation probability of state
$\left|  i\right\rangle $, while the off-diagonal elements $\rho_{ij}$
represent the coherence between $\left|  i\right\rangle $ and $\left|
j\right\rangle $. The density operator is positive semidefinite which
guarantees that any pure system state $\left|  \Psi\right\rangle $ has a
non-negative probability, i.e., $\left\langle \Psi\right|  \hat{\rho}\left|
\Psi\right\rangle \geq0$. This also implies Hermiticity, i.e., $\hat{\rho
}=\hat{\rho}^{\dagger}\,$\ and thus $\rho_{ij}=\rho_{ji}^{\ast}$, where the
dagger and asterisk denote the adjoint and the complex conjugate,
respectively. Furthermore, at least for closed systems
\cite{schirmer2004constraints}, the trace must remain constant to ensure
particle conservation and is usually normalized to unity, $\mathrm{Tr}\left\{
\hat{\rho}\right\}  =1$. The coherent time evolution of the density operator
is in the Schr\"{o}dinger picture described by the von Neumann equation%
\begin{equation}
\mathrm{i}\hbar\partial_{t}\hat{\rho}=\left[  \hat{H},\hat{\rho}\right]  .
\label{eq:neumann}%
\end{equation}

\begin{figure}[ptb]
\includegraphics{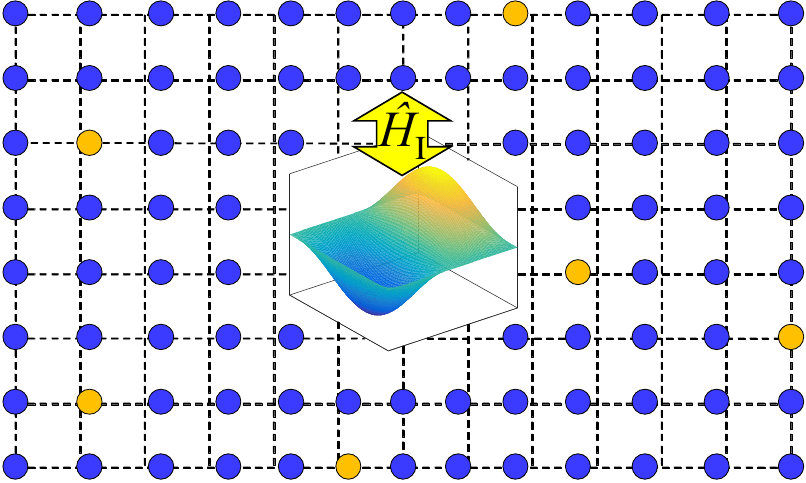}
\caption{Schematic illustration of a quantum system interacting with
impurities and thermal vibrations in the semiconductor lattice.}%
\label{fig:environment}%
\end{figure}

In realistic scenarios, often many degrees of freedom are relevant for the
time evolution and must thus be considered. Usually, only part of these
degrees of freedom are of direct interest for the application in mind, and
solving the full Eq.\thinspace(\ref{eq:neumann}) is typically also too
demanding. This issue can be addressed by performing a division into a system
containing the degrees of freedom which are of primary interest, and a second
one with the remaining degrees of freedom which then constitute the
environment. In semiconductor quantum devices, the degrees of freedom of
interest may be quantized states in a nanostructure such as a quantum well or
dot,\ while decoherence typically arises from interaction with the
semiconductor lattice itself, thus acting as the environment. This situation
is schematically illustrated in Fig.\thinspace\ref{fig:environment}. There are
various types of interactions, also referred to as scattering mechanisms,
which can induce decoherence in the quantum system of interest. These include
the interaction with phonons due to (longitudinal- and transverse-optical and
-acoustic) thermal lattice vibrations, lattice imperfections in form of
impurities (such as dopants), interface roughness or atomic disorder in
alloys, as well as piezoelectric fields. Also carrier-carrier interaction can
enter the single-particle picture as an additional scattering mechanism
\cite{rosati2014derivation,steinhoff2012treatment}. For a quantum system
interacting with the environment, $\hat{\rho}$ and $\hat{H}$ in Eq.\thinspace
(\ref{eq:neumann}) refer to the full dynamics of the combined system and environment.

The Hamiltonian $\hat{H}$ can be written as $\hat{H}=\hat{H}_{\mathrm{S}%
}\otimes\hat{I}_{\mathrm{E}}+\hat{I}_{\mathrm{S}}\otimes\hat{H}_{\mathrm{E}%
}+\hat{H}_{\mathrm{I}}$, where the Hamiltonians $\hat{H}_{\mathrm{S}}$,
$\hat{H}_{\mathrm{E}}$ and $\hat{H}_{\mathrm{I}}$ describe the system
$\mathrm{S}$, the environment $\mathrm{E}$ and the system-environment
interaction, $\hat{I}_{\mathrm{S}}$ and $\hat{I}_{\mathrm{E}}$ are the unit
operators in the respective Hilbert spaces, and $\otimes$ denotes the tensor
product \cite{breuer2002theory}. The reduced density matrix of the system of
interest is simply obtained by tracing over the environmental degrees of
freedom, $\hat{\rho}_{\mathrm{S}}=\mathrm{Tr}_{\mathrm{E}}\left\{  \hat{\rho
}\right\}  $. This step by itself does obviously not eliminate the dependence
of Eq.\thinspace(\ref{eq:neumann}) on the environment. Thus, additional
assumptions are necessary to arrive at a model for the non-unitary time
evolution of $\hat{\rho}_{\mathrm{S}}$, which is a consequence of eliminating
the environmental degrees of freedom. The resulting equation is expected to be
similar in structure as Eq.\thinspace(\ref{eq:neumann}), i.e., a first-order
linear differential equation in time for $\hat{\rho}_{\mathrm{S}}$, where the
linearity ensures consistency with the ensemble interpretation of the density
matrix \cite{preskill1998lecture}. The resulting time-local and Markovian
description of the reduced density matrix dynamics is commonly referred to as
(quantum) master equation. Its general form can be inferred by posing
additional requirements to avoid unphysical behavior. In particular, this
includes conservation of unit trace and positive semidefiniteness of the
density matrix, as discussed above. Closer inspection reveals that if there
exists another system $\mathrm{S}^{\prime}$, an evolution equation for
$\mathrm{S}$ which ensures positive semidefiniteness of $\hat{\rho
}_{\mathrm{S}}$ can still lead to unphysical time evolution of the combined
density matrix for $\mathrm{S}$ and $\mathrm{S}^{\prime}$, even if
$\mathrm{S}^{\prime}$ does not evolve and is completely decoupled from
$\mathrm{S}$ \cite{preskill1998lecture}. This problem is cured by demanding
complete positivity of the evolution, rather than only the preservation of
positive semidefiniteness of $\hat{\rho}_{\mathrm{S}}$.

From above requirements, the general form of the evolution equation can be
inferred by invoking the Kraus theorem \cite{kraus1971general}, characterizing
completely positive trace preserving maps. The resulting master equation is
called Lindblad equation \cite{lindblad1976generators,gorini1976completely}.
Dropping the subscript $\mathrm{S}$ from here on for ease of notation, it can
be written as%

\begin{align}
\partial_{t}\hat{\rho}  &  =-\frac{\mathrm{i}}{\hbar}\left[  \hat{H},\hat
{\rho}\right]  +\sum_{k}\left(  \hat{L}_{k}\hat{\rho}\hat{L}_{k}^{\dagger
}-\frac{1}{2}\hat{L}_{k}^{\dagger}\hat{L}_{k}\hat{\rho}-\frac{1}{2}\hat{\rho
}\hat{L}_{k}^{\dagger}\hat{L}_{k}\right) \nonumber\\
&  =\mathcal{L}\left(  \hat{\rho}\right)  +\sum_{k}\mathcal{D}_{k}\left(
\hat{\rho}\right)  =\mathcal{L}\left(  \hat{\rho}\right)  +\mathcal{D}\left(
\hat{\rho}\right)  , \label{eq:lindbl}%
\end{align}
where $\hat{H}=\hat{H}_{0}+\hat{H}_{1}$ is the effective Hamiltonian of the
reduced system. Here, the Hamiltonian $\hat{H}_{1}$ describes externally
induced perturbations, e.g., due to an incident optical field. The description
of light-matter interaction requires a time dependent Hamiltonian, which,
although lifting the originally assumed time-homogeneity of the Lindblad
equation, still gives a valid density matrix evolution
\cite{breuer2004genuine,kropf2016effective}. In addition, $\hat{H}$ may
contain non-dissipative contributions stemming from the interaction with the
environment, such as energy shifts \cite{breuer2002theory}. The dissipation is
described by the sum term, where the linear operators $\hat{L}_{k}$ are called
Lindblad or (quantum) jump operators, which can in principle be chosen without
further restrictions in the Hilbert space of the reduced system. Equation
(\ref{eq:lindbl}) now includes both the coherent dynamics due to the Liouville
superoperator $\mathcal{L}\left(  \hat{\rho}\right)  $, corresponding to
Eq.\thinspace(\ref{eq:neumann}), and the incoherent dynamics induced by the
dissipation superoperator $\mathcal{D}\left(  \hat{\rho}\right)  $ which
contains the interaction with the environment. Besides inferring the Lindblad
equation from the requirements given above, Eq.\thinspace(\ref{eq:lindbl}) can
also be microscopically derived, assuming that the quantum system is weakly
coupled to a large Markovian environment
\cite{davies1974markovian,breuer2002theory,le2011quantum}.

As mentioned above, we allow for a time dependent Hamiltonian in
Eq.\thinspace(\ref{eq:lindbl}), which is required to include light-matter
interaction as envisaged in this paper, and constitutes a slight
generalization of the original equation \cite{lindblad1976generators}.
Occasionally, also time dependent Lindblad operators $\hat{L}_{k}\left(
t\right)  $ are used, for example to model time dependent pumping rates
\cite{kantner2015modeling}. This also does not affect the physical validity of
Eq.\thinspace(\ref{eq:lindbl}), since conservation of trace and complete
positivity are further guaranteed \cite{breuer2004genuine,kropf2016effective}.
Moreover, Eq.\thinspace(\ref{eq:lindbl}) with time dependent operators
$\hat{H}$ and $\hat{L}_{k}$\ is still time-local and also Markovian
\cite{chruscinski2010non}.

\subsection{Introduction of Basis States}

In principle, the $N$ basis states of the (reduced) quantum system can be
freely selected as long as they span the entire Hilbert space of the $N$-level
system. In most cases, an orthonormal basis is the preferred option, since it
results in more compact expressions and provides a clearer physical
interpretation. The choice of energy eigenstates has the distinct advantage
that the reduced system Hamiltonian $\hat{H}_{0}$ is diagonal. In certain
cases, other choices may be preferable, such as a localized (or tight-binding)
basis set for the description of tunneling, e.g., in double- or multiple-well
systems \cite{grifoni1998driven}.

Assuming an orthonormal basis so that the unit operator becomes $\hat{I}=\sum
_{\ell=1}^{N}\left|  \ell\right\rangle \left\langle \ell\right|  $,
Eq.\thinspace(\ref{eq:lindbl}) can be written as%
\begin{align}
\partial_{t}\rho_{ij}  &  =-\frac{\mathrm{i}}{\hbar}\sum_{\ell}\left(
H_{i\ell}\rho_{\ell j}-H_{\ell j}\rho_{i\ell}\right) \nonumber\\
&\,\quad +\sum_{k}\sum_{\ell p}\bigg[  L_{k}^{i\ell}\left(  L_{k}^{jp}\right)
^{\ast}\rho_{\ell p}-\frac{1}{2}\left(  L_{k}^{\ell i}\right)  ^{\ast}%
L_{k}^{\ell p}\rho_{pj} \nonumber\\
&\,\quad -\frac{1}{2}\left(  L_{k}^{p\ell}\right)  ^{\ast}%
L_{k}^{pj}\rho_{i\ell}\bigg] \nonumber\\
&  =-\frac{\mathrm{i}}{\hbar}\sum_{mn}\mathcal{L}_{ijmn}\rho_{mn}+\sum
_{mn}\mathcal{D}_{ijmn}\rho_{mn}. \label{eq:lindbl2}%
\end{align}
Here, $H_{ij}=\left\langle i\right|  \hat{H}\left|  j\right\rangle $ and
$L_{k}^{ij}=\left\langle i\right|  \hat{L}_{k}\left|  j\right\rangle $ are the
matrix elements of the operators $\hat{H}$ and $\hat{L}_{k}$. Also the
superoperators can be represented in form of a matrix, albeit of size
$N^{2}\times N^{2}$, with elements
\begin{align}
\mathcal{L}_{ijmn}  &  =H_{im}\delta_{jn}-H_{nj}\delta_{im},\label{eq:Lijmn}\\
\mathcal{D}_{ijmn}  &  =\sum_{k}\bigg\{  L_{k}^{im}\left(  L_{k}^{jn}\right)
^{\ast} \nonumber\\
&\,\quad -\frac{1}{2}\sum_{\ell}\left[  \left(  L_{k}^{\ell i}\right)  ^{\ast
}L_{k}^{\ell m}\delta_{jn}+\left(  L_{k}^{\ell n}\right)  ^{\ast}L_{k}^{\ell
j}\delta_{im}\right]  \bigg\}  , \label{eq:Dijmn}%
\end{align}
where $\delta$ denotes the Kronecker delta. We emphasize that while
Eq.\thinspace(\ref{eq:Dijmn}) ensures that there is a matrix representation
$\mathcal{D}_{ijmn}$ for any given set of Lindblad operators, the converse is
not necessarily true, and arbitrarily chosen $\mathcal{D}_{ijmn}$ can produce
unphysical results.

\subsection{\label{sec:Lindop}Choice of Lindblad Operators}

The choice of the $\hat{L}_{k}$ for generating a certain time evolution is not
unique. In particular, for a given set $\hat{L}_{k}$ with $k=1,\dots,K$, the
set $\hat{L}_{k}^{\prime}=\sum_{\ell}u_{k\ell}\hat{L}_{\ell}$ (also with
$\ell=1,\dots,K$) generates the same dynamics for an arbitrary unitary matrix
with dimension $K$\ and elements $u_{k\ell}$
\cite{oi2012limits,breuer2002theory}. This can easily be verified by
substituting the $\hat{L}_{k}$ in Eq.\thinspace(\ref{eq:lindbl}) with above
expression for $\hat{L}_{k}^{\prime}$, and considering that $\sum_{k}u_{k\ell
}u_{km}^{\ast}=\delta_{\ell m}$. Furthermore, the $\hat{L}_{k}$ might also
contain unitary contributions, which can alternatively be included into the
Hamiltonian $\hat{H}$. In particular, replacing an operator $\hat{L}_{k}$\ by
$\hat{L}_{k}^{\prime}=\hat{L}_{k}+\alpha_{k}\hat{I}$ where $\alpha_{k}$ is an
arbitrary complex constant with dimension of inverse square root of time, and
$\hat{H}$\ by $\hat{H}^{\prime}=\hat{H}+\left(  \mathrm{i}\hbar/2\right)
\left(  \alpha_{k}\hat{L}_{k}^{\dagger}-\alpha_{k}^{\ast}\hat{L}_{k}\right)  $
generates the same dynamics \cite{oi2012limits,breuer2002theory}.\ From the
Kraus theorem \cite{kraus1971general} it follows that it is always possible to
choose the Lindblad operators so that a given non-unitary evolution can be
represented by $K\leq N^{2}-1$ operators [in addition to $\hat{L}_{0}%
\propto\hat{I}$ which gives a vanishing contribution in Eq.\thinspace
(\ref{eq:lindbl})]. Formally, such a representation can be constructed by
starting from the Kossakowski--Sudarshan form \cite{gorini1976completely} of
the Lindblad equation and applying a unitary transformation to convert it to
Eq.\thinspace(\ref{eq:lindbl}) \cite{breuer2002theory}. However, it has been
pointed out that the resulting standard form does not give much insight into
the underlying physical processes \cite{oi2012limits}. From a practical point
of view, it is more natural to choose the $\hat{L}_{k}$\ so that they
represent certain physical effects. In the following, we will discuss the two
most relevant mechanisms, i.e., incoherent transitions between states
corresponding to hopping transport, and pure dephasing which affects the
coherence between two states but does not involve population transfer between them.

\subsubsection{\label{sec:trans}Incoherent Transitions}

For a transition from a given basis state $\left|  \alpha\right\rangle $ to
$\left|  \beta\right\rangle $ with a rate $r_{\alpha\rightarrow\beta}$, the
associated Lindblad operator is given by
\begin{equation}
\hat{L}_{\alpha\rightarrow\beta}=r_{\alpha\rightarrow\beta}^{1/2}\left|
\beta\right\rangle \left\langle \alpha\right|  , \label{eq:Lif}%
\end{equation}
and Eq.\thinspace(\ref{eq:Dijmn}) for the corresponding superoperator matrix
elements yields
\begin{align}
\mathcal{D}_{ijmn}^{\alpha\rightarrow\beta}=  &  r_{\alpha\rightarrow\beta}\bigg[
\delta_{i\beta}\delta_{j\beta}\delta_{m\alpha}\delta_{n\alpha} \nonumber\\
&  -\frac{1}%
{2}\left(  \delta_{i\alpha}\delta_{jn}\delta_{m\alpha}+\delta_{im}%
\delta_{j\alpha}\delta_{n\alpha}\right)  \bigg]  . \label{eq:Dijmn2}%
\end{align}
Inserting Eq.\thinspace(\ref{eq:Lif}) into Eq.\thinspace(\ref{eq:lindbl2}), we
obtain population changes $\left[  \partial_{t}\rho_{\beta\beta}\right]
_{\alpha\rightarrow\beta}=r_{\alpha\rightarrow\beta}\rho_{\alpha\alpha}$,
$\left[  \partial_{t}\rho_{\alpha\alpha}\right]  _{\alpha\rightarrow\beta
}=-r_{\alpha\rightarrow\beta}\rho_{\alpha\alpha}$. The population relaxation
is thus generally described by rate equation terms%
\begin{equation}
\left[  \partial_{t}\rho_{\alpha\alpha}\right]  _{\mathrm{relax}}=\sum
_{j\neq\alpha}r_{j\rightarrow\alpha}\rho_{jj}-r_{\alpha}\rho_{\alpha\alpha},
\label{eq:relax1}%
\end{equation}
where%
\begin{equation}
r_{\alpha}=\sum_{j\neq\alpha}r_{\alpha\rightarrow j} \label{eq:rn}%
\end{equation}
is the total outscattering rate from level $\alpha$. Furthermore, we see that
apart from the population changes, $\hat{L}_{\alpha\rightarrow\beta}$ also
contains the associated lifetime contribution to dephasing, with $\left[
\partial_{t}\rho_{\alpha n}\right]  _{\alpha\rightarrow\beta}=-\left(
r_{\alpha\rightarrow\beta}/2\right)  \rho_{\alpha n}$ and $\left[
\partial_{t}\rho_{n\alpha}\right]  _{\alpha\rightarrow\beta}=-\left(
r_{\alpha\rightarrow\beta}/2\right)  \rho_{n\alpha}$ where $n\neq\alpha$. This
means that population transfer from a state $\left|  \alpha\right\rangle $ to
$\left|  \beta\right\rangle $ induces dephasing not only for this transition,
but also for other transitions involving $\left|  \alpha\right\rangle $, and
ignoring this fact might lead to unphysical results
\cite{schirmer2004constraints}. On the other hand, this implies that the total
lifetime contribution to the dephasing rate for a transition $\alpha
\rightarrow\beta$ is with Eq.\thinspace(\ref{eq:rn}) given by\ $\left(
r_{\alpha}+r_{\beta}\right)  /2$, i.e., is obtained from the total
outscattering rates for levels $\alpha$ and $\beta$. We note that the operator
in Eq.\thinspace(\ref{eq:Lif}) provides an elementary description of
transitions, but does for example not take into account correlations between
different transition processes.

\subsubsection{\label{sec:dep}Pure Dephasing}

In addition to above discussed population changes, there can be additional
mechanisms which do not involve population transfer between the chosen basis
states, but cause additional decoherence, resulting in a decay of off-diagonal
density matrix elements only
\cite{rebentrost2009environment,fathololoumi2012terahertz,dinh2012extended,2010PhRvB..81t5311D}%
. This so-called pure dephasing contribution between two levels $\alpha$ and
$\beta\neq\alpha$ can be described as $\left[  \partial_{t}\rho_{\alpha\beta
}\right]  _{\mathrm{pure}}=-\gamma_{\alpha\beta}^{\prime}\rho_{\alpha\beta}$,
which also implies $\left[  \partial_{t}\rho_{\beta\alpha}\right]
_{\mathrm{pure}}=-\gamma_{\alpha\beta}^{\prime}\rho_{\beta\alpha}$ since
$\rho_{\alpha\beta}=\rho_{\beta\alpha}^{\ast}$. Here, $\gamma_{\alpha\beta
}^{\prime}=\gamma_{\beta\alpha}^{\prime}\geq0$ denotes the pure dephasing
rate. As can easily be seen, the corresponding dissipation superoperator in
Eq.\thinspace(\ref{eq:lindbl2}) can be represented by the matrix elements%
\begin{equation}
\mathcal{D}_{ijmn}^{\alpha\beta}=-\gamma_{\alpha\beta}^{\prime}\left(
\delta_{i\alpha}\delta_{j\beta}\delta_{m\alpha}\delta_{n\beta}+\delta_{i\beta
}\delta_{j\alpha}\delta_{m\beta}\delta_{n\alpha}\right)  . \label{eq:Dijmn3}%
\end{equation}

The Lindblad operators for pure dephasing must be diagonal in the chosen basis
\cite{oi2012limits}. However, Eq.\thinspace(\ref{eq:Dijmn3}) does not
generally ensure physical behavior, and thus a representation in terms of
Lindblad operators does not always exist \cite{oi2012limits}. Notably, for
$N\geq3$ there are constraints on how to select the pure dephasing rates
$\gamma_{\alpha\beta}^{\prime}\geq0$ to ensure compatibility with
Eq.\thinspace(\ref{eq:lindbl}), and an ill-considered choice can for example
easily result in a violation of positive semidefiniteness for $\hat{\rho}$
\cite{schirmer2004constraints,oi2012limits}. For example, $\gamma_{12}%
^{\prime}+\gamma_{13}^{\prime}+\gamma_{23}^{\prime}\leq2\left(  \gamma
_{12}^{\prime}\gamma_{13}^{\prime}+\gamma_{12}^{\prime}\gamma_{23}^{\prime
}+\gamma_{13}^{\prime}\gamma_{23}^{\prime}\right)  ^{1/2}$ must hold in
three-level systems, which is already violated if only one of the three pure
dephasing rates is non-zero.

In two-level systems, pure dephasing is described by a single rate
$\gamma_{12}^{\prime}=\gamma_{21}^{\prime}=\gamma^{\prime}\geq0$, and can for
example be represented by a Lindblad operator $\hat{L}=\left(  2\gamma
^{\prime}\right)  ^{1/2}\left|  1\right\rangle \left\langle 1\right|  $ or
$\hat{L}=\left(  2\gamma^{\prime}\right)  ^{1/2}\left|  2\right\rangle
\left\langle 2\right|  $, or also by the set $\hat{L}_{1}=\left(
\gamma^{\prime}\right)  ^{1/2}\left|  1\right\rangle \left\langle 1\right|  $,
$\hat{L}_{2}=\left(  \gamma^{\prime}\right)  ^{1/2}\left|  2\right\rangle
\left\langle 2\right|  $. More generally, if the same (typically empirical)
pure dephasing rate $\gamma^{\prime}$ is assumed for all transitions of an
$N$-level system \cite{burnett2014density,2005JAP....98j4505C}, this case can
always be represented by Lindblad operators, for example by the set $\hat
{L}_{k}=\left(  \gamma^{\prime}\right)  ^{1/2}\left|  k\right\rangle
\left\langle k\right|  $, $k=1..N$ \cite{rebentrost2009environment}.

Taking into account the results of Section \ref{sec:trans}, the total phase
relaxation due to pure dephasing plus lifetime broadening associated with
incoherent transitions is described by the dissipation term%
\begin{equation}
\left[  \partial_{t}\rho_{\alpha\beta}\right]  _{\mathrm{relax}}%
=-\gamma_{\alpha\beta}\rho_{\alpha\beta}=-\left[  \left(  r_{\alpha}+r_{\beta
}\right)  /2+\gamma_{\alpha\beta}^{\prime}\right]  \rho_{\alpha\beta},
\label{eq:relax2}%
\end{equation}
where $\gamma_{\alpha\beta}=\left(  r_{\alpha}+r_{\beta}\right)
/2+\gamma_{\alpha\beta}^{\prime}$ is the total dephasing rate and the
$r_{\alpha,\beta}$ are given by Eq.\thinspace(\ref{eq:rn})
\cite{schirmer2004constraints}.

\subsubsection{General Case}

While physical dissipation channels can often be represented by either
incoherent transitions or pure dephasing \cite{pfanner2008entangled}, see
Sections \ref{sec:trans} and \ref{sec:dep}, the Lindblad operators should not
a priori be restricted to these two forms, but rather be found based on
physical considerations
\cite{palmieri2009lindblad,kirvsanskas2018phenomenological}. Even more, the
representation of a dissipative channel as, e.g., incoherent transition or
pure dephasing, only applies for the chosen basis
\cite{oi2012limits,burnett2014density}. For illustration, let's assume an
$N$-level system with orthonormal basis states\ $\left|  n\right\rangle $ and
dissipative channels described by a set of Lindblad operators $\hat{L}_{k}$.
Alternatively, an orthonormal basis with states $\left|  n^{\prime
}\right\rangle $ can be used, with $\left|  n\right\rangle =\hat{I}\left|
n\right\rangle =\sum_{n^{\prime}}\left\langle n^{\prime}\right.  \left|
n\right\rangle \left|  n^{\prime}\right\rangle $, which changes the character
of the Lindblad operators in the new basis system. As an illustrative example,
we restrict ourselves to two relevant levels $\left|  1\right\rangle $ and
$\left|  2\right\rangle $, which are assumed to be localized in adjacent
potential wells, and between\ which tunneling through the separating barrier
occurs. This mechanism plays for example an important role in QCLs, which are
frequently modeled with a density matrix approach for a discrete quantum
system, using localized states to describe the tunneling transport across
thick barriers
\cite{2009PhRvB..80x5316K,2010PhRvB..81t5311D,2010NJPh...12c3045T,dinh2012extended,tzenov2016time,tzenov2017analysis}%
. This tunneling process is critically affected by dephasing between the two
states involved, which can be modeled by Eq.\thinspace(\ref{eq:relax2})
\cite{2005JAP....98j4505C,2009PhRvB..80x5316K,2010PhRvB..81t5311D,jirauschek2017density}%
. We exemplarily focus on the pure dephasing contribution, which can for a
two-level system be described by the Lindblad operator $\hat{L}=\left(
2\gamma^{\prime}\right)  ^{1/2}\left|  1\right\rangle \left\langle 1\right|  $
as discussed in Section \ref{sec:dep}. Changing to energy eigenstates $\left|
1^{\prime}\right\rangle $ and $\left|  2^{\prime}\right\rangle $ and for
simplicity assuming near-degeneracy, we obtain $\left|  1\right\rangle
=2^{-1/2}\left(  \left|  1^{\prime}\right\rangle +\left|  2^{\prime
}\right\rangle \right)  $ and $\left|  2\right\rangle =2^{-1/2}\left(  \left|
1^{\prime}\right\rangle -\left|  2^{\prime}\right\rangle \right)  $
\cite{2009PhRvB..80s5317G}. In the energy basis, above Lindblad operator then
becomes $\hat{L}=\left(  \gamma^{\prime}\right)  ^{1/2}\left(  \left|
1^{\prime}\right\rangle \left\langle 1^{\prime}\right|  +\left|  1^{\prime
}\right\rangle \left\langle 2^{\prime}\right|  +\left|  2^{\prime
}\right\rangle \left\langle 1^{\prime}\right|  +\left|  2^{\prime
}\right\rangle \left\langle 2^{\prime}\right|  \right)  $ which is not
diagonal, i.e., does not represent pure dephasing in that basis.

To summarize, the frequently used classification of dissipation channels in
incoherent transitions and pure dephasing is not always possible and
additionally depends on the chosen basis system, but is frequently used since
it allows for an intuitive physical interpretation. Thus, this classification
might also be helpful for determining the corresponding dissipative rates
based on compact models or by comparison to experimental data
\cite{2005JAP....98j4505C,dubi2008thermoelectric,freeman2016self}.
Consequently, for a given system a criterion for a convenient choice of basis
states might be that the dissipation channels can reasonably well be described
in terms of incoherent transitions and pure dephasing, which for example
motivates the frequent use of localized states to describe tunneling transport
through thick barriers.

\subsection{\label{sec:val}Conditions for Validity}

As discussed in Section \ref{sec:Lindop},\ the dissipation parameters must
fulfill certain conditions to ensure physical behavior of the density matrix,
which is exactly true if a representation of the dissipation process in terms
of Lindblad operators exists. For example, the total dephasing rate of a given
transition cannot be smaller than the lifetime broadening contribution due to
incoherent transitions, as can be seen from Eq.\thinspace(\ref{eq:relax2}).
Also, as discussed in Section \ref{sec:dep}, the pure dephasing rates cannot
be independently chosen for each transition, but must fulfill certain
conditions for $N\geq3$ levels. Thus, if the experimentally obtained
dissipation rates for a system do not satisfy above conditions, this might
indicate that the chosen model is not adequate, for example that not enough
levels are considered \cite{schirmer2004constraints}.

As noted above, the Lindblad equation can also be microscopically derived for
a quantum system weakly coupled to a large Markovian environment
\cite{davies1974markovian,breuer2002theory,le2011quantum}. These assumptions
require in particular that the coherent system dynamics and relaxation
processes occur on a slower timescale than the memory decay of the environment
\cite{breuer2002theory,le2011quantum}. These additional microscopic
constraints are not required to ensure completely positive and trace
preserving evolution of the density matrix, which is guaranteed by the
Lindblad form of Eq.\thinspace(\ref{eq:lindbl}). However, disregarding the
microscopic validity criteria might result in a violation of other laws such
as Onsager's relation \cite{kirvsanskas2018phenomenological}. On the other
hand, it has been pointed out that some of the assumptions usually invoked in
microscopic derivations, such as the secular approximation, might be
unnecessarily restrictive \cite{kirvsanskas2018phenomenological}. Eventually,
for a description of realistic quantum systems where many degrees of freedom
affect the time evolution, there will always be a trade-off between exactness
and manageability of the model \cite{kirvsanskas2018phenomenological}. From a
practical point of view, Lindblad-type master equations, such as the MB
system, often still yield useful results on the verge of the microscopic
validity range, for example in semiconductor structures interacting with
high-intensity fields
\cite{ziolkowski1995ultrafast,hughes1998breakdown,kalosha1999formation,mucke2001signatures,freeman2013laser}%
.

\section{\label{sec:Bloch}Optical Bloch Equations}

The most basic quantum system is the two-level system with only $N=2$ relevant
states. This can be a natural two-level system with only two eigenstates such
as a spin 1/2 particle, or a quasi-two-level system with two strongly coupled
states, such as an optical transition in resonance with an electromagnetic
field \cite{allen1987optical} or a driven double-well potential
\cite{grifoni1998driven}. In an early application of this model, Rabi
investigated the interaction of a spin 1/2 particle with a rotating magnetic
field by solving the time dependent Schr\"{o}dinger equation
\cite{rabi1937space}. The term ''Bloch equations'', in the narrow sense,
refers to evolution equations for a dissipative two-level system, first
devised to describe the evolution of the nuclear magnetic moment in a magnetic
field \cite{bloch1946nuclear}. Here, the interaction with the environment was
taken into account by two phenomenological relaxation time constants. This
concept was extended to other two-level systems, such as a pair of levels in
resonance with a classical optical field
\cite{feynman1957geometrical,abella1966photon}. The resulting evolution
equations are occasionally called optical Bloch equations for distinction
\cite{allen1987optical}. The optical propagation can be considered by coupling
the Bloch model to Maxwell's equations
\cite{feynman1957geometrical,allen1987optical,mccall1967self}, resulting in
the so-called Maxwell-Bloch (MB) equations. In the following, we focus on the
interaction of a quantum system with an optical field, where the coupled MB
equations have to be used for a combined description of the system dynamics
and optical propagation. Here, we will not restrict ourselves to two-level
systems, but rather consider the more general case of $N\geq2$ discrete
levels. The resulting equations are for $N\geq3$ states occasionally also
referred to as a multilevel Bloch/MB model \cite{mukamel1990femtosecond}.
Furthermore, for the description of dissipative effects due to the system
interaction with the environment, the Lindblad formalism introduced in Section
\ref{sec:lindbl} will serve as a framework. Sometimes the Lindblad equation,
Eq.\thinspace(\ref{eq:lindbl}), is already referred to as Bloch equations
\cite{gisin1992quantum}. In the following, the (optical) Bloch equations will
be regarded as a special form of Eq.\thinspace(\ref{eq:lindbl}) containing an
interaction Hamiltonian $\hat{H}_{1}\left(  t\right)  $ to describe
light-matter coupling.

\subsection{Dipole Approximation}

We consider a Hamiltonian of the form $\hat{H}=\left(  \mathbf{\hat{p}%
}-q\mathbf{A}\right)  ^{2}/\left(  2m\right)  +q\varphi+V$, which models the
system's interaction with a classical optical field, represented by a time and
space dependent magnetic vector potential $\mathbf{A}=\mathbf{A}\left(
\mathbf{\hat{r}},t\right)  $ and electric potential $\varphi=\varphi\left(
\mathbf{\hat{r}},t\right)  $. Here, $\mathbf{\hat{r}}$ and $\mathbf{\hat{p}}$
denote the position and (canonical) momentum operators of the quantum system
with the commutator $\left[  \hat{r}_{i},\hat{p}_{j}\right]  =\mathrm{i}%
\hbar\delta_{ij}$, which are in position representation given by
$\mathbf{\hat{r}}=\mathbf{r}$ and $\mathbf{\hat{p}}=-\mathrm{i}\hbar
\mathbf{\nabla}$, and $V=V\left(  \mathbf{\hat{r}}\right)  $ represents the
system's potential energy. Furthermore, $m$ and $q$ denote the carrier mass
and charge, which are for electrons given by $m=m_{\mathrm{e}}$ and $q=-e$,
with the elementary charge $e$. Using the Coulomb gauge $\mathbf{\nabla A}=0$,
we have $\left[  \mathbf{A},\mathbf{p}\right]  =0$. Furthermore assuming a
radiation field without free charge contributions gives $\varphi=0$, and
$\mathbf{E}=-\partial_{t}\mathbf{A}$ for the corresponding electric field
\cite{meystre2013elements}. Under these assumptions, we obtain
$\hat{H}=\hat{H}_{0}+\hat{H}_{1}$ with the Hamiltonian of the unperturbed
system $\hat{H}_{0}=\mathbf{\hat{p}}^{2}/\left(  2m\right)  +V$, and the time
dependent interaction Hamiltonian \cite{scully1999quantum}\textrm{ }%
\begin{equation}
\hat{H}_{1}=-\left(  q/m\right)  \mathbf{A\hat{p}}+q^{2}\mathbf{A}^{2}/\left(
2m\right)  \mathrm{.} \label{eq:HA1}%
\end{equation}
\textrm{ }The Bloch equations are then obtained from Eq.\thinspace
(\ref{eq:lindbl2}) by choosing the energy eigenstates of the system
Hamiltonian $\hat{H}_{0}$ as basis, resulting in matrix elements
$H_{0,ij}=E_{i}\delta_{ij}$ where $E_{i}$ is the eigenenergy of state $i$, and%
\begin{equation}
H_{1,ij}=-\frac{q}{m}\left\langle i\right|  \mathbf{A\hat{p}}\left|
j\right\rangle +\frac{q^{2}}{2m}\left\langle i\right|  \mathbf{A}^{2}\left|
j\right\rangle . \label{eq:HA}%
\end{equation}
Typically, the field varies on the scale of the optical wavelengths involved,
and the system dimensions are much smaller. The carriers do then not
experience a spatial field variation across the quantum system, and
$\mathbf{A}$ in Eqs.\thinspace(\ref{eq:HA1}) and (\ref{eq:HA}) can be
represented by a space independent vector potential, evaluated at the
macroscopic position of the quantum system. In this case, it can be shown by a
gauge transformation that the interaction Hamiltonian in Eq.\thinspace
(\ref{eq:HA1}) is equivalent to
\begin{equation}
\hat{H}_{1}=-\mathbf{\hat{d}E}\left(  t\right)  , \label{eq:dE}%
\end{equation}
which corresponds to the interaction Hamiltonian in the widely used (electric)
dipole approximation \cite{meystre2013elements,scully1999quantum}. Here,
$\mathbf{\hat{d}}=q\mathbf{\hat{r}}$\ denotes the system's dipole operator,
and the electric field $\mathbf{E}\left(  t\right)  $ is taken at the system
position. Intuitively, the Hamiltonian in Eq.\thinspace(\ref{eq:dE})
corresponds to the potential energy associated with the force $q\mathbf{E}%
\left(  t\right)  $ exerted by the electric field on the carriers.

We note that under some special conditions, such as high harmonic generation
\cite{walser2000high} or strong plasmonic confinement in nanophotonic
structures \cite{stobbe2012spontaneous,lodahl2015interfacing}, the field
gradient may become so large that the dipole approximation is not applicable.
In this context, we re-emphasize that Eq.\thinspace(\ref{eq:dE}) only assumes
a spatially constant field within a given quantum system, but does not neglect
the term $\propto\mathbf{A}^{2}$ in Eq.\thinspace(\ref{eq:HA1}) and is thus
not restricted to weak fields, as is sometimes believed. For the interaction
Hamiltonian in Eq.\thinspace(\ref{eq:dE}), the mechanical and canonical
momentum operators coincide, $m\mathbf{\hat{v}}=\mathbf{\hat{p}}$. The
Hamiltonian of the unperturbed system $\hat{H}_{0}=\mathbf{\hat{p}}%
^{2}/\left(  2m\right)  +V$ thus corresponds to the instantaneous energy
operator, and a matrix element\ $\rho_{ii}$ in the eigenstate basis $\left|
i\right\rangle $ of $H_{0}$ can be interpreted as the measurable probability
of finding the system in the corresponding energy eigenstate
\cite{scully1999quantum}. For the interaction Hamiltonian in Eq.\thinspace
(\ref{eq:HA1}), the mechanical momentum operator is $m\mathbf{\hat{v}%
}=\mathbf{\hat{p}}-q\mathbf{A}$. This complicates the physical interpretation
of results, since, e.g., the instantaneous energy operator $m\mathbf{\hat{v}%
}^{2}/2+V$ is different from $\hat{H}_{0}$, which prohibits an interpretation
of $\rho_{ii}$ as a measurable probability
\cite{scully1999quantum,rzkazewski2004equivalence,bauer2005strong}. These
differences also explain why\ the matrix elements $H_{1,ij}$ in Eq.\thinspace
(\ref{eq:HA}) for spatially constant $\mathbf{A}$ and those obtained from
Eq.\thinspace(\ref{eq:dE}) deviate from each other \cite{scully1999quantum}.
While both versions of the interaction Hamiltonian lead to identical results
for observable quantities as expected, it has been pointed out that the use of
approximations, such as the rotating wave approximation discussed in Section
\ref{sec:RWA}, can cause deviations between the two formulations
\cite{rzkazewski2004equivalence,bauer2005strong}. In the following, we will
use the interaction operator of the form Eq.\thinspace(\ref{eq:dE}).

\subsection{Optical Bloch Equations in Standard Form}

From Eq.\thinspace(\ref{eq:lindbl2}), we obtain with Eq.\thinspace
(\ref{eq:dE}) in the dipole approximation the (multilevel) Bloch equations%
\begin{align}
\partial_{t}\rho_{ij}=  &  -\mathrm{i}\omega_{ij}\rho_{ij}+\frac{\mathrm{i}}{\hbar
}\sum_{\ell}\left(  \mathbf{d}_{i\ell}\rho_{\ell j}-\mathbf{d}_{\ell j}%
\rho_{i\ell}\right)  \mathbf{E} \nonumber\\
&  +\sum_{mn}\mathcal{D}_{ijmn}\rho_{mn},
\label{eq:MB}%
\end{align}
with the transition frequencies $\omega_{ij}=\left(  E_{i}-E_{j}\right)
/\hbar$. If we furthermore restrict the description of dissipative effects to
incoherent transitions and dephasing, Eqs.\thinspace(\ref{eq:relax1}) and
(\ref{eq:relax2}), Eq.\thinspace(\ref{eq:MB}) simplifies to%
\begin{subequations}%
\label{eq:MB2}
\begin{align}
\partial_{t}\rho_{ij}  &  =-\mathrm{i}\omega_{ij}\rho_{ij}+\frac{\mathrm{i}%
}{\hbar}\sum_{\ell}\left(  \mathbf{d}_{i\ell}\rho_{\ell j}-\mathbf{d}_{\ell
j}\rho_{i\ell}\right)  \mathbf{E} \nonumber\\
&\,\quad -\gamma_{ij}\rho_{ij},i\neq j,\label{eq:MB2a}%
\\
\partial_{t}\rho_{ii}  &  =\frac{\mathrm{i}}{\hbar}\sum_{\ell}\left(
\mathbf{d}_{i\ell}\rho_{\ell i}-\mathbf{d}_{\ell i}\rho_{i\ell}\right)
\mathbf{E}+\sum_{j\neq i}r_{j\rightarrow i}\rho_{jj}-r_{i}\rho_{ii}.
\label{eq:MB2b}%
\end{align}%
\end{subequations}%
Although quantum optoelectronic devices can in principle comprise a single
isolated quantum system, for example a QD
\cite{yoshie2004vacuum,englund2010resonant}, in general they are based on
extended nanostructures such as quantum well structures, or an ensemble of
many quantum systems such as multi-quantum-dot structures. This requires a
position resolved model, where the device is described by a representative
quantum system with density matrix $\rho_{ij}\left(  \mathbf{x},t\right)  $ at
each device position $\mathbf{x}$. Furthermore, also the parameters
$\omega_{ij}$, $\mathbf{d}_{ij}$, $\gamma_{ij}$, $r_{j\rightarrow i}$ and
$r_{i}$ in Eqs.\thinspace(\ref{eq:MB}) and\ (\ref{eq:MB2}) generally depend on
$\mathbf{x}$ for inhomogeneous device structures
\cite{hess1996maxwell,jirauschek2014modeling}, such as multi-section lasers
\cite{talukder2014quantum,arkhipov2016self,tzenov2018passive}.

\subsection{\label{sec:dip}Optical Dipole Matrix Element}

The Hamiltonian part of the Bloch equations, Eqs.\thinspace(\ref{eq:MB})
and\ (\ref{eq:MB2}), requires the dipole matrix element vectors $\mathbf{d}%
_{i\ell}$ of the optical transitions and the eigenenergies of the quantized
states as an input. These can be computed from models derived from the
stationary Schr\"{o}dinger equation, such as the effective mass or
\textbf{k.p} approach, as shortly discussed in the following.

\begin{figure}[ptb]
\includegraphics{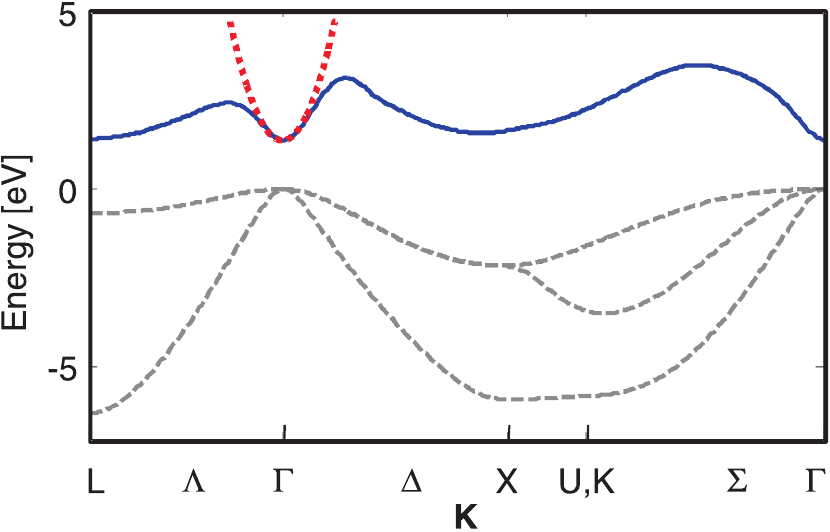}
\caption{Band structure of gallium arsenide (GaAs), obtained based on a simple
pseudo-potential tight-binding method without spin-orbit coupling
\cite{1980PhRvB..22.3886C}. Shown is the valence band (dashed lines), the
conduction band (solid line) and the parabolic dispersion relation assumed for
the $\Gamma$ valley in effective mass approximation (dotted line).}%
\label{fig:banddiagram}%
\end{figure}

In Fig.\thinspace\ref{fig:banddiagram}, the band structure of GaAs as an
exemplary direct bandgap semiconductor material is displayed. Shown is the
conduction band (solid line) and the valence band (dashed lines), consisting
of heavy hole, light hole and split-off band. The holes tend to accumulate
near the valence band maximum which is always at the $\Gamma$ point where the
crystal wavevector is $\mathbf{k}=\mathbf{0}$. For direct bandgap
semiconductors, the global conduction band minimum where the electrons
accumulate happens to be also at the $\Gamma$ point, and thus conservation of
crystal momentum can be satisfied for radiative electron-hole recombination.
This process is much less likely in indirect bandgap semiconductors, where the
global conduction band minimum is not at the $\Gamma$ point and the process
must additionally involve a phonon or crystal defect to achieve momentum conservation.

Assuming a direct bandgap semiconductor, it is practical to write the full
wavefunction $F_{i}\left(  \mathbf{r}\right)  $ of the initial and final state
as a product of periodic Bloch function $u_{v_{i}}\left(  \mathbf{r}\right)  $
at the $\Gamma$ point of band $v_{i}$ and an envelope wavefunction
$\varphi_{i}\left(  \mathbf{r}\right)  $ describing the slowly varying spatial
modulation of the full wavefunction across the nanostructure \cite{Bastard:88}%
. While a quantized state in a given band generally also contains
contributions from neighboring bands, in a first approximation only the
contribution of the dominant band is considered \cite{Bastard:88},
\begin{equation}
F_{i}\left(  \mathbf{r}\right)  =u_{v_{i}}\left(  \mathbf{r}\right)
\varphi_{i}\left(  \mathbf{r}\right)  . \label{eq:Fi}%
\end{equation}

\begin{figure}[ptb]
\includegraphics{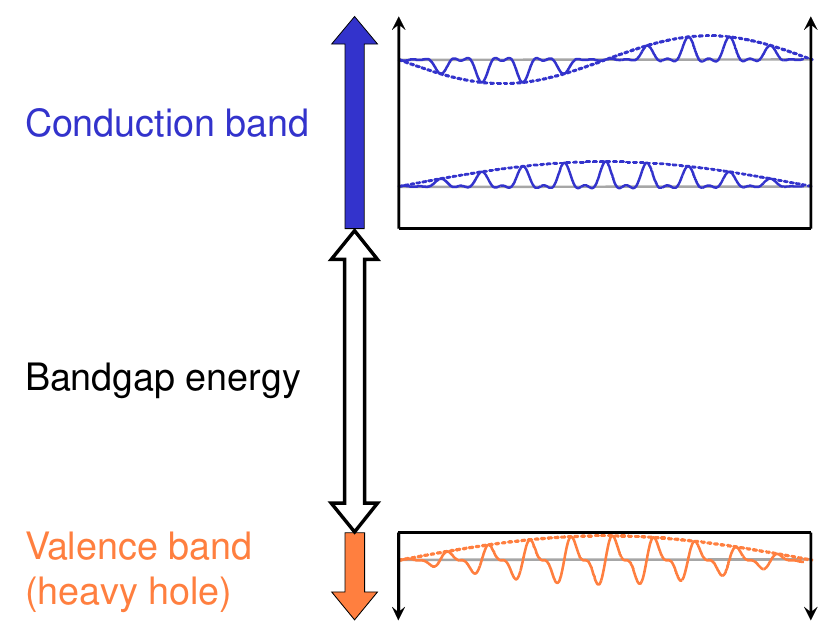}
\caption{Schematic illustration of full bound state wavefunctions
$F_{i}=u_{v_{i}}\varphi_{i}$ (solid) and envelope wavefunctions $\varphi_{i}$
(dashed) in the conduction and valence band of a quantum well.}%
\label{fig:blochwave}%
\end{figure}

In quantum well structures, the material composition changes only along the
growth direction $x$. Here, quantum confinement only occurs in $x$ direction,
while the carriers can move freely in the $yz$-plane. Thus, we can make the
ansatz
\begin{equation}
\varphi_{i}\left(  \mathbf{r}\right)  =S^{-1/2}\psi_{i}\left(  x\right)
\exp\left(  \mathrm{i}k_{y}y+\mathrm{i}k_{z}z\right)  . \label{eq:psi3D}%
\end{equation}
Here, $S$ is the in-plane cross section area, $\mathbf{k}=\left[  k_{y}%
,k_{z}\right]  ^{\mathrm{T}}$ denotes the in-plane wavevector in the
$yz$-plane where $\mathrm{T}$ indicates the transpose, and $\psi_{i}\left(
x\right)  $ is the (generally $\mathbf{k}$\ dependent) one-dimensional
envelope wavefunction in confinement direction. In Fig.\thinspace
\ref{fig:blochwave}, the full wavefunctions $F_{i}$ and corresponding envelope
wavefunctions $\varphi_{i}$ are schematically illustrated for the two lowest
conduction band states and the valence band ground state of a quantum well.
Similar considerations apply to quantum wires, where quantum confinement
occurs in two dimensions while the carriers can move freely along the third
coordinate. In QDs, the carriers are confined in all three dimensions.

\subsubsection{\label{sec:env}Computation of Envelope Wavefunction}

Neglecting the coupling between conduction and valence bands, the simplest
model for computing $\varphi_{i}\left(  \mathbf{r}\right)  $ in a quantum
structure is the Ben Daniel-Duke model, which works well for low-lying
conduction band states in the $\Gamma$ valley and also generally at the heavy
hole valence band maximum \cite{Bastard:88}. Here, we describe the dispersion
relation between energy and wavevector around the $\Gamma$ point by $E\left(
\mathbf{k}\right)  =V+\hbar^{2}\left|  \mathbf{k}\right|  ^{2}/\left(
2m^{\ast}\right)  $ which corresponds to a second order expansion, as
illustrated by the dotted line in Fig.\thinspace\ref{fig:banddiagram}. The
position dependent material composition in nanostructures causes the effective
mass $m^{\ast}$ and band edge energy $V$ to depend on $\mathbf{r}$, where $V$
additionally contains the externally applied bias. Within this model, the
stationary effective mass Schr\"{o}dinger equation is given by
\cite{Bastard:88}
\begin{equation}
0=\bigg\{-\frac{\hbar^{2}}{2}\left[  \nabla\frac{1}{m^{\ast}\left(
\mathbf{r}\right)  }\nabla\right]  +V\left(  \mathbf{r}\right)  -E_{i}\bigg
\}\varphi_{i}\left(  \mathbf{r}\right)  , \label{eq:s3D}%
\end{equation}
where $E_{i}$ denotes the eigenenergy of state $i$. For the valence band,
commonly the hole picture is adopted to avoid a negative effective mass in
Eq.\thinspace(\ref{eq:s3D}). For the transition between a conduction band
electron state with eigenenergy $E_{i}$ and a valence band hole state with
energy $E_{j}$, the transition energy is then given by $E_{i}+E_{j}%
+E_{\mathrm{g}}$ where $E_{\mathrm{g}}$ denotes the bandgap energy, i.e., the
energy difference between valence band maximum and conduction band minimum. In
quantum well systems, $m^{\ast}$ and $V$ only depend on the $x$ coordinate,
and Eq.\thinspace(\ref{eq:s3D}) can be reduced to the one-dimensional
effective mass equation by inserting Eq.\thinspace(\ref{eq:psi3D}). Similar
considerations apply to quantum wires where $m^{\ast}$ and $V$ only depend on
two coordinates.

The Ben Daniel-Duke model in Eq.\thinspace(\ref{eq:s3D}) can be extended,
e.g., by accounting for band bending due to space charge effects in the
potential, which is self-consistently included by solving Eq.\thinspace
(\ref{eq:s3D}) together with the Poisson equation
\cite{datta2005quantum,2009IJQE...45..1059J}. Furthermore, an energy dependent
effective mass can be introduced to include nonparabolicity effects associated
with the deviation of the dispersion relation from the parabolic form assumed
above \cite{1989PhRvB..40.7714E,1987PhRvB..35.7770N}.

A further refined treatment of the conduction and valence bands, which
accounts for band coupling, is usually performed based on \textbf{k.p} theory,
initially proposed by Kane \cite{kane1957band,kane1982energy} and Luttinger
and Kohn \cite{luttinger1955motion}. Here the envelope wavefunctions are not
scalar, but a multicomponent vector containing contributions from all the
bands considered. In many structures, strain arising from the lattice mismatch
between the different semiconductor compounds plays an important role, and can
be considered based on the Bir-Pikus model \cite{bir1974symmetry}. For
modeling interband devices, eight-band \textbf{k.p} is a common option which
considers the top three valence bands and the lowest conduction band, along
with spin orientation \cite{pidgeon1966interband,Bastard:88}. This approach is
routinely applied to nanostructures, such as quantum dots
\cite{stier1999electronic}, wires \cite{baraff1991eigenfunction} and wells
\cite{paul20088}. If only valence band states are considered, a restriction to
six bands is possible \cite{foreman1993effective}. This approach is sometimes
also combined with Eq.\thinspace(\ref{eq:s3D}) for the conduction band,
assuming that it is decoupled from the valence bands. On the other hand, it
has been found that for certain cases, eight-band \textbf{k.p} is not accurate
enough. For example, a 14-band \textbf{k.p} approach which also includes the
second conduction band in III-V semiconductors has been developed to obtain a
more accurate conduction band dispersion relation at higher energies
\cite{rossler1984nonparabolicity}, and 14-band \textbf{k.p} has also yielded
improved results for SiGe/Si heterostructures \cite{ridene2001infrared}.

\subsubsection{\label{sec:interdip}Inter- and Intraband Dipole Matrix Elements}

The dipole matrix element is best evaluated by computing the expectation value
of the momentum operator $\mathbf{\hat{p}}=-\mathrm{i}\hbar\mathbf{\nabla}$.
Employing the product rule and exploiting the fact that the periodic Bloch
functions and envelope wavefunctions vary on two different length scales, we
can with Eq.\thinspace(\ref{eq:Fi}) write \cite{1997plds.book.....D}
\begin{equation}
\left\langle F_{i}\right|  \mathbf{\hat{p}}\left|  F_{j}\right\rangle
\approx\left\langle \varphi_{i}\right.  \left|  \varphi_{j}\right\rangle
\left\langle u_{v_{i}}\right|  \mathbf{\hat{p}}\left|  u_{v_{j}}\right\rangle
+\left\langle u_{v_{i}}\right.  \left|  u_{v_{j}}\right\rangle \left\langle
\varphi_{i}\right|  \mathbf{\hat{p}}\left|  \varphi_{j}\right\rangle .
\label{eq:mom}%
\end{equation}

For transitions between conduction and valence band states, the first term
dominates because the Bloch functions vary much more rapidly than the envelope
wavefunctions. Using $\left\langle F_{i}\right|  \mathbf{\hat{p}}\left|
F_{j}\right\rangle =\mathrm{i}m_{\mathrm{e}}E_{\mathrm{g}}\left\langle
F_{i}\right|  \mathbf{\hat{r}}\left|  F_{j}\right\rangle /\hbar$ with the
electron mass $m_{\mathrm{e}}$ and band gap energy $E_{\mathrm{g}}$
\cite{burt1993evaluation}, the interband dipole matrix element can then in a
first approximation be written as%
\begin{equation}
\left\langle F_{i}\right|  \mathbf{\hat{r}}\left|  F_{j}\right\rangle
=-\mathrm{i}\hbar m_{\mathrm{e}}^{-1}E_{\mathrm{g}}^{-1}\left\langle u_{v_{i}%
}\right|  \mathbf{\hat{p}}\left|  u_{v_{j}}\right\rangle \left\langle
\varphi_{i}\right.  \left|  \varphi_{j}\right\rangle . \label{eq:d_interb}%
\end{equation}
For intraband optical transitions, we have $v_{i}=v_{j}$, $\left\langle
u_{v_{i}}\right|  \mathbf{\hat{p}}\left|  u_{v_{i}}\right\rangle =0$ and
$\left\langle u_{v_{i}}\right.  \left|  u_{v_{i}}\right\rangle =1$, and thus
Eq.\thinspace(\ref{eq:mom}) yields $\left\langle F_{i}\right|  \mathbf{\hat
{p}}\left|  F_{j}\right\rangle \approx\left\langle \varphi_{i}\right|
\mathbf{\hat{p}}\left|  \varphi_{j}\right\rangle $ and analogously%
\begin{equation}
\left\langle F_{i}\right|  \mathbf{\hat{d}}\left|  F_{j}\right\rangle
\approx\left\langle \varphi_{i}\right|  \mathbf{\hat{d}}\left|  \varphi
_{j}\right\rangle . \label{eq:d_intrab}%
\end{equation}

\begin{figure}[ptb]
\includegraphics{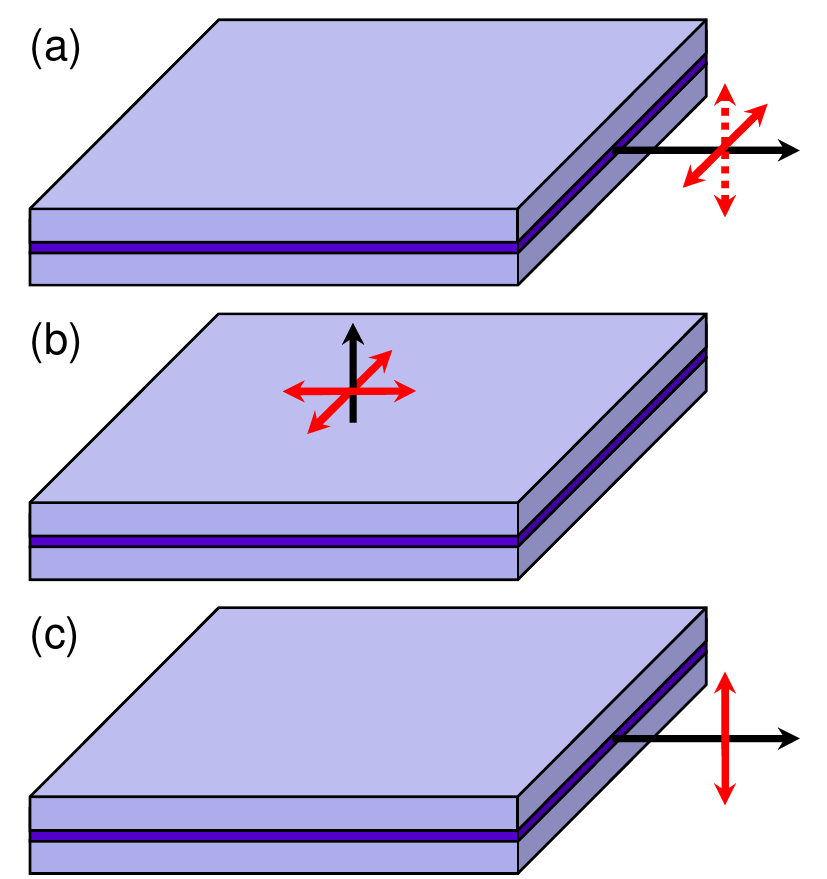}
\caption{Polarizations with non-vanishing dipole matrix elements for (a) and
(b) interband and (c) intraband transitions. The dashed arrow in (a) indicates
that for this polarization, only transitions between conduction band and light
hole states contribute.}%
\label{fig:polarization}%
\end{figure}

In quantum well systems, confinement only occurs in the growth direction $x$,
and the envelope wavefunction has the form given by Eq.\thinspace
(\ref{eq:psi3D}). For transitions between a conduction band state $\left|
\psi_{i},\mathbf{k}\right\rangle $ and a valence band state $\left|  \psi
_{j},\mathbf{k}^{\prime}\right\rangle $, $\left\langle \varphi_{i}\right.
\left|  \varphi_{j}\right\rangle =\left\langle \psi_{i}\right.  \left|
\psi_{j}\right\rangle \delta_{\mathbf{k},\mathbf{k}^{\prime}}$, i.e., the
optical transition is $\mathbf{k}$\ conserving. The absolute value of the
dipole matrix element can be approximately written as%
\begin{equation}
\left|  \mathbf{e}\left\langle F_{i\mathbf{k}}\right|  \mathbf{\hat{r}}\left|
F_{j\mathbf{k}^{\prime}}\right\rangle \right|  =c_{ij}E_{\mathrm{g}}%
^{-1}P_{\mathrm{cv}}\left\langle \psi_{i}\right.  \left|  \psi_{j}%
\right\rangle \delta_{\mathbf{k},\mathbf{k}^{\prime}}, \label{eq:dqw}%
\end{equation}
where $\mathbf{e}$ denotes the polarization direction of the electric field,
and $P_{\mathrm{cv}}\approx0.85..1\,\mathrm{nm}\times\mathrm{eV}%
\text{\textsl{\/}}{}$ for most common semiconductors
\cite{1997plds.book.....D}. For transitions between conduction band and heavy
hole states, $c_{ij}=2^{-1/2}$ for polarization in in-plane direction and
$c_{ij}=0$ for polarization in growth direction. For transitions between
conduction band and light hole states, $c_{ij}=6^{-1/2}$ for polarization in
in-plane direction and $c_{ij}=2\times6^{-1/2}$ for polarization in growth
direction. For intraband transitions occurring between quantized levels in the
conduction band of quantum wells, as are for example employed for QCLs, the
envelope wavefunctions again assume the form Eq.\thinspace(\ref{eq:psi3D}).
The dipole matrix element between an initial state $\left|  \psi
_{i},\mathbf{k}\right\rangle $ and a final state $\left|  \psi_{j}%
,\mathbf{k}^{\prime}\right\rangle $ is then with Eq.\thinspace
(\ref{eq:d_intrab}) given by $\mathbf{d}_{i\mathbf{k},j\mathbf{k}^{\prime}%
}=\mathbf{d}_{ij}\delta_{\mathbf{k},\mathbf{k}^{\prime}}$, where
\begin{equation}
\mathbf{d}_{ij}=\left\langle \psi_{i}\right|  \mathbf{\hat{d}}\left|  \psi
_{j}\right\rangle =-e\mathbf{e}_{x}\int\psi_{i}^{\ast}x\psi_{j}\mathrm{d}x.
\label{eq:dint}%
\end{equation}
Here, $\mathbf{e}_{x}$ denotes the unit vector in $x$ direction, and only the
dipole matrix element for polarization in growth direction $x$ is nonzero.
Notably, this is different from transitions between conduction band and heavy
hole states in quantum wells where the $x$ component of $\mathbf{d}_{ij}$ is
zero, as discussed above. In Fig.\thinspace\ref{fig:polarization}, the
possible field polarization directions for interband [Fig.\thinspace
\ref{fig:polarization}(a), (b)] and intraband [Fig.\thinspace
\ref{fig:polarization}(c)] transitions are indicated. For quantum well lasers,
Fig.\thinspace\ref{fig:polarization}(a), (b) and (c) correspond to the
standard edge-emitting, vertical-cavity surface-emitting and quantum cascade laser.

In quantum dots, the uppermost valence band eigenstates usually exhibit heavy
hole character
\cite{gerardot2008optical,karlsson2006optical,niquet2008quantum,cortez2001polarization}%
. Thus, band coupling effects can often be neglected in Eq.\thinspace
(\ref{eq:d_interb}) for interband transitions between the heavy-hole-like
states and low-lying conduction band states. Within the framework of these
assumptions, only optical dipole transitions between hole and electron states
with equal quantum numbers are allowed, and the envelope wavefunction overlap
in Eq.\thinspace(\ref{eq:d_interb}) typically approaches $\left\langle
\varphi_{i}\right.  \left|  \varphi_{j}\right\rangle \approx1$ for the allowed
transitions \cite{bimberg1999quantum,bimberg1999quantum2}. The symmetry of the
wavefunctions can however be affected by inhomogeneities in shape and
composition of the quantum dots as well as piezoelectric fields, resulting in
additional weakly allowed transitions
\cite{woggon1997optical,finley2008quantum}. Moreover, due to the strong
confinement in quantum dots, Coulomb interactions tend to play a pronounced
role, causing energy shifts as well as somewhat altered selection rules. Such
effects can be included in a more complete description based on the
electron-hole-pair picture, which replaces the single-carrier envelope
wavefunctions $\varphi_{i}$ and $\varphi_{j}$ in Eq.\thinspace
(\ref{eq:d_interb}) by expressions for the excited electron-hole pair state
and the corresponding ground state \cite{woggon1997optical,bimberg1999quantum}%
. Intraband transitions, which are mainly relevant in the context of quantum
dot infrared photodetectors, are again described by Eq.\thinspace
(\ref{eq:d_intrab}).

\subsection{\label{sec:coh}Non-Redundant Density Matrix Representation}

For a discrete-level system with $N$ eigenstates $\left|  j\right\rangle $,
the density matrix contains $N$ real diagonal elements and $N^{2}-N$ complex
off-diagonal elements which are related by $\rho_{ij}=\rho_{ji}^{\ast}$.
Furthermore considering the trace condition $\mathrm{Tr}\left\{  \hat{\rho
}\right\}  =1$, the density matrix can be represented by $N^{2}-1$
non-redundant, real-valued elements, which are conveniently written as a
vector $\mathbf{S}$. This non-redundant representation is for example achieved
by the coherence vector (or pseudospin) representation \cite{hioe1981n}, which
has also been found useful for numerically efficient implementations of the MB
equations
\cite{slavcheva2002coupled,slavcheva2003ultrashort,slavcheva2008model,slavcheva2010nonlinear}%
. For this purpose, the density matrix operator $\hat{\rho}$ is composed as
\begin{equation}
\hat{\rho}=\frac{1}{N}\hat{I}+\frac{1}{2}\sum_{j=1}^{N^{2}-1}S_{j}\hat{s}_{j}.
\label{eq:comp_rho}%
\end{equation}
Here, $\hat{s}_{j}$ are generators of the Lie algebra of SU($N$) which are
traceless Hermitian operators fulfilling the condition $\mathrm{Tr}\left\{
\hat{s}_{j}\hat{s}_{k}\right\}  =2\delta_{jk}$, and $\hat{I}$ is the identity
operator. $\hat{I}$ and $\hat{s}_{j}$ can be represented by corresponding
$N\times N$ matrices. A possible choice for the generators $\hat{s}=\left\{
\hat{u}_{12},\hat{u}_{13},\dots,\hat{v}_{12},\dots,\hat{w}_{1},\dots
,\hat{w}_{N-1}\right\}  $ consists of $N\left(  N-1\right)  /2$ generator
pairs
\begin{align}
\hat{u}_{jk}  &  =\hat{t}_{jk}+\hat{t}_{kj},\nonumber\\
\hat{v}_{jk}  &  =-\mathrm{i}\left(  \hat{t}_{jk}-\hat{t}_{kj}\right)  ,
\label{eq:ps1}%
\end{align}
and $N-1$ generators
\begin{equation}
\hat{w}_{l}=-\left[  2l^{-1}\left(  l+1\right)  ^{-1}\right]  ^{1/2}\left(
\sum_{\ell=1}^{l}\hat{t}_{\ell\ell}-l\hat{t}_{l+1,l+1}\right)  ,
\label{eq:ps2}%
\end{equation}
where $\hat{t}_{jk}=\left|  j\right\rangle \left\langle k\right|  $ is the
transition-projection operator, and the indices satisfy $1\leq j<k\leq N$ and
$1\leq l\leq N-1$~\cite{hioe1981n}. For $N=2$ and $N=3$ these generators
produce the Pauli and the Gell-Mann matrices, respectively.

The elements of the coherence vector $\mathbf{S}$ are defined as
$S_{j}=\mathrm{Tr}\left\{  \hat{\rho}\hat{s}_{j}\right\}  $ using the
Hilbert-Schmidt inner product. Since both $\hat{\rho}$ and the generators
$\hat{s}_{j}$ are Hermitian, the vector elements are real. A similar transform
can be applied to the Lindblad equation. Inserting Eq.\thinspace
(\ref{eq:comp_rho}) into Eq.\thinspace(\ref{eq:lindbl}) and applying
$\mathrm{Tr}\{\cdot\hat{s}_{k}\}$ yields
\begin{align}
\mathrm{Tr}\left\{  \partial_{t}\hat{\rho}\hat{s}_{k}\right\}  &  =\mathrm{Tr}%
\left\{  \frac{1}{2}\sum_{j=1}^{N^{2}-1}\partial_{t}S_{j}\hat{s}_{j}%
\hat{s}_{k}\right\} \nonumber\\
&  =\frac{1}{2}\sum_{j=1}^{N^{2}-1}\partial_{t}%
S_{j}\mathrm{Tr}\left\{  \hat{s}_{j}\hat{s}_{k}\right\}  =\partial_{t}S_{k}
\label{eq:cvrright}%
\end{align}
for the left hand side. For the right hand side we can write
\begin{subequations}
\begin{align}
&  \mathrm{Tr}\left\{  \mathcal{L}\left(  \hat{\rho}\right)  \hat{s}_{k}%
+\mathcal{D}\left(  \hat{\rho}\right)  \hat{s}_{k}\right\}  =\mathrm{Tr}%
\left\{  \mathcal{L}\left(  \hat{\rho}\right)  \hat{s}_{k}\right\}
+\mathrm{Tr}\left\{  \mathcal{D}\left(  \hat{\rho}\right)  \hat{s}_{k}%
\right\}  ,\label{eq:cvrlefta}\\
&  \mathrm{Tr}\left\{  \mathcal{L}\left(  \hat{\rho}\right)  \hat{s}_{k}%
\right\}  =\underbrace{\mathrm{Tr}\left\{  N^{-1}\mathcal{L}%
(\hat{I})\hat{s}_{k}\right\}  }_{=0}+\sum_{j=1}^{N^{2}-1}\underbrace{\frac
{1}{2}\mathrm{Tr}\left\{  \mathcal{L}\left(  \hat{s}_{j}\right)
\hat{s}_{k}\right\}  }_{=L_{jk}}S_{j},\label{eq:cvrleftb}\\
&  \mathrm{Tr}\left\{  \mathcal{D}\left(  \hat{\rho}\right)  \hat{s}_{k}%
\right\}  =\underbrace{\mathrm{Tr}\left\{  N^{-1}\mathcal{D}%
(\hat{I})\hat{s}_{k}\right\}  }_{S_{j}^{\mathrm{eq}}}+\sum_{j=1}^{N^{2}%
-1}\underbrace{\frac{1}{2}\mathrm{Tr}\left\{  \mathcal{D}\left(
\hat{s}_{j}\right)  \hat{s}_{k}\right\}  }_{D_{jk}}S_{j}, \label{eq:cvrleftc}%
\end{align}
\end{subequations}
since both superoperators $\mathcal{L}$ and $\mathcal{D}$ are linear. Noting
that $\mathcal{L}(\hat{I})=0$ and arranging Eqs.\thinspace(\ref{eq:cvrright})
and (\ref{eq:cvrlefta})-(\ref{eq:cvrleftc}) in matrix-vector form yields
\begin{equation}
\partial_{t}\mathbf{S}=\left(  L+D\right)  \mathbf{S}+\mathbf{S}^{\mathrm{eq}%
}, \label{eq:liouville-adj1}%
\end{equation}
where $L$ and $D$ are $\left(  N^{2}-1\right)  \times\left(  N^{2}-1\right)  $
real matrices and $\mathbf{S}^{\mathrm{eq}}$ denotes the equilibrium coherence vector.

Alternatively, one can start from Eq.\thinspace(\ref{eq:lindbl2}), where the
superoperators $\mathcal{L}$ and $\mathcal{D}$ are represented as $N^{2}\times
N^{2}$ matrices. This Liouville space representation was used for example
in~\cite{marskar2011}, where column-major order was applied to map the indices
$(i,j)\mapsto k$ and $(m,n)\mapsto l$. In this case, the density matrix is
represented as $N^{2}$ column vector $\mathbf{R}$, and the Lindblad equation
reads~\cite{marskar2011}
\begin{equation}
\partial_{t}\mathbf{R}=\left(  -\frac{\mathrm{i}}{\hbar}\mathcal{L}%
+\mathcal{D}\right)  \mathbf{R}, \label{eq:lindbl-liouv}%
\end{equation}
where $\mathcal{L}=\hat{H}^{\mathrm{T}}\otimes\hat{I}-\hat{I}\otimes\hat{H}$
and
\begin{equation}
\mathcal{D}=\sum_{k}\left[  \hat{L}_{k}^{\mathrm{T}}\otimes\hat{L}_{k}%
-\frac{1}{2}\left(  \hat{L}_{k}^{\mathrm{T}}\hat{L}_{k}^{\ast}\otimes
\hat{I}+\hat{I}\otimes\hat{L}_{k}^{\dagger}\hat{L}_{k}\right)  \right]  .
\end{equation}
Here, $\otimes$ denotes the Kronecker product. Then, since the Hilbert-Schmidt
inner product reads $\mathrm{Tr}\{\hat{a}^{\dagger}\hat{b}\}=\mathbf{a}%
^{\dagger}\mathbf{b}=\mathbf{b}^{\dagger}\mathbf{a}$ in this representation,
where the vectors $\mathbf{a}$ and $\mathbf{b}$ are the matrices $\hat{a}$ and
$\hat{b}$ in column-major order, we can write the transform from Liouville
space to the coherence vector representation as
\begin{equation}
S_{j}=\mathrm{Tr}\left\{  \hat{\rho}\hat{s}_{j}\right\}  =\mathbf{s}%
_{j}^{\dagger}\mathbf{R},\qquad\mathbf{S}=T^{\dagger}\mathbf{R},
\end{equation}
where the columns of the transformation matrix $T$ are the generators
$\mathbf{s}_{j}$ in column-major order. Conversely, the vector $\mathbf{R}$
can be recovered by
\begin{equation}
\mathbf{R}=\frac{1}{N}\mathbf{I}+\frac{1}{2}T\mathbf{S},
\end{equation}
where $\mathbf{I}$ is the identity matrix in vectorized form. Using these
transform relations, we can rewrite Eq.\thinspace(\ref{eq:lindbl-liouv}) as
\begin{equation}
\frac{1}{2}T\partial_{t}\mathbf{S}=\left(  -\frac{\mathrm{i}}{\hbar
}\mathcal{L}+\mathcal{D}\right)  \frac{1}{2}T\mathbf{S}+\left(  -\frac
{\mathrm{i}}{\hbar}\mathcal{L}+\mathcal{D}\right)  \frac{1}{N}\mathbf{I},
\end{equation}
and simplify the result by left-multiplication with $T^{\dagger}$ to
\begin{equation}
\partial_{t}\mathbf{S}=\left(  -\frac{\mathrm{i}}{2\hbar}T^{\dagger
}\mathcal{L}T+\frac{1}{2}T^{\dagger}\mathcal{D}T\right)  \mathbf{S}+\frac
{1}{N}T^{\dagger}\mathcal{D}\mathbf{I},
\end{equation}
where we used the orthogonality of the generators ($\frac{1}{2}T^{\dagger}%
T=I$, where $I$ is the $N^{2}\times N^{2}$ identity matrix) and the fact that
the commutator of the identity is zero ($\mathcal{L}\mathbf{I}=0$). This
corresponds to Eq.\thinspace(\ref{eq:liouville-adj1}).

As we shall see in Section~\ref{sec:maxwellbloch}, the derivative of the
macroscopic polarization $\partial_{t}\mathbf{P}_{\mathrm{q}}$ has to be
calculated for the Maxwell-Bloch equations. Naturally, it must be expressed as
function of the vector $\mathbf{S}$. By replacing the trace operation and
inserting the transformation rule, we can write for this term
\begin{align}
\partial_{t}\mathbf{P}_{\mathrm{q}}  &  =n_{\mathrm{3D}}\mathrm{Tr}\left\{
\partial_{t}\hat{\rho}\mathbf{\hat{d}}\right\}  =n_{\mathrm{3D}}%
\mathbf{u}^{\dagger}\partial_{t}\mathbf{R} \nonumber\\
&  =\frac{1}{2}n_{\mathrm{3D}%
}\mathbf{u}^{\dagger}T\left[  \left(  L+D\right)  \mathbf{S}+\mathbf{S}%
^{\mathrm{eq}}\right]  , \label{eq:cvr-polarization1}%
\end{align}
where $\mathbf{u}$ is the vectorized dipole moment operator and
$n_{\mathrm{3D}}$ denotes the carrier number density. Note that the elements
of $\mathbf{u}$ could be vectors themselves, depending on whether one ore more
dimensions are considered.

Using the dipole approximation, Eq.\thinspace(\ref{eq:dE}), we plug in the
Hamiltonian $\hat{H}=\hat{H}_{0}-\mathbf{\hat{d}}\mathbf{E}(t)$, which can be
represented with two matrices $\mathcal{L}_{0}$ and $\mathcal{L}_{1}$ in
Liouville space (and two matrices $L_{0}$ and $L_{1}$ in coherence vector
representation, respectively). Since
\begin{align}
\mathbf{u}^{\dagger}\mathcal{L}_{1}  &  =-\mathbf{u}^{\dagger}\left[
\mathbf{\hat{d}}^{\mathrm{T}}\mathbf{E}(t)\otimes\hat{I}-\hat{I}\otimes
\mathbf{\hat{d}}\mathbf{E}(t)\right] \nonumber\\
&  =-\left[  \mathbf{\hat{d}}^{\mathrm{T}}\mathbf{E}(t)\otimes
\hat{I}\mathbf{u}-\hat{I}\otimes\mathbf{\hat{d}}\mathbf{E}(t)\mathbf{u}%
\right]  ^{\dagger}\nonumber\\
&  =-\left[  \mathrm{vec}\left(  \hat{I}\mathbf{\hat{d}}\mathbf{\hat{d}%
}\mathbf{E}(t)\right)  -\mathrm{vec}\left(  \mathbf{\hat{d}}\mathbf{E}%
(t)\mathbf{\hat{d}}\hat{I}\right)  \right]  ^{\dagger}=0,
\end{align}
where $\mathrm{vec}$ denotes the vectorization of an operator, $\mathrm{vec}%
(\mathbf{\hat{d}})=\mathbf{u}$, and the Hermitian property of the operators
involved as well as the properties of the Kronecker product have been
exploited, the polarization does not depend on the electric field and
Eq.\thinspace(\ref{eq:cvr-polarization1}) can be refined as
\begin{equation}
\partial_{t}\mathbf{P}_{\mathrm{q}}=\frac{1}{2}n_{\mathrm{3D}}\mathbf{u}%
^{\dagger}T\left[  \left(  L_{0}+D\right)  \mathbf{S}+\mathbf{S}^{\mathrm{eq}%
}\right]  . \label{eq:cvr-polarization2}%
\end{equation}

\subsection{\label{sec:RWA}Rotating Wave Approximation}

The Bloch equations (\ref{eq:MB2}) are solvable only under special conditions,
like $\left|  \Delta M\right|  =1$ transitions in hydrogen-like atoms excited
with circularly polarized light \cite{allen1987optical,rabi1937space}. In
particular, closed analytical solutions do not exist for the basic and very
important case of excitation with a monochromatic, linearly polarized field
\cite{bloch1940magnetic}. Furthermore, the numerical solution of the
Maxwell-Bloch equations requires high spatiotemporal resolution since the
fields as well as the off-diagonal density matrix elements in Eq.\thinspace
(\ref{eq:MB2}) oscillate with the optical period. For these reasons, the
rotating wave approximation (RWA) is commonly invoked, which significantly
reduces the numerical burden and enables an analytical treatment of the Bloch
equations, at least for incident monochromatic radiation and some other
relevant cases.

The RWA is only applicable for not too broadband optical fields, which can
then be separated into a slowly varying amplitude, given in complex notation
by $\underline{\mathbf{E}}\left(  \mathbf{x},t\right)  =\left|  \underline
{\mathbf{E}}\left(  \mathbf{x},t\right)  \right|  \exp\left[  \mathrm{i}%
\phi\left(  \mathbf{x},t\right)  \right]  $, and a rapidly oscillating carrier
with frequency $\omega_{\mathrm{c}}>0$. We note that there is no unique
definition of $\omega_{\mathrm{c}}$, but rather any choice which ensures that
all relevant spectral components are close to $\omega_{\mathrm{c}}$ will
suffice (for optical fields with symmetric power spectra, it obviously makes
sense to pick the center frequency). In complex notation, the electric field
can then be written as
\begin{equation}
\mathbf{E}=\frac{1}{2}\underline{\mathbf{E}}\exp\left(  -\mathrm{i}%
\omega_{\mathrm{c}}t\right)  +c.c., \label{eq:E_RWA}%
\end{equation}
where c.c. denotes the complex conjugate. Furthermore assuming that all
transitions between pairs of states $i$ and $j$ with non-negligible coupling
to the optical field are in near-resonance, $\left|  \omega_{ij}\right|
\approx\omega_{\mathrm{c}}$, the corresponding off-diagonal density matrix
elements are transformed into a rotating reference frame,
\begin{equation}
\rho_{ij}=\eta_{ij}\exp\left[  -\mathrm{sgn}\left(  \omega_{ij}\right)
\mathrm{i\,}\omega_{\mathrm{c}}t\right]  , \label{eq:rho_RWA}%
\end{equation}
where $\mathrm{sgn}$ denotes the sign function. Inserting Eqs.\thinspace
(\ref{eq:E_RWA}) and (\ref{eq:rho_RWA}) in Eq.\thinspace(\ref{eq:MB2}),
multiplying both sides of Eq.\thinspace(\ref{eq:MB2a}) with $\exp\left[
\mathrm{sgn}\left(  \omega_{ij}\right)  \mathrm{i\,}\omega_{\mathrm{c}%
}t\right]  $ and applying the RWA, i.e., discarding all rapidly oscillating
terms $\propto\exp\left(  \pm\mathrm{i\,}\omega_{\mathrm{c}}t\right)  $ and
$\exp\left(  \pm2\mathrm{i\,}\omega_{\mathrm{c}}t\right)  $
\cite{allen1987optical}, we obtain%
\begin{subequations}%
\label{eq:RWA}
\begin{align}
\partial_{t}\eta_{ij}  &  =\mathrm{i}\Delta_{ij}\eta_{ij}+\frac{\mathrm{i}%
}{2\hbar}\left(  \rho_{jj}-\rho_{ii}\right)  \mathbf{d}_{ij}\left\{
\begin{array}
[c]{c}%
\underline{\mathbf{E}}\\
\underline{\mathbf{E}}^{\ast}%
\end{array}
\right\} \nonumber\\
&\,\quad  -\gamma_{ij}\eta_{ij},\omega_{ij}\left\{
\begin{array}
[c]{c}%
>0\\
<0
\end{array}
\right\}  ,\label{eq:RWA1}\\
\partial_{t}\rho_{ii}  &  =\frac{1}{\hbar}\sum_{\substack{j\\\omega_{ij}%
>0}}\Im\left\{  \mathbf{d}_{ji}\eta_{ij}\underline{\mathbf{E}}^{\ast}\right\}
+\frac{1}{\hbar}\sum_{\substack{j\\\omega_{ij}<0}}\Im\left\{  \mathbf{d}%
_{ji}\eta_{ij}\underline{\mathbf{E}}\right\} \nonumber\\
&\,\quad  +\sum_{j\neq i}r_{j\rightarrow
i}\rho_{jj}-r_{i}\rho_{ii}, \label{eq:RWA2}%
\end{align}%
\end{subequations}%
with $\Delta_{ij}=\mathrm{sgn}\left(  \omega_{ij}\right)  \left(
\omega_{\mathrm{c}}-\left|  \omega_{ij}\right|  \right)  $. As discussed
above, the RWA is only applicable if the near-resonance condition is
fulfilled, i.e., all significant spectral components $E\left(  \omega\right)
$ of the field are close to resonance with all relevant optical transitions at
frequencies $\omega_{ij}$, $\left|  \omega-\left|  \omega_{ij}\right|
\right|  \ll\left|  \omega_{ij}\right|  $. As a second condition, the
interaction energy must be so small that the eigenfrequencies of the quantum
system are not considerably perturbed \cite{allen1987optical}, i.e., $\left|
\mathbf{d}_{ij}\underline{\mathbf{E}}\right|  /\hbar\ll\left|  \omega
_{ij}\right|  $.

\section{\label{sec:maxwellbloch}Maxwell-Bloch Equations}

\begin{figure}[ptb]
\includegraphics{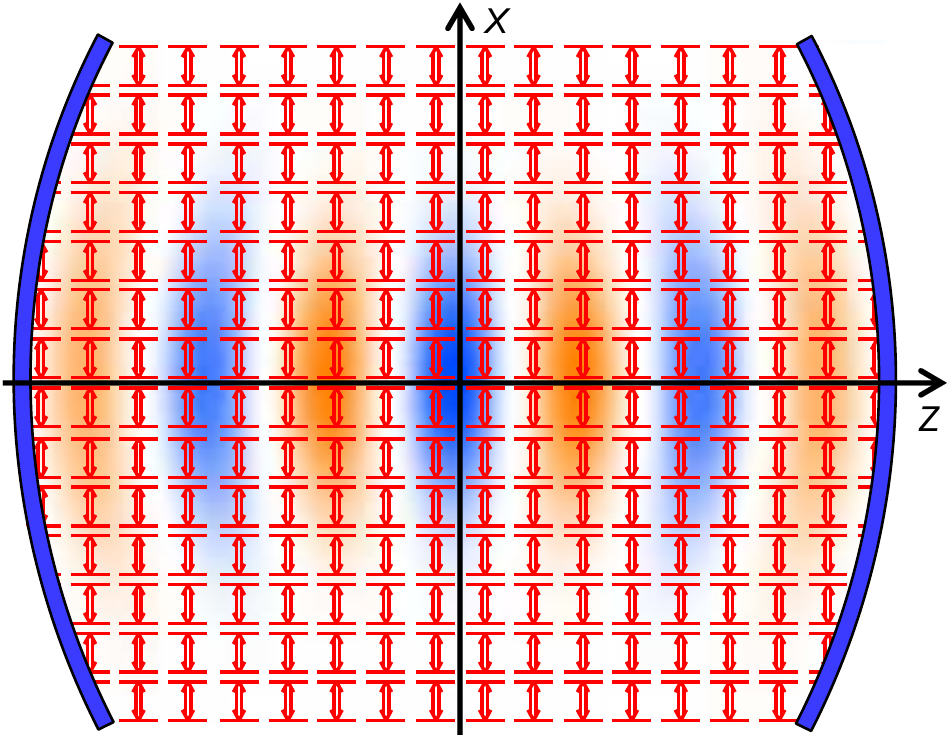}
\caption{Schematic illustration of modeling based on Maxwell-Bloch equations.
For an exemplary optical resonator, the distribution of representative
quantum systems is shown along with the cavity field.}%
\label{fig:mb}%
\end{figure}

The optical field propagation in the device is classically described in terms
of Maxwell's equations. Assuming that the magnetization is negligible at
optical frequencies, we can write Faraday's and Amp\`{e}re's law for the
electric field $\mathbf{E}$ and magnetic field $\mathbf{H}$ as
\begin{subequations}%
\label{eq:maxw}
\begin{align}
\mathbf{\nabla}\times\mathbf{E}  &  =-\mu_{0}\partial_{t}\mathbf{H}%
,\label{eq:maxw1}\\
\mathbf{\nabla}\times\mathbf{H}  &  =\epsilon_{0}\epsilon_{\mathrm{r}}%
\partial_{t}\mathbf{E}+\sigma\mathbf{E\mathbf{+}J}_{\mathrm{q}}\nonumber\\
&  =\epsilon_{0}\epsilon_{\mathrm{r}}\partial_{t}\mathbf{E}+\sigma
\mathbf{E\mathbf{+}J}_{\mathrm{f}}+\partial_{t}\mathbf{P}_{\mathrm{q}}.
\label{eq:maxw2}%
\end{align}%
\end{subequations}%
$\mathbf{E}$, $\mathbf{H}$, and $\mathbf{J}_{\mathrm{q}}$ are functions of
both $t$ and $\mathbf{x}$. $\mathbf{J}_{\mathrm{q}}=\mathbf{J}_{\mathrm{f}%
}+\partial_{t}\mathbf{P}_{\mathrm{q}}$ denotes the total current density
contribution of the quantum systems. Here, $\mathbf{J}_{\mathrm{f}}$ and
$\partial_{t}\mathbf{P}_{\mathrm{q}}$ correspond to the current density due to
free carrier motion and the polarization current density, respectively, where
$\mathbf{P}_{\mathrm{q}}$ is the macroscopic polarization. In Fig.\thinspace
\ref{fig:mb}, the coupled modeling of the field propagation and the quantum
system dynamics is schematically illustrated. For extended nanostructures such
as quantum well structures or ensembles of QDs, the medium must be described
by a representative quantum system at each position $\mathbf{x}$. The Bloch
equations, Eq.\thinspace(\ref{eq:MB2}), are coupled to Eq.\thinspace
(\ref{eq:maxw}) via $\mathbf{E}$. On the other hand, Eq.\thinspace
(\ref{eq:maxw}) is coupled to Eq.\thinspace(\ref{eq:MB2}) via $\mathbf{J}%
_{\mathrm{q}}$. For practical reasons, we consider the background polarization
due to the host medium separately by the (generally $\mathbf{x}$ dependent)
dielectric constant $\epsilon_{\mathrm{r}}\left(  \mathbf{x}\right)  $, where
we have for now neglected any frequency dependence, and assumed linearity and
isotropy of the host. Likewise, we include the absorption of the host medium
by a scalar conductivity $\sigma\left(  \mathbf{x}\right)  $, which gives rise
to an ohmic current contribution $\sigma\mathbf{E}$ in Eq.\thinspace
(\ref{eq:maxw2}). Furthermore, $\epsilon_{0}$ and $\mu_{0}$\ denote the vacuum
permittivity and permeability, respectively.

\subsection{\label{sec:pol}Macroscopic Polarization And Current Density}

Here, as above, we use the position variable $\mathbf{r}$ to resolve
microscopic behavior, while the variable $\mathbf{x}$ describes the position
in the modeled device or geometry and refers to macroscopic dependencies,
obtained from microscopic models by adequate ensemble averaging. In Maxwell's
equations (\ref{eq:maxw}), the total macroscopic current density contribution
of the quantum systems is given by $\mathbf{J}_{\mathrm{q}}\left(
\mathbf{x},t\right)  =\mathbf{J}_{\mathrm{f}}\left(  \mathbf{x},t\right)
+\partial_{t}\mathbf{P}_{\mathrm{q}}\left(  \mathbf{x},t\right)  $, where the
free charge current density $\mathbf{J}_{\mathrm{f}}$\ and polarization
current density $\partial_{t}\mathbf{P}_{\mathrm{q}}$ contain the
contributions due to free and bound charges, respectively. In optoelectronic
devices, $\mathbf{J}_{\mathrm{f}}$\ is for example induced by electrical
pumping or generated by the photovoltaic effect, while $\partial_{t}%
\mathbf{P}_{\mathrm{q}}$ is associated with the bound charge oscillations
induced by the optical field. Microscopically, in nanostructured devices the
carriers in bound or quasi-bound states may contribute to $\partial
_{t}\mathbf{P}_{\mathrm{q}}$ by coherent or incoherent interaction with the
optical field, as well as to $\mathbf{J}_{\mathrm{f}}$ via coherent transport
such as tunneling and incoherent transport such as scattering-induced hopping.
Thus, it makes sense to treat polarization and free current density together.

The macroscopic polarization $\mathbf{P}_{\mathrm{q}}$ can be obtained from
the dipole moment of the quantum system, given by the expectation value of the
dipole moment operator $\left\langle \mathbf{\hat{d}}\right\rangle \left(
t\right)  =\mathrm{Tr}\left\{  \hat{\rho}\left(  t\right)  \mathbf{\hat{d}%
}\right\}  $. $\mathbf{P}_{\mathrm{q}}$ at a position $\mathbf{x}$ is then
obtained by summing over the quantum systems in a volume $V_{\mathrm{p}}$
around $\mathbf{x}$,
\begin{equation}
\mathbf{P}_{\mathrm{q}}=V_{\mathrm{p}}^{-1}\sum_{i}\left\langle \mathbf{\hat
{d}}_{i}\right\rangle , \label{eq:Ps}%
\end{equation}
where $V_{\mathrm{p}}$ is chosen big enough to obtain a smooth dependence of
$\mathbf{P}_{\mathrm{q}}$, but small enough so that spatial variations on
classical length scales can still be resolved. For a large ensemble of
identical systems with carrier number density $n_{\mathrm{3D}}$, the
polarization is then given by
\begin{equation}
\mathbf{P}_{\mathrm{q}}=n_{\mathrm{3D}}\left\langle \mathbf{\hat{d}%
}\right\rangle =n_{\mathrm{3D}}q\mathrm{Tr}\left\{  \mathbf{\hat{r}}\hat{\rho
}\right\}  , \label{eq:P}%
\end{equation}
where $\hat{\rho}$ is the density operator of a representative quantum system
at position $\mathbf{x}$. On the other hand, the electric current in the
quantum system can be computed from the expectation value of carrier velocity
$\left\langle \mathbf{\hat{v}}\right\rangle $ in the system, where the
velocity operator is defined in the Heisenberg picture by the time derivative
of the position operator $\mathbf{\hat{r}}_{\mathrm{H}}$, $\mathbf{\hat{v}%
}_{\mathrm{H}}=\mathrm{d}_{t}\mathbf{\hat{r}}_{\mathrm{H}}$. For the coherent
contribution corresponding to the Hamiltonian part in Eq.\thinspace
(\ref{eq:lindbl}), we then obtain with the Ehrenfest equation $\left\langle
\mathbf{\hat{v}}\right\rangle =\mathrm{d}_{t}\left\langle \mathbf{\hat{r}%
}\right\rangle $ \cite{breuer2002theory}, where we have dropped the index
$\mathrm{H}$ since expectation values for physical observables are independent
of the chosen picture. In the Schr\"{o}dinger picture, we thus obtain
$\left\langle \mathbf{\hat{v}}\right\rangle =\mathrm{d}_{t}\mathrm{Tr}\left\{
\mathbf{\hat{r}}\hat{\rho}\right\}  =\mathrm{Tr}\left\{  \mathbf{\hat{r}%
}\mathrm{d}_{t}\hat{\rho}\right\}  $, which is also valid for the incoherent
contribution induced by the Lindblad operator term in Eq.\thinspace
(\ref{eq:lindbl}) \cite{burnett2014density}. Thus, $I=q\left|  \left\langle
\mathbf{\hat{v}}\right\rangle \right|  /L$ corresponds to the current through
an individual (single-carrier) quantum system, where $L$ here indicates the
system length in the direction of current flow, and $L/\left|  \left\langle
\mathbf{\hat{v}}\right\rangle \right|  $ is the transit time of the carrier
through the system. Again averaging over a large ensemble of identical
systems, we obtain the macroscopic current density $\mathbf{J}_{\mathrm{f}%
}=n_{\mathrm{3D}}q\mathrm{Tr}\left\{  \mathbf{\hat{r}}\mathrm{d}_{t}\hat{\rho
}\right\}  $. We note that this result is the same as for the polarization
current density, obtained by taking the time derivative of Eq.\thinspace
(\ref{eq:P}), which reflects the fact that the carriers of the quantum system
are responsible for both the free charge current and polarization current.
Even more, from a microscopic standpoint, this distinction is inappropriate
for our case. Thus we can write
\begin{equation}
\mathbf{J}_{\mathrm{q}}=n_{\mathrm{3D}}q\mathrm{Tr}\left\{  \mathbf{\hat{r}%
}\partial_{t}\hat{\rho}\right\}  , \label{eq:Jt}%
\end{equation}
where $\hat{\rho}$ again describes a representative quantum system at position
$\mathbf{x}$, and $\partial_{t}\hat{\rho}$, given by Eq.\thinspace
(\ref{eq:lindbl}), contains both the coherent and incoherent dynamics.
Assuming an $N$-level system with orthonormal basis states\ $\left|
i\right\rangle $, Eq.\thinspace(\ref{eq:Jt}) can with the dipole matrix
element $\mathbf{d}_{ij}=q\left\langle i\right|  \mathbf{\hat{r}}\left|
j\right\rangle $ be written as%
\begin{equation}
\mathbf{J}_{\mathrm{q}}=n_{\mathrm{3D}}\sum_{i,j}\mathbf{d}_{ji}\partial
_{t}\rho_{ij}, \label{eq:Jt2}%
\end{equation}
with $\partial_{t}\rho_{ij}$ given by Eq.\thinspace(\ref{eq:lindbl2}) or
(\ref{eq:MB2}).

A widely used criterion to distinguish between the macroscopic free charge and
polarization current contributions in Eq.\thinspace(\ref{eq:Jt2}) is the
frequency range, where commonly $\partial_{t}\mathbf{P}_{\mathrm{q}}$ is
expected to contain frequencies in the range of the driving optical field
spectrum, while $\mathbf{J}_{\mathrm{f}}$\ covers the low-frequency and direct
current contributions. In this context, we point out that due to nonlinear
optical mixing, the polarization generally contains up- and down-converted
components \cite{bloembergen1996nonlinear,Boyd,wacker,wegener2005extreme}.
This especially applies to nanostructured optoelectronic devices where giant
optical nonlinearities can be artificially engineered, and are actively
exploited in both the optical and terahertz regime
\cite{2003PhRvL..90d3902O,englund2007controlling,srinivasan2007linear,belkin2008room}%
. On the other hand, the electric current can contain components up to tens of
GHz due to external modulation or back-coupling of the optical dynamics to the
electrical circuitry. Notably, in QCLs embedded into a micro-strip line,
strong coupling of the co-propagating microwave current modulation and optical
waveform has recently been found
\cite{maineult2010microwave,calvar2013high,st2014injection,wang2015generating,faist2016quantum}%
, indicating that a clear differentiation between free and polarization
current contributions is not always possible. However, as pointed out above,
such a distinction is also not necessary since the current density and
polarization appear as $\mathbf{J}_{\mathrm{q}}=\mathbf{J}_{\mathrm{f}%
}+\partial_{t}\mathbf{P}_{\mathrm{q}}$ in Maxwell's equations. Ultimately, the
frequency range of the measured electrical current will be limited by both the
measurement setup itself and the electrical properties of the device, such as
its intrinsic capacitance.

\subsubsection{Coherent Contribution}

Using the Ehrenfest equation $\mathrm{d}_{t}\left\langle \mathbf{\hat{r}%
}\right\rangle =\mathrm{i}\hbar^{-1}\left\langle \left[  \hat{H},\mathbf{\hat
{r}}\right]  \right\rangle $ \cite{breuer2002theory}, we can write the
coherent part of the current density as%
\begin{equation}
\mathbf{J}_{\mathrm{coh}}=\mathrm{i}n_{\mathrm{3D}}q\hbar^{-1}\left\langle
\left[  \hat{H},\mathbf{\hat{r}}\right]  \right\rangle .
\end{equation}
In the following, we assume an effective mass Hamiltonian of the form
$\hat{H}=\left(  1/2\right)  \mathbf{\hat{p}}\left[  m^{\ast}\left(
\mathbf{\hat{r}}\right)  \right]  ^{-1}\mathbf{\hat{p}+V}\left(
\mathbf{\hat{r}},t\right)  $ as used in Eq.\thinspace(\ref{eq:s3D}), yielding%
\begin{equation}
\mathbf{J}_{\mathrm{coh}}=\left(  1/2\right)  n_{\mathrm{3D}}q\left\langle
\left[  m^{\ast}\left(  \mathbf{\hat{r}}\right)  \right]  ^{-1}\mathbf{\hat
{p}}+\mathbf{\hat{p}}\left[  m^{\ast}\left(  \mathbf{\hat{r}}\right)  \right]
^{-1}\right\rangle . \label{eq:Jcoh2}%
\end{equation}
Using an orthonormal basis as for Eq.\thinspace(\ref{eq:Jt2}), and inserting
the unit operator $\hat{I}=\int\mathrm{d}r^{3}\,\left|  \mathbf{r}%
\right\rangle \left\langle \mathbf{r}\right|  $ in Eq.\thinspace
(\ref{eq:Jcoh2}), we can express the result in terms of wavefunctions
$\varphi_{i}\left(  \mathbf{r}\right)  =\left\langle \mathbf{r}\right.
\left|  i\right\rangle $,
\begin{equation}
\mathbf{J}_{\mathrm{coh}}=\hbar n_{\mathrm{3D}}q\sum_{i,j}\int\Im\left\{
\rho_{ji}\varphi_{i}^{\ast}\left(  \mathbf{r}\right)  \left[  m^{\ast}\left(
\mathbf{r}\right)  \right]  ^{-1}\mathbf{\nabla}\varphi_{j}\left(
\mathbf{r}\right)  \right\}  \mathrm{d}^{3}r. \label{eq:Jcoh3}%
\end{equation}
Equation\ (\ref{eq:Jcoh3}) can also be interpreted as the current density
contribution of a representative individual (single-carrier) quantum system at
the corresponding position, averaged over the associated volume $V_{\mathrm{p}%
}=n_{\mathrm{3D}}^{-1}$, where the microscopically resolved current density is
given by the familiar expression \cite{kirvsanskas2018phenomenological}
\begin{equation}
\mathbf{J}_{\mathrm{coh}}=\hbar q\sum_{i,j}\Im\left\{  \rho_{ji}\varphi
_{i}^{\ast}\left(  \mathbf{r}\right)  \left[  m^{\ast}\left(  \mathbf{r}%
\right)  \right]  ^{-1}\mathbf{\nabla}\varphi_{j}\left(  \mathbf{r}\right)
\right\}  . \label{eq:Jcoh4}%
\end{equation}

\subsubsection{Incoherent Contribution}

The incoherent contribution to the current density is given by $\mathbf{J}%
_{\mathrm{inc}}=n_{\mathrm{3D}}q\left[  \left\langle \mathbf{\hat{v}%
}\right\rangle \right]  _{\mathrm{inc}}=n_{\mathrm{3D}}q\mathrm{Tr}\left\{
\mathbf{\hat{r}}\left[  \mathrm{d}_{t}\hat{\rho}\right]  _{\mathrm{inc}%
}\right\}  $\ \cite{burnett2014density}, which yields with Eqs.\thinspace
(\ref{eq:lindbl}) and (\ref{eq:lindbl2})
\begin{align}
\mathbf{J}_{\mathrm{inc}}  &  =n_{\mathrm{3D}}q\mathrm{Tr}\left\{
\mathbf{\hat{r}}\sum_{k}\left(  \hat{L}_{k}\hat{\rho}\hat{L}_{k}^{\dagger
}-\frac{1}{2}\hat{L}_{k}^{\dagger}\hat{L}_{k}\hat{\rho}-\frac{1}{2}\hat{\rho
}\hat{L}_{k}^{\dagger}\hat{L}_{k}\right)  \right\} \nonumber\\
&  =n_{\mathrm{3D}}\sum_{ij}\mathbf{d}_{ji}\sum_{mn}\mathcal{D}_{ijmn}%
\rho_{mn}. \label{eq:Jinc}%
\end{align}
For an incoherent transition from a state $\alpha$ to $\beta\neq\alpha$, we
obtain with the corresponding Lindblad operator given in Eq.\thinspace
(\ref{eq:Lif})%
\begin{equation}
\mathbf{J}_{\mathrm{inc}}^{\alpha\rightarrow\beta}=r_{\alpha\rightarrow\beta
}n_{\mathrm{3D}}\left[  \left(  \mathbf{d}_{\beta\beta}-\mathbf{d}%
_{\alpha\alpha}\right)  \rho_{\alpha\alpha}-\sum_{i\neq\alpha}\Re\left\{
\mathbf{d}_{i\alpha}\rho_{\alpha i}\right\}  \right]  . \label{eq:Jtrans}%
\end{equation}
Furthermore, inserting Eq.\thinspace(\ref{eq:relax2}) in Eq.\thinspace
(\ref{eq:Jinc}) yields the pure dephasing contribution between two levels
$\alpha$ and $\beta$%
\begin{equation}
J_{\mathrm{inc}}^{\alpha\beta}=-2\gamma_{\alpha\beta}^{\prime}n_{\mathrm{3D}%
}\Re\left\{  \mathbf{d}_{\beta\alpha}\rho_{a\beta}\right\}  , \label{eq:Jpure}%
\end{equation}
with the pure dephasing rate $\gamma_{\alpha\beta}^{\prime}$. The current
contributions from incoherent transitions due to Eqs.\thinspace
(\ref{eq:Jtrans}) and (\ref{eq:Jpure}) can also be rearranged so that%
\begin{equation}
J_{\mathrm{hop}}^{\alpha\beta}=\left(  r_{\alpha\rightarrow\beta}\rho
_{\alpha\alpha}-r_{\beta\rightarrow\alpha}\rho_{\beta\beta}\right)
n_{\mathrm{3D}}\left(  \mathbf{d}_{\beta\beta}-\mathbf{d}_{\alpha\alpha
}\right)  \label{eq:Jhop}%
\end{equation}
is the net current due to the hopping transport between states $\alpha$ and
$\beta$ which corresponds to the classical rate equation description, and%
\begin{equation}
J_{\mathrm{dep}}^{\alpha\beta}=-2\gamma_{\alpha\beta}n_{\mathrm{3D}}%
\Re\left\{  \mathbf{d}_{\beta\alpha}\rho_{a\beta}\right\}  \label{eq:Jdep}%
\end{equation}
is the dephasing contribution due to the decay of the corresponding
off-diagonal matrix elements $\rho_{a\beta}$ and $\rho_{\beta a}$. Here,
$\gamma_{\alpha\beta}=\left(  r_{\alpha}+r_{\beta}\right)  /2+\gamma
_{\alpha\beta}^{\prime}$ is the total dephasing rate, including lifetime
broadening and pure dephasing, and $r_{\alpha,\beta}$ is given by
Eq.\thinspace(\ref{eq:rn}). The total incoherent current density, resulting
from incoherent transitions and pure dephasing, is then obtained by summing
over all transitions. With Eqs.\thinspace(\ref{eq:Jhop})\ and (\ref{eq:Jdep}),
we obtain%
\begin{align}
J_{\mathrm{inc}}  &  =\sum_{\alpha=1}^{N-1}\sum_{\beta=\alpha+1}^{N}\left(
J_{\mathrm{hop}}^{\alpha\beta}+J_{\mathrm{dep}}^{\alpha\beta}\right)
\nonumber\\
&  =n_{\mathrm{3D}}\sum_{\alpha}\sum_{\beta\neq\alpha}\big[  r_{\alpha
\rightarrow\beta}\rho_{\alpha\alpha}\left(  \mathbf{d}_{\beta\beta}%
-\mathbf{d}_{\alpha\alpha}\right) \nonumber\\
&  \phantom{==}-\gamma_{\alpha\beta}\Re\left\{
\mathbf{d}_{\beta\alpha}\rho_{a\beta}\right\}  \big]  .
\end{align}

\subsection{\label{sec:SVAA}Slowly Varying Amplitude Approximation}

Although the Bloch equations in RWA, Eq.\thinspace(\ref{eq:RWA}), are
sometimes solved in combination with the full Maxwell's equations,
Eq.\thinspace(\ref{eq:maxw}), typically the RWA is combined with an envelope
propagation equation, derived from Maxwell's equations under the assumption of
a slowly varying field amplitude. In this way, above mentioned advantages of
the RWA, namely a significantly reduced numerical burden and a larger number
of analytical solutions, also applies to the coupled Maxwell-Bloch system.
Taking the curl of Eq.\thinspace(\ref{eq:maxw1}) and eliminating $\mathbf{H}$
using Eq.\thinspace(\ref{eq:maxw2}) yields
\begin{align}
\mathbf{\nabla\times\nabla\times E}=  &  -\frac{n_{0}^{2}\left(  1-2\Delta
_{n}\right)  }{c^{2}}\partial_{t}^{2}\mathbf{E}-\mu_{0}\sigma\partial
_{t}\mathbf{E}-\mu_{0}\partial_{t}\mathbf{J}_{\mathrm{f}} \nonumber\\
&  -\mu_{0}\partial
_{t}^{2}\mathbf{P}_{\mathrm{q}}\mathbf{.} \label{eq:wavemax}%
\end{align}
Here, $c=\left(  \mu_{0}\epsilon_{0}\right)  ^{-1/2}$ is the vacuum speed of
light. Furthermore, the background permittivity of the host material
$\epsilon_{\mathrm{r}}\left(  \mathbf{x}\right)  $ is here modeled as
$\epsilon_{\mathrm{r}}=n_{0}^{2}\left(  1-2\Delta_{n}\right)  $, where
$\Delta_{n}\left(  x,y\right)  \ $(with the minimum value $0$) describes a
transverse refractive index profile, as widely employed in waveguiding
structures \cite{1991ONT}.

For no free space charges, Gauss's law dictates that $\mathbf{\nabla D}=0$
where $\mathbf{D}=\mathbf{P}_{\mathrm{q}}+\epsilon_{0}n_{0}^{2}\left(
1-2\Delta_{n}\right)  \mathbf{E}$ is the displacement field in Eq.\thinspace
(\ref{eq:wavemax}). Assuming an isotropic medium, we can thus set
$\mathbf{\nabla}\left(  \mathbf{\nabla E}\right)  \approx0$ in the case of
weak nonlinearity \cite{Boyd} and weak inhomogeneity \cite{1991ONT,yar89}, or
generally if the field intensity transverse to the propagation direction is
slowly varying over an optical wavelength \cite{sar87}. This assumption is
only fulfilled for weak waveguiding, i.e., if the relative changes of the
refractive index $\left|  \Delta n_{\mathrm{b}}\right|  /n_{\mathrm{b}}$ and
its gradient $\left|  \Delta\left(  \mathbf{\nabla}n_{\mathrm{b}}\right)
\right|  /\left|  \mathbf{\nabla}n_{\mathrm{b}}\right|  $ over the distance of
a wavelength in the medium is small against unity \cite{1991ONT}, where
$n_{\mathrm{b}}=n_{0}\left(  1-2\Delta_{n}\right)  ^{1/2}$ in Eq.\thinspace
(\ref{eq:wavemax}). Furthermore, also the polarization contribution
$\mathbf{P}_{\mathrm{q}}$ of the quantum structure must be compatible with the
assumption of weak inhomogeneity. As discussed in Section \ref{sec:dip},
quantum structures, as modeled by the Bloch equations, can be highly
anisotropic; e.g., the dipole moment element vector $\mathbf{d}_{ij}$ of
inter-conduction band transitions in quantum wells only has a nonzero
component in growth direction. If the optical field is however also polarized
in this direction, which is for example often the case in lasers since only
the corresponding field component gets amplified, then $\mathbf{\nabla
E}\approx0$ can still hold for weak nonlinearity and inhomogeneity. Using
$\mathbf{\nabla\times}\left(  \mathbf{\nabla\times E}\right)  =\mathbf{\nabla
}\left(  \mathbf{\nabla E}\right)  -\mathbf{\nabla}^{2}\mathbf{E}$ and
subsequently neglecting the term $\mathbf{\nabla}\left(  \mathbf{\nabla
E}\right)  $, we obtain the generalized inhomogeneous wave equation%
\begin{equation}
\mathbf{\nabla}^{2}\mathbf{E}=\frac{n_{0}^{2}\left(  1-2\Delta_{n}\right)
}{c^{2}}\partial_{t}^{2}\mathbf{E}+\mu_{0}\sigma\partial_{t}\mathbf{E}+\mu
_{0}\partial_{t}\mathbf{J}_{\mathrm{f}}+\mu_{0}\partial_{t}^{2}\mathbf{P}%
_{\mathrm{q}}. \label{eq:waveinh}%
\end{equation}

For deriving the slowly varying amplitude approximation (SVAA), $\mathbf{E}$
and $\mathbf{P}_{\mathrm{q}}$ are written as a product of its envelope and
carrier, as done above for the derivation of the RWA. However, in contrast to
Eq.\thinspace(\ref{eq:E_RWA}), we also take into account the spatial
dependence of the carrier, where we assume that the direction of the optical
energy flow at every position\ is close to a reference direction defined by
the carrier wavevector $\mathbf{k}_{\mathrm{c}}$, which corresponds to the
paraxial approximation. This assumption is for example typically fulfilled in
laser resonators or optical fibers. Introducing the complex-valued field and
polarization amplitudes, $\underline{\mathbf{E}}\left(  \mathbf{x},t\right)  $
and $\underline{\mathbf{P}}\left(  \mathbf{x},t\right)  $, and assuming
propagation along the $z$ direction, we have%
\begin{subequations}%
\label{eq:EP_SVAA}
\begin{align}
\mathbf{E}\left(  \mathbf{x},t\right)   &  =\frac{1}{2}\underline{\mathbf{E}%
}\left(  \mathbf{x},t\right)  \exp\left(  \mathrm{i}k_{\mathrm{c}}%
z-\mathrm{i}\omega_{\mathrm{c}}t\right)  +c.c.,\label{eq:E_SVAA}\\
\mathbf{P}_{\mathrm{q}}\left(  \mathbf{x},t\right)   &  =\frac{1}{2}%
\underline{\mathbf{P}}\left(  \mathbf{x},t\right)  \exp\left(  \mathrm{i}%
k_{\mathrm{c}}z-\mathrm{i}\omega_{\mathrm{c}}t\right)  +c.c.,
\label{eq:P_SVAA}%
\end{align}%
\end{subequations}%
with $\left|  k_{\mathrm{c}}\right|  =n_{0}\omega_{\mathrm{c}}/c$. We note
that although Eq.\thinspace(\ref{eq:E_SVAA}) contains the term $\exp\left(
\mathrm{i}k_{\mathrm{c}}z\right)  $ not included in Eq.\thinspace
(\ref{eq:E_RWA}), the Bloch equations in RWA, Eq.\thinspace(\ref{eq:RWA}),
remain unchanged since $\exp\left(  \mathrm{i}k_{\mathrm{c}}z\right)  $
cancels out. To apply the SVAA, we insert Eq.\thinspace(\ref{eq:EP_SVAA}%
)\thinspace in Eq.\thinspace(\ref{eq:waveinh}). Just as for the RWA, we assume
that all significant spectral components of the field are close to
$\omega_{\mathrm{c}}$, i.e., at frequencies $\omega_{\mathrm{c}}%
+\Delta_{\omega}$ with $\left|  \Delta_{\omega}\right|  \ll\omega_{\mathrm{c}%
}$. This implies that $\partial_{t}^{2}\underline{\mathbf{E}}$ can be
neglected against $-2\mathrm{i}\omega_{\mathrm{c}}\partial_{t}\underline
{\mathbf{E}}$, as can be seen in Fourier domain where the two terms become
$-\Delta_{\omega}^{2}\underline{\mathbf{E}}\left(  \Delta_{\omega}\right)  $
and $-2\omega_{\mathrm{c}}\Delta_{\omega}\underline{\mathbf{E}}\left(
\Delta_{\omega}\right)  $. Similarly, also $\sigma\partial_{t}\underline
{\mathbf{E}}$ and $\Delta_{n}\partial_{t}\mathbf{E}$ can be dropped against
$-\mathrm{i}\omega_{\mathrm{c}}\sigma\underline{\mathbf{E}}$ and
$-\mathrm{i}\omega_{\mathrm{c}}\Delta_{n}\mathbf{E}$. The polarization
amplitude $\underline{\mathbf{P}}$, introduced in Eq.\thinspace
(\ref{eq:P_SVAA}), couples the optical propagation equation to the Bloch
equations, Eq.\thinspace(\ref{eq:RWA}), as further discussed in Section
\ref{sec:pol_RWA}. The RWA implies that also $\underline{\mathbf{P}}$ is
narrowband, which means that for example harmonic or difference frequency
generation cannot be included. Thus, similarly as for the field, $\partial
_{t}^{2}\underline{\mathbf{P}}$ and $-2\mathrm{i}\omega_{\mathrm{c}}%
\partial_{t}\underline{\mathbf{P}}$ can be neglected against $-\omega
_{\mathrm{c}}^{2}\underline{\mathbf{P}}$. In addition, the paraxial
approximation implies that $\partial_{z}^{2}\underline{\mathbf{E}}$ can be
neglected against $\mathrm{i}k_{\mathrm{c}}\partial_{z}\underline{\mathbf{E}}%
$. Finally multiplying all terms with $\exp\left(  \mathrm{i}\omega
_{\mathrm{c}}t-\mathrm{i}k_{\mathrm{c}}z\right)  $ and discarding all rapidly
oscillating terms, which also eliminates $\mathbf{J}_{\mathrm{f}}$ since it is
assumed to contain only low frequency components (see Section \ref{sec:pol}),
we arrive at
\begin{align}
\partial_{t}\underline{\mathbf{E}}\pm\frac{c}{n_{0}}\partial_{z}%
\underline{\mathbf{E}}=  &  -\mathrm{i}\omega_{\mathrm{c}}\Delta_{n}\underline
{\mathbf{E}} \nonumber\\
&  +\frac{1}{2n_{0}^{2}}\left(  \mathrm{i}\frac{c^{2}}{\omega
_{\mathrm{c}}}\nabla_{\mathrm{T}}^{2}\underline{\mathbf{E}}+\mathrm{i}%
\frac{\omega_{\mathrm{c}}}{\epsilon_{0}}\underline{\mathbf{P}}-\frac{\sigma
}{\epsilon_{0}}\underline{\mathbf{E}}\right)  . \label{eq:SVAA}%
\end{align}
Here, $\nabla_{\mathrm{T}}^{2}=\partial_{x}^{2}+\partial_{y}^{2}$ denotes the
transverse Laplace operator. The ''+'' and ''-'' signs in Eq.\thinspace
(\ref{eq:SVAA}) are for forward and backward propagation corresponding to
$k_{\mathrm{c}}>0$ and $k_{\mathrm{c}}<0$, respectively. For
counterpropagating fields which for example arise in Fabry-P\'{e}rot
resonators, the standing wave pattern causes a position dependent inversion
grating, also referred to as spatial hole burning. This effect is not yet
included in Eq.\thinspace(\ref{eq:SVAA}), and its implementation is discussed
in Section \ref{sec:shb}.

\subsubsection{\label{sec:pol_RWA}Polarization in Rotating Wave Approximation}

In the RWA, the off-diagonal density matrix elements\ $\rho_{ij}$ that are
associated with near-resonant optical transitions are represented in terms of
transformed elements\ $\eta_{ij}$ in a rotating reference frame, as obtained
with Eq.\thinspace(\ref{eq:rho_RWA}). Writing the total current as
$\mathbf{J}_{\mathrm{q}}=\mathbf{J}_{\mathrm{f}}+\partial_{t}\mathbf{P}%
_{\mathrm{q}}$ as in Eq.\thinspace(\ref{eq:Jt}), and assigning the
low-frequency contributions to $\mathbf{J}_{\mathrm{f}}$ and the optical
contributions to $\mathbf{P}_{\mathrm{q}}$, we see from Eq.\thinspace
(\ref{eq:Jt2}) that the transformation into the rotating frame only affects
the evaluation of the polarization $\mathbf{P}_{\mathrm{q}}$. With
Eq.\thinspace(\ref{eq:P}) and Eq.\thinspace(\ref{eq:rho_RWA}), we obtain%

\begin{equation}
\mathbf{P}_{\mathrm{q}}=n_{\mathrm{3D}}\sum_{\omega_{ij}>0}\mathbf{d}_{ji}%
\eta_{ij}\exp\left(  -\mathrm{i\,}\omega_{\mathrm{c}}t\right)
+c.c.+n_{\mathrm{3D}}\sum_{i}\mathbf{d}_{ii}\rho_{ii}. \label{eq:P_RWA}%
\end{equation}
For inclusion of optical propagation, the RWA is often not coupled to the full
Maxwell equations, but rather solved together with Eq.\thinspace
(\ref{eq:SVAA}) in SVAA which contains the polarization in terms of the
amplitude $\underline{\mathbf{P}}$. As discussed above, we have to replace
$\exp\left(  -\mathrm{i\,}\omega_{\mathrm{c}}t\right)  $\ by $\exp\left(
\mathrm{i}k_{\mathrm{c}}z-\mathrm{i}\omega_{\mathrm{c}}t\right)  $ in
Eq.\thinspace(\ref{eq:P_RWA}) since the SVAA also takes into account the
spatial dependence of the carrier. Comparing the resulting equation with
Eq.\thinspace(\ref{eq:P_SVAA}), and neglecting the quasi-static dipole moment
contribution $\sum_{i}\mathbf{d}_{ii}\rho_{ii}$ which does not oscillate at
the optical excitation frequency and thus drops out in the SVAA, we obtain%
\begin{equation}
\underline{\mathbf{P}}=2n_{\mathrm{3D}}\sum_{\omega_{ij}>0}\mathbf{d}_{ji}%
\eta_{ij}. \label{eq:P_SVAA2}%
\end{equation}

\subsection{Initial Conditions\label{sec:bocos}}

The Bloch equations without or with RWA, Eq.\thinspace(\ref{eq:MB2}) or
Eq.\thinspace(\ref{eq:RWA}), have to be supplemented by corresponding initial
conditions at time $t=t_{0}$. Apart from special cases where the quantum
system may be coherently prepared in a certain initial state such as a
coherent superposition \cite{brewer1972optical}, the system will be initially
in equilibrium. The corresponding density matrix elements are then obtained by
setting $\partial_{t}=0$ in Eq.\thinspace(\ref{eq:MB2}) or (\ref{eq:RWA}) and
assuming a vanishing optical field for $t\leq t_{0}$, which gives rise to a
mixed state with off-diagonal elements $\rho_{ij}\left(  t=t_{0}\right)  =0$
and $\eta_{ij}\left(  t=t_{0}\right)  =0$, respectively. The diagonal elements
$\rho_{ii}\left(  t=t_{0}\right)  =\rho_{ii}^{\mathrm{eq}}$ are given by the
equilibrium occupation probabilities $\rho_{ii}^{\mathrm{eq}}$, which can be
obtained by setting $\mathrm{d}_{t}=0$ in Eq.\thinspace(\ref{eq:relax1}). This
yields for a system with $N$ levels the linear equation system%
\begin{align}
0  &  =\sum_{j\neq i}r_{j\rightarrow i}\rho_{jj}^{\mathrm{eq}}-r_{i}\rho
_{ii}^{\mathrm{eq}},\,i=1..\left(  N-1\right)  ,\\
1  &  =\sum_{i=1}^{N}\rho_{ii}^{\mathrm{eq}},
\end{align}
where $r_{i}$ is given by Eq.\thinspace(\ref{eq:rn}). The $\rho_{ii}%
^{\mathrm{eq}}$ do not necessarily correspond to a thermal distribution, but
are rather determined by the transition rates which may for example include
the pumping process in lasers. For inhomogeneous device structures, the rates
$r_{i}$ and $r_{j\rightarrow i}$ generally depend on position $\mathbf{x}$,
giving rise to $\mathbf{x}$\ dependent $\rho_{ii}^{\mathrm{eq}}$.

Suitable initial conditions have also to be defined for the Maxwell equations,
Eq.\thinspace(\ref{eq:maxw}), or the propagation equations in SVAA derived
thereof, Eq.\thinspace(\ref{eq:SVAA}). Here we cannot choose identically
vanishing fields, since the optical field would then remain zero throughout
the MB simulation. Laser seeding by spontaneous emission is often mimicked by
initializing the electric field with white Gaussian amplitude noise, which can
also be added at every time step of the simulation to model spontaneous
emission noise \cite{slavcheva2004fdtd}. In the SVAA, Eq.\thinspace
(\ref{eq:SVAA}), the electric field is represented by its complex envelope
function, and thus complex white Gaussian noise is used in this case.

\subsection{Two-Level Approximation}

In most cases, the simplest model with only two relevant states (e.g., an
upper laser level $2$ and lower laser level $1$) is considered. The
corresponding density matrix contains the elements $\rho_{11}$, $\rho_{22}$
and $\rho_{21}=\rho_{12}^{\ast}$. Assuming a closed system, we obtain
$\rho_{11}+\rho_{22}=1$. The dissipation in the Bloch equations,
Eq.\thinspace(\ref{eq:MB2}), is then parametrized by the three rates
$\gamma_{21}$, $r_{1\rightarrow2}$ and $r_{2\rightarrow1}$, where $\gamma
_{12}=\gamma_{21}$ as described in Section \ref{sec:dep}, and Eq.\thinspace
(\ref{eq:rn}) gives $r_{1}=r_{1\rightarrow2}$, $r_{2}=r_{2\rightarrow1}$.
Introducing the population inversion $w=\rho_{22}-\rho_{11}$, we can
substitute $\rho_{11}=\left(  1-w\right)  /2$ and $\rho_{22}=\left(
1+w\right)  /2$ in Eq.\thinspace(\ref{eq:MB2}). Furthermore neglecting static
dipole moments, $\mathbf{d}_{22}-\mathbf{d}_{11}\approx\mathbf{0}$, we obtain%
\begin{subequations}%
\label{eq:bloch2l}
\begin{align}
\partial_{t}\rho_{21}  &  =-\mathrm{i}\omega_{21}\rho_{21}-\mathrm{i}%
w\Omega-\gamma_{2}\rho_{21},\label{eq:bloch2l1}\\
\partial_{t}w  &  =2\mathrm{i}\left(  \rho_{21}^{\ast}\Omega-\rho_{21}%
\Omega^{\ast}\right)  -\gamma_{1}w+\Gamma_{12}. \label{eq:bloch2l2}%
\end{align}%
\end{subequations}%
Here, $\Omega=\hbar^{-1}\mathbf{d}_{21}\mathbf{E}$ denotes the instantaneous
Rabi frequency, $\gamma_{1}=r_{1\rightarrow2}+r_{2\rightarrow1}$ and
$\gamma_{2}=\gamma_{21}$ are the population inversion relaxation and dephasing
rates, and $\Gamma_{12}=r_{1\rightarrow2}-r_{2\rightarrow1}$ represents the
(net) pumping rate from the lower to the upper level. Equation
(\ref{eq:bloch2l2}) is often also written as
\begin{equation}
\partial_{t}w=2\mathrm{i}\left(  \rho_{21}^{\ast}\Omega-\rho_{21}\Omega^{\ast
}\right)  -\gamma_{1}\left(  w-w_{\mathrm{eq}}\right)  , \label{eq:bloch2l3}%
\end{equation}
where $w_{\mathrm{eq}}=\Gamma_{12}/\gamma_{1}$ denotes the equilibrium
population inversion for $\Omega=0$.

A real-valued, redundance-free representation can be obtained by applying
Eqs.\thinspace(\ref{eq:ps1}) and (\ref{eq:ps2}), yielding the three
real-valued quantities $u\mathbf{:=}\left\langle \hat{u}_{12}\right\rangle
=\rho_{12}+\rho_{21}=2\Re\left\{  \rho_{21}\right\}  $, $v\mathbf{:=-}%
\left\langle \hat{v}_{12}\right\rangle =\mathbf{-}\mathrm{i}\left(  \rho
_{12}-\rho_{21}\right)  =-2\Im\left\{  \rho_{21}\right\}  $, and
$w\mathbf{:=}\left\langle \hat{w}_{1}\right\rangle =\rho_{22}-\rho_{11}$.
These are usually represented in terms of the so-called Bloch vector
$\mathbf{S}=\left[  u,v,w\right]  ^{\mathrm{T}}$, where a minus sign has been
added to the definition of $v$ in order to obtain the usual convention for the
Bloch vector \cite{allen1987optical}. Separating $\Omega$ in its real and
imaginary part $\Omega_{\mathrm{r}}+\mathrm{i}\Omega_{\mathrm{i}}$,
Eq.\thinspace(\ref{eq:bloch2l}) then becomes
\begin{subequations}%
\label{eq:blochr}%
\begin{align}
\partial_{t}u  &  =-\omega_{21}v+2\Omega_{\mathrm{i}}w-\gamma_{2}%
u,\label{eq:blochr1}\\
\partial_{t}v  &  =\omega_{21}u+2\Omega_{\mathrm{r}}w-\gamma_{2}%
v,\label{eq:blochr2}\\
\partial_{t}w  &  =-2\Omega_{\mathrm{i}}u-2\Omega_{\mathrm{r}}v-\gamma
_{1}w+\Gamma_{12}, \label{eq:blochr3}%
\end{align}%
\end{subequations}%
which can also be written as \cite{allen1987optical}%
\begin{equation}
\partial_{t}\mathbf{S}=\left(
\begin{array}
[c]{c}%
-2\Omega_{\mathrm{r}}\\
2\Omega_{\mathrm{i}}\\
\omega_{21}%
\end{array}
\right)  \times\mathbf{S}-\left(
\begin{array}
[c]{c}%
\gamma_{2}u\\
\gamma_{2}v\\
\gamma_{1}\left(  w-w_{\mathrm{eq}}\right)
\end{array}
\right)  .
\end{equation}
The polarization term in Eq.\thinspace(\ref{eq:maxw}) is then with
Eq.\thinspace(\ref{eq:Jt2}) obtained as%
\begin{align}
\partial_{t}\mathbf{P}_{\mathrm{q}}  &  =2n_{\mathrm{3D}}\Re\left\{
\mathbf{d}_{12}\partial_{t}\rho_{21}\right\} \nonumber\\
&  =n_{\mathrm{3D}}\left(  \Re\left\{  \mathbf{d}_{12}\right\}  \partial
_{t}u+\Im\left\{  \mathbf{d}_{12}\right\}  \partial_{t}v\right)  .
\label{eq:P2u}%
\end{align}

\begin{figure}[ptb]
\includegraphics{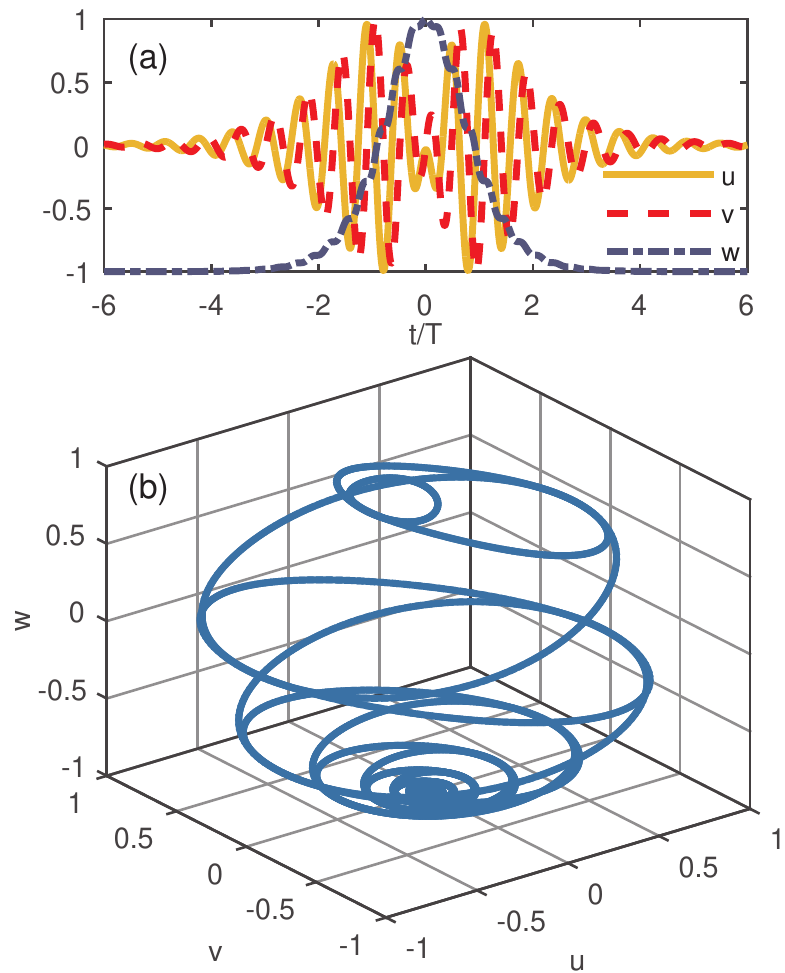}
\caption{(a)\ Bloch vector components and (b) Bloch vector trajectory for a
dissipationless two-level system excited with a sech pulse.}%
\label{fig:blochsphere}%
\end{figure}

The time evolution of the Bloch vector $\mathbf{S}(t)$\ can be visualized in
the Bloch sphere representation, where the Bloch vector trajectory is
displayed in a Cartesian coordinate system with axes $u$, $v$ and $w$
\cite{feynman1957geometrical,allen1987optical}. For $\gamma_{1}=\gamma_{2}=0$,
$\left|  \mathbf{S}\left(  t\right)  \right|  $ is conserved over time, as can
be seen by multiplying Eqs.\thinspace(\ref{eq:blochr1}), (\ref{eq:blochr2})
and (\ref{eq:blochr3}) with $u$, $v$ and $w$, respectively, and adding the
resulting equations, which yields $\partial_{t}\left(  u^{2}+v^{2}%
+w^{2}\right)  =0$. For pure states, $\left|  \mathbf{S}\left(  t\right)
\right|  =1$, i.e., the tip of the Bloch vector moves along the surface of a
unit sphere, the so-called Bloch sphere. For mixed states, the tip is located
within the Bloch sphere, corresponding to $\left|  \mathbf{S}\left(  t\right)
\right|  <1$. In Fig.\thinspace\ref{fig:blochsphere}, the time evolution of
the Bloch vector components and the corresponding Bloch vector trajectory are
shown for a two-level system with $\gamma_{1}=\gamma_{2}=0$ and initial
conditions $u=v=0$, $w=-1$. The optical field is assumed to be a sech pulse
$\Omega=2T^{-1}\mathrm{sech}\left(  t/T\right)  $, which corresponds to a
self-induced transparency soliton as further discussed in Section
\ref{sec:SIT}, with $T=10/\omega_{21}$ chosen for this example.

A further representation of the Bloch equations is obtained by assuming a real
$\mathbf{d}_{21}$ and thus $\Omega=\Omega_{\mathrm{r}}$. Solving
Eq.\thinspace(\ref{eq:blochr1}) for $v$ and using the result to eliminate $v$
in Eqs.\thinspace(\ref{eq:blochr2}) and (\ref{eq:blochr3}) yields with
Eq.\thinspace(\ref{eq:P2u}) \cite{Boyd}%
\begin{subequations}%
\label{eq:Pcomp}
\begin{gather}
\left[ \partial_{t}^{2}+2\gamma_{2}\partial_{t}%
+\left(  \omega_{21}^{2}+\gamma_{2}^{2}\right)\right]  \mathbf{P}%
_{\mathrm{q}}=-\frac{2\omega_{21}\mathbf{d}_{12}^{2}}{\hbar}n_{\mathrm{3D}%
}w\frac{\mathbf{d}_{12}}{\left|  \mathbf{d}_{12}\right|  }\mathbf{E,}%
\label{eq:Pcomp1}\\
\partial_{t}w=2\frac{\partial_{t}\mathbf{P}_{\mathrm{q}}+\gamma_{2}%
\mathbf{P}_{\mathrm{q}}}{\hbar n_{\mathrm{3D}}\omega_{21}}\mathbf{E}%
-\gamma_{1}w+\Gamma_{12}. \label{eq:Pcomp2}%
\end{gather}%
\end{subequations}%

This representation can be seen as an extension of the classical Lorentz model
for resonant polarization in dielectrics, assuming the same mathematical form
as Eq.\thinspace(\ref{eq:Pcomp1}) if we set $w$ constant. Accordingly,
Eq.\thinspace(\ref{eq:Pcomp}) is mainly used in computational electrodynamics,
especially in combination with the finite-difference time-domain method, as a
substitute for more basic classical polarization models \cite{taflove2005}.

\subsubsection{Rotating Wave/Slowly Varying Amplitude Approximation}

In the RWA, we obtain from Eq.\thinspace(\ref{eq:RWA}) with $\underline
{\Omega}=\hbar^{-1}\mathbf{d}_{21}\underline{\mathbf{E}}$
\begin{subequations}%
\label{eq:RWA2l}
\begin{align}
\partial_{t}\eta_{21}  &  =\mathrm{i}\Delta\eta_{21}-\frac{1}{2}%
\mathrm{i}w\underline{\Omega}-\gamma_{2}\eta_{21},\label{eq:RWA2l1}\\
\partial_{t}w  &  =\mathrm{i}\left(  \eta_{21}^{\ast}\underline{\Omega}%
-\eta_{21}\underline{\Omega}^{\ast}\right)  -\gamma_{1}\left(
w-w_{\mathrm{eq}}\right)  , \label{eq:RWA2l2}%
\end{align}%
\end{subequations}%
where $\Delta=\omega_{\mathrm{c}}-\omega_{21}$ denotes the detuning of the
optical field from the resonance frequency $\omega_{21}$. In analogy to above,
we can introduce the Bloch vector $\mathbf{s}$\ for the off-diagonal density
matrix elements in RWA, with components $s_{1}=\eta_{12}+\eta_{21}%
=2\Re\left\{  \eta_{21}\right\}  $, $s_{2}=\mathbf{-}\mathrm{i}\left(
\eta_{12}-\eta_{21}\right)  =-2\Im\left\{  \eta_{21}\right\}  $, $s_{3}=w$,
and obtain in analogy to Eq.\thinspace(\ref{eq:blochr}) with $\underline
{\Omega}=\underline{\Omega}_{\mathrm{r}}+\mathrm{i}\underline{\Omega
}_{\mathrm{i}}$
\begin{subequations}%
\label{eq:RWAu}%
\begin{align}
\partial_{t}s_{1}  &  =\Delta s_{2}+\underline{\Omega}_{\mathrm{i}}%
w-\gamma_{2}s_{1},\label{eq:RWAu1}\\
\partial_{t}s_{2}  &  =-\Delta s_{1}+\underline{\Omega}_{\mathrm{r}}%
w-\gamma_{2}s_{2},\label{eq:RWAu2}\\
\partial_{t}w  &  =-\underline{\Omega}_{\mathrm{i}}s_{1}-\underline{\Omega
}_{\mathrm{r}}s_{2}-\gamma_{1}\left(  w-w_{\mathrm{eq}}\right)  .
\label{eq:RWAu3}%
\end{align}%
\end{subequations}%
The polarization term in the SVAA propagation equation, Eq.\thinspace
(\ref{eq:SVAA}), is then with Eq.\thinspace(\ref{eq:P_SVAA2}) obtained as%

\begin{equation}
\underline{\mathbf{P}}=2n_{\mathrm{3D}}\mathbf{d}_{12}\eta_{21}=n_{\mathrm{3D}%
}\mathbf{d}_{12}\left(  s_{1}-\mathrm{i}s_{2}\right)  . \label{eq:pol2rwa}%
\end{equation}

\section{\label{sec:1D}Reduction to One-Dimensional Model}

\begin{figure}[ptb]
\includegraphics{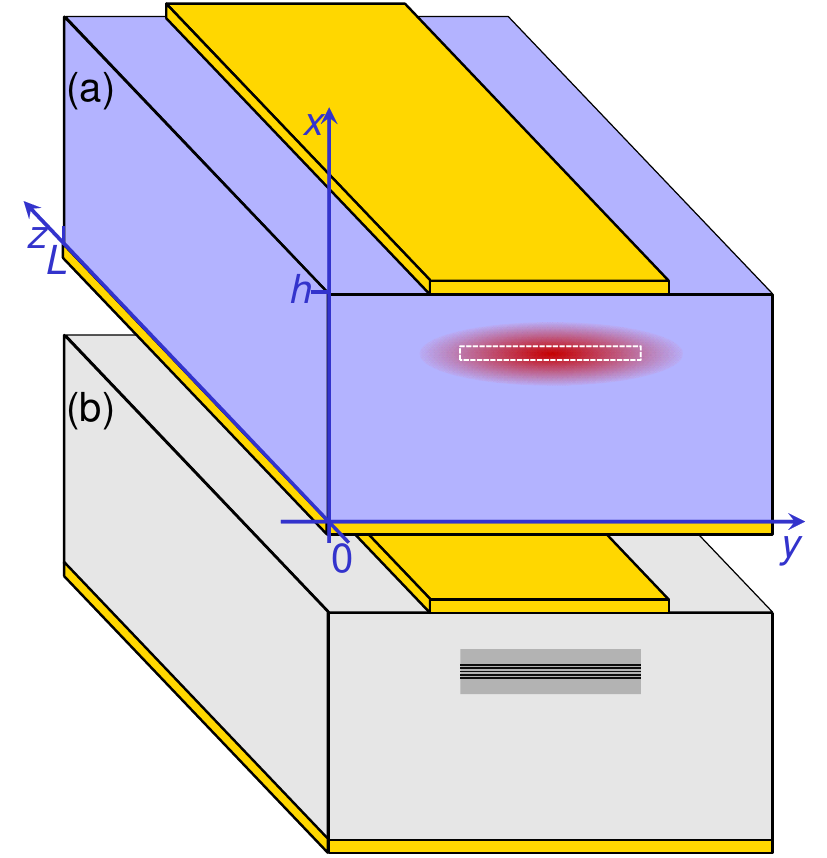}
\caption{Schematic illustration of one-dimensional propagation model for a
waveguide structure. (a) The white rectangle on the front facet denotes the
cross section of the quantum structure, and the optical intensity distribution
is indicated. (b) The refractive index profile is indicated, with darker
colors corresponding to higher refractive indices.}%
\label{fig:resonator1D}%
\end{figure}

Although the MB equations are sometimes solved in two or even three spatial
dimensions
\cite{slavcheva2002coupled,klaedtke2006ultrafast,sukharev2011,pusch2012coherent,lopata2009nonlinear,takeda2011self,dridi2013model,cartar2017}%
, the model is frequently reduced to a single spatial coordinate in order to
minimize the numerical load \cite{allen1987optical}. This is usually achieved
by assuming plane wave propagation in the Maxwell equations, Eq.\thinspace
(\ref{eq:maxw}), or the corresponding propagation equations in SVAA,
Eq.\thinspace(\ref{eq:SVAA}) \cite{siddons2014light}. For extended beams
propagating in a homogeneous medium such as a gas or bulk solid-state medium,
the plane wave approximation may be a reasonable assumption. For
optoelectronic devices which are the focus of this paper, the light is usually
strongly guided, often with sub-wavelength confinement in at least one
dimension. Here, the plane wave approximation is clearly too simplistic.
However, optoelectronic devices such as semiconductor-based lasers often
employ waveguiding structures which are invariant in propagation direction
$z$, in particular schemes with a suitable transverse refractive index profile
or metal cladding. Such geometries provide lateral field confinement and give
rise to guided mode solutions, i.e., field solutions which are at a given
frequency $\omega$ characterized by a propagation constant and a $z$
independent transverse field distribution. While some one-dimensional plane
wave treatments have included all transverse field components to describe
elliptically or circularly polarized light
\cite{slavcheva2005dynamical,slavcheva2010nonlinear,slavcheva2019ultrafast,slavcheva2008model,song2006propagation}%
, we assume linearly polarized waveguide modes in the following, and thus
consider a single transverse component of the electric and magnetic fields. In
Fig.\thinspace\ref{fig:resonator1D}, an exemplary waveguide structure is
schematically illustrated.

\subsection{\label{sec:full}Full Maxwell Equations}

We employ the full Maxwell equations, Eq.\thinspace(\ref{eq:maxw}), coupled to
the Bloch equations, Eq.\thinspace(\ref{eq:MB}) or (\ref{eq:MB2}), to describe
the carrier-light interaction and optical propagation in a waveguide geometry
which is invariant with respect to the propagation direction $z$. Our goal is
to extract a one-dimensional MB model with a single electric and magnetic
field component, as typically used in simulations due to the associated
computational burden. We focus on guided mode solutions, which are at a given
frequency $\omega$ characterized by a (generally complex) propagation constant
$\underline{\beta}$, and $z$ invariant transverse field dependencies
$\underline{E}_{x,y,z}^{\mathrm{t}}\left(  x,y\right)  $ and $\underline
{H}_{x,y,z}^{\mathrm{t}}\left(  x,y\right)  $\ for the electric and magnetic
field components. Thus, we can make the ansatz%
\begin{subequations}%
\label{eq:EH}
\begin{align}
E_{p}\left(  \mathbf{x},t\right)   &  =\Re\left\{  \underline{E}%
_{p}^{\mathrm{t}}\left(  x,y\right)  \exp\left(  \mathrm{i}\underline{\beta
}z-\mathrm{i}\omega t\right)  \right\} \nonumber\\
&  =\Re\left\{  \underline{E}_{p}%
^{\omega}\left(  \mathbf{x}\right)  \exp\left(  -\mathrm{i}\omega t\right)
\right\}  ,\label{eq:EH1}\\
H_{p}\left(  \mathbf{x},t\right)   &  =\Re\left\{  \underline{H}%
_{p}^{\mathrm{t}}\left(  x,y\right)  \exp\left(  \mathrm{i}\underline{\beta
}z-\mathrm{i}\omega t\right)  \right\} \nonumber\\
&  =\Re\left\{  \underline{H}_{p}%
^{\omega}\left(  \mathbf{x}\right)  \exp\left(  -\mathrm{i}\omega t\right)
\right\}  , \label{eq:EH2}%
\end{align}%
\end{subequations}%
with $p=x,y,z$. By inserting Eq.\thinspace(\ref{eq:EH}) into Eq.\thinspace
(\ref{eq:maxw}), the computation of the transverse mode profile in the
$xy$-plane can be decoupled from the $z$ coordinate and reduces to a
two-dimensional problem. For example, by eliminating the electric field, we obtain%

\begin{equation}
\left(  \partial_{p}^{2}+\partial_{q}^{2}\right)  \underline{H}_{p}%
^{\mathrm{t}}+\frac{\partial_{q}\underline{\epsilon}_{\mathrm{r}}}%
{\underline{\epsilon}_{\mathrm{r}}}\left(  \partial_{p}\underline{H}%
_{q}^{\mathrm{t}}-\partial_{q}\underline{H}_{p}^{\mathrm{t}}\right)  =\left(
\underline{\beta}^{2}-\frac{\omega^{2}}{c^{2}}\underline{\epsilon}%
_{\mathrm{r}}\right)  \underline{H}_{p}^{\mathrm{t}} \label{eq:Hpq}%
\end{equation}
with $p=x,y$ and $q=y,x$ \cite{1990ITMTT..38..722S}. Waveguiding is for
instance obtained by surrounding the optically active region with another
dielectric material featuring a lower refractive index as illustrated in
Fig.\thinspace\ref{fig:resonator1D}(b), or with a metal cladding. Both cases
can be described by a transversely dependent complex background permittivity%
\begin{equation}
\underline{\epsilon}_{\mathrm{r}}=\epsilon_{\mathrm{r}}+\mathrm{i}%
\sigma/\left(  \omega\epsilon_{0}\right)  , \label{eq:eps}%
\end{equation}
where $\epsilon_{\mathrm{r}}$ and $\sigma$ generally depend on $x$, $y$ and
$\omega$, and $\sigma$ accounts for the conductivity or dielectric losses.
Together with the boundary condition $\underline{H}_{p,q}^{\mathrm{t}%
}\rightarrow0$ for $x^{2}+y^{2}\rightarrow\infty$, Eq.\thinspace(\ref{eq:Hpq})
constitutes a complex eigenvalue problem. Equation (\ref{eq:Hpq}) can for
example be solved with the film mode matching method, which is especially
suitable for waveguides with a rectangular cross section
\cite{1994PApOp...3..381S}. The polarization contribution $\mathbf{P}%
_{\mathrm{q}}$ in Eq.\thinspace(\ref{eq:maxw}) due to the quantum systems is
not yet considered\ in Eq.\thinspace(\ref{eq:Hpq}) since it is assumed to be
small enough to be included in first-order perturbation theory, with
negligible influence on the transverse field distribution. Using
$\mathbf{\nabla H}=0$, we can calculate the longitudinal component
$\underline{H}_{z}^{\mathrm{t}}\left(  x,y\right)  $ from $\underline{H}%
_{x}^{\mathrm{t}}$ and $\underline{H}_{y}^{\mathrm{t}}$ as%
\begin{equation}
\underline{H}_{z}^{\mathrm{t}}=\mathrm{i}\underline{\beta}^{-1}\left(
\partial_{x}\underline{H}_{x}^{\mathrm{t}}+\partial_{y}\underline{H}%
_{y}^{\mathrm{t}}\right)  . \label{eq:Hz}%
\end{equation}
Furthermore, the electric field components are with Eq.\thinspace
(\ref{eq:maxw2}) obtained as%
\begin{subequations}%
\label{eq:Exyz}%
\begin{align}
\omega\epsilon_{0}\underline{\epsilon}_{\mathrm{r}}\underline{E}%
_{x}^{\mathrm{t}}  &  =\mathrm{i}\partial_{y}\underline{H}_{z}^{\mathrm{t}%
}+\underline{\beta}\underline{H}_{y}^{\mathrm{t}},\label{eq:Exyz1}\\
\omega\epsilon_{0}\underline{\epsilon}_{\mathrm{r}}\underline{E}%
_{y}^{\mathrm{t}}  &  =-\underline{\beta}\underline{H}_{x}^{\mathrm{t}%
}-\mathrm{i}\partial_{x}\underline{H}_{z}^{\mathrm{t}},\label{eq:Exyz2}\\
\omega\epsilon_{0}\underline{\epsilon}_{\mathrm{r}}\underline{E}%
_{z}^{\mathrm{t}}  &  =\mathrm{i}\partial_{x}\underline{H}_{y}^{\mathrm{t}%
}-\mathrm{i}\partial_{y}\underline{H}_{x}^{\mathrm{t}}. \label{eq:Exyz3}%
\end{align}%
\end{subequations}%

For general solutions of Eq.\thinspace(\ref{eq:Hpq}), the polarization varies
over the waveguide cross section. As indicated in Fig.\thinspace
\ref{fig:resonator1D}(b), in many optoelectronic devices, such as typical
standard edge-emitting and quantum cascade lasers, rectangular waveguides are
used where the width in lateral $y$ direction significantly exceeds its
thickness in $x$ direction. This allows an approximate treatment as a slab
waveguide structure, which is assumed to be infinitely extended in $y$
direction and thus can, to first order, be described by $\underline{\epsilon
}_{\mathrm{r}}\left(  x\right)  $. The field components are then assumed to be
constant in $y$ direction, which corresponds to setting $\partial_{y}=0$
\cite{yar89}. The guided field solutions can be divided into two classes:
Transverse electric (TE) modes are characterized by $\underline{E}_{z}%
^{\omega}=0$, where for $\partial_{y}=0$ all components except $\underline
{E}_{y}^{\omega}$, $\underline{H}_{x}^{\omega}$ and $\underline{H}_{z}%
^{\omega}$ vanish as can be seen from Eqs.\thinspace(\ref{eq:Hz}) and
(\ref{eq:Exyz}); similarly, transverse magnetic (TM) modes, characterized by
$\underline{H}_{z}^{\omega}=0$, have for $\partial_{y}=0$ only non-vanishing
$\underline{H}_{y}^{\omega}$, $\underline{E}_{x}^{\omega}$ and $\underline
{E}_{z}^{\omega}$ components \cite{yar89}. The $y$ dependence of the field
distribution may then be reintroduced using the effective refractive index
approximation method \cite{chiang1991performance}, which preserves the TE or
TM character of the solution. From the discussion of dipole matrix elements
for quantum well structures in Section \ref{sec:dip}, it follows that standard
edge-emitting lasers, which utilize interband transitions, preferably operate
in TE mode [see also Fig.\thinspace\ref{fig:polarization}(a)]. On the other
hand, QCLs, which rely on intraband transitions, only operate in TM\ mode [see
also Fig.\thinspace\ref{fig:polarization}(c)].

As pointed out above, simulations typically employ a plane-wave-type
propagation model which only depends on the propagation coordinate $z$ and
time $t$, and considers a single transverse electric and transverse magnetic
field component. Our goal is to derive such equations, with a form equivalent
to the Maxwell equations, for guided rather than plane-wave propagation, as
applies to many photonic devices and systems.

\subsubsection{\label{sec:TE}Transverse Electric Mode}

For TE modes in slab waveguides, Eq.\thinspace(\ref{eq:Hpq}) yields with
$\underline{H}_{y}^{\mathrm{t}}=0$ and $\partial_{y}=0$%
\begin{equation}
\partial_{x}^{2}\underline{H}_{x}^{\mathrm{t}}=\left(  \underline{\beta}%
^{2}-\frac{\omega^{2}}{c^{2}}\underline{\epsilon}_{\mathrm{r}}\right)
\underline{H}_{x}^{\mathrm{t}}, \label{eq:Hx}%
\end{equation}
and the boundary conditions are given by $\underline{H}_{x}^{\mathrm{t}%
}\left(  x\rightarrow\pm\infty\right)  \rightarrow0$. From Eq.\thinspace
(\ref{eq:maxw}), we furthermore obtain%
\begin{subequations}%
\label{eq:Maxw_TE}%
\begin{align}
\mathbf{-\partial}_{z}E_{y}  &  =-\mu_{0}\partial_{t}H_{x},
\label{eq:Maxw_TE1}\\
\mathbf{\partial}_{z}H_{x}-\partial_{x}H_{z}  &  =\epsilon_{0}\epsilon
_{\mathrm{r}}\partial_{t}E_{y}+\sigma E_{y}. \label{eq:Maxw_TE2}%
\end{align}%
\end{subequations}%

The polarization contribution of the quantum systems is not contained in
Eq.\thinspace(\ref{eq:Maxw_TE2}) since it will subsequently be included in a
perturbative manner. Equation (\ref{eq:Maxw_TE2}) does not yet have the
desired form since it contains an $x$ derivative and the longitudinal field
component in the term $\partial_{x}H_{z}$. With Eqs.\thinspace(\ref{eq:EH2}),
(\ref{eq:Hz}) and (\ref{eq:Hx}), we obtain%
\begin{equation}
\partial_{x}\underline{H}_{z}^{\omega}=\mathrm{i}\underline{\beta}^{-1}\left(
\underline{\beta}^{2}-\frac{\omega^{2}}{c^{2}}\underline{\epsilon}%
_{\mathrm{r}}\right)  \underline{H}_{x}^{\omega}, \label{eq:Hx2}%
\end{equation}
where $\underline{\epsilon}_{\mathrm{r}}=\epsilon_{\mathrm{r}}+\mathrm{i}%
\sigma/\left(  \omega\epsilon_{0}\right)  $. In the following, it is practical
to switch to the frequency domain, where Eq.\thinspace(\ref{eq:Maxw_TE1}) is
with Eq.\thinspace(\ref{eq:EH}) given by%
\begin{equation}
\partial_{z}\underline{E}_{y}^{\omega}=\mathrm{i}\underline{\beta}%
\underline{E}_{y}^{\omega}=-\mathrm{i}\omega\mu_{0}\underline{H}_{x}^{\omega}.
\label{eq:TE2}%
\end{equation}
From Eq.\thinspace(\ref{eq:TE2}), we see that the electric and magnetic fields
have the same transverse distribution. Inserting Eq.\thinspace(\ref{eq:Hx2})
into Eq.\thinspace(\ref{eq:Maxw_TE2}) in frequency domain, and employing
Eq.\thinspace(\ref{eq:TE2}), we arrive at
\begin{equation}
\mathbf{\partial}_{z}\underline{H}_{x}^{\omega}=\mathbf{-}\mathrm{i}%
\omega^{-1}\mu_{0}^{-1}\underline{\beta}^{2}\underline{E}_{y}^{\omega}.
\label{eq:TE3}%
\end{equation}

In the following, the polarization contribution of the quantum system will be
included as a perturbation \cite{agr01}. Re-deriving Eq.\thinspace
(\ref{eq:Hx}) from Maxwell's equations Eq.\thinspace(\ref{eq:maxw}), but now
with the polarization contribution due to the quantum systems included, we see
that the perturbation generated by the polarization on the $\underline{E}%
_{y}^{\omega}$ field component is formally equivalent to an additional
background permittivity $\Delta\epsilon_{\mathrm{r}}=\epsilon_{0}%
^{-1}\underline{P}_{y}^{\omega}/\underline{E}_{y}^{\omega}$, where
$\underline{P}_{y}^{\omega}$ contains the polarization contribution of the
quantum systems in frequency domain. In the following, we assume that the
device operates in a single transverse mode with the magnetic field
distribution $\underline{H}_{x}^{\mathrm{t}}$, possibly the fundamental mode.
Using the similarity of Eq.\thinspace(\ref{eq:Hx}) to the Schr\"{o}dinger
equation in quantum mechanics, we can apply perturbation theory in an
analogous matter \cite{schrodinger1926quantisierung}. To first order,
$\underline{H}_{x}^{\mathrm{t}}$ remains unchanged, and for the eigenvalue
$\underline{\beta}^{2}$ we obtain the correction
\begin{equation}
\Delta\underline{\beta}^{2}=\frac{\omega^{2}}{\epsilon_{0}\underline{E}%
_{y}^{\omega}c^{2}}\frac{\iint_{-\infty}^{\infty}\left|  \underline{H}%
_{x}^{\mathrm{t}}\right|  ^{2}\underline{P}_{y}^{\omega}\mathrm{d}%
x\mathrm{d}y}{\iint_{-\infty}^{\infty}\left|  \underline{H}_{x}^{\mathrm{t}%
}\right|  ^{2}\mathrm{d}x\mathrm{d}y}. \label{eq:Db2}%
\end{equation}
For completeness, we also include integration over the $y$ coordinate in
Eq.\thinspace(\ref{eq:Db2}) since the $y$ dependence of $\underline{H}%
_{x}^{\mathrm{t}}$ may be reintroduced based on above mentioned effective
refractive index method. It should be mentioned that if $\underline{\epsilon
}_{\mathrm{r}}$ in Eq.\thinspace(\ref{eq:Hx}) has a non-vanishing imaginary
part, the eigenvalue problem is non-Hermitian and strictly speaking, a
biorthogonal basis set must be used \cite{sternheim1972non}. In this case,
Eq.\thinspace(\ref{eq:Db2}) serves as an approximation to the exact
perturbation term. Furthermore, it is practical to split the unperturbed
propagation constant $\underline{\beta}$ into a real and an imaginary part,
$\underline{\beta}=\beta+\mathrm{i}\beta^{\prime}$. Here, $\beta^{\prime}$\ is
related to the power loss coefficient $a$ by $\beta^{\prime}=\mathrm{sgn}%
\left(  \beta\right)  a/2$, with the sign function $\mathrm{sgn}$. Assuming
$\left|  \beta\right|  \gg\left|  \beta^{\prime}\right|  $, we can write
$\underline{\beta}^{2}\approx\beta^{2}+\mathrm{i}\left|  \beta\right|
a+\Delta\underline{\beta}^{2}$. Introducing the effective waveguide refractive
index $n_{\mathrm{eff}}\left(  \omega\right)  $ defined by $\beta
=\mathrm{sgn}\left(  \beta\right)  \omega n_{\mathrm{eff}}/c$, we then obtain%
\begin{align}
\partial_{z}\underline{H}_{x}^{\omega}=  &  -\mathrm{i}\omega\epsilon
_{0}n_{\mathrm{eff}}^{2}\underline{E}_{y}^{\omega}+\epsilon_{0}%
cn_{\mathrm{eff}}a\underline{E}_{y}^{\omega} \nonumber\\
&  -\mathrm{i}\omega\frac
{\iint_{-\infty}^{\infty}\left|  \underline{H}_{x}^{\mathrm{t}}\right|
^{2}\underline{P}_{y}^{\omega}\mathrm{d}x\mathrm{d}y}{\iint_{-\infty}^{\infty
}\left|  \underline{H}_{x}^{\mathrm{t}}\right|  ^{2}\mathrm{d}x\mathrm{d}y}.
\label{eq:TE5}%
\end{align}

\paragraph{\label{sec:conf}Field Confinement Factor}

\begin{figure}[ptb]
\includegraphics{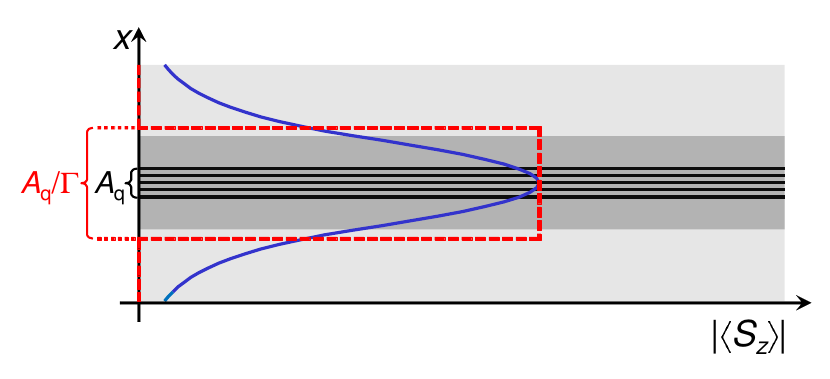}
\caption{Illustration of the transverse electric mode profile of the
waveguiding structure in Fig.\thinspace\ref{fig:resonator1D}. The solid curve
shows the mode profile $\left|  \underline{E}_{y}^{\mathrm{t}}\right|  ^{2}$,
and the dashed curve represents the equivalent rectangular mode profile.}%
\label{fig:Gamma}%
\end{figure}

Equations (\ref{eq:TE2}) and (\ref{eq:TE5}) effectively reduce the complexity
of the propagation problem from three spatial dimensions to a single
coordinate $z$. However, for computing the integral in the polarization term
of Eq.\thinspace(\ref{eq:TE5}), $\underline{P}_{y}^{\omega}$ must be obtained
by solving the Bloch equations in the whole device volume, using the full
spatial field dependence given by Eq.\thinspace(\ref{eq:EH}). This greatly
impedes the numerical efficiency of the one-dimensional propagation model. As
indicated in Fig.\thinspace\ref{fig:resonator1D}(a), frequently the transverse
field distribution does not vary significantly across the quantum
nanostructure, e.g., because the nanostructure covers only part of the
waveguide cross section, preferably at the position of maximum intensity.
Consequently, also $\underline{P}_{y}^{\omega}$ is approximately constant over
the quantum system cross section and can be taken out of the integral in
Eq.\thinspace(\ref{eq:Db2}), which can then be written as $\Gamma\underline
{P}_{y}^{\omega}$. Here, $\Gamma$ denotes the field confinement factor, which
gives the overlap of the quantum nanostructure with the mode profile and is
thus also referred to as overlap factor. With Eq.\thinspace(\ref{eq:TE2}),
$\Gamma$ can be written as%
\begin{equation}
\Gamma=\frac{\iint_{A_{\mathrm{q}}}\left|  \underline{H}_{x}^{\mathrm{t}%
}\right|  ^{2}\mathrm{d}x\mathrm{d}y}{\iint_{-\infty}^{\infty}\left|
\underline{H}_{x}^{\mathrm{t}}\right|  ^{2}\mathrm{d}x\mathrm{d}y}=\frac
{\iint_{A_{\mathrm{q}}}\left|  \underline{E}_{y}^{\mathrm{t}}\right|
^{2}\mathrm{d}x\mathrm{d}y}{\iint_{-\infty}^{\infty}\left|  \underline{E}%
_{y}^{\mathrm{t}}\right|  ^{2}\mathrm{d}x\mathrm{d}y}. \label{eq:over}%
\end{equation}
Here, the enumerator contains an integration over the cross section area
$A_{\mathrm{q}}$ of the active region formed by the quantum systems. The
intensity distribution in the waveguide is given by the time-averaged
magnitude of the $z$ component of the Poynting vector, which is with
Eq.\thinspace(\ref{eq:TE2}) obtained as
\begin{equation}
I=\left|  \left\langle S_{z}\right\rangle \right|  =\left|  \Re\left\{
\underline{E}_{y}^{\omega}\left(  \underline{H}_{x}^{\omega}\right)  ^{\ast
}\right\}  \right|  /2=\epsilon_{0}cn_{\mathrm{eff}}\left|  \underline{E}%
_{y}^{\omega}\right|  ^{2}/2. \label{eq:I}%
\end{equation}
With Eqs.\thinspace(\ref{eq:TE2}) and (\ref{eq:over}), we then arrive at the
usual definition \cite{visser1997confinement}%
\begin{equation}
\Gamma=\frac{\iint_{A_{\mathrm{q}}}\left|  \left\langle S_{z}\right\rangle
\right|  \mathrm{d}x\mathrm{d}y}{\iint_{-\infty}^{\infty}\left|  \left\langle
S_{z}\right\rangle \right|  \mathrm{d}x\mathrm{d}y}. \label{eq:Gamma}%
\end{equation}
The meaning of $\Gamma$ is visualized in Fig.\thinspace\ref{fig:Gamma}: We can
represent the field confinement factor as $\Gamma=A_{\mathrm{q}}%
/A_{\mathrm{eff}}$ where $A_{\mathrm{eff}}$ is the area covered by an
equivalent mode which conserves $\iint_{-\infty}^{\infty}\left|  \left\langle
S_{z}\right\rangle \right|  \mathrm{d}x\mathrm{d}y$, but has a rectangular
intensity distribution with $\left|  \left\langle S_{z}\right\rangle \right|
$ fixed to the value in the quantum nanostructure. Thus, the optical power can
with Eq.\thinspace(\ref{eq:I}) be written as%
\begin{equation}
P=\left[  I\right]  _{\mathrm{q}}A_{\mathrm{eff}}=\left|  \left[
\underline{E}_{y}^{\omega}\right]  _{\mathrm{q}}\right|  ^{2}A_{\mathrm{eff}%
}\epsilon_{0}cn_{\mathrm{eff}}/2, \label{eq:P2}%
\end{equation}
where $\left[  I\right]  _{\mathrm{q}}$ and $\left[  \underline{E}_{y}%
^{\omega}\right]  _{\mathrm{q}}$ refer to the values of $I$ and $\underline
{E}_{y}^{\omega}$ in the quantum nanostructure.

\paragraph{\label{sec:TE1D}One-Dimensional Maxwell Equations}

In the following, we regard the $\underline{E}_{y}^{\omega}$ and
$\underline{H}_{x}^{\omega}$ fields at a frequency $\omega$ as spectral
components of time dependent fields $E_{y}$ and $H_{x}$, and transform
Eq.\thinspace(\ref{eq:TE5}) into time domain. For convenience, we do not
consider the $\omega$ dependence of $\underline{E}_{y}^{\mathrm{t}}$ and
$\underline{H}_{x}^{\mathrm{t}}$ in Eq.\thinspace(\ref{eq:over}), but rather
evaluate $\Gamma$ at the center frequency $\omega_{\mathrm{c}}$ of the optical
field. Furthermore, to obtain a form compatible with Eq.\thinspace
(\ref{eq:Maxw_TE2}), we divide $n_{\mathrm{eff}}^{2}\left(  \omega\right)  $
into a constant part, e.g., the value $n_{\mathrm{eff}}^{2}\left(
\omega_{\mathrm{c}}\right)  $ at $\omega=\omega_{\mathrm{c}}$, and a frequency
dependent part $\Delta\epsilon_{\mathrm{eff}}=n_{\mathrm{eff}}^{2}\left(
\omega\right)  -n_{\mathrm{eff}}^{2}\left(  \omega_{\mathrm{c}}\right)  $
which describes chromatic waveguide dispersion and gives rise to an extra
polarization contribution. Considering that multiplications with $\omega$ in
frequency domain correspond to operators $\mathrm{i}\partial_{t}$ in time
domain, we obtain from Eqs.\thinspace(\ref{eq:TE2}), (\ref{eq:TE5}) and
(\ref{eq:Jt2})%
\begin{subequations}%
\label{eq:MaxwTE}%
\begin{align}
\partial_{z}E_{y}  &  =\mu_{0}\partial_{t}H_{x},\label{eq:MaxwTE1}\\
\partial_{z}H_{x}  &  =\epsilon_{0}n_{\mathrm{eff}}^{2}\left(  \omega
_{\mathrm{c}}\right)  \partial_{t}E_{y}+\sigma\left(  \mathrm{i}\partial
_{t}\right)  E_{y}\label{eq:MaxwTE2}\\
&  +\Gamma n_{\mathrm{3D}}\sum_{i,j}d_{y,ji}\partial_{t}\rho_{ij}+\epsilon
_{0}\partial_{t}\left[  \Delta\epsilon_{\mathrm{eff}}\left(  \mathrm{i}%
\partial_{t}\right)  E_{y}\right]  .\nonumber
\end{align}%
\end{subequations}%
Here, the generally frequency dependent conductivity
\begin{equation}
\sigma\left(  \omega\right)  =\epsilon_{0}cn_{\mathrm{eff}}\left(
\omega\right)  a\left(  \omega\right)  \label{eq:sig}%
\end{equation}
is\ often approximated by $\sigma=\sigma\left(  \omega_{\mathrm{c}}\right)  $.
Obviously, $\Delta\epsilon_{\mathrm{eff}}$ and $\sigma$ must be even functions
$f\left(  -\omega\right)  =f\left(  \omega\right)  $ to preserve the
real-valued character of Eq.\thinspace(\ref{eq:MaxwTE}). Furthermore,
causality requires that the real and imaginary parts of the complex
permittivity defined in Eq.\thinspace(\ref{eq:eps}) fulfill the Kramers-Kronig
relation \cite{Jackson}, which is, strictly speaking, already violated when
modeling a medium as a lossless, frequency independent dielectric with
$\underline{\epsilon}_{\mathrm{r}}=\epsilon_{\mathrm{r}}\neq1$ \cite{1991ONT}.

Notably Eq.\thinspace(\ref{eq:MaxwTE}) does not explicitly depend on $x$ and
$y$ anymore. Thus, it is practical to identify $H_{x}\left(  z,t\right)  $
and\ $E_{y}\left(  z,t\right)  $ with the field strengths at the transverse
position of the nanostructure, because then $E_{y}$ can directly be used in
Eq.\thinspace(\ref{eq:MB}) or (\ref{eq:MB2}) to evaluate $\partial_{t}%
\rho_{ji}\left(  z,t\right)  $. For completeness, we mention that the
longitudinal magnetic field component $H_{z}$ can be obtained from
$\mathbf{\nabla H}=0$, i.e., $\partial_{z}H_{z}=-\partial_{x}H_{x}$.

\subsubsection{Transverse Magnetic Mode}

\begin{figure}[ptb]
\includegraphics{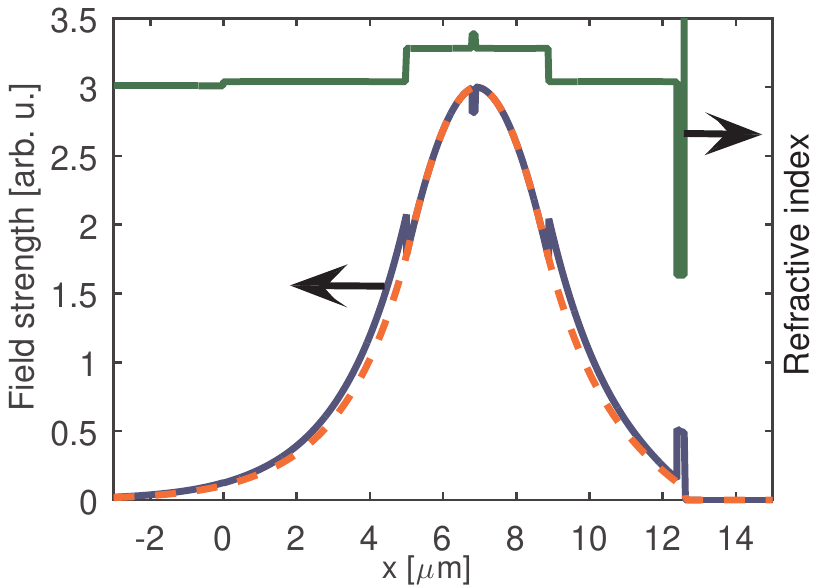}
\caption{Fundamental mode of a QCL\ waveguide structure at $28\,\mathrm{THz}$
\cite{2011ApPhL..99m1106L}. The solid and dashed curves show $\left|
\underline{E}_{x}^{\mathrm{t}}\right|  $ and $\left|  \underline{H}%
_{y}^{\mathrm{t}}\right|  $, respectively. Furthermore, the refractive index
profile, i.e., $\Re\left\{  \underline{\epsilon}_{\mathrm{r}}^{1/2}\left(
x\right)  \right\}  $, is displayed.}%
\label{fig:TM}%
\end{figure}

For TM modes in slab waveguides, Eq.\thinspace(\ref{eq:Hpq}) yields with
$\underline{H}_{x}^{\mathrm{t}}=0$ and $\partial_{y}=0$%
\begin{equation}
\underline{\epsilon}_{\mathrm{r}}\partial_{x}\left(  \underline{\epsilon
}_{\mathrm{r}}^{-1}\partial_{x}\underline{H}_{y}^{\mathrm{t}}\right)  =\left(
\underline{\beta}^{2}-\frac{\omega^{2}}{c^{2}}\underline{\epsilon}%
_{\mathrm{r}}\right)  \underline{H}_{y}^{\mathrm{t}}, \label{eq:TM}%
\end{equation}
and the boundary conditions are given by $\underline{H}_{y}^{\mathrm{t}%
}\left(  x\rightarrow\pm\infty\right)  \rightarrow0$. The mutual dependence of
the field components is given by Eq.\thinspace(\ref{eq:Exyz}), which yields
with Eq.\thinspace(\ref{eq:EH})%
\begin{align}
\mathrm{i}\omega\epsilon_{0}\underline{\epsilon}_{\mathrm{r}}\underline{E}%
_{x}^{\omega}  &  =\partial_{z}\underline{H}_{y}^{\omega}=\mathrm{i}%
\underline{\beta}\underline{H}_{y}^{\omega},\label{eq:TM1}\\
\mathrm{i}\omega\epsilon_{0}\underline{\epsilon}_{\mathrm{r}}\underline{E}%
_{z}^{\omega}  &  =-\partial_{x}\underline{H}_{y}^{\omega}. \label{eq:TM2}%
\end{align}
From Eq.\thinspace(\ref{eq:maxw1}), we furthermore obtain with Eq.\thinspace
(\ref{eq:EH})%
\begin{equation}
\mathbf{\partial}_{z}\underline{E}_{x}^{\omega}=\mathrm{i}\omega\mu
_{0}\underline{H}_{y}^{\omega}+\partial_{x}\underline{E}_{z}^{\omega}.
\label{eq:Maxw_TM1}%
\end{equation}
Using Eq.\thinspace(\ref{eq:TM2}) to eliminate $\underline{E}_{z}^{\omega}$ in
Eq.\thinspace(\ref{eq:Maxw_TM1}) yields with Eq.\thinspace(\ref{eq:TM})%
\begin{equation}
\mathbf{\partial}_{z}\underline{E}_{x}^{\omega}=\frac{\mathrm{i}}%
{\omega\epsilon_{0}\underline{\epsilon}_{\mathrm{r}}}\underline{\beta}%
^{2}\underline{H}_{y}^{\omega}. \label{eq:TM3}%
\end{equation}
In Fig.\thinspace\ref{fig:TM}, the fundamental TM mode of a QCL waveguide
structure is shown. The magnetic field distribution $\underline{H}%
_{y}^{\mathrm{t}}$ is continuous, while $\underline{E}_{x}^{\mathrm{t}}%
$\ exhibits jumps at interfaces of layers with different $\underline{\epsilon
}_{\mathrm{r}}$, as can also be seen from Eq.\thinspace(\ref{eq:TM1}).

In the following, we treat the background refractive index $n_{0}$ of the host
material for the quantum systems as a real constant, assuming that the main
frequency dependence and gain/loss in the nanostructure is provided by the
quantum systems rather than the host material. Furthermore, although for
example in a quantum well structure the barrier and well materials will have
different refractive indices as indicated in Figs.\thinspace
\ref{fig:resonator1D}(b) and \ref{fig:Gamma}, the nanostructured region can
still be approximately described by a single effective $\underline{\epsilon
}_{\mathrm{r}}$ since the individual layers are too thin to be resolved by the
optical field. For TM modes, the effective permittivity is then obtained as
the harmonic mean of the individual permittivity values, i.e., we obtain in
the host material $n_{0}^{-2}=\left(  \Delta_{1}\epsilon_{\mathrm{r,}1}%
^{-1}+\Delta_{2}\epsilon_{\mathrm{r,}2}^{-1}\right)  /\left(  \Delta
_{1}+\Delta_{2}\right)  $ where $\Delta_{i}$ and $\epsilon_{\mathrm{r,}i}$
denote the total thicknesses and permittivities of the regions made from
material $i=1,2$ \cite{1999poet.book.....B,1994ApOpt..33.7875B}.

Generally, the treatment of TM modes is known to be more complex than for TE
modes \cite{visser1997confinement}. Specifically, in contrast to the TE case
the derivation of one-dimensional Maxwell-type equations for the transverse
field components is not as straightforward as in Section \ref{sec:TE}. This
can for example be seen from Eq.\thinspace(\ref{eq:Maxw_TM1}), which only
assumes a Maxwell-type form in analogy to Eq. \thinspace(\ref{eq:MaxwTE1}) if
$\partial_{x}\underline{E}_{z}^{\omega}$ can be neglected. Similarly as in
Section \ref{sec:TE}, we assume that the fields are approximately constant
over the transverse cross section of the nanostructured region, and identify
in the following $\underline{E}_{x}^{\omega}$ and\ $\underline{H}_{y}^{\omega
}$ with the field strengths at the transverse position of the nanostructure.
With $\underline{\epsilon}_{\mathrm{r}}=n_{0}^{2}$, we obtain from
Eq.\thinspace(\ref{eq:TM3})%
\begin{equation}
\mathbf{\partial}_{z}\underline{E}_{x}^{\omega}=\mathrm{i}\omega\mu_{0}%
\frac{n_{\mathrm{eff}}^{2}\left(  \omega_{\mathrm{c}}\right)  }{n_{0}^{2}%
}\underline{H}_{y}^{\omega}, \label{eq:TM4}%
\end{equation}
where we have assumed that $\left|  \beta^{\prime}\right|  \ll\left|
\beta\right|  $ and approximated $\underline{\beta}^{2}\approx\beta^{2}$
$=n_{\mathrm{eff}}^{2}\omega^{2}/c^{2}$ as in Section \ref{sec:TE}, with the
effective waveguide refractive index $n_{\mathrm{eff}}$. Furthermore, we have
neglected the frequency dependence of $n_{\mathrm{eff}}$, evaluating it at the
center frequency $\omega=\omega_{\mathrm{c}}$ of the optical field, so as to
formally obtain a Maxwell-type equation with a frequency independent effective
relative permeability $\mu_{\mathrm{eff}}=n_{\mathrm{eff}}^{2}\left(
\omega_{\mathrm{c}}\right)  /n_{0}^{2}$. In order to complete our model,
Eq.\thinspace(\ref{eq:TM4}) must be complemented by a second Maxwell-type
equation with a form similar to Eq.\thinspace(\ref{eq:TE3}) in frequency
domain, i.e., Eq.\thinspace(\ref{eq:MaxwTE2}) in time domain. Importantly,
this equation has to include the losses and frequency dependence omitted in
Eq.\thinspace(\ref{eq:TM4}), so that the correct field propagation dynamics is
obtained. Specifically, from Eq.\thinspace(\ref{eq:EH}) we obtain the field
propagation equations in frequency domain $\mathbf{\partial}_{z}^{2}%
\underline{E}_{x}^{\omega}=-\underline{\beta}^{2}\underline{E}_{x}^{\omega}$,
$\mathbf{\partial}_{z}^{2}\underline{H}_{y}^{\omega}=-\underline{\beta}%
^{2}\underline{H}_{y}^{\omega}$. This requires that%
\begin{equation}
\partial_{z}\underline{H}_{y}^{\omega}=\mathrm{i}\frac{n_{0}^{2}}{\omega
\mu_{0}n_{\mathrm{eff}}^{2}\left(  \omega_{\mathrm{c}}\right)  }%
\underline{\beta}^{2}\underline{E}_{x}^{\omega}, \label{eq:TM5}%
\end{equation}
as can be verified by differentiating Eq.\thinspace(\ref{eq:TM4}) with respect
to $z$ and eliminating $\underline{H}_{y}^{\omega}$ with Eq.\thinspace
(\ref{eq:TM5}), or alternatively eliminating $\underline{E}_{x}^{\omega}$ in
an analogous way.

As in Section \ref{sec:TE}, the polarization due to the quantum systems is
again perturbatively included in terms of a change $\Delta\underline{\beta
}^{2}$ to $\underline{\beta}^{2}$. To this end, we re-derive Eq.\thinspace
(\ref{eq:TM}) from Maxwell's equations, Eq.\thinspace(\ref{eq:maxw}), but now
keep the polarization contribution, which yields on the left side of
Eq.\thinspace(\ref{eq:TM}) the perturbation term $\underline{\beta}%
\omega\underline{P}_{x}^{\omega}$. Here, $\underline{P}_{x}^{\omega}$ is the
$x$ component of $\mathbf{P}_{\mathrm{q}}$ in frequency domain, while a
possible additional $z$ component has been neglected. With Eq.\thinspace
(\ref{eq:TM1}) and $\underline{\beta}^{2}\approx\beta^{2}$ $=n_{\mathrm{eff}%
}^{2}\omega^{2}/c^{2}$, the perturbation term can then be written as
$\underline{\beta}\omega\underline{P}_{x}^{\omega}\approx\Delta L\underline
{H}_{y}^{\mathrm{t}}$ with
\begin{equation}
\Delta L\approx\frac{\omega^{2}}{c^{2}}\frac{n_{\mathrm{eff}}^{2}}{n_{0}^{2}%
}\frac{\underline{P}_{x}^{\omega}}{\epsilon_{0}\underline{E}_{x}^{\omega}},
\label{eq:DL}%
\end{equation}
and Eq.\thinspace(\ref{eq:TM}) becomes $\left(  \hat{L}+\Delta L\right)
\underline{H}_{y}^{\mathrm{t}}=\underline{\beta}^{2}\underline{H}%
_{y}^{\mathrm{t}}$ with $\hat{L}=\underline{\epsilon}_{\mathrm{r}}\partial
_{x}\underline{\epsilon}_{\mathrm{r}}^{-1}\partial_{x}+\omega^{2}%
c^{-2}\underline{\epsilon}_{\mathrm{r}}$. Similarly as in Section
\ref{sec:TE}, we assume that the device operates in a given transverse mode
with propagation constant $\underline{\beta}$, and use that first order
perturbation theory does not affect the corresponding eigenfunction
$\underline{H}_{y}^{\mathrm{t}}$. Since $\hat{L}$ is non-Hermitian, a
biorthogonal basis set must be used, and the change of $\underline{\beta}^{2}$
is given by $\Delta\underline{\beta}^{2}=\big\langle \tilde{\phi}\big|
\Delta L\big|  \phi\big\rangle /\big\langle \tilde{\phi}
\big|  \phi\big\rangle $ \cite{sternheim1972non}. Here $\phi=H_{y}%
^{\mathrm{t}}$, while $\tilde{\phi}$ denotes the corresponding eigenfunction
of the adjoint problem $\hat{L}^{\dagger}\tilde{\phi}\left(  x\right)
=\left(  \underline{\beta}^{2}\right)  ^{\ast}\tilde{\phi}\left(  x\right)  $,
with $\hat{L}^{\dagger}=\partial_{x}\left(  \underline{\epsilon}_{\mathrm{r}%
}^{-1}\right)  ^{\ast}\partial_{x}\left(  \underline{\epsilon}_{\mathrm{r}%
}^{\ast}\dots\right)  +\omega^{2}c^{-2}\underline{\epsilon}_{\mathrm{r}}%
^{\ast}$ and $\tilde{\phi}\left(  x\rightarrow\pm\infty\right)  \rightarrow0$.
As can be seen by inserting Eq.\thinspace(\ref{eq:TM1}) into Eq.\thinspace
(\ref{eq:TM}), $\tilde{\phi}$ simply corresponds to the conjugate complex
electric field distribution of the mode $\left(  \underline{E}_{x}%
^{\mathrm{t}}\right)  ^{\ast}$, and in analogy to Section \ref{sec:conf} we
then obtain the field confinement factor $\Gamma=\iint_{A_{\mathrm{q}}%
}\underline{E}_{x}^{\mathrm{t}}\underline{H}_{y}^{\mathrm{t}}\mathrm{d}%
x\mathrm{d}y/\iint_{-\infty}^{\infty}\underline{E}_{x}^{\mathrm{t}}%
\underline{H}_{y}^{\mathrm{t}}\mathrm{d}x\mathrm{d}y$. Generally, this
expression is complex, but for real $\underline{\epsilon}_{\mathrm{r}}$ it
coincides with the previous result Eq.\thinspace(\ref{eq:Gamma}). Thus, the
expression for the confinement factor given in Eq.\thinspace(\ref{eq:Gamma})
corresponds to the perturbative expression for both TE and TM\ modes in the
case of real $\underline{\epsilon}_{\mathrm{r}}$, and is commonly also used
for complex $\underline{\epsilon}_{\mathrm{r}}$ where it can be seen as a
real-valued approximation to the perturbative result. Similarly as in Section
\ref{sec:TE}, we insert $\underline{\beta}^{2}\approx\beta^{2}+\mathrm{i}%
\left|  \beta\right|  a+\Delta\underline{\beta}^{2}$ with $\beta\left(
\omega\right)  =\mathrm{sgn}\left(  \beta\right)  \omega n_{\mathrm{eff}%
}\left(  \omega\right)  /c$ into Eq.\thinspace(\ref{eq:TM5}), which yields%
\begin{align}
\partial_{z}\underline{H}_{y}^{\omega}=  &  \mathrm{i}\omega\frac{\epsilon_{0}%
n_{0}^{2}}{n_{\mathrm{eff}}^{2}\left(  \omega_{\mathrm{c}}\right)
}n_{\mathrm{eff}}^{2}\left(  \omega\right)  \underline{E}_{x}^{\omega}%
-\frac{\epsilon_{0}n_{0}^{2}}{n_{\mathrm{eff}}^{2}\left(  \omega_{\mathrm{c}%
}\right)  }cn_{\mathrm{eff}}\left(  \omega\right)  a\underline{E}_{x}^{\omega
} \nonumber\\
&  +\mathrm{i}\omega\Gamma\underline{P}_{x}^{\omega}. \label{eq:TM6}%
\end{align}
Here, we have neglected a possible frequency dependence of $n_{\mathrm{eff}}$
and $\Gamma$ in the last term. Obviously, our two derived Maxwell-type
equations, Eqs.\thinspace(\ref{eq:TM4}) and (\ref{eq:TM6}), have the same form
as Eqs.\thinspace(\ref{eq:TE2}) and (\ref{eq:TE5}) for TE\ modes, as can be
seen by substituting $\underline{H}_{x}^{\omega}\rightarrow-\underline{H}%
_{y}^{\omega}$, $\underline{E}_{y}^{\omega}\rightarrow\underline{E}%
_{x}^{\omega}$, $\epsilon_{0}\rightarrow\epsilon_{0}n_{0}^{2}/n_{\mathrm{eff}%
}^{2}\left(  \omega_{\mathrm{c}}\right)  $, $\mu_{0}\rightarrow\mu
_{0}n_{\mathrm{eff}}^{2}\left(  \omega_{\mathrm{c}}\right)  /n_{0}^{2}$. Thus,
the Maxwell-type equations in time domain can be obtained in the same way as
Eq.\thinspace(\ref{eq:MaxwTE}), yielding%
\begin{subequations}%
\label{eq:MaxwTM}%
\begin{align}
\partial_{z}E_{x}  &  =-\mu_{0}\frac{n_{\mathrm{eff}}^{2}\left(
\omega_{\mathrm{c}}\right)  }{n_{0}^{2}}\partial_{t}H_{y},\label{eq:MaxwTM1}\\
\partial_{z}H_{y}  &  =-\epsilon_{0}n_{0}^{2}\partial_{t}E_{x}-\sigma\left(
\mathrm{i}\partial_{t}\right)  E_{x}\label{eq:MaxwTM2}\\
&  -\Gamma n_{\mathrm{3D}}\sum_{i,j}d_{x,ji}\partial_{t}\rho_{ij}-\epsilon
_{0}\partial_{t}\left[  \Delta\epsilon_{\mathrm{eff}}\left(  \mathrm{i}%
\partial_{t}\right)  E_{x}\right]  ,\nonumber
\end{align}%
\end{subequations}%
where the generally frequency dependent conductivity
\begin{equation}
\sigma\left(  \omega\right)  =\epsilon_{0}n_{0}^{2}cn_{\mathrm{eff}}\left(
\omega\right)  a\left(  \omega\right)  /n_{\mathrm{eff}}^{2}\left(
\omega_{\mathrm{c}}\right)  \label{eq:sig2}%
\end{equation}
is\ often approximated by $\sigma=\sigma\left(  \omega_{\mathrm{c}}\right)  $.
Furthermore, in the last term describing chromatic waveguide dispersion, we
now have $\Delta\epsilon_{\mathrm{eff}}=n_{0}^{2}n_{\mathrm{eff}}^{2}\left(
\omega\right)  /n_{\mathrm{eff}}^{2}\left(  \omega_{\mathrm{c}}\right)
-n_{0}^{2}$. As discussed below Eq.\thinspace(\ref{eq:sig}), certain
conditions apply to $\Delta\epsilon_{\mathrm{eff}}$ and $\sigma$. In
particular, they must be even functions $f\left(  -\omega\right)  =f\left(
\omega\right)  $ to preserve the real-valued character of Eq.\thinspace
(\ref{eq:MaxwTM}). Identifying $E_{x}\left(  z,t\right)  $ and\ $H_{y}\left(
z,t\right)  $ with the field strengths at the transverse position of the
nanostructure, $E_{x}$ can directly be used in Eq.\thinspace(\ref{eq:MB}) or
(\ref{eq:MB2}) to evaluate $\partial_{t}\rho_{ij}\left(  z,t\right)  $. As
stated above, Eq.\thinspace(\ref{eq:MaxwTM}) has been constructed to assume
the form of one-dimensional Maxwell equations and to yield the correct
propagation behavior for $E_{x}$ and $H_{y}$. On the other hand, the relation
between $E_{x}$ and $H_{y}$, given by Eq.\thinspace(\ref{eq:TM1}) for
$\underline{P}_{x}^{\omega}=0$, is in Eq.\thinspace(\ref{eq:MaxwTM}) for the
general case of waveguide loss and dispersion only approximately fulfilled.

\subsection{\label{sec:SVAA2}Slowly Varying Amplitude Approximation}

As in Section \ref{sec:full}, we assume a waveguiding structure which is
invariant in propagation direction $z$, and now employ the slowly varying
amplitude approximation. The guided mode solutions at a given frequency
$\omega$ are characterized by the propagation constant and a $z$ independent
transverse field distribution $F\left(  x,y\right)  $. Here $F$ is induced by
the refractive index profile $\Delta_{n}\left(  x,y\right)  $, while the
polarization of the quantum systems and other nonlinear effects are assumed to
act as perturbations which do not significantly affect the transverse field
distribution \cite{agr01}.

We start from Eq.\thinspace(\ref{eq:waveinh}) and introduce the slowly varying
field envelopes by inserting Eq.\thinspace(\ref{eq:EP_SVAA}), where we however
replace $k_{\mathrm{c}}$\ by the propagation constant of the guided mode
$\beta_{0}=\beta\left(  \omega_{\mathrm{c}}\right)  $. In the following, it is
advantageous to switch to the spectral domain, where the slowly varying
envelopes depend on the frequency variable $\Delta_{\omega}=\omega
-\omega_{\mathrm{c}}$, corresponding to the frequency offset from
$\omega_{\mathrm{c}}$. Neglecting higher order derivatives of $t$ and $z$ in
the spirit of the SVAA and the paraxial approximation as described in Section
\ref{sec:SVAA}, and considering that time derivatives of the envelopes are
replaced by multiplications with $-\mathrm{i}\Delta_{\omega}$ in frequency
domain, we obtain%
\begin{equation}
-2\mathrm{i}\beta_{0}\mathbf{\partial}_{z}\underline{\mathbf{E}}+\beta_{0}%
^{2}\underline{\mathbf{E}}=\nabla_{\mathrm{T}}^{2}\underline{\mathbf{E}}%
+\frac{\omega^{2}}{c^{2}}\underline{n}^{2}\underline{\mathbf{E}}%
+\omega_{\mathrm{c}}^{2}\mu_{0}\underline{\mathbf{P}}. \label{eq:SVAA3}%
\end{equation}
For the term $\omega^{2}c^{-2}\underline{n}^{2}\underline{\mathbf{E}}$, we
have retained the full frequency dependence of the complex refractive index
$\underline{n}\left(  x,y,\omega\right)  $ to include chromatic dispersion, as
described further below. More specifically, $\underline{n}^{2}=n_{0}%
^{2}\left(  1-2\Delta_{n}\right)  +\mathrm{i}\sigma/\left(  \omega\epsilon
_{0}\right)  $ contains the refractive index profile via $\Delta_{n}\left(
x,y\right)  $ and losses via the conductivity $\sigma$, where $n_{0}$,
$\Delta_{n}$\ and $\sigma$ may be treated as frequency dependent. Assuming a
guided mode solution, we can use the separation ansatz
\begin{equation}
\underline{\mathbf{E}}\left(  \mathbf{x},t\right)  =\mathbf{e}\underline
{E}\left(  z,t\right)  F\left(  x,y\right)  , \label{eq:EF}%
\end{equation}
with the polarization direction of the electric field $\mathbf{e}$ and modal
distribution $F$. Inserting Eq.\thinspace(\ref{eq:EF}) into Eq.\thinspace
(\ref{eq:SVAA3}), multiplying by $\mathbf{e}$ and introducing the separation
constant $\underline{\beta}^{2}$, the right side of the resulting equation
becomes%
\begin{equation}
\left(  \underline{\beta}^{2}-\frac{\omega^{2}}{c^{2}}\underline{n}%
^{2}\right)  F=\nabla_{\mathrm{T}}^{2}F, \label{eq:F}%
\end{equation}
which does not depend on the propagation coordinate $z$. As in Section
\ref{sec:TE}, $\underline{\mathbf{P}}$ is subsequently included based on first
order perturbation theory \cite{agr01}. Equation (\ref{eq:F}), together with
the boundary condition that $F\rightarrow0$ for $x^{2}+y^{2}\rightarrow\infty
$, constitutes an eigenvalue equation for $F$ with complex eigenvalues
$\underline{\beta}^{2}$, featuring multiple eigensolutions which correspond to
the different transverse waveguide modes. In the following, we assume that the
device operates in a single transverse mode, possibly the fundamental mode. As
in Section \ref{sec:TE}, we split the complex propagation constant
$\underline{\beta}$ into a real and an imaginary part $\underline{\beta}%
=\beta+\mathrm{sgn}\left(  \beta\right)  \mathrm{i}a/2$ with power loss
coefficient $a$, and assume that $\left|  a\right|  \ll\left|  \beta\right|
$. Including $\underline{\mathbf{P}}$ in first order perturbation theory in
analogy to Section \ref{sec:TE} does not alter $F$, but yields a modified
propagation constant $\beta+\Delta\beta$ with \cite{agr01}%
\begin{equation}
\Delta\beta=\frac{\omega_{\mathrm{c}}^{2}}{2c^{2}\beta_{0}}\frac
{\iint_{-\infty}^{\infty}\Delta\epsilon_{\mathrm{r}}\left|  F\right|
^{2}\mathrm{d}x\mathrm{d}y}{\iint_{-\infty}^{\infty}\left|  F\right|
^{2}\mathrm{d}x\mathrm{d}y}\approx\frac{\omega_{\mathrm{c}}^{2}}{2c^{2}%
\beta_{0}}\Gamma\Delta\epsilon_{\mathrm{r}}, \label{eq:pert}%
\end{equation}
where $\Delta\epsilon_{\mathrm{r}}=\mathbf{e}\underline{\mathbf{P}}%
\mathbf{/}\left(  \epsilon_{0}\underline{E}\right)  $ and $\Gamma
=\iint_{A_{\mathrm{q}}}\left|  F\right|  ^{2}\mathrm{d}x\mathrm{d}%
y/\iint_{-\infty}^{\infty}\left|  F\right|  ^{2}\mathrm{d}x\mathrm{d}y$ in
agreement with Eq.\thinspace(\ref{eq:over}). Here we have evaluated
$\Delta\beta$ at the carrier frequency $\omega_{\mathrm{c}}$, in accordance
with the SVAA. For a realistic description of guided mode propagation, the
frequency dependence of $\beta$ itself should however be retained, giving rise
to chromatic dispersion \cite{agr01}. This effect is commonly described in
terms of a Taylor series, $\beta\left(  \Delta_{\omega}\right)  =\sum
_{n}\left(  \beta_{n}/n!\right)  \Delta_{\omega}^{n}$ with $\beta_{n}=\left[
\mathrm{d}_{\omega}^{n}\beta\right]  _{\omega=\omega_{\mathrm{c}}}$. While
frequency dependent waveguide loss can be included in a similar manner by an
$\omega$ dependent coefficient $a$, we ignore this effect since usually the
spectral gain or loss profile is dominated by the contribution of the quantum
systems, contained in $\underline{\mathbf{P}}$. Furthermore assuming that
$\left|  \Delta\beta\right|  \ll\left|  \beta\right|  $ and $\beta\approx
\beta_{0}$, we can approximate $\underline{\beta}^{2}-\beta_{0}^{2}%
\approx2\beta_{0}\left(  \beta-\beta_{0}\right)  +\mathrm{i}\left|  \beta
_{0}\right|  a+2\beta_{0}\Delta\beta$. With this result and Eq.\thinspace
(\ref{eq:pert}), the separation ansatz yields for the left-hand side of
Eq.\thinspace(\ref{eq:SVAA3}) in time domain%
\begin{align}
\frac{1}{v_{\mathrm{g}}}\partial_{t}\underline{E}+\mathbf{\partial}%
_{z}\underline{E}=  &  \mathrm{i}\sum_{n\geq2}\frac{\beta_{n}}{n!}\left(
\mathrm{i}\partial_{t}\right)  ^{n}\underline{E}-\mathrm{sgn}\left(  \beta
_{0}\right)  \frac{a}{2}\underline{E} \nonumber\\
&  +\mathrm{i}\frac{\omega_{\mathrm{c}}^{2}%
}{2\epsilon_{0}\beta_{0}c^{2}}\Gamma\mathbf{e}\underline{\mathbf{P}},
\label{eq:prop1D}%
\end{align}
where $v_{\mathrm{g}}=\beta_{1}^{-1}$ denotes the group velocity at
$\omega_{\mathrm{c}}$.

The guided field solution is characterized by a linearly polarized field
distribution, with the electric field pointing in direction $\mathbf{e}$ as
reflected by the ansatz for the electric field,\ Eq.\thinspace(\ref{eq:EF})
\cite{agr01,1991ONT}. The corresponding modes are transverse electromagnetic,
i.e., with transverse, perpendicular electric and magnetic fields. Notably,
this approach always yields two degenerate modes, orthogonally polarized in
transverse $x$ and $y$ directions. In reality, this applies for example to an
ideal, cylindrically symmetric single-mode fiber, while irregularities such as
random variations in the core shape already break the degeneracy. Within the
assumptions of weak waveguiding, the optical power is given by the
corresponding expression for the TE mode,\ Eq.\thinspace(\ref{eq:P2}).

Above approach is commonly used to model coherent propagation effects in
optical fibers like self-induced transparency, where the dopants, such as
erbium ions, take the role of the quantum systems modeled by the Bloch
equations, and the host material is for example glass
\cite{maimistov1983propagation,doktorov1983optical,nakazawa1991coexistence,guo2014breathers}%
. Here, in addition to the refractive index profile, fiber loss and chromatic
dispersion, also other effects related to the host material are commonly
considered. This in particular includes optical nonlinearity due to an
intensity dependent refractive index of the host material, which induces an
intensity dependent phase shift of the optical field and is thus referred to
as self-phase modulation. This effect can be included in Eq.\thinspace
(\ref{eq:SVAA3}) by substituting $\underline{n}^{2}$ with $\left(
\underline{n}+n_{2}\left|  \underline{\mathbf{E}}\right|  ^{2}\right)
^{2}\approx\underline{n}^{2}+2n_{0}n_{2}\left|  \underline{E}F\right|  ^{2}$
\cite{agr01}. Treating the nonlinear component as a perturbation, we can again
use Eq.\thinspace(\ref{eq:pert}) with $\Delta\epsilon_{\mathrm{r}}=2n_{0}%
n_{2}\left|  \underline{E}F\right|  ^{2}$ and include this effect in a similar
manner as discussed above. With Eqs.\thinspace(\ref{eq:prop1D}),
(\ref{eq:P_SVAA2}) and (\ref{eq:over}), we finally obtain the propagation
equation
\begin{align}
\frac{1}{v_{\mathrm{g}}}\partial_{t}\underline{E}+\mathbf{\partial}%
_{z}\underline{E}  &  =\mathrm{i}\sum_{n\geq2}\frac{\beta_{n}}{n!}\left(
\mathrm{i}\partial_{t}\right)  ^{n}\underline{E}-\mathrm{sgn}\left(  \beta
_{0}\right)  \frac{a}{2}\underline{E}\nonumber\\
&  +\mathrm{i}\gamma\left|  \underline{E}\right|  ^{2}\underline{E}%
+\mathrm{i}\frac{n_{\mathrm{3D}}\omega_{\mathrm{c}}^{2}}{\epsilon_{0}\beta
_{0}c^{2}}\Gamma\sum_{\omega_{ij}>0}d_{ji}\eta_{ij}, \label{eq:SVAA4}%
\end{align}
with the self-phase modulation coefficient%
\begin{equation}
\gamma=\frac{n_{0}n_{2}\omega_{\mathrm{c}}^{2}}{\beta_{0}c^{2}}\frac
{\iint_{-\infty}^{\infty}\left|  F\right|  ^{4}\mathrm{d}x\mathrm{d}y}%
{\iint_{-\infty}^{\infty}\left|  F\right|  ^{2}\mathrm{d}x\mathrm{d}y}.
\label{eq:SPM}%
\end{equation}
We note that in Eq.\thinspace(\ref{eq:SPM}), the nonlinearity is assumed to
extend over the whole fiber cross section since both the core and cladding
typically consist of the same host material. The MB equations are then
obtained by coupling Eq.\thinspace(\ref{eq:SVAA4}) to the Bloch equations in
RWA, Eq.\thinspace(\ref{eq:RWA}). Here it is practical to normalize $F$ in
Eq.\thinspace(\ref{eq:EF}) so that $F=1$ at the transverse position of the
dopants acting as quantum systems; then the field in Eq.\thinspace
(\ref{eq:RWA}) is directly given by $\underline{\mathbf{E}}\mathbf{=e}%
\underline{E}\left(  z,t\right)  $.

Typically, the MB equations are stepped in time to obtain the temporal
evolution of the optical field in a given geometry. For the case of
unidirectional propagation along a fiber where the input at $z=0$ is a given
time-limited optical waveform such as a pulse, it is more practical to
propagate the field in $z$ direction. It is then convenient to introduce the
retarded time variable $\tau=t-z/v_{\mathrm{g}}$, which is defined with
respect to a time frame which co-propagates with the waveform. Denoting the
position variable in the new coordinate system as $\zeta=z$, we then obtain
the partial derivatives $\partial_{z}=\partial_{\zeta}-\left(  1/v_{\mathrm{g}%
}\right)  \partial_{\tau}$, $\partial_{t}=\partial_{\tau}$. Thus,
Eq.\thinspace(\ref{eq:SVAA4}) becomes%
\begin{align}
\partial_{z}\underline{E}=  &  \mathrm{i}\sum_{n\geq2}\frac{\beta_{n}}{n!}\left(
\mathrm{i}\partial_{\tau}\right)  ^{n}\underline{E}-\mathrm{sgn}\left(
\beta_{0}\right)  \frac{a}{2}\underline{E}+\mathrm{i}\gamma\left|
\underline{E}\right|  ^{2}\underline{E} \nonumber\\
&  +\mathrm{i}\frac{n_{\mathrm{3D}}%
\omega_{\mathrm{c}}^{2}}{\epsilon_{0}\beta_{0}c^{2}}\Gamma\sum_{\omega_{ij}%
>0}d_{ji}\eta_{ij}, \label{eq:P_SVAA5}%
\end{align}
where we have resubstituted $\zeta$ with $z$. For very short pulses with
durations of only a few optical cycles, additional corrections might have to
be included on the right side of Eq.\thinspace(\ref{eq:P_SVAA5}). In
particular, this includes the self-steepening term $-\gamma\omega_{\mathrm{c}%
}^{-1}\partial_{\tau}\left(  \left|  \underline{E}\right|  ^{2}\underline
{E}\right)  $ \cite{demartini1967self,agr01} which is a higher order term
dropped in the SVAA, and the Raman-induced frequency shift term $-\mathrm{i}%
\gamma T_{\mathrm{R}}\underline{E}\partial_{\tau}\left(  \left|  \underline
{E}\right|  ^{2}\right)  $ with Raman-response time $T_{\mathrm{R}}$
\cite{mitschke1986discovery,gordon1986theory,agr01}. In Eqs.\thinspace
(\ref{eq:SVAA4}) and (\ref{eq:P_SVAA5}), we have assumed that $d_{ji}$ and
$\underline{E}$ are aligned in the same direction or, as is more realistic for
an optical fiber, $d_{ji}$ are effective dipole moments which average over the
different orientations of the dopant ions with respect to the field. Equation
(\ref{eq:P_SVAA5}) is solved together with the Bloch equations in RWA,
Eq.\thinspace(\ref{eq:RWA}), which are expressed in the retarded time frame
simply by substituting $t$ with $\tau$ in the density matrix elements and
derivative operators. The resulting equation system is sometimes also referred
to as Hirota--Maxwell--Bloch system \cite{nakkeeran1995solitons}.

\subsection{Fabry-P\'{e}rot Type Resonator}

For lasers, optical feedback has to be provided, which is in semiconductor
lasers typically achieved by using a Fabry-P\'{e}rot type waveguide resonator.
Here, the cleaved end facets provide natural reflection due to the refractive
index jump between the semiconductor material and air.

\subsubsection{\label{sec:bound}Boundary Conditions at the End Facets}

In the Fabry-P\'{e}rot type resonator, the ansatz for the optical field,
Eqs.\thinspace(\ref{eq:E_SVAA}), is extended to include a forward and a
backward propagating component, with amplitudes $\underline{E}_{+}$ and
$\underline{E}_{-}$, respectively. Furthermore assuming a guided mode solution
as in Eq.\thinspace(\ref{eq:EF}), we obtain
\begin{align}
E\left(  z,t\right)  =  &  \frac{1}{2}\big[  \underline{E}_{+}\left(  z,t\right)
\exp\left(  \mathrm{i}\beta_{0}z-\mathrm{i}\omega_{\mathrm{c}}t\right) \nonumber\\
&  +\underline{E}_{-}\left(  z,t\right)  \exp\left(  -\mathrm{i}\beta
_{0}z-\mathrm{i}\omega_{\mathrm{c}}t\right)  +c.c.\big]  , \label{eq:SHB1}%
\end{align}
where $\beta_{0}$ is the real part of the propagation constant at
$\omega=\omega_{\mathrm{c}}$. With the (generally complex) field reflection
coefficients $\underline{r}_{1}$ and $\underline{r}_{2}$\ of the facets,
assumed to be located at $z=0$ and $z=L$ where $L$ is the resonator length, we
obtain%
\begin{align}
\underline{E}_{+}\left(  z=0,t\right)   &  =\underline{r}_{1}\underline{E}%
_{-}\left(  z=0,t\right)  ,\nonumber\\
\underline{E}_{-}\left(  z=L,t\right)   &  =\underline{r}_{2}\underline{E}%
_{+}\left(  z=L,t\right)  , \label{eq:refl}%
\end{align}
where we have neglected a possible frequency dependence of $\underline
{r}_{1,2}$.

For the full Maxwell equations, a decomposition of the field into a forward
and a backward propagating component is not practical. Here, reflecting
boundary conditions can in principle be implemented by position dependent
parameters, e.g., by setting $n_{\mathrm{eff}}=1$, $\Delta\epsilon
_{\mathrm{eff}}=0$, $\sigma=0$ and $d_{y,ji}=0$ in Eq.\thinspace
(\ref{eq:MaxwTE2}) for $z<0$ and $z>L$ if we assume air outside of the
resonator region and neglect modal effects. In this context, care has to be
taken to suppress unwanted spurious reflections at the simulation domain
boundaries, which can be achieved by implementing absorbing boundary
conditions. However, this is not quite trivial, and various methods with
different degrees of complexity have been developed \cite{taflove2005}. A
simplified treatment, which works best for highly reflecting facets, is to use
perfectly reflecting boundary conditions by setting the transverse electric
field component at the facet positions to zero. The mirror loss, i.e., the
decay of the optical field in the cavity due to outcoupling through the
mirrors, can then be considered by a distributed power loss coefficient
$a_{\mathrm{m}}$, which is is obtained from $\left|  \underline{r}%
_{1}\underline{r}_{2}\right|  ^{2}=\exp\left(  -2a_{\mathrm{m}}L\right)  $ as%
\begin{equation}
a_{\mathrm{m}}=-\ln\left(  \left|  \underline{r}_{1}\underline{r}_{2}\right|
\right)  /L. \label{eq:am}%
\end{equation}
Using Eq.\thinspace(\ref{eq:sig}) or (\ref{eq:sig2}), $\sigma$ in
Eq.\thinspace(\ref{eq:MaxwTE2}) or (\ref{eq:MaxwTM2}) can then be determined
from the total power loss coefficient $a=a_{\mathrm{m}}+a_{\mathrm{w}}$, where
$a_{\mathrm{w}}=2\mathrm{sgn}\left(  \Re\left\{  \underline{\beta}\right\}
\right)  \Im\left\{  \underline{\beta}\right\}  $ denotes the waveguide loss.

\paragraph{Reflection Coefficient}

Using special reflective structures, such as reflection/antireflection
coatings or distributed Bragg reflectors, $\underline{r}_{1,2}$ can be
custom-tailored. In the following, we focus on the highly relevant case where
the bare end facets are used as reflective elements. For sufficiently large
transverse waveguide dimensions, Fresnel's formula for normal incidence can be
used to estimate the field reflection coefficient at the facet as%
\begin{equation}
\underline{r}=\frac{\underline{n}_{\mathrm{eff}}-1}{\underline{n}%
_{\mathrm{eff}}+1}=\frac{\underline{\beta}-k_{0}}{\underline{\beta}+k_{0}},
\label{eq:fresnel}%
\end{equation}
with $\underline{n}_{\mathrm{eff}}=\underline{\beta}/k_{0}$ and $k_{0}%
=\omega/c$. While Eq.\thinspace(\ref{eq:fresnel}) is usually valid for weak
waveguiding assumed in the derivation of Eqs.\thinspace(\ref{eq:waveinh}) and
(\ref{eq:SVAA4}), modal effects can result in increased reflection at the
facets \cite{2005JAP....97e3106K}. Various methods are available to compute
the reflectance $R=\left|  r\right|  ^{2}$ from the transverse mode profile
\cite{butler1974radiation,1972IJQE....8..470I,1993IPTL....5..148K,1971JAP....42.4466R}%
.

For TE polarization, it is practical to decompose the waveguide mode,
characterized by its complex propagation constant $\underline{\beta}$ and
magnetic field distribution $\underline{H}_{x}^{\mathrm{t}}$ which can be
computed from Eq.\thinspace(\ref{eq:Hx}), into plane waves, using the Fourier
transform
\begin{equation}
\Phi_{x}\left(  k_{x}\right)  =\int_{-\infty}^{\infty}\underline{H}%
_{x}^{\mathrm{t}}\left(  x\right)  \exp\left(  -\mathrm{i}k_{x}x\right)
\mathrm{d}x. \label{eq:Fourier}%
\end{equation}
Then, a generalized version of Eq.\thinspace(\ref{eq:fresnel}) for tilted
incidence is applied to each plane wave in order to calculate the reflection
coefficient \cite{butler1974radiation}. The reflectance $R$, i.e., the ratio
of the optical power reflected at the facet to the incident power, is obtained
by integrating over all components, yielding%
\begin{equation}
R=\frac{1}{2\pi}\left[  \int_{-\infty}^{\infty}\left|  \underline{H}%
_{x}^{\mathrm{t}}\right|  ^{2}\mathrm{d}x\right]  ^{-1}\int_{-\infty}^{\infty
}\left|  \frac{\underline{\beta}-\kappa}{\underline{\beta}+\kappa}\right|
^{2}\left|  \Phi_{x}\right|  ^{2}\mathrm{d}k_{x}, \label{eq:TER}%
\end{equation}
where $\kappa\left(  k_{x}\right)  =\sqrt{k_{0}^{2}-k_{x}^{2}}$ and the square
root is chosen so that $\Im\left\{  \kappa\right\}  >0$. From Eq.\thinspace
(\ref{eq:TE2}), we see that Eq.\thinspace(\ref{eq:TER}) can also be evaluated
by replacing $\underline{H}_{x}^{\mathrm{t}}$ with $\underline{E}%
_{y}^{\mathrm{t}}$ in Eqs.\thinspace(\ref{eq:TER}) and (\ref{eq:Fourier}).
Making the reasonable assumption that $\left|  \Re\left\{  \underline{\beta
}\right\}  \right|  \gg\left|  \Im\left\{  \underline{\beta}\right\}  \right|
$, we can approximately treat the reflection coefficient $r$ as real-valued.
Furthermore assuming that the share of reflected power going into other
waveguide modes is negligible \cite{butler1974radiation}, we obtain
$r=R^{1/2}$.

The case of TM polarization is somewhat more complex and can be treated based
on the boundary value method \cite{butler1974radiation}. Starting from the
magnetic field distribution $\underline{H}_{y}^{\mathrm{t}}$ given by
Eq.\thinspace(\ref{eq:TM}), we first evaluate the power transmittance $T$
through the facet \cite{butler1974radiation,jirauschek2014modeling},%
\begin{align}
T  &  =\frac{2}{\pi}\left|  \underline{\beta}\right|  ^{2}\left[
\int_{-\infty}^{\infty}\left|  \underline{\epsilon}_{\mathrm{r}}\right|
^{-2}\Re\left\{  \underline{\beta}^{\ast}\underline{\epsilon}_{\mathrm{r}%
}\right\}  \left|  \underline{H}_{y}^{\mathrm{t}}\right|  ^{2}\mathrm{d}%
x\right]  ^{-1}\nonumber\\
&  \times\int_{-k_{0}}^{k_{0}}\frac{\kappa\left|  \Phi_{y}\right|  ^{2}\left|
\Phi_{y}^{\prime}\right|  ^{2}}{\left|  \kappa\Phi_{y}+\underline{\beta}%
\Phi_{y}^{\prime}\right|  ^{2}}\mathrm{d}k_{x}. \label{eq:TMT}%
\end{align}
Here, $\Phi_{y}\left(  k_{x}\right)  $ denotes the Fourier transform,
Eq.\thinspace(\ref{eq:Fourier}), of $\underline{H}_{y}^{\mathrm{t}}\left(
x\right)  $, and $\Phi_{y}^{\prime}\left(  k_{x}\right)  $ is the Fourier
transform applied to the function $\underline{\epsilon}_{\mathrm{r}}%
^{-1}\left(  x\right)  \underline{H}_{y}^{\mathrm{t}}\left(  x\right)  $, with
the complex relative permittivity profile of the slab waveguide structure
$\underline{\epsilon}_{\mathrm{r}}\left(  x\right)  $. Again neglecting modal
effects and assuming a real $r$, the reflection coefficient is then given by
$r=\left(  1-T\right)  ^{1/2}$.

Since the field distributions and $\underline{\beta}$\ in Eqs.\thinspace
(\ref{eq:TER}) and\ (\ref{eq:TMT}) depend on $\omega$, this also applies to
the obtained reflection coefficients. Usually, this frequency dependence is
neglected in the formulation of the boundary conditions, and $r$ is taken at
the center frequency $\omega_{\mathrm{c}}$ of the optical field.

\subsubsection{\label{sec:shb}Spatial Hole Burning}

\begin{figure}[ptb]
\includegraphics{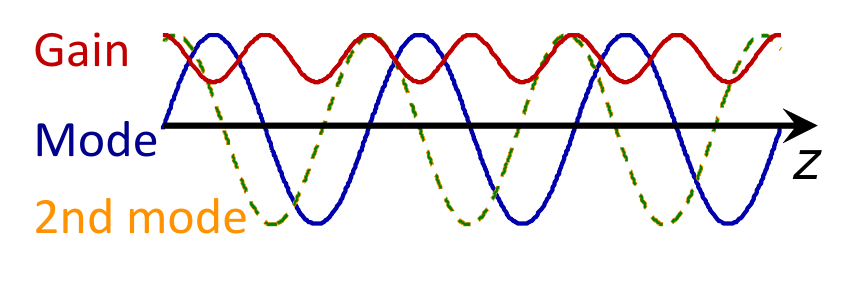}
\caption{Illustration of spatial hole burning in a Fabry-P\'{e}rot cavity.}%
\label{fig:shb}%
\end{figure}

In a Fabry-P\'{e}rot resonator, the reflection at the end facets gives rise to
counterpropagating waves, which produce a standing wave pattern with a
periodicity corresponding to the wavelength. At the field node positions,
there is no interaction of the optical field with the quantum systems. This
also implies that the population inversion and resulting optical gain, as
provided by the quantum systems in the active region of a semiconductor laser,
do not get saturated at those positions. Thus, other modes at slightly
different frequencies which have their maxima close to these unsaturated
regions can also start lasing. In Fig.\thinspace\ref{fig:shb}, this effect is
illustrated, which is referred to as (longitudinal) spatial hole burning
(SHB). The resulting multimode lasing can be desired or undesired, depending
on the envisaged application. For example, the broadening of the lasing
spectrum is beneficial in applications such as the generation of frequency
combs in QCLs, which are comb-like optical spectra used for precision
metrology and sensing \cite{tzenov2017analysis}. On the other hand, spatial
hole burning tends to introduce optical instabilities in form of irregular
variations in the mode amplitudes and phases
\cite{2007PhRvA..75c1802W,tzenov2017analysis}. In a similar way as just
discussed for the propagation direction, SHB can also occur along the
transverse directions, and has been shown to affect the spatiotemporal
dynamics especially in broad-area semiconductor lasers \cite{hess1996maxwell2}%

The inversion grating is smoothed out by carrier diffusion processes, and SHB
can even be neglected in a first approximation if diffusion is strong enough
\cite{hess1996maxwell}. Diffusion can be generically described by adding a
term $\left.  \partial_{t}\rho_{ii}\right|  _{\mathrm{diff}}=\mathbf{\nabla
}D_{i}\mathbf{\nabla}\rho_{ii}$ to the Bloch equations, Eq.\thinspace
(\ref{eq:MB}) or (\ref{eq:MB2}) \cite{torrey1956bloch,hess1996maxwell}. Here
$D_{i}\left(  \mathbf{x}\right)  $ is the diffusion coefficient associated
with level $i$. In the following, we focus on longitudinal SHB. Furthermore
assuming constant coefficients $D_{i}$, the diffusion term added to the Bloch
equations, Eq.\thinspace(\ref{eq:MB}) or (\ref{eq:MB2}), becomes%
\begin{equation}
\left[  \partial_{t}\rho_{ii}\right]  _{\mathrm{diff}}=D_{i}\mathbf{\partial
}_{z}^{2}\rho_{ii}, \label{eq:diff}%
\end{equation}
with zero-flux boundary conditions $\partial_{z}\rho_{ii}=0$ at the resonator
ends. In a two-level description of bulk semiconductors, the levels $i$
correspond to the conduction and valence bands, and the diffusion process is
largely mediated by carrier-phonon and carrier-carrier scattering between the
$\mathbf{k}$ states in the bands \cite{hess1996maxwell}. In quantum wells, the
levels correspond to the subbands formed by one-dimensional carrier
confinement, and the diffusion process is mediated by scattering between the
$\mathbf{k}$ states in the subbands \cite{ando1982electronic}. In
multi-quantum-dot structures, the SHB dynamics is often modeled by taking into
account the carrier diffusion in the wetting layer, as well as carrier capture
and escape processes to and from the quantum dots
\cite{asryan2000longitudinal,capua2013finite}. Since these processes
effectively reduce the diffusion length, SHB can have a strong effect,
similarly as in intersubband devices like QCLs which typically feature a very
fast gain recovery dynamics \cite{gordon2008multimode}. On the other hand, the
inversion grating is usually eliminated in interband bulk and quantum well
lasers due to effective diffusion \cite{capua2013finite}.

\paragraph{Slowly Varying Amplitude Approximation}

As for Eq.\thinspace(\ref{eq:rho_RWA}), we assume that all transitions between
pairs of states $i$ and $j$ with non-negligible coupling to the optical field
are in near-resonance, $\left|  \omega_{ij}\right|  \approx\omega_{\mathrm{c}%
}$. For the corresponding off-diagonal density matrix elements, we now make
the ansatz%
\begin{align}
\rho_{ij}  &  =\eta_{ij}^{+}\left(  z,t\right)  \exp\left[  \mathrm{sgn}%
\left(  \omega_{ij}\right)  \mathrm{i}\left(  \beta_{0}z-\omega_{\mathrm{c}%
}t\right)  \right] \nonumber\\
&  +\eta_{ij}^{-}\left(  z,t\right)  \exp\left[  \mathrm{sgn}\left(
\omega_{ij}\right)  \mathrm{i}\left(  -\beta_{0}z-\mathrm{\,}\omega
_{\mathrm{c}}t\right)  \right]  . \label{eq:SHB2}%
\end{align}
The periodicity of the inversion grating corresponds to that of the optical
intensity, i.e., half the wavelength. Thus, for the populations we make the
ansatz%
\begin{align}
\rho_{ii}=  &  \rho_{ii}^{0}\left(  z,t\right)  +\rho_{ii}^{+}\left(  z,t\right)
\exp\left(  2\mathrm{i}\beta_{0}z\right) \nonumber\\
&  +\rho_{ii}^{-}\left(  z,t\right)
\exp\left(  -2\mathrm{i}\beta_{0}z\right)  , \label{eq:SHB3}%
\end{align}
where\ $\rho_{ii}^{+}=\left(  \rho_{ii}^{-}\right)  ^{\ast}$ correspond to the
inversion grating's amplitudes. An analogous ansatz with $\rho_{ij}^{0}$ and
$\rho_{ij}^{\pm}$ is also chosen for off-diagonal density matrix elements
which are associated with two closely aligned levels and are thus not treated
in RWA, such as resonant tunneling transitions in QCLs \cite{tzenov2016time}.
Inserting Eqs.\thinspace(\ref{eq:SHB1}), (\ref{eq:SHB2}) and (\ref{eq:SHB3})
into Eq.\thinspace(\ref{eq:MB2}) with the diffusion term Eq.\thinspace
(\ref{eq:diff}) added to Eq.\thinspace(\ref{eq:MB2b}), we obtain in a similar
way as described in Section \ref{sec:RWA}%
\begin{align}
\partial_{t}\eta_{ij}^{\pm}  &  =\mathrm{i}\Delta_{ij}\eta_{ij}^{\pm}%
+\frac{\mathrm{i}}{2\hbar}\left(  \rho_{jj}^{0}-\rho_{ii}^{0}\right)
\mathbf{d}_{ij}\left\{
\begin{array}
[c]{c}%
\underline{\mathbf{E}}_{\pm}\\
\underline{\mathbf{E}}_{\pm}^{\ast}%
\end{array}
\right\} \nonumber\\
&\,\quad  +\frac{\mathrm{i}}{2\hbar}\mathbf{d}_{ij}\left\{
\begin{array}
[c]{c}%
\left(  \rho_{jj}^{\pm}-\rho_{ii}^{\pm}\right)  \underline{\mathbf{E}}_{\mp}\\
\left(  \rho_{jj}^{\mp}-\rho_{ii}^{\mp}\right)  \underline{\mathbf{E}}_{\mp
}^{\ast}%
\end{array}
\right\}  -\gamma_{ij}\eta_{ij}^{\pm},\omega_{ij}\left\{
\begin{array}
[c]{c}%
>0\\
<0
\end{array}
\right\}  ,\nonumber\\
\partial_{t}\rho_{ii}^{0}  &  =\frac{1}{\hbar}\sum_{\substack{j\\\omega
_{ij}>0}}\Im\left\{  \mathbf{d}_{ji}\left(  \eta_{ij}^{+}\underline
{\mathbf{E}}_{+}^{\ast}+\eta_{ij}^{-}\underline{\mathbf{E}}_{-}^{\ast}\right)
\right\} \nonumber\\
&\,\quad  +\frac{1}{\hbar}\sum_{\substack{j\\\omega_{ij}<0}}\Im\left\{
\mathbf{d}_{ji}\left(  \eta_{ij}^{+}\underline{\mathbf{E}}_{+}+\eta_{ij}%
^{-}\underline{\mathbf{E}}_{-}\right)  \right\} \nonumber\\
&\,\quad +\sum_{j\neq i}%
r_{j\rightarrow i}\rho_{jj}^{0}-r_{i}\rho_{ii}^{0},\nonumber\\
\partial_{t}\rho_{ii}^{+}  &  =\frac{\mathrm{i}}{2\hbar}\sum
_{\substack{j\\\omega_{ij}>0}}\left(  \mathbf{d}_{ij}\eta_{ji}^{-}%
\underline{E}_{+}-\mathbf{d}_{ji}\eta_{ij}^{+}\underline{E}_{-}^{\ast}\right)
\nonumber\\
&\,\quad  +\frac{\mathrm{i}}{2\hbar}\sum_{\substack{j\\\omega_{ij}%
<0}}\left(  \mathbf{d}_{ij}\eta_{ji}^{+}\underline{E}_{-}^{\ast}%
-\mathbf{d}_{ji}\eta_{ij}^{-}\underline{E}_{+}\right) \nonumber\\
&\,\quad +\sum_{j\neq
i}r_{j\rightarrow i}\rho_{jj}^{+}-r_{i}\rho_{ii}^{+}-4\beta_{0}^{2}D_{i}%
\rho_{ii}^{+}, \label{eq:SHB4}%
\end{align}
with $\Delta_{ij}=\mathrm{sgn}\left(  \omega_{ij}\right)  \left(
\omega_{\mathrm{c}}-\left|  \omega_{ij}\right|  \right)  $. Furthermore, we
have $\rho_{ii}^{-}=\left(  \rho_{ii}^{+}\right)  ^{\ast}$, $\eta_{ji}^{\pm
}=\left(  \eta_{ij}^{\pm}\right)  ^{\ast}$.

Assuming that the coefficients $\beta_{n}$ in Eq.\thinspace(\ref{eq:SVAA4})
refer to forward propagation, i.e., $\beta_{0}>0$ and $\beta_{1}%
=v_{\mathrm{g}}^{-1}>0$, the backward propagating field is described by
coefficients $-\beta_{n}$. Furthermore deriving the polarization term
analogously to Eq.\thinspace(\ref{eq:P_SVAA2}), we can summarize the equations
for the forward and backward propagating fields as
\begin{align}
\frac{1}{v_{\mathrm{g}}}\partial_{t}\underline{E}_{\pm}\pm\mathbf{\partial
}_{z}\underline{E}_{\pm}=  &  \mathrm{i}\sum_{n\geq2}\frac{\beta_{n}}{n!}\left(
\mathrm{i}\partial_{t}\right)  ^{n}\underline{E}_{\pm}-\frac{a}{2}\underline{E}_{\pm} \nonumber\\
&  +\mathrm{i}\frac{n_{\mathrm{3D}}\omega_{\mathrm{c}}^{2}%
}{\epsilon_{0}\beta_{0}c^{2}}\Gamma\sum_{\omega_{ij}>0}d_{ji}\eta_{ij}^{\pm},
\label{eq:SHB5}%
\end{align}
where we have neglected the self-phase modulation term. Equation
(\ref{eq:SHB5}) has to be complemented by the boundary conditions
Eq.\thinspace(\ref{eq:refl}), which together with Eq.\thinspace(\ref{eq:SHB4})
constitute the MB model in RWA and SVAA for a waveguide resonator.

This treatment of SHB can be extended by considering higher spatial
frequencies of the inversion grating in Eq.\thinspace(\ref{eq:SHB3})
\cite{gordon2008multimode}. Furthermore, it has been suggested to consider the
formation of a grating, and its relaxation due to diffusion, also for the
off-diagonal density matrix elements in Eq.\thinspace(\ref{eq:SHB2})
\cite{vukovic2016multimode}.

\section{\label{sec:Ana}Analytical Solutions}

\subsection{\label{sec:Ana_RWA}Rotating Wave Approximation}

\subsubsection{Monochromatic Excitation}

The Bloch equations in RWA, Eq.\thinspace(\ref{eq:RWA}), are in principle
analytically solvable for monochromatic excitation, corresponding to a
time-constant field envelope \cite{torrey1949transient,allen1987optical}.
This is usually achieved by using the Laplace transform, which takes a time
dependent function $f\left(  t\right)  $ to a function $\mathcal{L}\left\{
f\right\}  \left(  s\right)  $ of a complex frequency variable $s$. The main
advantage is that differentiation becomes a multiplication with $s$, i.e.,
$\mathcal{L}\left\{  \partial_{t}f\right\}  =s\mathcal{L}\left\{  f\right\}
-f\left(  t=0+\right)  $. Restricting ourselves to a two-level system with
initial conditions $w_{0}=w\left(  t=0\right)  $, $\eta_{21}^{0}=\eta
_{21}\left(  t=0\right)  $, and considering that the transform is linear and
$\mathcal{L}\left\{  1\right\}  =s^{-1}$, Eq.\thinspace(\ref{eq:RWA2l})
becomes in Laplace domain
\begin{subequations}%
\label{eq:lapl}
\begin{align}
\mathcal{L}\left\{  \eta_{21}\right\}   &  =s_{-}^{-1}\left(  -\mathrm{i}%
\mathcal{L}\left\{  w\right\}  \underline{\Omega}/2+\eta_{21}^{0}\right)
,\label{eq:lapl1}\\
\mathcal{L}\left\{  \eta_{12}\right\}   &  =s_{+}^{-1}\left[  \mathrm{i}%
\mathcal{L}\left\{  w\right\}  \underline{\Omega}^{\ast}/2+\left(  \eta
_{21}^{0}\right)  ^{\ast}\right]  ,\label{eq:lapl2}\\
\mathcal{L}\left\{  w\right\}   &  =\frac{2ss_{0}+s_{+}s_{-}\left(  \gamma
_{1}w_{\mathrm{eq}}+sw_{0}\right)  }{s\left[  ss_{+}s_{-}+\left(  s+\gamma
_{2}\right)  \left|  \underline{\Omega}\right|  ^{2}+\gamma_{1}s_{+}%
s_{-}\right]  }, \label{eq:lapl3}%
\end{align}%
\end{subequations}%
with $s_{\pm}=\left(  s+\gamma_{2}\pm\mathrm{i}\Delta\right)  $ and $s_{0}%
=\Im\left\{  \eta_{21}^{0}\underline{\Omega}^{\ast}\right\}  \left(
s+\gamma_{2}\right)  +\Delta\Re\left\{  \eta_{21}^{0}\underline{\Omega}^{\ast
}\right\}  $. Here, the off-diagonal matrix elements have already been
eliminated in Eq.\thinspace(\ref{eq:lapl3}) by inserting Eqs.\thinspace
(\ref{eq:lapl1}) and (\ref{eq:lapl2}). Back-transformation of Eq.\thinspace
(\ref{eq:lapl3}) is achieved by performing a partial fraction decomposition,
which results in a sum of simpler fractions with known inverse Laplace
transforms. This requires finding the poles of the rational function in
Eq.\thinspace(\ref{eq:lapl3}), given by $s=0$ and the roots of the cubic
function of $s$ in the square brackets of the denominator, for which closed
analytical expressions are readily available. The solution for $f\left(
t\right)  =w\left(  t\right)  $ and $f\left(  t\right)  =\eta_{21}\left(
t\right)  $ is of the form \cite{allen1987optical}
\[
f\left(  t\right)  =A+B\exp\left(  -at\right)  +\left[  C\cos\left(  \omega
t\right)  +D\sin\left(  \omega t\right)  \right]  \exp\left(  -bt\right)  ,
\]
where the decay constants $a$, $b$ and the oscillation frequency $\omega$ are
the same for $w$ and $\eta_{21}$, but the coefficients $A$, $B$, $C$ and $D$
are different and also depend on the initial conditions.

\paragraph{Rabi Oscillations}

\begin{figure}[ptb]
\includegraphics{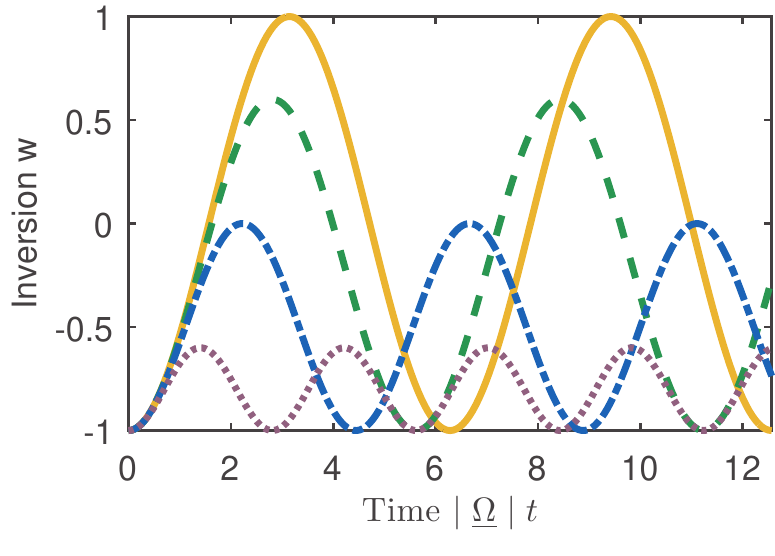}
\caption{Time dependent inversion for monochromatic excitation with a field of
amplitude $\underline{\Omega}$ and a detuning of $\Delta=0$ (solid line),
$\Delta=\left|  \underline{\Omega}\right|  /2$ (dashed line), $\Delta=\left|
\underline{\Omega}\right|  $ (dash-dotted line), and $\Delta=2\left|
\underline{\Omega}\right|  $ (dotted line).}%
\label{fig:Rabi}%
\end{figure}

In the following, we consider the case of dissipationless light-matter
interaction, i.e., $\gamma_{1}=\gamma_{2}=0$. Then, Eq.\thinspace
(\ref{eq:lapl3}) becomes%
\begin{align}
\mathcal{L}\left\{  w\right\}   &  =\frac{2s_{0}+\left(  s^{2}+\Delta
^{2}\right)  w_{0}}{s\left(  s+\mathrm{i}\Omega_{\mathrm{g}}\right)  \left(
s-\mathrm{i}\Omega_{\mathrm{g}}\right)  }\nonumber\\
&  =\frac{A}{s}+\frac{\left(  w_{0}-A\right)  s+2\Im\left\{  \eta_{21}%
^{0}\underline{\Omega}^{\ast}\right\}  }{s^{2}+\Omega_{\mathrm{g}}^{2}},
\label{eq:RabiL}%
\end{align}
where $\Omega_{\mathrm{g}}=\left(  \Delta^{2}+\left|  \underline{\Omega
}\right|  ^{2}\right)  ^{1/2}$ denotes the generalized Rabi frequency for
detuned excitation, and $A=\Omega_{\mathrm{g}}^{-2}\left(  2\Delta\Re\left\{
\eta_{21}^{0}\underline{\Omega}^{\ast}\right\}  +\Delta^{2}w_{0}\right)  $.
Inverse Laplace transformation of Eq.\thinspace(\ref{eq:RabiL}) yields%
\begin{equation}
w\left(  t\right)  =A+\left(  w_{0}-A\right)  \cos\left(  \Omega_{\mathrm{g}%
}t\right)  +2\Omega_{\mathrm{g}}^{-1}\Im\left\{  \eta_{21}^{0}\underline
{\Omega}^{\ast}\right\}  \sin\left(  \Omega_{\mathrm{g}}t\right)  .
\label{eq:Rabi}%
\end{equation}
As can be seen from Eq.\thinspace(\ref{eq:Rabi}), a monochromatic,
near-resonant light field interacting with an ideal, dissipationless two-level
system causes Rabi flopping, i.e., an oscillation of the population between
states $1$ and $2$ with frequency $\Omega_{\mathrm{g}}$, as predicted by I. I.
Rabi for the analogous case of a two-level system in a rotating magnetic field
\cite{rabi1937space}. In Fig.\thinspace\ref{fig:Rabi}, $w\left(  t\right)  $
is shown for the initial conditions $w_{0}=-1$, $\eta_{21}^{0}=0$ and
different detunings. As can be seen, complete population inversion with $w=1$
is achieved only for resonant excitation.

Due to the presence of dissipation, above presented analytical treatment of
Rabi flopping can rarely be directly used for the description of
optoelectronic device operation. However, under favorable conditions,
signatures of Rabi oscillations have been observed in nanostructured
optoelectronic systems and devices. These includes quantum well structures
\cite{cundiff1994rabi,schulzgen1999direct}, QCLs \cite{choi2010ultrafast},
single quantum dots \cite{stievater2001rabi,kamada2001exciton}, and nanowire
lasers \cite{mayer2017long} at cryogenic temperatures, as well as quantum dot
\cite{kolarczik2013quantum,karni2013rabi} and quantum dash \cite{capua2014}
amplifiers at room temperature. Besides the usually strong influence of
dissipation, effects beyond the two-level dynamics and inherent restrictions
of models based on macroscopic Maxwell-Bloch equations, the applicability of
Eq.\thinspace(\ref{eq:RabiL}) is also limited by the validity range of the
RWA. In particular, for very strong optical excitation where the Rabi
frequency approaches the optical resonance frequency, effects beyond the RWA
have been observed in bulk and nanostructured semiconductors
\cite{mucke2001signatures,muller2007resonance,xu2007coherent,wagner2010observation}%
.

\paragraph{Steady State Solution}

In the following, we consider the steady state behavior for a dissipative
two-level system under monochromatic optical excitation, i.e., for a field at
frequency $\omega_{\mathrm{c}}$ with constant amplitude. In the presence of
dissipation, the coherent transients associated with above discussed Rabi
oscillation, Eq.\thinspace(\ref{eq:Rabi}), decay, and the system approaches
the steady state for $t\rightarrow\infty$. The steady state solution can be
obtained by setting $\partial_{t}=0$ in Eq.\thinspace(\ref{eq:RWA2l})
\cite{Boyd}. Alternatively, we can apply the final value theorem to
Eq.\thinspace(\ref{eq:lapl}), stating that if $\lim_{t\rightarrow\infty
}f\left(  t\right)  $ exists, it is identical to $\lim_{s\rightarrow
0}s\mathcal{L}\left\{  f\right\}  $. Introducing the relaxation times
$T_{1,2}=\gamma_{1,2}^{-1}$, we then obtain%
\begin{subequations}%
\label{eq:stat}
\begin{align}
\eta_{21}  &  =\frac{1}{2}\frac{\underline{\Omega}T_{2}\left(  \Delta
T_{2}-\mathrm{i}\right)  w_{\mathrm{eq}}}{1+\Delta^{2}T_{2}^{2}+T_{1}%
T_{2}\left|  \underline{\Omega}\right|  ^{2}},\label{eq:stat1}\\
w  &  =\frac{\left(  1+\Delta^{2}T_{2}^{2}\right)  w_{\mathrm{eq}}}%
{1+\Delta^{2}T_{2}^{2}+T_{1}T_{2}\left|  \underline{\Omega}\right|  ^{2}}.
\label{eq:stat2}%
\end{align}%
\end{subequations}%

From Eq.\thinspace(\ref{eq:stat}), an expression for the relative permittivity
$\underline{\epsilon}_{\mathrm{r}}$ and susceptibility $\underline{\chi
}=\underline{\epsilon}_{\mathrm{r}}-1$ due to the quantum systems can be
derived. Setting the classical expression for the complex polarization
amplitude $\underline{\mathbf{P}}=\epsilon_{0}\underline{\chi}\underline
{\mathbf{E}}$ equal to Eq.\thinspace(\ref{eq:pol2rwa}), we obtain with
Eq.\thinspace(\ref{eq:stat1}), $\underline{\Omega}=\hbar^{-1}\mathbf{d}%
_{21}\underline{\mathbf{E}}$ and Eq.\thinspace(\ref{eq:I}) the frequency and
intensity dependent susceptibility \cite{Boyd,yar89}%
\begin{equation}
\underline{\chi}=\frac{n_{\mathrm{3D}}\left|  d_{21}\right|  ^{2}T_{2}%
}{\epsilon_{0}\hbar}\frac{\left(  \Delta T_{2}-\mathrm{i}\right)
w_{\mathrm{eq}}}{1+\Delta^{2}T_{2}^{2}+I/I_{\mathrm{s}}}, \label{eq:chi}%
\end{equation}
with the saturation intensity at zero detuning%
\begin{equation}
I_{\mathrm{s}}=\frac{\hbar^{2}\epsilon_{0}n_{\mathrm{eff}}c}{2T_{1}%
T_{2}\left|  d_{21}\right|  ^{2}}. \label{eq:Is}%
\end{equation}
For arbitrary detuning, the saturation intensity is then given by
$I_{\mathrm{s}}\left(  1+\Delta^{2}T_{2}^{2}\right)  $ \cite{Boyd}, i.e.,
non-resonant fields interact less strongly with the two-level system, and thus
saturation occurs for higher intensities. As in Section \ref{sec:SVAA2}, it is
here assumed that $d_{21}$ and $\underline{E}$ are aligned in the same
direction, or that $d_{21}$ is an effective value averaged over the different
orientations of, e.g., dopant ions in an optical fiber.

\begin{figure}[ptb]
\includegraphics{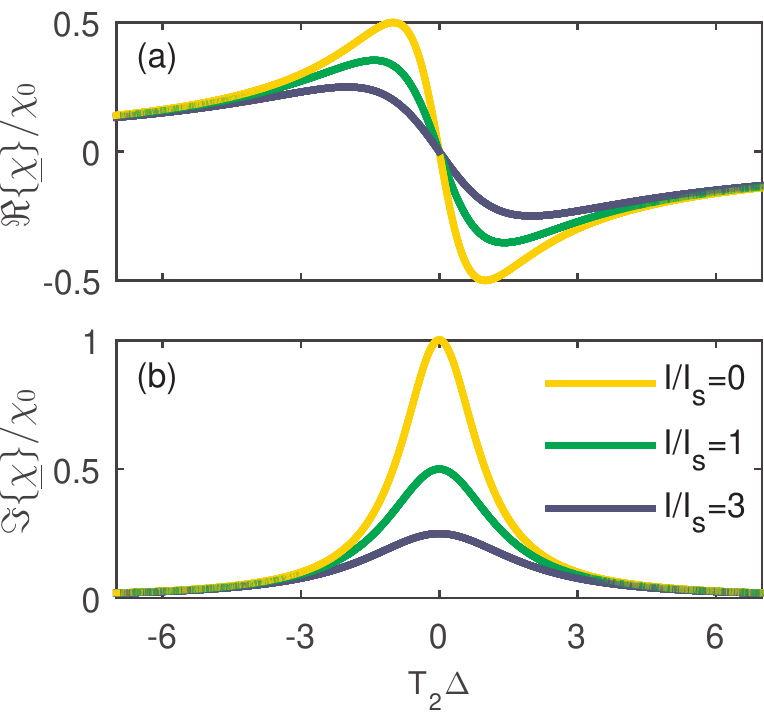}
\caption{Complex susceptibility of the two-level system, normalized to
$\chi_{0}=-n_{\mathrm{3D}}\left|  d_{21}\right|  ^{2}T_{2}\epsilon_{0}%
^{-1}\hbar^{-1}w_{\mathrm{eq}}$, as a function of the detuning frequency
$\Delta$. Shown is the (a) real and (b) imaginary part for various optical
intensities.}%
\label{fig:susc}%
\end{figure}

In Fig.\thinspace\ref{fig:susc}, the real and imaginary parts of
$\underline{\chi}$ are shown as a function of detuning from the optical
resonance frequency for various optical intensities. $\Re\left\{
\underline{\chi}\right\}  $, which contains chromatic dispersion, changes sign
at the resonance frequency. $\Im\left\{  \underline{\chi}\right\}  $ describes
gain for $w_{\mathrm{eq}}>0$, i.e., positive population inversion, and loss
for $w_{\mathrm{eq}}<0$, where the frequency dependence is given by a
Lorentzian profile with the full width at half-maximum (FWHM) bandwidth
$2\gamma_{2}\left(  1+I/I_{\mathrm{s}}\right)  ^{1/2}$. For increased
intensities, the profile thus gets broadened which is known as power
broadening, and also the peak value at resonance frequency is reduced by a
factor of $1+I/I_{\mathrm{s}}$, which corresponds to gain saturation for
$w_{\mathrm{eq}}>0$ and saturable absorption for\ $w_{\mathrm{eq}}<0$.

In the following, we investigate optical power amplification or absorption by
two-level systems. For TE modes or generally in the limit of weak waveguiding,
the power is given by Eq.\thinspace(\ref{eq:P2}). Multiplying Eq.\thinspace
(\ref{eq:SVAA4}) from left with $\underline{E}^{\ast}$ and adding the complex
conjugate, we obtain with Eqs.\thinspace(\ref{eq:stat1}) and (\ref{eq:P2}) and
$\beta_{0}=\omega_{\mathrm{c}}n_{\mathrm{eff}}/c$
\begin{equation}
\mathbf{\partial}_{z}P=-aP+gP, \label{eq:Pz}%
\end{equation}
where $a$ is the waveguide loss coefficient, and the two-level power gain
coefficient is given by%
\begin{equation}
g=\alpha\frac{w_{\mathrm{eq}}}{1+\Delta^{2}T_{2}^{2}+P/P_{\mathrm{s}}},
\label{eq:gP}%
\end{equation}
with%
\begin{equation}
\alpha=\Gamma\frac{\omega_{\mathrm{c}}n_{\mathrm{3D}}\left|  d_{21}\right|
^{2}T_{2}}{\epsilon_{0}c\hbar n_{\mathrm{eff}}} \label{eq:alpha}%
\end{equation}
and the saturation power at zero detuning $P_{\mathrm{s}}=A_{\mathrm{eff}%
}I_{\mathrm{s}}$. With the help of the Lambert $\mathrm{W}$ function, defined
by $x=\mathrm{W}\left(  x\right)  \exp\left[  \mathrm{W}\left(  x\right)
\right]  $, we can write the solution of Eqs.\thinspace(\ref{eq:Pz}) and
(\ref{eq:gP}) for zero waveguide loss, $a=0$, as $P\left(  z\right)
=P_{0}G\left(  z\right)  $ with the power gain factor%
\begin{equation}
G\left(  z\right)  =\frac{P_{\mathrm{s}}^{\prime}}{P_{0}}\mathrm{W}\left[
\frac{P_{0}}{P_{\mathrm{s}}^{\prime}}\exp\left(  \frac{P_{0}}{P_{\mathrm{s}%
}^{\prime}}\right)  \exp\left(  \frac{\alpha w_{\mathrm{eq}}}{1+\Delta
^{2}T_{2}^{2}}z\right)  \right]  , \label{eq:Pz2}%
\end{equation}
where $P_{0}=P\left(  z=0\right)  $ and $P_{\mathrm{s}}^{\prime}=\left(
1+\Delta^{2}T_{2}^{2}\right)  P_{\mathrm{s}}$. With Eq.\thinspace
(\ref{eq:Pz2}), the steady state field solution of Eq.\thinspace
(\ref{eq:SVAA4}) can then for $a=\gamma=0$ be written as%
\begin{equation}
\underline{E}\left(  z\right)  =\underline{E}\left(  z=0\right)  \left[
G\left(  z\right)  \right]  ^{\left(  1+\mathrm{i}\Delta T_{2}\right)  /2},
\label{eq:stead}%
\end{equation}
and the density matrix elements are with $\underline{\Omega}=\hbar^{-1}%
d_{21}\underline{E}$ given by Eq.\thinspace(\ref{eq:stat}). In the exponent of
Eq.\thinspace(\ref{eq:stead}), $\alpha_{\mathrm{H}}=-\Delta T_{2}$ corresponds
to the Henry or linewidth enhancement factor \cite{1982IJQE...18..259H}, which
relates phase changes to changes in the optical gain.

\begin{figure}[ptb]
\includegraphics{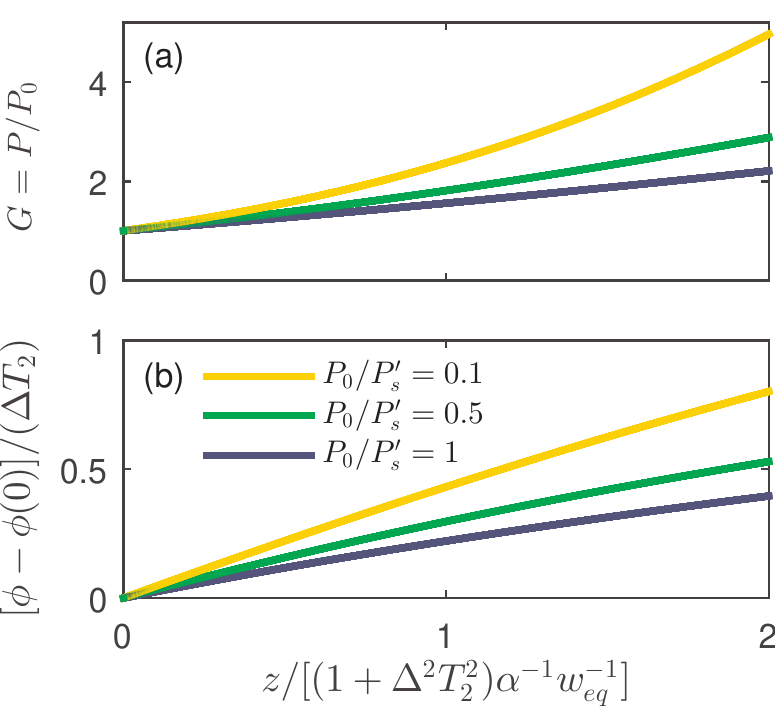}
\caption{(a) Optical power gain and (b) phase shift of the electric field as a
function of propagation distance for different values of initial power.}%
\label{fig:amp}%
\end{figure}

In Fig.\thinspace\ref{fig:amp}, the optical power gain and phase shift of the
electric field are shown as a function of propagation distance for different
values of initial power $P_{0}$ for $w_{\mathrm{eq}}>0$, i.e., amplification.
In the small signal limit, $P\ll P_{\mathrm{s}}^{\prime}$, the typical
exponential increase in power is observed. This can also be seen from
Eq.\thinspace(\ref{eq:Pz2}), which yields with $\mathrm{W}\left(  x\right)
\approx x$ for $x\ll1$ the usual exponential amplification (for $G>1$) or loss
(for $G<1$) characteristics. In the saturation regime, the power increases
only linearly. Physically, this is a consequence of the fact that the growth
in optical power is ultimately limited by the supplied pump power.

The expressions Eqs.\thinspace(\ref{eq:chi}) and (\ref{eq:gP}) for the
susceptibility $\underline{\chi}$ and optical gain $g$ are widely used to
model the optical properties of homogeneously broadened atomic
\cite{Boyd,yar89} and nanostructured
\cite{bimberg1997ingaas,jirauschek2010monte} optical media. In particular for
interband transitions in bulk semiconductor and quantum well media, the
electron wavevector must be explicitly considered, along with additional
corrections due to Coulomb interactions
\cite{chow2012semiconductor,ning1997effective}. Above derivation of an
expression for the susceptibility from the Bloch equations can be extended to
more than two levels, which is for example relevant for the investigation of
slow light propagation. This is usually achieved based on electromagnetically
induced transparency (EIT), where a control laser beam induces a narrow
transparency window with an extremely low group velocity in the absorption
spectrum of a suitable medium
\cite{boller1991observation,fleischhauer2005electromagnetically,kasapi1995}.
EIT requires a three-level configuration, and expressions for the
susceptibility have been derived in a similar way as above
\cite{khurgin2005optical,kasapi1995}. A reduction of the group velocity to
subsonic speeds, as well as complete halting of light, has been demonstrated
in an ultra-cold atomic
vapor~\cite{hau1999light,liu2001observation,phillips2001storage}. Possible
applications include optical buffers~\cite{khurgin2005optical},
imaging~\cite{camacho2007,firstenberg2009elimination}, and quantum
memory~\cite{lukin2000entanglement}. In view of a future commercialization of
these technologies, a compact solid-state based implementation is desirable,
and slow light propagation as well as light trapping has meanwhile been
demonstrated in doped crystals
\cite{turukhin2001observation,bigelow2003observation}. The realization of slow
light in suitably engineered semiconductor structures is especially
attractive. Here, the exploitation of tunneling induced transparency is highly
promising, which differs from EIT in that it does not require an optical
control field, but utilizes strong tunneling coupling between a pair of
states. For this case, the susceptibility has been analytically derived for
quantum dot and intersubband quantum well systems
\cite{ginzburg2006slow,borges2012tunneling,tzenov2017slow}. Furthermore,
nonlinear optical mixing effects which involve optical field contributions at
two or more frequencies, and often rely on more than two energy levels, are
exploited in many semiconductor-based applications, requiring a description by
higher order susceptibilities
\cite{agrawal1988population,hu1990theory,sugawara2004theory,khurgin1989second,belkin2007terahertz,burnett2016origins,jirauschek2015monte,2011IJQE...47..691V}%
. The corresponding expressions can for example be obtained from the Bloch
equations by employing time dependent perturbation theory
\cite{butcher1991elements,shen1984principles}.

\subsubsection{\label{sec:SIT}Self-Induced Transparency}

In addition to above presented steady state solution to the Maxwell-Bloch
equations, also dynamic solutions are available for some special cases. An
important example is self-induced transparency (SIT), where a special optical
pulse solution exists which can propagate through the two-level medium without
being attenuated or disturbed. This effect was theoretically predicted, and
first experimentally demonstrated in ruby, by McCall and Hahn
\cite{mccall1967self,mccall1969self}. SIT is based on coherent interaction
with the medium, which requires that the pulse duration must be much shorter
than the relaxation processes described by $\gamma_{1}$ and $\gamma_{2}$, and
thus we can set $\gamma_{1}=\gamma_{2}=0$. Furthermore we assume that the
field envelope $\underline{\Omega}=\hbar^{-1}d_{21}\underline{E}$ is
real-valued, and initially restrict ourselves to resonant excitation, i.e.,
$\Delta=0$ . Then, the solution of Eq.\thinspace(\ref{eq:RWAu}) for the
initial condition $s_{1}\left(  z,-\infty\right)  =s_{2}\left(  z,-\infty
\right)  =0$, $w\left(  z,-\infty\right)  =-1$ can be written as $s_{1}=0$,
$s_{2}=-\sin\vartheta$ and $w=-\cos\vartheta$ with $\vartheta\left(
z,t\right)  =\int_{-\infty}^{t}\underline{\Omega}\left(  z,t^{\prime}\right)
\mathrm{d}t^{\prime}$, as can easily be verified by re-insertion of the
solution into Eq.\thinspace(\ref{eq:RWAu}). This analysis can be extended to
incorporate inhomogeneous broadening in media consisting of quantum systems
with slightly different resonance frequencies
\cite{allen1987optical,meystre2013elements}. Assuming that non-resonant
systems with $\Delta\neq0$ essentially respond in the same way to
$\underline{\Omega}$ as the resonant ones, apart from a change in amplitude,
we can make the factorization ansatz $s_{2,\Delta}\left(  z,t\right)
=F\left(  \Delta\right)  s_{2}\left(  z,t\right)  $, which again yields closed
analytical solutions to Eq.\thinspace(\ref{eq:RWAu}),
\begin{align}
s_{2,\Delta}\left(  z,t\right)   &  =-F\left(  \Delta\right)  \sin
\vartheta\left(  z,t\right)  ,\label{eq:Vw1}\\
w_{\Delta}\left(  z,t\right)   &  =-F\left(  \Delta\right)  \left[
\cos\vartheta\left(  z,t\right)  -1\right]  -1. \label{eq:Vw2}%
\end{align}
Taking the second derivative of Eq.\thinspace(\ref{eq:RWAu2}) for $\Delta
\neq0$ and inserting Eqs.\thinspace(\ref{eq:RWAu1}), (\ref{eq:Vw1}) and
(\ref{eq:Vw2}) yields%
\begin{equation}
\partial_{t}^{2}\vartheta=\frac{\Delta^{2}F\left(  \Delta\right)  }{1-F\left(
\Delta\right)  }\sin\vartheta:=T^{-2}\sin\vartheta. \label{eq:pend}%
\end{equation}
Since the electric field envelope, and hence also $\vartheta$, is $\Delta$
independent, this must also apply to $\Delta^{2}F\left(  \Delta\right)
/\left[  1-F\left(  \Delta\right)  \right]  $ which we have thus set equal to
a constant $T^{-2}$ in Eq.\thinspace(\ref{eq:pend}). This yields a Lorentzian
dependence $F=1/\left(  1+T^{2}\Delta^{2}\right)  $. Equation (\ref{eq:pend})
corresponds to the pendulum problem, where the solutions are given by elliptic
functions. Here we require $\underline{\Omega}=\partial_{t}\underline{\Omega
}=0$\ and thus $\partial_{t}\vartheta=\partial_{t}^{2}\vartheta=0$ at
$t=\pm\infty$, yielding the unique solution $\vartheta=4\arctan\left\{
\exp\left[  \left(  t-t_{0}\right)  /T\right]  \right\}  $. Introducing
$t_{0}=z/v$ with the pulse propagation velocity $v$, we thus obtain%
\begin{equation}
\underline{\Omega}\left(  z,t\right)  =2T^{-1}\mathrm{sech}\left(
\tau/T\right)  \label{eq:SIT}%
\end{equation}
with the retarded time variable $\tau=t-z/v$.

In the optical propagation equation, inhomogeneous broadening can
approximately be included by substituting $s_{1}-\mathrm{i}s_{2}$ with
$\int_{-\infty}^{\infty}g\left(  \Delta\right)  \left(  s_{1,\Delta
}-\mathrm{i}s_{2,\Delta}\right)  \mathrm{d}\Delta$ in the polarization,
Eq.\thinspace(\ref{eq:pol2rwa}), where $g\left(  \Delta\right)  $ with
$\int_{-\infty}^{\infty}g\left(  \Delta\right)  \mathrm{d}\Delta=1$ gives the
distribution of quantum systems as a function of the detuning $\Delta$ from
$\omega_{\mathrm{c}}$. Here, a possible dependence of the dipole matrix
element on $\Delta$ has been neglected. Using $s_{2,\Delta}=F\left(
\Delta\right)  s_{2}$ and above result for $F\left(  \Delta\right)  $, we
obtain $s_{2,-\Delta}=s_{2,\Delta}$, and with Eq.\thinspace(\ref{eq:RWAu1}) we
see that then $s_{1,-\Delta}=-s_{1,\Delta}$ for $\underline{\Omega
}_{\mathrm{i}}=\gamma_{2}=0$. Often $g\left(  \Delta\right)  $ is an even
function as further discussed in Section \ref{sec:inh}, and then the
contribution of $s_{1,\Delta}$ cancels out. Under this assumption,
Eq.\thinspace(\ref{eq:SVAA4}) becomes without dispersion\ ($\beta_{n}=0$ for
$n\geq2$), loss ($a=0$) and self-phase modulation ($\gamma=0$)%
\begin{equation}
\partial_{t}\underline{\Omega}+v_{\mathrm{g}}\mathbf{\partial}_{z}%
\underline{\Omega}=2^{-1}\alpha\gamma_{2}v_{\mathrm{g}}s_{2}\int_{-\infty
}^{\infty}g\left(  \Delta\right)  F\left(  \Delta\right)  \mathrm{d}\Delta,
\label{eq:sitprop}%
\end{equation}
where $\alpha$ is given by Eq.\thinspace(\ref{eq:alpha}). Inserting
Eqs.\thinspace(\ref{eq:SIT}) and (\ref{eq:Vw1}) into Eq.\thinspace
(\ref{eq:sitprop}) yields the pulse propagation velocity%
\begin{equation}
v=v_{\mathrm{g}}\frac{2}{2+\alpha\gamma_{2}v_{\mathrm{g}}T^{2}\int_{-\infty
}^{\infty}g\left(  \Delta\right)  \left(  1+T^{2}\Delta^{2}\right)
^{-1}\mathrm{d}\Delta}. \label{eq:v}%
\end{equation}

\begin{figure}[ptb]
\includegraphics{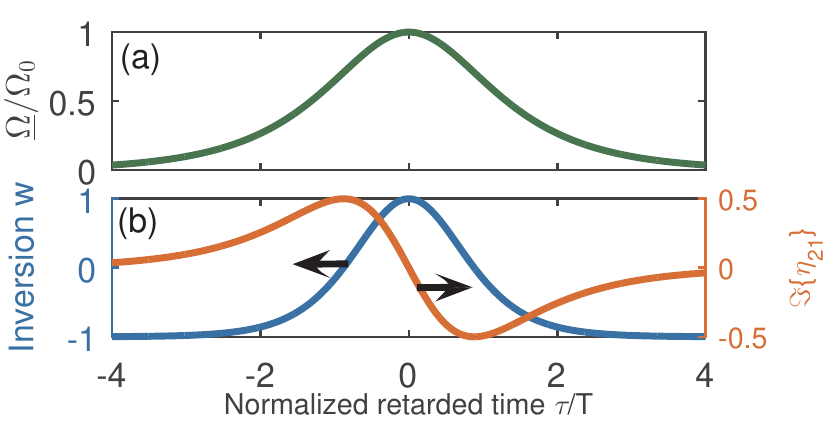}
\caption{Self-induced transparency soliton. Shown are (a) the normalized
electromagnetic field and (b) inversion $w$ and imaginary part of the
off-diagonal matrix element $\eta_{21}$, where $\Re\left\{  \eta_{21}\right\}
=0$.}%
\label{fig:sit}%
\end{figure}

In Fig.\thinspace\ref{fig:sit}, the pulse shape, inversion $w$ and the
imaginary part of the off-diagonal matrix element $\Im\left\{  \eta
_{21}\right\}  =-s_{2}/2$ is shown. Notably, for $w_{\mathrm{eq}}<0$ when the
two-level medium normally absorbs light [see Eq.\thinspace(\ref{eq:gP})], the
optical energy absorbed during the first half of the SIT pulse and stored in
the inversion, is re-emitted during the second half, which delays the pulse so
that $v\ $is smaller than the group velocity $v_{\mathrm{g}}$ without the
coherent interaction as can be seen from Eq.\thinspace(\ref{eq:v}), but does
not change its shape or amplitude. This is accompanied by a Rabi flop of the
population inversion from $w=-1$ to $w=1$ and back again.

Generally, based on the area theorem it was found that for coherent
propagation, the pulse area $\Theta=\int\underline{\Omega}\mathrm{d}t$ evolves
towards the closest even multiple of $\pi$ ($\Theta=0,2\pi,4\pi,\dots$) for
absorbing media ($w_{\mathrm{eq}}<0$), and to the closest odd multiple of
$\pi$ for gain media ($w_{\mathrm{eq}}>0$) \cite{mccall1967self}. Importantly,
the area theorem only makes a statement about $\Theta$, but does not indicate
if the pulse envelope changes. As discussed above, the SIT pulse
Eq.\thinspace(\ref{eq:SIT}), which has a pulse area $\Theta=2\pi$, is the only
finite energy solution of Eq.\thinspace(\ref{eq:pend}) where the pulse
envelope is preserved. However, analytical solutions of the MB equations with
changing pulse shapes can also be obtained for other cases of coherent
propagation \cite{lamb1971analytical,maimistov1990present}.

With regards to novel practical applications, SIT is for example a highly
interesting candidate for the generation of ultrashort optical pulses in
various types of lasers with sufficiently long coherence times, such as
quantum dot and quantum cascade lasers
\cite{kozlov1997self,kalosha1999theory,kozlov2011obtaining,2009PhRvL.102b3903M,2009ApPhL..95g1109T,2010OExpr..18.5639T,arkhipov2016self,arkhipov2016self2}%
. This SIT (or coherent) mode-locking approach requires a laser design with
one or multiple gain and absorber regions, where an SIT soliton with a pulse
area of $d_{21}\int\underline{E}\mathrm{d}t/\hbar=2\pi$ is approximately
realized in the absorber sections. In order to obtain a stable pulse area of
$\pi$ in the gain regions, they are engineered to have half the dipole moment
$d_{21}$ of the absorber sections. Instead of sequential gain and absorber
regions, another option is to stack the gain and loss regions in transverse
direction, i.e., perpendicular to the propagation axis. This approach is for
example compatible with the manufacturing process of QCLs, and an analytical
solution has been derived for the steady state pulse solution
\cite{2009PhRvL.102b3903M}. Despite its great promise, SIT mode-locking has
not been experimentally demonstrated to date.

\subsection{Full-Wave Bloch Equations}

Without employing the RWA, the Bloch equations (\ref{eq:MB2}) are solvable
only for very special conditions. In particular, for $\left|  \Delta M\right|
=1$ transitions in hydrogen-like atoms where the dipole matrix element is
given by $\mathbf{d}=2^{-1/2}\left|  \mathbf{d}\right|  \left(  \mathbf{e}%
_{x}-\mathrm{i}\mathbf{e}_{y}\right)  $, excitation with circularly polarized
light where $E_{x}=E\left(  t\right)  \cos\left(  \omega t\right)  $ and
$E_{y}=E\left(  t\right)  \sin\left(  \omega t\right)  $ leads to
$\mathbf{dE}\left(  t\right)  =2^{-1/2}\left|  \mathbf{d}\right|  E\left(
t\right)  \exp\left(  -\mathrm{i}\omega t\right)  $
\cite{allen1987optical,rabi1937space}. Furthermore using Eq.\thinspace
(\ref{eq:rho_RWA}) to substitute the off-diagonal density matrix elements in
Eq.\thinspace(\ref{eq:MB2}), the resulting equation formally corresponds to
the RWA Bloch equation, with the analytical solutions discussed in Section
\ref{sec:Ana_RWA}. Closed analytical solutions are not available for the
relatively simple, but very important case of monochromatic excitation with a
linearly polarized wave. Some approximate corrections have been derived, such
as the Bloch-Siegert shift describing the change in the system's resonance
frequency for strong driving \cite{bloch1940magnetic}, and the Mollow triplet
which refers to the emergence of satellite peaks in the spectrum of resonantly
excited systems \cite{mollow1969power}. Interestingly, the full Bloch
equations can be solved analytically if the linearly polarized electric field
has the form of an $N$-soliton. This is also true for the so-called reduced MB
equations, which combine the full-wave Bloch equations with a first-order
unidirectional optical propagation equation
\cite{bullough1979solitons,kujawski1986coherent}.

\section{\label{sec:Num}Numerical Schemes}

As discussed in Section {\ref{sec:Ana}}, the full-wave MB equations have known
analytical solutions only for very special cases, and also in the RWA/SVAA
approximation, no general analytical solution exists. Therefore computer
simulations are in general necessary. From a practical point of view, the
numerical scheme should be stable, accurate, and efficient, and a naive
discretization will often fail. The goal of this section is to introduce
well-established approaches which are straightforward to implement, and give a
critical discussion of their properties. Furthermore, an overview of recent
developments in the field will be given. Since the RWA/SVAA problem and the
full MB equations are not of the same mathematical form, their numerical
implementation has to be treated separately.

Several software projects have been published that are able to solve the
Maxwell-Bloch equations. For example, the Freetwm tool~\cite{freetwm} is an
open-source MATLAB code that simulates the dynamics of semiconductor lasers
using the 1D MB equations in rotating wave approximation. The Electromagnetic
Template Library (EMTL) is a free C++ library with Message Passing Interface
(MPI) support~\cite{emtl}, which has for example been used to model quantum
emitters with the full-wave MB equations in two
dimensions~\cite{deinega2014self}. Another solver library for the full-wave MB
equations is the open-source MEEP project~\cite{oskooi2010}, using a similar
representation of the Bloch equations as given in Eq.\thinspace(\ref{eq:Pcomp}%
). The mbsolve project~\cite{mbsolve-github} solves the full-wave MB equations
using different parallel acceleration techniques and features an open-source
codebase. Finally, a commercial MB solver has been announced
\cite{quantillion}.

\subsection{Rotating Wave/Slowly Varying Amplitude Approximation}

\subsubsection{\label{sec:RNFD}Finite Difference Discretization of the
One-Dimensional Propagation Equation}

In the following, the numerical solution of the one-dimensional optical
propagation equation in the SVAA is discussed. Neglecting chromatic
dispersion, i.e., setting $\beta_{n}=0$ for $n\geq2$, we write Eq.\thinspace
({\ref{eq:SHB5})} in the form
\begin{equation}
\partial_{t}E^{\pm}=\mp v_{\mathrm{g}}\partial_{z}E^{\pm}+f^{\pm}(z,t)-\ell
E^{\pm}. \label{chapter:numerics:eq:genericwave}%
\end{equation}
An obvious choice is to use a finite difference discretization approach where
a full spatiotemporal discretization of $E^{\pm}$ onto an equidistant grid
with $z_{m}=m\Delta_{z}$, $t_{n}=n\Delta_{t}$ is imposed. In the following,
$E_{m,n}^{\pm}$ and $f_{m,n}^{\pm}$ denote the numerical solution of $E^{\pm}$
and $f^{\pm}$ on the grid. The starting point is a Taylor series expansion of
$E^{\pm}(z_{m},t_{n+1})$ around the point $(z_{m},t_{n})$, yielding up to
second order
\begin{equation}
E_{m,n+1}^{\pm}=E_{m,n}^{\pm}+\Delta_{t}\partial_{t}E_{m,n}^{\pm}+\frac
{\Delta_{t}^{2}}{2}\partial_{t}^{2}E_{m,n}^{\pm}.
\label{chapter:numerics:eq:laxtaylor}%
\end{equation}
Then $\partial_{t}E_{m,n}^{\pm}$ and $\partial_{t}^{2}E_{m,n}^{\pm}$ are
replaced by space derivatives: {Differentiating Eq.\thinspace
(\ref{chapter:numerics:eq:genericwave}) with respect to }$z$, multiplying the
result {by $\mp v_{\mathrm{g}}$ and adding it to the time derivative of
Eq.\thinspace(\ref{chapter:numerics:eq:genericwave})} yields $\big(
\partial_{t}^{2}-v_{\mathrm{g}}^{2}\partial_{z}^{2}\big)  E^{\pm}=\big(
\partial_{t}\mp v_{\mathrm{g}}\partial_{z}\big)  (f^{\pm}-\ell E^{\pm})$.
Plugging the result into Eq.\thinspace({\ref{chapter:numerics:eq:laxtaylor}})
and furthermore using Eq.\thinspace(\ref{chapter:numerics:eq:genericwave}){,
we obtain}
\begin{align}
E_{m,n+1}^{\pm}=  &  E_{m,n}^{\pm}+\Delta_{t}\left\{  \mp v_{\mathrm{g}%
}\left[  \partial_{z}E^{\pm}\right]  _{m,n}+f_{m,n}^{\pm}-\ell E_{m,n}^{\pm
}\right\} \nonumber\\
&  +\frac{\Delta_{t}^{2}}{2}\Big\{  v_{\mathrm{g}}^{2}\left[  \partial
_{z}^{2}E^{\pm}\right]  _{m,n}+\left[  \partial_{t}f^{\pm}\right]  _{m,n} \nonumber\\
&  \mp
v_{\mathrm{g}}\left[  \partial_{z}f^{\pm}\right]  _{m,n}\Big\} \nonumber\\
&  +\frac{\ell\Delta_{t}^{2}}{2}\left\{  \pm2v_{\mathrm{g}}\left[
\partial_{z}E^{\pm}\right]  _{m,n}-f_{m,n}^{\pm}+\ell E_{m,n}^{\pm}\right\}  .
\label{chapter:numerics:eq:fulllaxwendroff}%
\end{align}
\ \ \

For finite difference discretization, there are different possibilities such
as the well known and widely used $2^{nd}$ order Lax-Wendroff method
\cite{laxwendroff60}, or the Risken-Nummedal finite differences (RNFD) scheme
which was specifically developed in the context of MB simulations
\cite{risken1968self}. In both cases, $\left[  \partial_{z}^{2}E^{\pm}\right]
_{m,n}$ is approximated by the standard finite difference approximation
$\big(  E_{m+1,n}^{\pm}-2E_{m,n}^{\pm}+E_{m-1,n}^{\pm}\big)  /\Delta
_{z}^{2}$. Here we will treat in detail the RNFD scheme, since it has some
advantageous properties as discussed further below. The main difference as
compared to the Lax-Wendroff method is that rather than employing centered
differences, depending on the propagation direction backward/forward finite
differences are used, with $\left[  \partial_{z}E^{\pm}\right]  _{m,n}%
\approx\pm\big(  E_{m,n}^{\pm}-E_{m\mp1,n}^{\pm}\big)  /\Delta_{z}$,
$\left[  \partial_{z}f^{\pm}\right]  _{m,n}\approx\pm\big(  f_{m,n}^{\pm
}-f_{m\mp1,n}^{\pm}\big)  /\Delta_{z}$. Furthermore, a time step of
$\Delta_{t}=\Delta_{z}/v_{\mathrm{g}}$ is imposed. From Eq.\thinspace
({\ref{chapter:numerics:eq:fulllaxwendroff}}), we then obtain the RNFD scheme
\begin{align}
E_{m,n+1}^{\pm}  &  =E_{m\mp1,n}^{\pm}+\Delta_{t}\left(  \frac{1}{2}%
f_{m,n}^{\pm}+\frac{1}{2}f_{m\mp1,n}^{\pm}-\ell E_{m\mp1,n}^{\pm}\right)
\nonumber\\
&  +\frac{\Delta_{t}^{2}}{2}\left\{  \left[  \partial_{t}f^{\pm}\right]
_{m,n}-\ell f_{m,n}^{\pm}+\ell^{2}E_{m,n}^{\pm}\right\}  . \label{eq:RNFD}%
\end{align}
The term $\left[  \partial_{t}f^{\pm}\right]  _{m,n}$ is not substituted with
a corresponding finite difference approximation since $\partial_{t}\eta_{ij}${
}can directly be obtained from the Bloch equations, Eq.\thinspace
({\ref{eq:RWA})}. In a Fabry-P\'{e}rot type resonator, Eq.\thinspace
({\ref{eq:RNFD}) is complemented by the boundary conditions }Eq.\thinspace
({\ref{eq:refl})}.

\paragraph{Numerical Properties of the RNFD Scheme}

For a numerical scheme to be useful, an important requirement is that
round-off and truncation errors do not get amplified during the computation,
since this will eventually lead to numerical instability.~The stability of
finite difference discretization schemes can be investigated based on a von
Neumann stability analysis \cite{issacson1994}. It turns out that the RNFD
scheme is stable for $\ell\geq0$, which is also true for the Lax-Wendroff
method for a sufficiently small Courant number $v_{\mathrm{g}}\Delta
_{t}/\Delta_{z}$. On the other hand, for positive linear gain, i.e., $\ell<0$,
we obtain unconditionally unstable behavior for both schemes. Furthermore,
like the Lax-Wendroff method, the RNFD scheme is second order accurate in
space and time \cite{risken1968self}. This guarantees that the numerical
scheme converges to the original partial differential equation as the grid
spacing approaches zero, with a convergence order of two. However, this does
not yet guarantee that the numerical solution for finite grid spacing has a
physically meaningful behavior, e.g., satisfies certain physical conservation
laws. Thus, additional conditions might be desirable for a finite difference
discretization of Eq.\thinspace({\ref{chapter:numerics:eq:genericwave}}),
which has the form of an inhomogeneous scalar convection equation and thus
allows us to draw from related work \cite{harten1983high}. Specifically, it
has been established that second and higher order linear finite difference
schemes tend to introduce artificial numerical dispersion, yielding phase
errors and numerical oscillations near extrema or discontinuities of the
solution \cite{godunov1959}. The numerical solution is less prone to phase
errors for monotonicity preserving schemes, which guarantee that for every
non-decreasing (non-increasing) initial condition $E_{m,n=0}^{\pm}$, the
numerical solution at all later instants $n>0$ is also non-decreasing
(non-increasing). A sufficient condition for the RNFD scheme to be
monotonicity preserving for the homogeneous propagation equation, i.e.,
Eq.\thinspace(\ref{chapter:numerics:eq:genericwave}) with a vanishing source
term $f^{\pm}(z,t)\equiv0$, can be easily derived: Formulating Eq.\thinspace
(\ref{eq:RNFD}) for $E_{m+1,n+1}^{\pm}$ and subtracting the resulting
expression from Eq.\thinspace(\ref{eq:RNFD}), we arrive at
\begin{align*}
E_{m+1,n+1}^{\pm}-E_{m,n+1}^{\pm}  &  =\left(  1-\ell\Delta_{t}\right)
\left(  E_{m+1\mp1,n}^{\pm}-E_{m\mp1,n}^{\pm}\right) \\
&  +\frac{\Delta_{t}^{2}}{2}\ell^{2}\left(  E_{m+1,n}^{\pm}-E_{m,n}^{\pm
}\right)  ,
\end{align*}
which yields $\Delta_{t}\leq1/\ell$ as sufficient condition for monotonicity
preservation in the stability regime $\ell\geq0$. This is a unique feature for
a second order finite difference propagation scheme which is directly related
to the choice of time step $\Delta_{t}=\Delta_{z}/v_{\mathrm{g}}$. Also, this
constitutes an important advantage of the RNFD scheme over the Lax-Wendroff
method, which does not have this property in numerically stable regions, as
can be shown in a similar way as above or directly from Godunov's order
barrier theorem~\cite{godunov1959}.

\begin{figure}[ptb]
\includegraphics{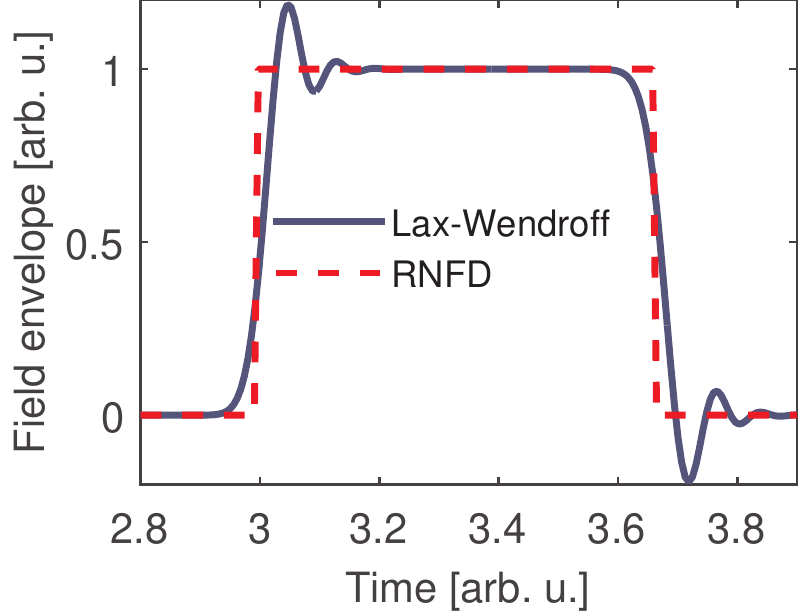}
\caption{Comparison
between Lax-Wendroff and RNFD schemes for the forward propagation of a
rectangular pulse with $f^{+}=\ell=0$.}%
\label{fig:comp_LW_RNFD}%
\end{figure}

In Fig.\thinspace\ref{fig:comp_LW_RNFD}, the Lax-Wendroff and the RNFD scheme
are compared for lossless propagation of an initially rectangular pulse
without interaction with a quantum system. For the Lax-Wendroff scheme,
spurious oscillatory features arise in the vicinity of the field
discontinuities, which are absent in the RNFD scheme due to its monotonicity
preserving nature.

\subsubsection{Density Matrix Equations}

\label{chapter:numerics:subsec:density-matrix-equations} The numerical scheme
for the optical propagation equation has to be coupled to a time-propagation
scheme for the Bloch equations, Eq.\thinspace(\ref{eq:RWA}). These constitute
an ordinary differential equation (ODE) system describing the temporal
evolution of the density matrix, which has to be solved for each spatial grid
point. In principle, most standard methods should do the job although they
will differ in numerical stability, accuracy and efficiency, and
well-established schemes such as Runge-Kutta
\cite{nielsen2007numerical,xiong2008numerical,demeter2013solving} and
Adams-Bashforth \cite{tzenov2016time} have successfully been used. Most
research on the suitability of different numerical schemes in literature has
focused on the full-wave Bloch equations, as detailed in Section
\ref{sec:fullwave}. Since they are identical in structure to the RWA Bloch
equations [compare for example Eqs.\thinspace(\ref{eq:bloch2l1}) and
(\ref{eq:bloch2l3}) to Eq.\thinspace(\ref{eq:RWA2l})], the obtained insights
should in principle also be valid for the RWA Bloch equations. The Runge-Kutta
method is further described in Section \ref{sec:RK} in the context of
full-wave Bloch equations. Here we exemplarily discuss the explicit
Adams-Bashforth scheme as an especially straightforward to implement and
numerically highly efficient method. The RNFD scheme for the MB equations,
Eq.\thinspace(\ref{eq:RNFD}), is strongly coupled, i.e., requires an
evaluation of the density matrix and the electric field at the same time
value. The $k-$step Adams-Bashforth method for the solution of an ODE system
is given by \cite{butcher2003,arieh1996,gear1971}
\begin{equation}
\hat{\rho}^{n+1}=\hat{\rho}^{n}+\Delta_{t}\sum\limits_{m=0}^{k-1}%
c_{m}\mathcal{F}^{n-m}\left(  \hat{\rho}^{n-m}\right)  . \label{eq:AB}%
\end{equation}
Here, $n$ corresponds to the time $t_{n}=n\Delta_{t}$, $\Delta_{t}$ is the
time step size, and $\mathcal{F}(\hat{\rho})=\mathcal{L}(\hat{\rho
})+\mathcal{D}(\hat{\rho})$ represents the right hand side of the Lindblad
equation~(\ref{eq:lindbl}), and specifically in our case of the RWA Bloch
equations, Eq.\thinspace(\ref{eq:RWA}). Furthermore, the $c_{m}$ are suitably
chosen coefficients~\cite{butcher2003} so that maximal accuracy is reached in
the approximation. A $k-$step Adams-Bashforth method has a global numerical
error on the order of $O(\Delta_{t}^{k})$~\cite{arieh1996}. In this context,
it must be considered that the overall numerical accuracy cannot be
arbitrarily improved by choosing a high value of $k$, since it is also limited
by the numerical discretization of the optical propagation equation, e.g.,
based on the RNFD method. As discussed in Section \ref{sec:bocos}, the Bloch
equations are initialized by the starting values of the density matrix
elements at a given time, while the Adams-Bashforth method would require $k$
initial values as can be seen from Eq.\thinspace(\ref{eq:AB}). This problem
can for example be solved by doing the first $k-1$ time steps with a different
numerical scheme such as the Runge-Kutta method, or by initializing the
simulations with two-step Adams-Bashforth on a finer grid. In simulations of
laser operation which are typically started from noise
\cite{slavcheva2004fdtd}, the exact choice of initial conditions is not
critical and thus the initialization steps required by Adams-Bashforth do not
pose a problem. The main advantage is the reduced numerical load as compared
to the Runge-Kutta method (see Section \ref{sec:RK}), which however requires
initialization only at a single time point.

\subsubsection{Generalizations and Alternative Methods}

In Section \ref{sec:RNFD}, one-dimensional propagation has been assumed,
neglecting the transverse coordinates in the SVAA propagation equation,
Eq.\thinspace(\ref{eq:SVAA}). In reality, the field dependence, and thus also
the temporal evolution of the quantum systems, is varying along the $x$ and
$y$ coordinates, which must be explicitly considered for an inclusion of
diffraction and other effects
\cite{xiong2008numerical,arve2004propagation,siddons2014light}. As long as no
transverse boundary conditions or material dependencies have to be considered,
i.e., $\Delta_{n}$ and $\sigma$ in Eq.\thinspace(\ref{eq:SVAA}) are constant,
the most straightforward approach is to Fourier-transform Eq.\thinspace
(\ref{eq:SVAA}) with respect to $x$ and $y$ before the time propagation step
is carried out \cite{xiong2008numerical}. The resulting equation then depends
on $z$, $t$ and the spatial Fourier frequencies $k_{x}$ and $k_{y}$,
converting the derivative operator $\nabla_{\mathrm{T}}^{2}$ into a
multiplication with $-\left(  k_{x}^{2}+k_{y}^{2}\right)  $. Thus a
one-dimensional propagation method can be used, such as the one discussed in
Section \ref{sec:RNFD}. Since this procedure requires a Fourier transform
before and an inverse transform after each propagation step, the numerically
efficient fast Fourier transform method is usually employed.

As discussed in the context of Eq.\thinspace(\ref{eq:P_SVAA5}), for the
modeling of unidirectional fiber or beam propagation often the initial field
at $z=0$ is given, and the solution at a certain distance $z=L$ is required.
Then it is more practical to propagate the field in $z$ direction rather than
in time, and to introduce the retarded time variable $\tau=t-z/v_{\mathrm{g}}$
which simplifies the propagation operator $\left(  v_{\mathrm{g}}^{-1}%
\partial_{t}\underline{E}+\mathbf{\partial}_{z}\underline{E}\right)  $ to
$\partial_{z}\underline{E}$. In the absence of other time derivatives, e.g.,
due to chromatic dispersion, this effectively reduces the propagation equation
to an ODE. The solution is then marched in $z$ direction in dependence of
$\tau$ (and $k_{x,y}$ if applicable), and the density matrix is updated after
every propagation step \cite{xiong2008numerical,demeter2013solving}. The
propagation along $z$ can be performed with a conventional ODE scheme where
for example the Adams-Moulton method (with the trapezoidal rule as a widely
used special case) or Adams-Bashforth method, Eq.\thinspace(\ref{eq:AB}), have
been employed, in both cases combined with fourth-order Runge-Kutta for the
Bloch equations \cite{xiong2008numerical,demeter2013solving}. In the more
general case where time derivatives have to be considered in Eq.\thinspace
(\ref{eq:P_SVAA5}), for example to incorporate chromatic dispersion, these can
be handled in Fourier domain, similarly as for the $x$ and $y$ derivatives
discussed in the previous paragraph. One option is to process all terms in
Fourier domain \cite{gross1992numerical}, which however complicates the
treatment of expressions which are nonlinear in the field, such as the
self-phase modulation term in Eq.\thinspace(\ref{eq:P_SVAA5}). Another
strategy might be to couple the Bloch equations to the split-step Fourier
method, which treats only the terms containing time derivatives in Fourier
domain, and the others in time domain \cite{agr01}.

\subsection{\label{sec:fullwave}Full-Wave Simulation}

While the RWA significantly reduces the computational workload, care must be
taken in cases where its basic assumptions are not fulfilled. For example, the
RWA assumes that the electric field intensity is small and the field spectrum
narrow. However, in a scenario where ultrashort pulse generation is simulated
(e.g., mode-locking operation in quantum cascade lasers), the electric field
features high peak intensity and a broad spectrum. In such cases, the full
electromagnetic wave might have to be considered in the simulation, and a
suitable numerical scheme has to be used. In the following, we describe the
methods for the Maxwell and full-wave Bloch equations, Eqs.\thinspace
(\ref{eq:maxw}) and (\ref{eq:MB2}), which are most widely used in related
literature, and address the coupling between the updates of the electric field
and the density matrix. Finally, we assess the advantages and drawbacks of the
different methods.

\subsubsection{Numerical Schemes for Maxwell's Equations}

Out of the many numerical methods that solve Maxwell's Equations, mainly two
-- namely the finite-difference time-domain (FDTD) and the pseudo-spectral
time-domain (PSTD) method -- are used in the context of Maxwell-Bloch equations.

\begin{figure}[ptb]
\includegraphics{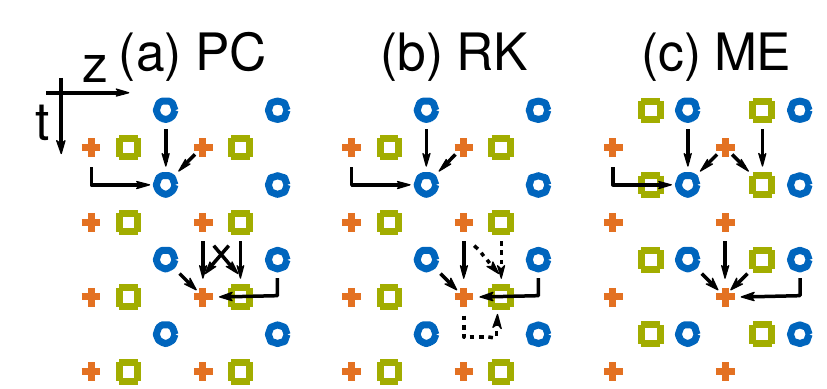}
\caption{Discretization
and data dependencies of the finite-difference time-domain (FDTD) method
combined with the (a) predictor-corrector (PC), (b) Runge-Kutta (RK), and
(c)\ matrix exponential (ME) method for the density matrix updates. The
crosses and circles denote the electric and magnetic fields, respectively,
while the squares represent the density matrix. (a)\textbf{ }The PC approach
updates the electric field and the density matrix in parallel. (b)\textbf{
}The RK method first updates the electric field (solid line) and then the
density matrix (dashed line), where in the latter both the old and the new
electric field values are used. (c) For the ME scheme,\textbf{ }the updates of
density matrix and the magnetic field are performed in parallel. The electric
field discretization is shifted by a half time step.}%
\label{fig:methods}%
\end{figure}

The FDTD method is one of the standard methods for Maxwell's
equations~\cite{taflove2005}, and is widely used in combination with the
optical Bloch equations~\cite{ziolkowski1995ultrafast,hughes1998breakdown,
bidegaray2001, bidegaray2003, slavcheva2002coupled,
slavcheva2005dynamical,sukharev2011,
cartar2017,klaedtke2006ultrafast,pusch2012coherent,lopata2009nonlinear,takeda2011self,dridi2013model}%
. Here, the derivatives with respect to time and space are approximated using
central differences. Hence, the method has second order accuracy. In order to
facilitate the calculation of the central differences, the Yee grid is used
where the discretization points are staggered by half of the respective step
size~\cite{yee1966}. Figure~\ref{fig:methods} depicts an example of a Yee grid
in one spatial dimension. The main advantage of the FDTD scheme is its
simplicity. The implementation of the method as well as boundary conditions or
sources is straightforward~\cite{taflove2005}. Additionally, it can be
executed efficiently in parallel, although the naive implementation will not
yield the maximum performance and a more advanced approach must be
used~\cite{krishnamoorthy2007, riesch2018oqel}. The major drawback is the
introduced numerical dispersion which can only be avoided by using very fine
discretization sizes. Otherwise, artifacts in the simulation results could be
the consequence. In the context of MB simulations, different values (and value
ranges) for the maximal spatial discretization size $\Delta_{z}$ have been
found adequate for the FDTD scheme. Namely, $\lambda$/20 to $\lambda
$/100~\cite{bidegaray2003}, $\lambda$/50~\cite{slavcheva2003ultrashort},
$\lambda$/100~\cite{schlottau2005modeling}, and $\lambda$%
/200~\cite{ziolkowski1995ultrafast} have been used, where $\lambda$ represents
the smallest occuring wavelength. The maximum time step $\Delta_{t}$ is,
similarly as in Section \ref{sec:RNFD}, determined by the Courant number,
which leads for the FDTD scheme to the condition $v\Delta_{t}<\Delta_{z}$ [or
$v\Delta_{t}\leq\left(  \Delta_{x}^{-2}+\Delta_{y}^{-2}+\Delta_{z}%
^{-2}\right)  ^{-1/2}$ for three spatial dimensions]~\cite{taflove2005}. Here,
the velocity is obtained from the parameters in Eq.\thinspace(\ref{eq:maxw})
as $v=\left(  \mu_{0}\epsilon_{0}\epsilon_{\mathrm{r}}\right)  ^{-1/2}$. In
related literature, choosing $v\Delta_{t}=\Delta_{z}/2$ was found to be
adequate \cite{schlottau2005modeling, ziolkowski1995ultrafast}.

To reduce the numerical burden, different approaches using the pseudo-spectral
time-domain method~\cite{liu1997} have been presented~\cite{saut2006,
marskar2011}. This method calculates the spatial derivatives using the fast
Fourier transform in space. As long as Nyquist-Shannon theorem is not
violated, the method is exact in space (and the introduced numerical
dispersion minimal). However, the time derivative is still approximated with
finite differences that cause numerical error and dispersion. Nevertheless,
fewer spatial grid points are required to achieve reasonable accuracy (for
example, the spatial discretization size $\Delta_{z}=\lambda/10$ has been used
\cite{marskar2011}). Thereby, the computational workload is reduced. These
advantages come at the price of a more complicated implementation. In
particular, absorbing boundary conditions must be implemented in order to
avoid the wrap-around effect. Furthermore, sharp material parameter changes
and the implementation of sources are not trivial anymore~\cite{liu1997}.

\subsubsection{Coupling Electric Field Updates and Density Matrix Updates}

Since the electric field in Maxwell's equations and the density matrix in the
Bloch equations depend on each other, this coupling must be treated
appropriately for any numerical method that solves Maxwell's equations.
Bid\'{e}garay distinguishes between strongly and weakly coupled
methods~\cite{bidegaray2003}. The difference is the discretization of the
density matrix in time and in relation to the electric field. Strongly coupled
methods discretize the density matrix and the electric field at the same time
value, weakly coupled methods apply a discretization which is staggered (a
half time step difference between density matrix and electric field). In the
following, we discuss various approaches to update the density matrix with
different forms of coupling.

\subsubsection{Crank-Nicolson Scheme/Predictor-Corrector Method}

The pioneering work by Ziolkowski et al.~\cite{ziolkowski1995ultrafast,
slavcheva2002coupled} treats the Bloch equations with the Crank-Nicolson
scheme
\begin{equation}
\hat{\rho}^{n+1}=\hat{\rho}^{n}+\frac{\Delta_{t}}{2}\left[  \mathcal{F}%
^{n+1}\left(  \hat{\rho}^{n+1}\right)  +\mathcal{F}^{n}\left(  \hat{\rho}%
^{n}\right)  \right]  ,
\end{equation}
where $n$ corresponds to the time $t_{n}=n\Delta_{t}$, $\Delta_{t}$ is the
time step size, and $\mathcal{F}(\hat{\rho})=\mathcal{L}(\hat{\rho
})+\mathcal{D}(\hat{\rho})$ represents the right hand side of the Lindblad
equation~(\ref{eq:lindbl}). Since this implicit scheme requires solving a
linear system of equations at every time step, usually modifications are
employed to reduce the numerical load, such as keeping the field at a fixed
value while advancing the density matrix by a time step~\cite{capua2013finite}%
. A widely used variant is based on the predictor-corrector technique, where
the update step first initializes $\hat{\rho}_{\mathrm{PC}}=\hat{\rho}_{n}$,
then executes the procedure
\begin{equation}
\hat{\rho}_{\mathrm{PC}}\leftarrow\hat{\rho}_{n}+\Delta_{t}\mathcal{F}\left(
\frac{1}{2}\hat{\rho}_{\mathrm{PC}}+\frac{1}{2}\hat{\rho}_{n}\right)
\end{equation}
four times, and finally assigns the result to the value $\hat{\rho}_{n+1}%
=\hat{\rho}_{\mathrm{PC}}$~\cite{ziolkowski1995ultrafast,
slavcheva2002coupled}.

In Fig.\thinspace\ref{fig:methods}(a), the coupling of the method to the FDTD
scheme is illustrated. It should be noted that this is a strongly coupled
method and the electric field is updated with the same procedure (of course,
$\mathcal{F}$ is replaced with the right hand side of Ampere's law) and in
parallel to the density matrix update.

\subsubsection{\label{sec:RK}Runge-Kutta Method}

Several research groups use the fourth-order Runge-Kutta (RK) method to solve
the Bloch equations \cite{sukharev2011, cartar2017, deinega2014self,
garraway1994}. As illustrated in Fig.\thinspace\ref{fig:methods}(b), the
method is strongly coupled since electric field and density matrix are
discretized at the same time steps. The exact procedure is not always
described in related work, but can be outlined as
follows~\cite{riesch2018oqel}: First, the electric field is updated using the
standard FDTD update step. Then, the update of the density matrix using the
rule
\begin{equation}
\hat{\rho}^{n+1}=\hat{\rho}^{n}+\Delta_{t}\left(  k_{1}+2k_{2}+2k_{3}%
+k_{4}\right)  /6
\end{equation}
follows, where $k_{1}=\mathcal{F}^{n}(\hat{\rho}^{n})$, $k_{2}=\mathcal{F}%
^{n+1/2}(\hat{\rho}^{n}+\Delta_{t}k_{1}/2)$, $k_{3}=\mathcal{F}^{n+1/2}%
(\hat{\rho}^{n}+\Delta_{t}k_{2}/2)$, and $k_{4}=\mathcal{F}^{n+1}(\hat{\rho
}^{n}+\Delta_{t}k_{3})$~\cite{hairer1993}. Since the $\mathcal{F}^{n}$
contains the electric field $\mathbf{E}^{n}$ at time $t_{n}=n\Delta_{t}$, not
only the old and updated field values are required, but also the value at the
half time step. The latter can be approximated by averaging between the old
and the updated field value, i.e., $\mathbf{E}^{n+1/2}\approx\left(
\mathbf{E}^{n}+\mathbf{E}^{n+1}\right)  /2$.

\subsubsection{Matrix Exponential Methods}

The methods of this group aim to solve the Bloch equations exactly for one
time step. As illustrated in Fig.\thinspace\ref{fig:methods}(c), the updates
of electric field and density matrix are weakly coupled, i.e., their updates
are performed alternatingly. The density matrix update reads
\begin{equation}
\hat{\rho}^{n+1/2}=\exp\left(  \mathcal{F}^{n}\Delta_{t}\right)  \hat{\rho
}^{n-1/2},
\end{equation}
where $\mathcal{F}^{n}$ may depend on the electric field $\mathbf{E}^{n}$ and
$\exp(\mathcal{F}^{n}\Delta_{t})$ represents the exact solution of the
Lindblad equation. After that, the standard FDTD update rule calculates
$\mathbf{E}^{n+1}$ using $\mathbf{E}^{n}$ and $\hat{\rho}^{n+1/2}$.

If an analytical expression for the solution superoperator $\exp
(\mathcal{F}t)$ exists, this method is clearly the most accurate one. However,
finding such an analytical expression is far from trivial. In fact, the exact
form of the exponential depends on the representation. In Liouville or
coherence vector representation described in Section \ref{sec:coh}, the
solution superoperator has the form $\exp(Ft)$, where $F$ is a matrix. While
this is straightforward to solve, the size of the matrix is in the order
$N^{2}\times N^{2}$ for a $N\times N$ density matrix. Since the exponential of
a $N\times N$ matrix would generally need $\mathcal{O}(N^{3})$ operations,
calculating the exponential in Liouville representation requires
$\mathcal{O}(N^{6})$ operations and becomes unfeasible for large $N$.

In regular representation, a solution for the Lindblad equation must be found
first. The Strang splitting technique~\cite{strang1968} can help here to
separate the effects of the Liouvillian $\mathcal{L}$ and the dissipation
superoperator $\mathcal{D}$. The solution for the Liouvillian requires the
calculation of $\exp(-\mathrm{i}\hbar^{-1}\hat H t)$, where the Hamiltonian
$\hat H$ is a $N \times N$ Hermitian matrix. The calculation requires
$\mathcal{O}(N^{3})$ operations, which is still quite intensive.

The Strang splitting introduces an additional error of $\mathcal{O}(\Delta
_{t}^{2})$ in general. Furthermore, $\mathcal{F}$ is generally time dependent
due to its dependence on the time-varying electric field, in which case the
resulting matrix exponentials contain an integral in the exponent. Commonly,
the integral is approximated using the midpoint rule. This leads to the
conclusion that in reality the accuracy of matrix exponential methods is
comparable to other approaches. Nevertheless, this group of methods preserves
certain matrix properties and despite their limited performance they have
attracted the focus of many research groups.

Several techniques have been applied in order to improve the performance of
matrix exponential methods. The already mentioned Strang splitting has not
only been used to allow analytical solutions, but also to separate the time
dependent and time independent part of $\mathcal{F}$~\cite{bidegaray2001,
bidegaray2003, saut2006, marskar2011}. This has the advantage that a part of
the solution can be precalculated and applied at every time step, while for
the remaining part efficient evaluation techniques exist in some cases. For
example, we discovered that the coherence vector representation leads to a
real skew-symmetric matrix in the exponential. This expression can be
evaluated efficiently using the generalized Rodrigues'
formula~\cite{gallier2003}. Other techniques to calculate the matrix
exponential~\cite{moler2003} have been applied in related work: An
approximation based on the Cayley transform~\cite{bidegaray2001,
bidegaray2003, saut2006}, Magnus expansion via Sylvester's
formula~\cite{hailu2016}, diagonalization of the matrix~\cite{weninger2013},
the scaling and squaring method as well as a Krylov subspace
method~\cite{guduff2017}, and Chebyshev polynomials~\cite{kosloff1994}.

\subsubsection{Comparison of Numerical Methods for the Bloch Equations}

As already outlined above, the matrix exponential methods are the most
computationally expensive ones. In fact, this was confirmed in a detailed
investigation~\cite{riesch2018oqel}, where both the Runge-Kutta and the
predictor-corrector implementation outperformed the matrix exponential method.
In this comparison, the predictor-corrector method demonstrated the best performance.

In terms of accuracy, Runge-Kutta methods have the highest order. However, the
accuracy alone is not the crucial criterion. In particular, it was
demonstrated that the Crank-Nicolson scheme does not preserve the positivity
of the density matrix in the general case (at least when more than two energy
levels are considered) and therefore might yield unrealistic results, e.g.,
negative populations~\cite{bidegaray2001}. Furthermore, it was found that both
the predictor-corrector and Runge-Kutta method yield negative populations in
certain cases (e.g., long simulation end time combined with unfortunate
choices for the time step size), while the matrix exponential method preserves
the properties of the density matrix independent of the simulation
settings~\cite{riesch2017e}.

\subsubsection{Alternative Methods}

Besides the full-wave Bloch equations of the form Eq.\thinspace(\ref{eq:MB2}),
also related formalisms are used to model quantum systems interacting with a
semiclassical optical field, requiring adapted numerical schemes which are
often combined with the FDTD method for Maxwell's equations. For example, MB
simulations which replace the Bloch equations by an equivalent evolution
equation for the polarization vector, Eq.\thinspace(\ref{eq:Pcomp}), require
modified schemes adapted to the second-order differential form of
Eq.\thinspace(\ref{eq:Pcomp}) \cite{taflove2005,dridi2013model,chua2011}.

If the dissipation term in the Bloch equations can be neglected and the
quantum system is in a pure state, the time evolution can be described in a
simplified manner with the time dependent Schr\"{o}dinger equation,
Eq.\thinspace(\ref{eq:schr}), for which suitable numerical schemes have been
developed
\cite{leforestier1991comparison,peskin1994solution,tremblay2004using,gordon2006numerical,blanes2017symplectic}%
. In analogy to the MB equations, the Schr\"{o}dinger and Maxwell's equations
can be combined to the Maxwell-Schr\"{o}dinger approach, which is for example
used~to model nanoelectronic systems \cite{pierantoni2008new,ahmed2010hybrid},
or to describe the interaction of atoms with intense laser
fields~\cite{christov1998generation,lorin2007numerical}. As for the MB
equations, such a coupled simulation complicates the numerical treatment, and
various numerical schemes have been developed, e.g., combining FDTD or
transmission line matrix simulations of Maxwell's equations with a spatial
grid representation or eigenstate expansion of the wavefunction
\cite{lorin2007numerical,pierantoni2008new,ahmed2010hybrid,chen2017canonical,chen2017hamiltonian,masiello2005dynamics}%
. In this context, a recent interest has been on algorithms preserving the
symplectic structure of the Maxwell-Schr\"{o}dinger equations, thus ensuring
energy conservation of the coupled system
\cite{chen2017canonical,chen2017hamiltonian,masiello2005dynamics}.

\section{\label{sec:eff}Inclusion of Further Effects}

\subsection{Local-Field Correction}

In principle, the current/polarization contribution of an individual quantum
system at a given position can be directly represented in Maxwell's equations
by a point source \cite{deinega2014self}, without using ensemble averaging as
in Section \ref{sec:pol}. However, the complexity of such an approach
increases significantly with the number of quantum systems to be included
\cite{deinega2014self}. Moreover, care must be taken that the field which
drives the quantum system does not contain the divergent self-field
contribution of the system itself, which further adds to the numerical load
\cite{deinega2014self,schelew2017self}. An alternative approach, which is
especially suitable for a large ensemble of quantum systems as considered in
this paper, is based on macroscopic MB equations. Rather than setting up Bloch
equations for each of the quantum systems, the ensemble is here modeled by
representative density matrices $\rho_{ij}\left(  \mathbf{x},t\right)  $
distributed over the device volume, e.g., placed on the spatial grid points,
where $\mathbf{x}$ is the macroscopic position coordinate. Likewise, Maxwell's
equations then contain the macroscopic current densities (see Section
\ref{sec:pol}) and fields, defined as ensemble averages over the individual
microscopic contributions. From the macroscopic electric field $\mathbf{E}$
which is averaged over local variations associated with the individual
dipoles, the local microscopic field which interacts with a given physical
quantum system can be determined based on a compact Clausius-Mosotti type
model. This local-field correction can for example lead to frequency shifts,
and becomes relevant for densely spaced quantum systems as discussed in
Section \ref{sec:NDD} \cite{wegener1990line,bowden1993near}. A related effect
emerges in tightly localized artificial quantum systems consisting of multiple
semiconductor atoms, see Section \ref{sec:depol} \cite{slepyan2004rabi}. It
has been pointed out that local-field effects can be exploited as an
additional design degree of freedom in nanostructures
\cite{dolgaleva2012local}.

\subsubsection{\label{sec:NDD}Near-Dipole-Dipole Effects in Dense Media}

\begin{figure}[ptb]
\includegraphics{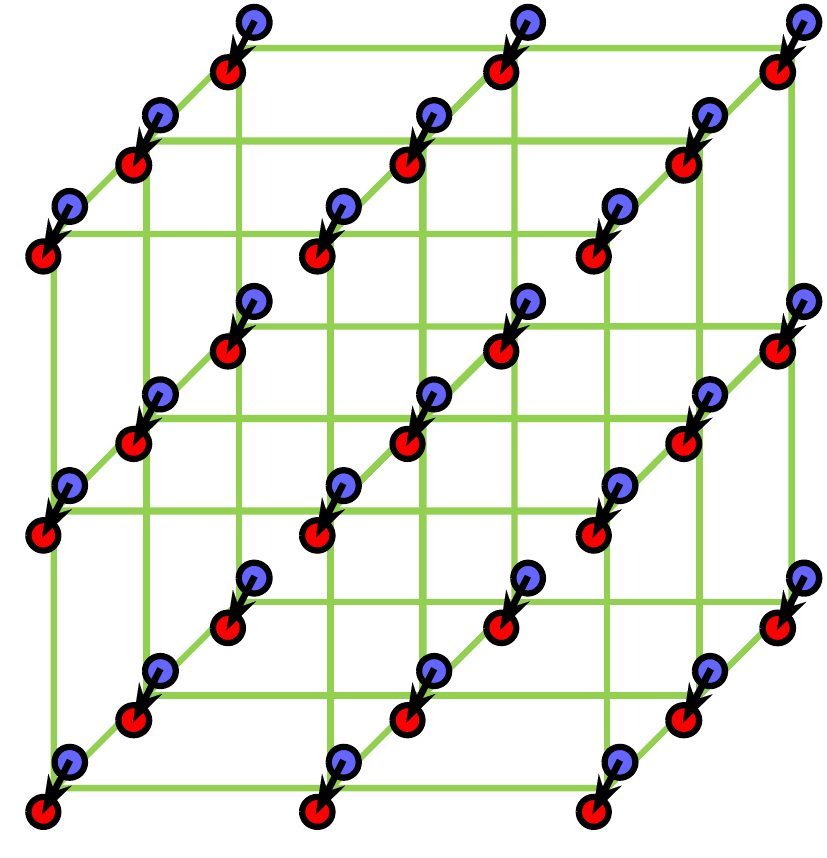}
\caption{Illustration of dipoles arranged in a cubic lattice.}%
\label{fig:crystal}%
\end{figure}

The macroscopic field $\mathbf{E}$ comprises contributions of external sources
as well as an internal contribution $\mathbf{E}_{\mathrm{p}}$ due to the
induced dipoles in the material, which is related to $\mathbf{P}_{\mathrm{q}}%
$. In the following, we consider a medium such as a gas or crystal lattice
which consists of a dense collection of atoms, molecules or other quantum
systems, as illustrated in Fig.\thinspace\ref{fig:crystal}. The local field
$\mathbf{E}_{\mathrm{L}}$ at the position of the considered quantum system is
determined by replacing the volume-averaged field $\mathbf{E}_{\mathrm{p}}$
with the microscopic contribution $\mathbf{E}_{\mathrm{near}}$ due to the
nearby dipole moments, $\mathbf{E}_{\mathrm{L}}=\mathbf{E-E}_{\mathrm{p}%
}+\mathbf{E}_{\mathrm{near}}$ \cite{Jackson}. Based on a microscopically
large, but macroscopically small probe volume, which is conveniently chosen to
be spherical, the macroscopic polarization contribution is obtained as
$\mathbf{E}_{\mathrm{p}}=-\mathbf{P}_{\mathrm{q}}/\left(  3\epsilon
_{0}\right)  $ \cite{Jackson}. On the other hand, it can be shown that for
dipoles arranged in a cubic lattice as illustrated in Fig.\thinspace
\ref{fig:crystal}, $\mathbf{E}_{\mathrm{near}}\ $vanishes for a particle on a
lattice site, where the particle's self-field is not included
\cite{Jackson,bowden1993near}. This yields $\mathbf{E}_{\mathrm{L}%
}=\mathbf{E+P}_{\mathrm{q}}/\left(  3\epsilon_{0}\right)  $, which is also
approximately true for other reasonably isotropic media and completely random
arrangements, such as amorphous media or gases \cite{Jackson,bowden1993near}.
Retardation effects are here negligible since the probe volume diameter is
assumed to be much smaller than the optical wavelength
\cite{bowden1993near,bowden1994erratum}. For two-level systems where the
polarization is given by Eq.\thinspace(\ref{eq:P2u}), the local-field
corrections can thus be included in the Maxwell-Bloch equations by formally
substituting $\Omega=\mathbf{d}_{12}\mathbf{E/\hbar}\ $with
\begin{equation}
\Omega_{\mathrm{L}}=\mathbf{d}_{12}\mathbf{E}_{\mathrm{L}}\mathbf{/\hbar
}=\Omega+2\omega_{\mathrm{L}}\Re\left\{  \rho_{21}\right\}  \label{eq:loc1}%
\end{equation}
in Eq.\thinspace(\ref{eq:bloch2l}) \cite{xia2005near}, where the static
Lorentz shift
\[
\omega_{\mathrm{L}}=d_{12}^{2}n_{\mathrm{3D}}/\left(  3\hbar\epsilon
_{0}\right)
\]
has the dimension of frequency. Here we have assumed a real-valued
$\mathbf{d}_{12}$ for simplicity. Applying the RWA, we obtain Eq.\thinspace
(\ref{eq:RWA2l}) where we have to substitute $\underline{\hat{\Omega}}$ by
\begin{equation}
\underline{\hat{\Omega}}_{\mathrm{L}}=\underline{\hat{\Omega}}+2\omega
_{\mathrm{L}}\eta_{21}, \label{eq:loc2}%
\end{equation}
which changes Eq.\thinspace(\ref{eq:RWA2l1}) but not Eq.\thinspace
(\ref{eq:RWA2l2}) since there the local-field correction term cancels out
\cite{bowden1993near}.

If the quantum systems are embedded in a host medium, such as dopant ions in a
crystal, they interact not only with particles of the same species, but also
with those of the host material, and the local-field correction must be
suitably extended. Accordingly, above approach has been generalized to a dense
collection of two-level atoms embedded in a linear, potentially dispersive and
absorptive host medium \cite{crenshaw1996local}, and to multicomponent media
in general \cite{crenshaw1997local}. Furthermore, above concept can be
straightforwardly extended to more than two levels \cite{dowling1993near}.

We note that above correction to the MB equations has mainly been considered
for ensembles of atoms or molecules, where typically much higher number
densities are obtained than for artificial systems such as quantum dots. This
model has enabled the analytical \cite{afanas2000optical} and numerical
\cite{crenshaw1992ultrafast,afanas2002coherent,novitsky2011femtosecond}
investigation of numerous effects, such as solitonic and ultrashort pulse
propagation or optical switching. For artificial, tightly localized quantum
systems, local-field effects are typically governed by the depolarization
field, as discussed below.

\subsubsection{\label{sec:depol}Depolarization Field in Tightly Localized
Quantum Systems}

\begin{figure}[ptb]
\includegraphics{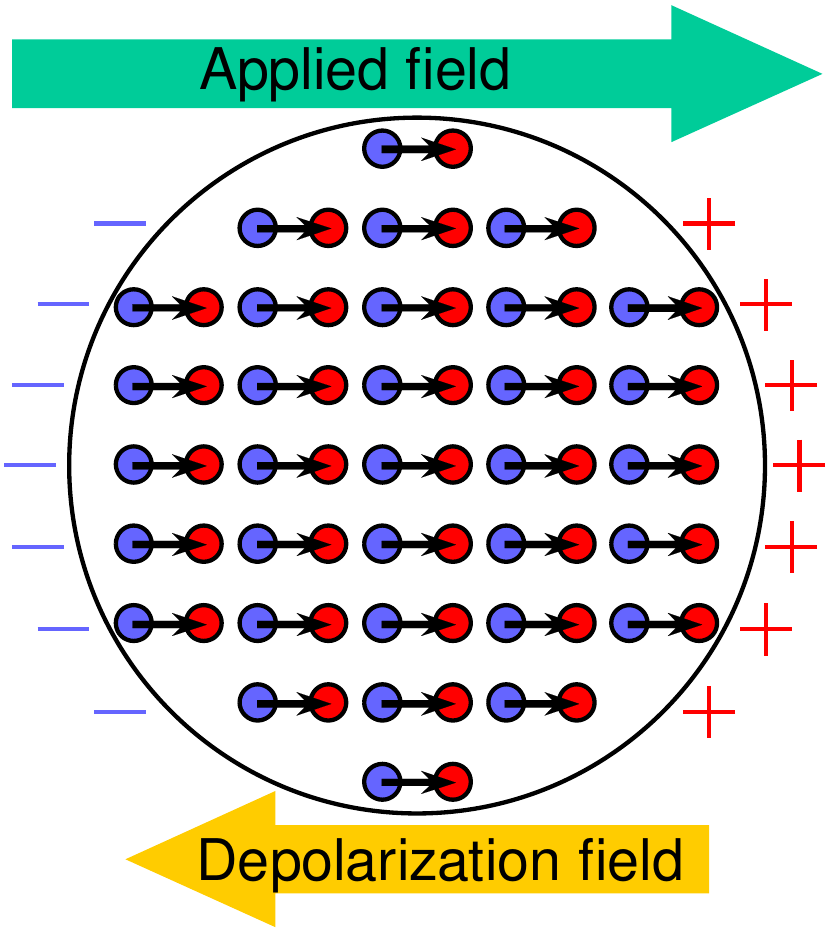}
\caption{Illustration of a localized polarizable object in an electric field.}%
\label{fig:depol}%
\end{figure}

Artificial quantum systems, such as quantum dots, are nanostructures built
from a larger number of semiconductor atoms. Thus, the quantum systems only
''feel'' an averaged polarization contribution of the individual atoms, which
can be described by the background dielectric constant of the host material
$\epsilon_{\mathrm{r}}$. Here, the local-field correction accounts for the
deviation of the mesoscopic average field inside the localized polarizable
nanostructure from the macroscopic average field $\mathbf{E}$ in the entire
composite material, which enters the macroscopic Maxwell equations
\cite{dolgaleva2012local}. In the following, we assume that $\epsilon
_{\mathrm{r}}$ is real-valued and frequency independent, and is identical for
the nanostructure and the surrounding material. As illustrated in
Fig.\thinspace\ref{fig:depol}, the field $\mathbf{E}$ generates a polarization
inside the polarizable object, and the uncompensated surface dipoles give rise
to bound surface charges which induce an electric field in the object, the
so-called depolarization field $\mathbf{E}_{\mathrm{d}}$, thus altering the
local field inside to $\mathbf{E}_{\mathrm{L}}=\mathbf{E+E}_{\mathrm{d}}$. We
assume that the object's dimensions are much smaller than the wavelength of
the exciting field. The (spatially averaged) depolarization field is then
obtained as $\mathbf{E}_{\mathrm{d}}=-\mathbf{NP}_{\mathrm{q}}/\left(
\epsilon_{0}\epsilon_{\mathrm{r}}\right)  $, where $\mathbf{N}$ is the
depolarization tensor which only depends on the geometry of the polarizable
object \cite{slepyan2002quantum}. For a localized two-level system of volume
$V_{\mathrm{s}}$, the (volume-averaged) polarization is given by
$\mathbf{P}_{\mathrm{q}}=\mathbf{d}_{12}\left(  \rho_{12}+\rho_{21}\right)
/V_{\mathrm{s}}$. Proceeding in the same way as in Section \ref{sec:NDD}, we
include the local-field correction into the MB equations by using
Eq.\thinspace(\ref{eq:loc1}) or Eq.\thinspace(\ref{eq:loc2}), where we now
have \cite{slepyan2002quantum,slepyan2004rabi}
\[
\omega_{\mathrm{L}}=-d_{12}^{2}L_{\mathrm{d}}/\left(  \mathbf{\hbar}%
\epsilon_{0}\epsilon_{\mathrm{r}}V_{\mathrm{s}}\right)  .
\]
The depolarization factor $L_{\mathrm{d}}=\mathbf{d}_{12}\mathbf{Nd}%
_{12}/d_{12}^{2}$ with $0\leq L_{\mathrm{d}}\leq1$ accounts for the anisotropy
of the object, such as an ellipsoidal quantum dot, and becomes $L_{\mathrm{d}%
}=1/3$ for spherical geometries \cite{slepyan2002quantum}. The resulting
equations have been used for both analytical \cite{paspalakis2006local} and
numerical \cite{slepyan2004rabi,paspalakis2006local,mitsumori2018effect}
studies of local field effects in quantum dots.

\subsection{\label{sec:inh}Inhomogeneous Broadening}

Homogeneous broadening is naturally considered in the Bloch equations. This
can best be seen from the steady state solution in RWA, Eq.\thinspace
(\ref{eq:stat}), yielding a Lorentzian lineshape with the width given by the
dephasing rate of the corresponding transition, see Eqs.\thinspace
(\ref{eq:chi}) and (\ref{eq:gP}) as well as Fig.\thinspace\ref{fig:susc}(b).
In addition, inhomogeneous broadening arises if the optically active medium
consists of quantum systems with slightly different resonance frequencies
\cite{allen1987optical,meystre2013elements}, as for example frequently arises
in ensembles of quantum dots due to size fluctuations. Another example is
Doppler broadening in a gas \cite{1986lase.book.....S}, caused by the Doppler
shift due to the thermal motion of the atoms or molecules. For a given
transition between two states $\left|  i\right\rangle $\ and $\left|
j\right\rangle $, the distribution of the resonance frequency $\omega_{ij}$ is
commonly described by a distribution function $g_{ij}\left(  \omega
_{ij}-\overline{\omega}_{ij}\right)  $ with $\int_{-\infty}^{\infty}%
g_{ij}\left(  \omega\right)  \mathrm{d}\omega=1$, where $\omega_{ij}%
-\overline{\omega}_{ij}$ is the deviation from the average resonance frequency
$\overline{\omega}_{ij}$. For thermal Doppler broadening, $g_{ij}$ is given by
a Gaussian distribution%
\begin{equation}
g_{ij}\left(  \omega_{ij}-\overline{\omega}_{ij}\right)  =\frac{1}{\sqrt{2\pi
}\sigma_{ij}}\exp\left[  -\frac{\left(  \omega_{ij}-\overline{\omega}%
_{ij}\right)  ^{2}}{2\sigma_{ij}^{2}}\right]  . \label{eq:Gauss}%
\end{equation}
In this case, the standard deviation becomes $\sigma_{ij}=\left(  \omega
_{ij}/c\right)  \left(  k_{\mathrm{B}}T/m\right)  ^{1/2}$, where
$k_{\mathrm{B}}$ is the Boltzmann constant, $T$ indicates the temperature, and
$m$ is the mass of the atom or molecule \cite{1986lase.book.....S}. The
Gaussian distribution, Eq.\thinspace(\ref{eq:Gauss}), is also frequently used
as a generic model if the distribution of resonance frequencies is not exactly
known, e.g., to describe above mentioned inhomogeneous broadening in ensembles
of quantum dots due to size fluctuations \cite{majer2010cascading}. If the
individual quantum systems contain more than one relevant optical transition
with distributed resonance frequencies, in principle joint distribution
functions have to be used.

Numerically, the full-wave or RWA Bloch equations, Eq.\thinspace(\ref{eq:MB2})
or (\ref{eq:RWA}), have to be solved separately for each possible value of the
resonance frequency (or each possible combination of resonance frequency
values if the individual quantum systems contain more than one relevant
optical transition) \cite{schlottau2005modeling}. This requires discretizing
the distribution function into a finite number of $N_{\mathrm{inh}}$ bins with
resonance frequencies $\omega_{ij}^{s}$, $s=1..N_{\mathrm{inh}}$. Each of
these bins is represented by a corresponding quantum system subensemble with
density matrix $\rho_{ij}^{s}$, where the fraction of carriers $p_{s}$ is
proportional to the weight of that bin and $\sum_{s}p_{s}=1$. The polarization
current density is then obtained from a generalized version of Eq.\thinspace
(\ref{eq:Jt2}),%
\begin{equation}
\mathbf{J}_{\mathrm{q}}=n_{\mathrm{3D}}\sum_{i,j}\mathbf{d}_{ji}\sum_{s}%
p_{s}\partial_{t}\rho_{ij}^{s}. \label{eq:dtP}%
\end{equation}
In SVAA, the polarization amplitude is given by a generalized form of
Eq.\thinspace(\ref{eq:P_SVAA2}),
\begin{equation}
\underline{\mathbf{P}}=2n_{\mathrm{3D}}\sum_{\omega_{ij}>0}\mathbf{d}_{ji}%
\sum_{s}p_{s}\eta_{ij}^{s}. \label{eq:Pinh}%
\end{equation}
For certain broadening mechanisms, such as fluctuations in quantum dot size,
also $\mathbf{d}_{ij}$ can in principle vary which could be considered by
introducing a quantity $\mathbf{d}_{ij}^{s}$ in Eqs.\thinspace(\ref{eq:dtP})
and (\ref{eq:Pinh}) in analogy to $\omega_{ij}^{s}$, but this effect is
usually neglected.

In certain cases, inhomogeneous broadening can also be considered in
analytical solutions of the MB equations based on the RWA. These usually
invoke the factorization ansatz, which assumes that non-resonant systems with
a finite frequency detuning $\Delta=\omega_{\mathrm{c}}-\omega_{21}\neq0$
essentially respond in the same way to the optical field as the resonant ones,
apart from a detuning dependent amplitude $F\left(  \Delta\right)  $
\cite{allen1987optical}. In Section \ref{sec:SIT}, this approach has been
demonstrated in the context of self-induced transparency.

\subsection{\label{sec:noise}Noise}

Noise in optoelectronic devices arises for example from spontaneous emission
and from processes in the semiconductor host, such as lattice vibrations.
Noise and fluctuations can generally be included into the semiclassical MB
equations by adding stochastic terms
\cite{polder1979superfluorescence,wodkiewicz1979stochastic}. Numerically, the
stochastic terms are typically implemented by using a pseudorandom number
generator to obtain uncorrelated, Gaussian distributed random numbers for
every grid point \cite{slavcheva2004fdtd,andreasen2009finite}. The MB
equations, complemented by these additional stochastic terms, are then
numerically solved as discussed in Section \ref{sec:Num}, for example with an
FDTD-based approach \cite{slavcheva2004fdtd,andreasen2009finite}. The
stochastic terms are systematically obtained from quantum Langevin equations
\cite{gardiner2004quantum,sar87}, which are then represented by equivalent
stochastic c-number equations
\cite{lax1969quantum,drummond1991quantum,slavcheva2004fdtd}, i.e., evolution
equations for operator expectation values with additional stochastic terms.

Spontaneous emission obviously plays an important role in optoelectronic
devices. While the resulting recombination can simply be included by nonlinear
rate terms for the carrier occupations in Eq.\thinspace(\ref{eq:MB2b}) or
(\ref{eq:RWA2}) \cite{gehrig2002mesoscopic,majer2010cascading}, the noise
contribution is not included in the MB model due to its semiclassical nature.
This effect can however be considered in terms of a Gaussian white noise
source in the optical propagation equation
\cite{slavcheva2004fdtd,kim2010maxwell,wilkinson2013influence}. In a different
model, also dipole fluctuations are included by adding Langevin noise terms
not only to the propagation equation, but also to Eq.\thinspace(\ref{eq:MB2a})
for the off-diagonal density matrix elements
\cite{hofmann1999quantum,gehrig2002mesoscopic}. By virtue of the
fluctuation-dissipation theorem, a decay of populations, coherences or the
optical field is generally accompanied by fluctuations, and an MB equation
model which includes such decay-induced fluctuations has been presented
\cite{andreasen2009finite,pusch2012coherent}. Furthermore, an extension of the
stochastic c-number approach to incorporate nonclassical effects has been
discussed \cite{drummond1991quantum}.

\section{\label{sec:Appl}Application to Optoelectronic Devices}

\subsection{\label{sec:Bulk}Bulk and Quantum Well Interband Optoelectronic Devices}

For interband optoelectronic devices based on semiconductor bulk or
quantum-well media, the conduction and valence band states are given by
$\left|  \mathrm{c},\mathbf{k}\right\rangle $ and $\left|  \mathrm{v}%
,\mathbf{k}\right\rangle $, respectively. Here, $\mathbf{k}$ is the
three-dimensional crystal wavevector in bulk media or the two-dimensional
in-plane wavevector for quantum well structures, see Section \ref{sec:dip}. In
the following, we define $\rho_{\mathrm{c},\mathbf{k}}$ and $\left(
1-\rho_{\mathrm{v},\mathbf{k}}\right)  $ as the electron occupation
probability of a conduction band state $\left|  \mathrm{c},\mathbf{k}%
\right\rangle $ and a valence band state $\left|  \mathrm{v},\mathbf{k}%
\right\rangle $, respectively, i.e., $\rho_{\mathrm{v},\mathbf{k}}$
corresponds to the hole occupation probability of a valence band state.
Restricting ourselves to direct bandgap semiconductors in a two-band
approximation and using that the optical transitions are $\mathbf{k}$
conserving, as can for quantum wells be seen from Eq.\thinspace(\ref{eq:dqw})
\cite{chow2012semiconductor}, the Bloch equations, Eq.\thinspace
(\ref{eq:MB2}), become%
\begin{subequations}%
\label{eq:SB}%
\begin{align}
\partial_{t}\rho_{\mathrm{cv},\mathbf{k}}  &  =-\mathrm{i}\omega
_{\mathrm{cv},\mathbf{k}}\rho_{\mathrm{cv},\mathbf{k}}-\mathrm{i}%
\Omega_{\mathbf{k}}\left(  \rho_{\mathrm{c},\mathbf{k}}+\rho_{\mathrm{v}%
,\mathbf{k}}-1\right) \nonumber\\
&\,\quad +\left[  \partial_{t}\rho_{\mathrm{cv},\mathbf{k}%
}\right]  _{\mathrm{col}},\label{eq:SB1}\\
\partial_{t}\rho_{\alpha,\mathbf{k}}  &  =\mathrm{i}\left(  \Omega
_{\mathbf{k}}\rho_{\mathrm{cv},\mathbf{k}}^{\ast}-\Omega_{\mathbf{k}}^{\ast
}\rho_{\mathrm{cv},\mathbf{k}}\right)  +\left[  \partial_{t}\rho
_{\alpha,\mathbf{k}}\right]  _{\mathrm{col}}, \label{eq:SB2}%
\end{align}%
\end{subequations}%
with $\alpha=\mathrm{c},\mathrm{v}$. Here, the sum over $\mathbf{k}$
implicitly also includes summation over the two possible spin orientations.
This equation applies to quantum wells, where the carriers are treated as a
two-dimensional gas, as well as bulk media. The dissipation processes are here
included by general collision terms $\left[  \partial_{t}\rho_{\mathrm{cv}%
,\mathbf{k}}\right]  _{\mathrm{col}}$ and $\left[  \partial_{t}\rho
_{\alpha,\mathbf{k}}\right]  _{\mathrm{col}}$. These can be modeled on a
microscopic level, in particular accounting for carrier-carrier and
carrier-phonon scattering, or under certain approximations by relaxation rate
terms similar to those in Eq.\thinspace(\ref{eq:MB2})
\cite{chow2012semiconductor,haug2009quantum}. Many-body Coulomb interactions
can be taken into account based on the Hartree-Fock approximation, which
results in the so-called semiconductor Bloch equations, which have the form of
Eq.\thinspace(\ref{eq:SB}) but feature renormalized transition and Rabi
frequencies%
\begin{subequations}%
\label{eq:renorm}
\begin{align}
\omega_{\mathrm{cv},\mathbf{k}}  &  =\frac{1}{\hbar}E_{\mathrm{cv},\mathbf{k}%
}\mathbf{-}\frac{1}{\hbar}\sum_{\mathbf{q}\neq\mathbf{k}}V_{\left|
\mathbf{k}-\mathbf{q}\right|  }\left(  \rho_{\mathrm{c},\mathbf{q}}%
+\rho_{\mathrm{v},\mathbf{q}}\right)  ,\label{eq:renorm1}\\
\Omega_{\mathbf{k}}  &  =\frac{\mathbf{d}_{\mathrm{cv},\mathbf{k}}\mathbf{E}%
}{\hbar}\mathbf{+}\frac{1}{\hbar}\sum_{\mathbf{q}\neq\mathbf{k}}V_{\left|
\mathbf{k}-\mathbf{q}\right|  }\rho_{\mathrm{cv},\mathbf{q}},
\label{eq:renorm2}%
\end{align}%
\end{subequations}%
leading to a coupling of the states with different $\mathbf{k}$ through the
time dependent renormalization terms
\cite{vu2004light,chow2012semiconductor,haug2009quantum}. In Eq.\thinspace
(\ref{eq:renorm1}), $E_{\mathrm{cv},\mathbf{k}}$ is the energy of free
electron-hole pairs, which can in a simple model be described as%
\begin{equation}
E_{\mathrm{cv},\mathbf{k}}=\frac{1}{2}\hbar^{2}\mathbf{k}^{2}\left(  \frac
{1}{m_{\mathrm{e}}^{\ast}}+\frac{1}{m_{\mathrm{h}}^{\ast}}\right)
+E_{\mathrm{g}}. \label{eq:renorm3}%
\end{equation}
Here, $E_{\mathrm{g}}$ denotes the band gap energy, or for quantum wells the
energy difference between the electron and hole state energies. Furthermore,
$m_{\mathrm{e}}^{\ast}$ and $m_{\mathrm{h}}^{\ast}$\ are the electron and hole
effective masses, assuming a parabolic dispersion relation near the conduction
and valence band edges, respectively. $V_{\mathbf{q}}$ is the Coulomb
potential in Fourier representation, which is in a bulk\ with probe volume
$V_{\mathrm{p}}$ and background permittivity $\epsilon_{\mathrm{r}}$\ given by
$V_{q}=e^{2}/\left(  \epsilon_{0}\epsilon_{\mathrm{r}}V_{\mathrm{p}}%
q^{2}\right)  $, and in a quantum well structure with in-plane cross section
$S$ by $V_{q}=e^{2}/\left(  2\epsilon_{0}\epsilon_{\mathrm{r}}qS\right)  $.
Screening effects, which result from the response of the other carriers and
weaken the potential, are then incorporated as corrections to the Hartree-Fock
equations in the form of a modified $V_{q}$
\cite{chow2012semiconductor,haug2009quantum}. Summing in Eq.\thinspace
(\ref{eq:Jt2}) over the initial and final states $\left|  i\right\rangle
=\left|  \mathrm{c},\mathbf{k}\right\rangle $ and $\left|  j\right\rangle
=\left|  \mathrm{v},\mathbf{q}\right\rangle $ where we consider that the total
number density of electron-hole pairs is given by
\begin{equation}
n_{\mathrm{cv}}\left(  t\right)  =V_{\mathrm{p}}^{-1}\sum_{\mathbf{k}}%
\rho_{\mathrm{c},\mathbf{k}}\left(  t\right)  =V_{\mathrm{p}}^{-1}%
\sum_{\mathbf{k}}\rho_{\mathrm{v},\mathbf{k}}\left(  t\right)  , \label{eq:Ne}%
\end{equation}
and using the $\mathbf{k}$ conservation of optical transitions, $\mathbf{d}%
_{ij}=\mathbf{d}_{\mathrm{cv},\mathbf{k}}\delta_{\mathbf{k},\mathbf{q}}$, we
obtain for the polarization term
\begin{equation}
\partial_{t}\mathbf{P}_{\mathrm{q}}=V_{\mathrm{p}}^{-1}\sum_{\mathbf{k}%
}\left(  \mathbf{d}_{\mathrm{cv},\mathbf{k}}\partial_{t}\rho_{\mathrm{vc}%
,\mathbf{k}}+c.c.\right)  . \label{eq:SBP}%
\end{equation}
With Eq.\thinspace(\ref{eq:SBP}), Eqs.\thinspace(\ref{eq:SB}) and
(\ref{eq:renorm}) can be coupled to the Maxwell equations, Eq.\thinspace
(\ref{eq:maxw}), resulting in the semiconductor Maxwell-Bloch equations. These
equations have been extensively used for the simulation of semiconductor
lasers and related devices
\cite{hess1996maxwell,hess1996maxwell2,witzigmann2006microscopic}.
Furthermore, they have been adapted to the modeling of quantum wire structures
\cite{rossi1996, marti2005,golde2008}, as well as graphene
\cite{stroucken2012excitonic} and carbon nanotubes \cite{hirtschulz2008carbon}.

As stated in Section \ref{sec:intro}, a further discussion of this model is
beyond the scope of this paper. Rather, we will focus here on approaches which
reduce above two-band model to macroscopic two- or $N$-level Bloch equations.
As a first step, the collision terms in Eq.\thinspace(\ref{eq:SB}) are modeled
by relaxation rate terms similar to those in Eq.\thinspace(\ref{eq:MB2})
\cite{chow2012semiconductor,haug2009quantum,ning1997effective,yao1995semiconductor}%
. Extending the rate equation model Eq.\thinspace(\ref{eq:relax1}) to states
$\left|  \alpha,\mathbf{k}\right\rangle $ results in a Boltzmann-type
collision term for the populations \cite{chow2012semiconductor}, which can in
consideration of Pauli blocking be written as%
\begin{align}
\left[  \partial_{t}\rho_{\alpha,\mathbf{k}}\right]  _{\mathrm{col}}%
=  &  -\rho_{\alpha,\mathbf{k}}\left(  W_{\alpha\beta,\mathbf{k}}^{\alpha
}+W_{\alpha\alpha,\mathbf{k}}^{\alpha}\right) \nonumber\\
&  +\left(  1-\rho_{\alpha
,\mathbf{k}}\right)  \left(  W_{\alpha\beta,\mathbf{k}}^{\beta}+W_{\alpha
\alpha,\mathbf{k}}^{\beta}\right)  . \label{eq:Boltz}%
\end{align}
Here, $\beta\neq\alpha$, i.e., for the conduction band collision term
($\alpha=\mathrm{c}$) we have $\beta=\mathrm{v}$, while $\beta=\mathrm{c}%
$\ for $\alpha=\mathrm{v}$.\ The Boltzmann rates are related to the electron
transition rates of Section \ref{sec:trans} by $W_{\alpha\alpha^{\prime
},\mathbf{k}}^{\mathrm{c}}=\sum_{\mathbf{k}^{\prime}}r_{\alpha,\mathbf{k}%
\rightarrow\alpha^{\prime},\mathbf{k}^{\prime}}\left(  1-f_{\alpha^{\prime
},\mathbf{k}^{\prime}}\right)  $ and $W_{\alpha\alpha^{\prime},\mathbf{k}%
}^{\mathrm{v}}=\sum_{\mathbf{k}^{\prime}}r_{\alpha^{\prime},\mathbf{k}%
^{\prime}\rightarrow\alpha,\mathbf{k}}f_{\alpha^{\prime},\mathbf{k}^{\prime}}%
$, where $f_{\alpha,\mathbf{k}}$ denotes the electron occupation probability,
i.e., $f_{\mathrm{c},\mathbf{k}}=\rho_{\mathrm{c},\mathbf{k}}$ and
$f_{\mathrm{v},\mathbf{k}}=1-\rho_{\mathrm{v},\mathbf{k}}$. For example,
spontaneous emission and carrier-phonon scattering can be modeled by
Eq.\thinspace(\ref{eq:Boltz}) with adequately chosen transition rates
$r_{\alpha,\mathbf{k}\rightarrow\alpha^{\prime},\mathbf{k}^{\prime}}$, while
the inclusion of carrier-carrier interactions beyond Hartree-Fock effectively
requires rates which themselves depend on the carrier distribution
\cite{chow2012semiconductor}. Furthermore modeling the dephasing in analogy to
Eq.\thinspace(\ref{eq:relax2}), we obtain%
\begin{subequations}%
\textrm{ }%
\begin{align}
\left[  \partial_{t}\rho_{\mathrm{cv},\mathbf{k}}\right]  _{\mathrm{col}}  &
=-\gamma_{\mathrm{cv},\mathbf{k}}\rho_{\mathrm{cv},\mathbf{k}}%
,\label{eq:dissip1}\\
\left[  \partial_{t}\rho_{\alpha,\mathbf{k}}\right]  _{\mathrm{col}}  &
=\Gamma_{\beta\alpha,\mathbf{k}}-r_{\alpha\beta,\mathbf{k}}\rho_{\alpha
,\mathbf{k}}+\Gamma_{\alpha\alpha,\mathbf{k}}-r_{\alpha\alpha,\mathbf{k}}%
\rho_{\alpha,\mathbf{k}}. \label{eq:dissip2}%
\end{align}%
\end{subequations}%

In Eq.\thinspace(\ref{eq:dissip1}), $\gamma_{\mathrm{cv},\mathbf{k}}$
indicates the dephasing rate. Rearranging the contributions in Eq.\thinspace
(\ref{eq:Boltz}), Eq.\thinspace(\ref{eq:dissip2}) is obtained, where the first
two terms on the right hand side with $\beta\neq\alpha$ describe interband
processes, while the other two terms model the intraband transitions. Here,
$r_{\alpha\alpha^{\prime},\mathbf{k}}=W_{\alpha\alpha^{\prime},\mathbf{k}%
}^{\alpha}$ denotes the interband (for $\alpha^{\prime}=\beta$) or intraband
(for $\alpha^{\prime}=\alpha$) recombination rate due to nonradiative
transitions and spontaneous emission. Furthermore, $\Gamma_{\alpha^{\prime
}\alpha,\mathbf{k}}=\left(  1-\rho_{\alpha,\mathbf{k}}\right)  W_{\alpha
\alpha^{\prime},\mathbf{k}}^{\beta}$ describes the filling of state $\left|
\alpha,\mathbf{k}\right\rangle $, and can for $\alpha^{\prime}=\beta$ be
interpreted as a pump rate,\textrm{ }where carriers are induced for example by
an injection current or optical pumping. Summation of Eq.\thinspace
(\ref{eq:Boltz}) or Eq.\thinspace(\ref{eq:dissip2}) over $\mathbf{k}$ yields
$\sum_{\mathbf{k}}\left[  \partial_{t}\rho_{\mathrm{c},\mathbf{k}}\right]
_{\mathrm{col}}=\sum_{\mathbf{k}}\left[  \partial_{t}\rho_{\mathrm{v}%
,\mathbf{k}}\right]  _{\mathrm{col}}$, as expected from Eq.\thinspace
(\ref{eq:Ne}). In more detail, for the intraband contributions we have
\begin{equation}
\sum_{\mathbf{k}}\left(  \Gamma_{\alpha\alpha,\mathbf{k}}-r_{\alpha
\alpha,\mathbf{k}}\rho_{\alpha,\mathbf{k}}\right)  =0, \label{eq:intra}%
\end{equation}
while the interband terms fulfill
\begin{align}
\sum_{\mathbf{k}}\Gamma_{\mathrm{vc},\mathbf{k}}  &  =\sum_{\mathbf{k}}%
\Gamma_{\mathrm{cv},\mathbf{k}},\nonumber\\
\sum_{\mathbf{k}}r_{\mathrm{cv},\mathbf{k}}\rho_{\mathrm{c},\mathbf{k}}  &
=\sum_{\mathbf{k}}r_{\mathrm{vc},\mathbf{k}}\rho_{\mathrm{v},\mathbf{k}}.
\label{eq:inter}%
\end{align}

To obtain compact two-level Bloch equations, we assume a $\mathbf{k}$
independent dipole matrix element $\mathbf{d}_{\mathrm{cv},\mathbf{k}%
}=\mathbf{d}_{21}$ in Eq.\thinspace(\ref{eq:renorm2}) and ignore the
renormalization contribution, yielding the usual definition $\Omega
_{\mathbf{k}}=\hbar^{-1}\mathbf{d}_{21}\mathbf{E}$. Summing Eq.\thinspace
(\ref{eq:SB2}) over $\mathbf{k}$ yields\ with Eqs.\thinspace(\ref{eq:Ne}),
(\ref{eq:dissip2}), (\ref{eq:intra}) and (\ref{eq:inter}) and $P_{\mathrm{cv}%
}=V_{\mathrm{p}}^{-1}\sum_{\mathbf{k}}\rho_{\mathrm{cv},\mathbf{k}}$%
\begin{equation}
\partial_{t}n_{\mathrm{cv}}=\mathrm{i}\left(  \Omega P_{\mathrm{cv}}^{\ast
}-\Omega^{\ast}P_{\mathrm{cv}}\right)  +\Gamma_{12}-\gamma_{1}n_{\mathrm{cv}}.
\label{eq:eff1}%
\end{equation}
For electrical pumping, the injection rate $\Gamma_{12}=\sum_{\mathbf{k}%
}\Gamma_{\beta\alpha,\mathbf{k}}$ (with $\alpha=\mathrm{c}$, $\beta
=\mathrm{v}$ or $\alpha=\mathrm{v}$, $\beta=\mathrm{c}$) can be modeled as
$\Gamma_{12}=\eta V_{\mathrm{p}}^{-1}I/e$, where $\eta$ and $I$ denote the
injection efficiency and current, respectively. The recombination rate
$\gamma_{1}$, which includes nonradiative and spontaneous transitions, is
obtained by averaging over the carrier distribution, $\gamma_{1}%
=\sum_{\mathbf{k}}r_{\alpha\beta,\mathbf{k}}\rho_{\alpha,\mathbf{k}%
}/n_{\mathrm{cv}}$. Proceeding in a similar manner for Eq.\thinspace
(\ref{eq:SB1}) by neglecting the $\mathbf{k}$ dependence of $\omega
_{\mathrm{cv},\mathbf{k}}$ is not feasible, due to the problems arising from
the summation over $\left(  \rho_{\mathrm{c},\mathbf{k}}+\rho_{\mathrm{v}%
,\mathbf{k}}-1\right)  $. Various strategies have been developed to circumvent
this problem \cite{ning1997effective,yao1995semiconductor,balle1995effective}.
Here we follow the approach by Yao et al. \cite{yao1995semiconductor},
formulated in the framework of the RWA. Thus, we start by introducing the
slowly varying field envelope $\underline{\mathbf{E}}$ and the transformed
off-diagonal elements defined in Eqs.\thinspace(\ref{eq:E_RWA})
and\ (\ref{eq:rho_RWA}), respectively. Furthermore assuming a $\mathbf{k}$
independent Rabi frequency $\underline{\Omega}=\hbar^{-1}\mathbf{d}%
_{\mathrm{cv}}\underline{\mathbf{E}}$, Eq.\thinspace(\ref{eq:SB1}) yields with
Eq.\thinspace(\ref{eq:dissip1})
\begin{equation}
\partial_{t}\eta_{\mathrm{cv},\mathbf{k}}=\left(  \mathrm{i}\Delta
_{\mathbf{k}}-\gamma_{\mathrm{cv},\mathbf{k}}\right)  \eta_{\mathrm{cv}%
,\mathbf{k}}-\frac{\mathrm{i}}{2}\underline{\Omega}\left(  \rho_{\mathrm{c}%
,\mathbf{k}}+\rho_{\mathrm{v},\mathbf{k}}-1\right)  , \label{eq:eff2}%
\end{equation}
where $\Delta_{\mathbf{k}}=\omega_{\mathrm{c}}-\omega_{\mathrm{cv},\mathbf{k}%
}$ with $\omega_{\mathrm{cv},\mathbf{k}}$ given by Eq.\thinspace
(\ref{eq:renorm1}). In the framework of semi-phenomenological macroscopic MB
equation models, the renormalization term is often neglected
\cite{yao1995semiconductor,balle1995effective}. The $\mathbf{k}$\ dependence
can be modeled with Eq.\thinspace(\ref{eq:renorm3}) or a more sophisticated
description. Dividing Eq.\thinspace(\ref{eq:eff2}) by $\left(  \mathrm{i}%
\Delta_{\mathbf{k}}-\gamma_{\mathrm{cv},\mathbf{k}}\right)  $, summing over
$\mathbf{k}$ and defining $p_{\mathrm{cv}}=V_{\mathrm{p}}^{-1}\sum
_{\mathbf{k}}\eta_{\mathrm{cv},\mathbf{k}}$, we obtain%
\begin{equation}
\partial_{t}p_{\mathrm{cv}}=-\frac{\mathrm{i}}{2}\frac{\tau_{1}}{\tau_{2}%
}\underline{\Omega}n_{\mathrm{cv}}-\tau_{2}^{-1}p_{\mathrm{cv}},
\label{eq:eff3}%
\end{equation}
where we have furthermore used Eq.\thinspace(\ref{eq:Ne}) and introduced the
complex parameters $\tau_{1}$ and $\tau_{2}$ with%
\begin{align}
V_{\mathrm{p}}^{-1}\sum_{\mathbf{k}}\frac{\rho_{\mathrm{c},\mathbf{k}}%
+\rho_{\mathrm{v},\mathbf{k}}-1}{\gamma_{\mathrm{cv},\mathbf{k}}%
-\mathrm{i}\Delta_{\mathbf{k}}}  &  =\tau_{1}n_{\mathrm{cv}},\nonumber\\
V_{\mathrm{p}}^{-1}\sum_{\mathbf{k}}\frac{\eta_{\mathrm{cv},\mathbf{k}}%
}{\gamma_{\mathrm{cv},\mathbf{k}}-\mathrm{i}\Delta_{\mathbf{k}}}  &  =\tau
_{2}p_{\mathrm{cv}}. \label{eq:t12}%
\end{align}
Equation (\ref{eq:eff3}) and Eq.\thinspace(\ref{eq:eff1}) in RWA,%
\begin{equation}
\partial_{t}n_{\mathrm{cv}}=\frac{\mathrm{i}}{2}\left(  \underline{\Omega
}p_{\mathrm{cv}}^{\ast}-\underline{\Omega}^{\ast}p_{\mathrm{cv}}\right)
+\Gamma_{12}-\gamma_{1}n_{\mathrm{cv}}, \label{eq:eff4}%
\end{equation}
constitute macroscopic Bloch equations for interband transitions in bulk
semiconductor and quantum well systems. In order to obtain a Maxwell-Bloch
model, Eqs.\thinspace(\ref{eq:eff3}) and (\ref{eq:eff4}) can be coupled to the
optical propagation equation in SVAA, Eq.\thinspace(\ref{eq:SVAA}), where the
RWA polarization term is obtained from Eq.\thinspace(\ref{eq:pol2rwa}) as
$\underline{\mathbf{P}}=2\mathbf{d}_{\mathrm{cv}}p_{\mathrm{cv}}$. The
parameters $\tau_{1}$ and $\tau_{2}$ introduced in Eq.\thinspace(\ref{eq:t12})
can for example be evaluated numerically, or by fitting to experimental data
\cite{yao1995semiconductor}. In this context, it has been found that $\tau
_{1}$ and $\tau_{2}$ can be treated as independent of the optical intensity,
but that especially $\tau_{2}$ shows a pronounced dependence on
$n_{\mathrm{cv}}$ which should be considered in the model
\cite{yao1995semiconductor}, and also allows a phenomenological reintroduction
of renormalization effects.

\subsection{Quantum Well Intersubband Devices}

Intersubband devices, such as QCLs \cite{1994Sci...264..553F}, quantum cascade
detectors \cite{hofstetter2002quantum,gendron2004quantum} and quantum well
infrared photodetectors (QWIPs) \cite{levine1993quantum}, commonly utilize
optical intersubband transitions between quantized energy levels in the
conduction band $\Gamma$ valley of a multiple quantum well structure. The
Maxwell-Bloch model has been extensively applied to such devices, especially
for the dynamic modeling of QCLs. The quantized states $\left|  \psi
_{i},\mathbf{k}\right\rangle $, also referred to as subbands, are
characterized by their wavefunction $\psi_{i}\left(  x\right)  $ where $i$ is
the subband index, and in-plane wavevector $\mathbf{k}=\left[  k_{y}%
,k_{z}\right]  ^{\mathrm{T}}$. These states are commonly found by solving the
one-dimensional effective mass Schr\"{o}dinger equation, obtained from
inserting the ansatz Eq.\thinspace(\ref{eq:psi3D}) into Eq.\thinspace
(\ref{eq:s3D}), for the quantum well potential $V\left(  z\right)  $. As
mentioned in Section \ref{sec:env}, band bending due to space charge effects
is usually considered by solving the coupled Schr\"{o}dinger-Poisson equation
system \cite{datta2005quantum,2009IJQE...45..1059J}, and also nonparabolicity
effects, which play a role especially in mid-infrared devices, can be included
\cite{1989PhRvB..40.7714E,1987PhRvB..35.7770N}. These calculations yield the
eigenenergies $E_{i}$ and wavefunctions $\psi_{i}$, and thus the transition
frequencies $\omega_{ij}=\hbar^{-1}\left(  E_{i}-E_{j}\right)  $ and the
dipole matrix elements, Eq.\thinspace(\ref{eq:dint}), as input for the Bloch
equations. We again choose the semiconductor Bloch equations as a starting
point, with a form analogous to Eq.\thinspace(\ref{eq:SB}). Due to the
typically low doping levels of QCLs, the Hartree--Fock renormalization effects
in Eq.\thinspace(\ref{eq:renorm}) have been found to be relatively small
\cite{waldmueller2006nonequilibrium}, and also Pauli blocking only plays a
secondary role. Furthermore, we can assume $\mathbf{k}$ independent transition
frequencies $\omega_{ij}$ at least for terahertz QCLs, where the energetic
level spacings are smaller than in mid-infrared QCLs and thus the subbands
have nearly parallel dispersion relationships. Under these assumptions,
summing over $\mathbf{k}$ yields Bloch equations of the form Eq.\thinspace
(\ref{eq:MB2}), where we have used that the dipole matrix element is
$\mathbf{k}$ conserving, and introduced intersubband scattering rates
$r_{j\rightarrow i}$ which are related to the generally $\mathbf{k}$ dependent
rates $r_{j,\mathbf{q}\rightarrow i,\mathbf{k}}$ by $r_{j\rightarrow i}%
\rho_{jj}=\sum_{\mathbf{k},\mathbf{q}}r_{j,\mathbf{q}\rightarrow i,\mathbf{k}%
}\rho_{jj,\mathbf{q}}$. Assuming either moderate temporal variations of the
intrasubband electron distribution $\rho_{jj,\mathbf{k}}$ or a moderate
$\mathbf{k}$ dependence of the rates, the $r_{j\rightarrow i}$ are
approximately given by%
\begin{equation}
r_{j\rightarrow i}=\sum_{\mathbf{k},\mathbf{q}}r_{j,\mathbf{q}\rightarrow
i,\mathbf{k}}\rho_{jj,\mathbf{q}}^{0}/\sum_{\mathbf{k}}\rho_{jj,\mathbf{k}%
}^{0}. \label{eq:rate}%
\end{equation}
Here,\ $\rho_{jj,\mathbf{k}}^{0}$ describes the steady state electron
distribution in the subband \cite{jirauschek2017self}. Notably, the
intrasubband electron distributions in QCLs can often be reasonably well
approximated by Fermi--Dirac or Maxwell-Boltzmann distributions, parametrized
by subband electron temperatures which can significantly exceed the lattice
temperatures \cite{2005ApPhL..86k1115V,jirauschek2014modeling}. By contrast,
the off-diagonal density matrix elements generally vary strongly with time,
and no clearly defined concept exists how the $\mathbf{k}$ averaging should be
performed to obtain an effective dephasing rate $\gamma_{ij}$ from a
relaxation term of the form Eq.\thinspace(\ref{eq:dissip1}), $\left[
\partial_{t}\rho_{ij,\mathbf{k}}\right]  _{\mathrm{col}}=-\gamma
_{ij,\mathbf{k}}\rho_{ij,\mathbf{k}}$. This especially matters if the ratio
$\rho_{ii,\mathbf{k}}^{0}/\rho_{jj,\mathbf{k}}^{0}$ has a strong $\mathbf{k}$
dependence, as is the case for significantly different subband electron
temperatures or highly non-thermal distributions \cite{jirauschek2017self}.
Often, an average over the population inversion of the involved subbands is
applied \cite{2009JAP...106f3115N,freeman2016self},%
\begin{equation}
\gamma_{ij}=\sum_{\mathbf{k}}\gamma_{ij,\mathbf{k}}\left|  \rho_{ii,\mathbf{k}%
}^{0}-\rho_{jj,\mathbf{k}}^{0}\right|  /\sum_{\mathbf{k}}\left|
\rho_{ii,\mathbf{k}}^{0}-\rho_{jj,\mathbf{k}}^{0}\right|  , \label{eq:gamma}%
\end{equation}
and a comparison to an alternative way of averaging has yielded similar
results for terahertz QCLs \cite{jirauschek2017density}. Apart from very few
exceptions based on the full-wave MB equations
\cite{freeman2013laser,dietze2011terahertz,riesch2018dynamic}, usually the MB
equations in RWA and SVAA, Eqs.\thinspace(\ref{eq:SHB5})\ and (\ref{eq:RWA}),
are used for the modeling of QCLs and related devices. Also, apart from some
cases including multiple subbands
\cite{choi2010ultrafast,wang2015active,tzenov2016time,tzenov2017analysis,jirauschek2017self,tzenov2018passive}%
, typically a two-level model is employed. In the case of mid-infrared QCLs,
where nonparabolicity effects play a more important role, an approach similar
to Eq.\thinspace(\ref{eq:eff3}) can be envisioned. The transition and
dephasing rates are usually empirically chosen, or extracted from fits to
experimental data. Alternatively, they can be calculated from Eqs.\thinspace
(\ref{eq:rate}) and (\ref{eq:gamma}) based on the Hamiltonians of the relevant
scattering mechanisms, such as electron-electron interactions, scattering with
acoustic and longitudinal optical phonons, as well as impurity, interface
roughness and alloy scattering
\cite{jirauschek2014modeling,jirauschek2017self}. Here, the use of dissipation
rates derived from steady-state models is consistent with the Markovian and
time-homogeneous character of the Lindblad dissipator, which provides the
basis for the Bloch equations. The corresponding scattering rates are
typically evaluated based on Fermi's golden rule \cite{jirauschek2014modeling}%
, while the associated pure dephasing rates can be obtained from Ando's model
\cite{ando1978broadening,unuma2003intersubband,jirauschek2017density}.

For QCLs, the MB equations have primarily been used to study ultrashort pulse
generation by mode-locking
\cite{2010OExpr..1813616G,wojcik2013generation,revin2016active,columbo2018dynamics,2009PhRvL.102b3903M,2009ApPhL..95g1109T,2010OExpr..18.5639T,talukder2014quantum,tzenov2018passive,wang2015active}
and the closely related formation of coherent instabilities
\cite{2007PhRvA..75c1802W,gordon2008multimode,vukovic2017low}, as well as the
generation of frequency combs
\cite{khurgin2014coherent,villares2015quantum,tzenov2016time,tzenov2017analysis,jirauschek2017self}%
. In detail, it has been found that coherent multimode instabilities result in
the emergence of sidebands around the original longitudinal mode, giving rise
to broadband multimode operation
\cite{2007PhRvA..75c1802W,gordon2008multimode}. Furthermore, active
mode-locking has been investigated where short pulses are generated by
modulating the laser current at the cavity roundtrip frequency, yielding good
agreement between simulations and measurements
\cite{2010OExpr..1813616G,wojcik2013generation,revin2016active,columbo2018dynamics}%
. Also the possibility of realizing passive mode-locking in QCLs has been
theoretically explored
\cite{2009PhRvL.102b3903M,2009ApPhL..95g1109T,2010OExpr..18.5639T,talukder2014quantum,tzenov2018passive}%
. Here, pulse formation is obtained by adding saturable absorption regions,
where SIT mode-locking, discussed in Section \ref{sec:SIT}, constitutes a
special variant. Besides, frequency comb operation has been studied, where an
equidistant line spectrum is generated, which serves as a ruler in the
frequency domain for spectroscopic and sensing applications. Here, a
perturbative treatment of the MB equations imposing a comb-like spectrum has
been employed \cite{khurgin2014coherent,villares2015quantum}, as well as full
numerical simulations
\cite{tzenov2016time,tzenov2017analysis,jirauschek2017self}. In most of above
works, spatial hole burning has been considered based on Eqs.\thinspace
(\ref{eq:SHB4}) and (\ref{eq:SHB5}), as it considerably affects the
QCL\ dynamics. In addition, various other effects have been implemented which
can play an important role for mode-locked and frequency comb operation in
QCLs, such as tunneling across thick barriers
\cite{tzenov2016time,tzenov2017analysis,jirauschek2017self} and group velocity
dispersion due to the waveguide and bulk semiconductor material
\cite{2010OExpr..18.5639T,villares2015quantum,bai2016coherent,tzenov2016time,tzenov2017analysis,jirauschek2017self}%
. For optical excitation on very short timescales, memory effects become
important and the presuppositions of the Lindblad approach are too
restrictive, requiring the use of more complex models such as the
quantum-kinetic schemes \cite{iotti2016electronic,butscher2005ultrafast,Savic}.

\begin{figure}[ptb]
\includegraphics{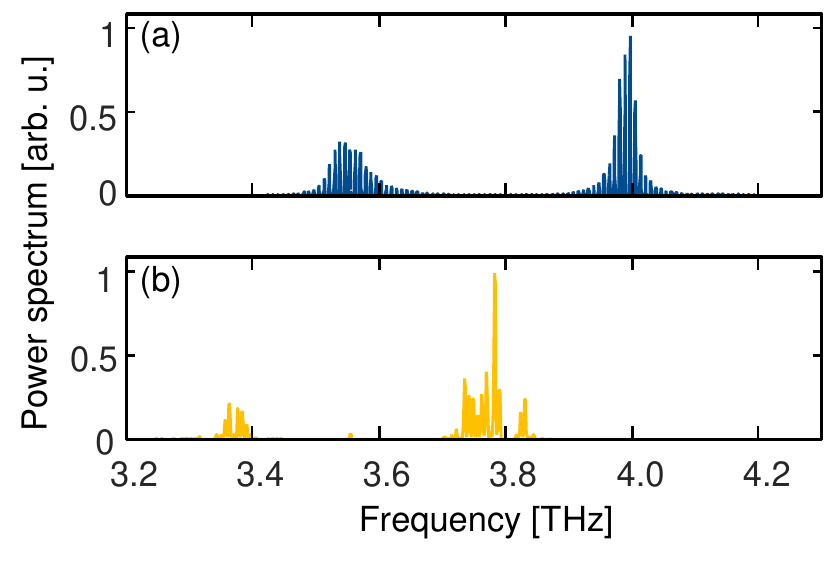}
\caption{Optical comb spectrum of a terahertz QCL, as obtained from (a)
simulation and (b) experiment.}%
\label{fig:comb}%
\end{figure}

\begin{figure}[ptb]
\includegraphics{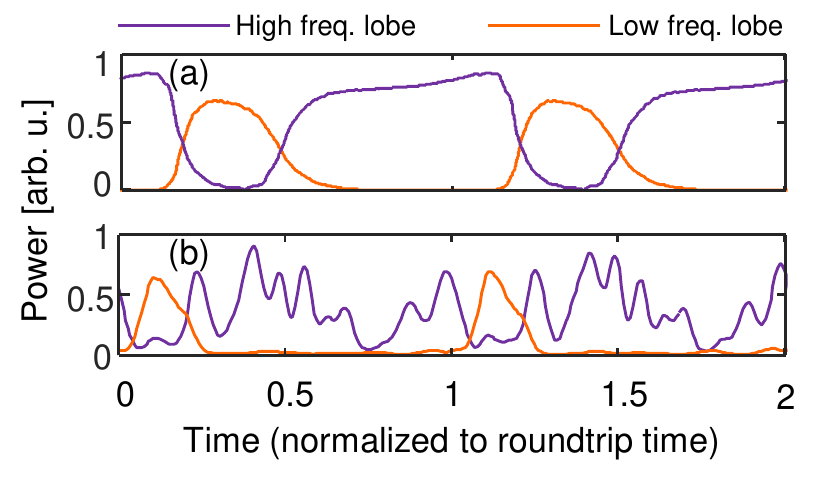}
\caption{Instantaneous optical power in the high and low frequency lobe as
obtained from (a) simulation and (b) experiment.}%
\label{fig:comb2}%
\end{figure}

As mentioned above, MB simulations have for example been used to model
frequency comb operation of QCLs, identifying four-wave mixing as the primary
comb forming mechanism and explaining experimentally observed features
\cite{khurgin2014coherent,villares2015quantum,tzenov2016time,tzenov2017analysis,jirauschek2017self}%
. In Fig.\thinspace\ref{fig:comb}, a comparison between simulation and
experiment is presented for the power spectrum \cite{tzenov2016time},
generated by a THz QCL for frequency comb generation
\cite{burghoff2014terahertz}. Good agreement is found; in particular, a
splitting of the comb spectrum into a high and a low frequency lobe is
observed in both simulation and experiment. Also the simulated temporal
dynamics agrees well with experiment. In Fig.\thinspace\ref{fig:comb2}, the
simulated and measured instantaneous optical power in the high and low
frequency lobe of the comb is shown \cite{tzenov2016time}. Again, good
agreement between theory and experiment is obtained, confirming the validity
of the MB model. In particular, the temporal switching behavior between the
two lobes is reproduced in the simulation.

\subsection{\label{sec:QD}Quantum Dot Devices}

Due to the strong carrier localization and discrete energy spectrum resulting
from the three-dimensional confinement in QDs, they enable lasers and laser
amplifiers with excellent gain, threshold, temperature, and dynamic
characteristics \cite{kirstaedter1994low,ledentsov1998quantum,huffaker19981}.
While these devices rely on interband optical transitions, also the
possibility has been studied to exploit intraband transitions similarly as in
QCLs to obtain lasing in the mid-infrared or terahertz regime
\cite{wingreen1997quantum,burnett2014density,zibik2009long,zhuo2014quantum}.
Furthermore, intraband transitions between bound electron or hole states (or
from bound to continuum states) have been employed for quantum dot infrared
photodetectors \cite{phillips1998far,lee1999bound,liu2001quantum}.

In contrast to bulk semiconductors and quantum well or wire structures which
feature a continuum of states due to the free carrier motion in at least one
dimension, the QD possesses a discrete set of energy eigenstates. Thus, the
application of phenomenological models based on generic discrete-level MB
equations appears to be especially justified for QD systems. Phenomenological
two-level MB equations have been employed for a large range of applications
based on QDs. This includes studies of the spatiotemporal dynamics
\cite{gehrig2004dynamic,sailliot2002filamentation,mukherjee2009spatial} and
SIT mode-locking \cite{arkhipov2016self} in QD lasers, FDTD-based MB
simulations of QD photonic-crystal-cavity lasers \cite{cartar2017}, and QDs
coupled to a nanoparticle or cavity
\cite{protsenko2005dipole,kulkarni2014cavity,waks2010cavity}. Furthermore,
three-level MB equations have been used, for example to study EIT
\cite{nielsen2007numerical}, soliton propagation \cite{adamashvili2007optical}
or all-optical switching \cite{schneebeli2010zeno} in QD structures, and also
four-level models have been developed
\cite{slavcheva2008model,slavcheva2019ultrafast}.

\begin{figure}[ptb]
\includegraphics{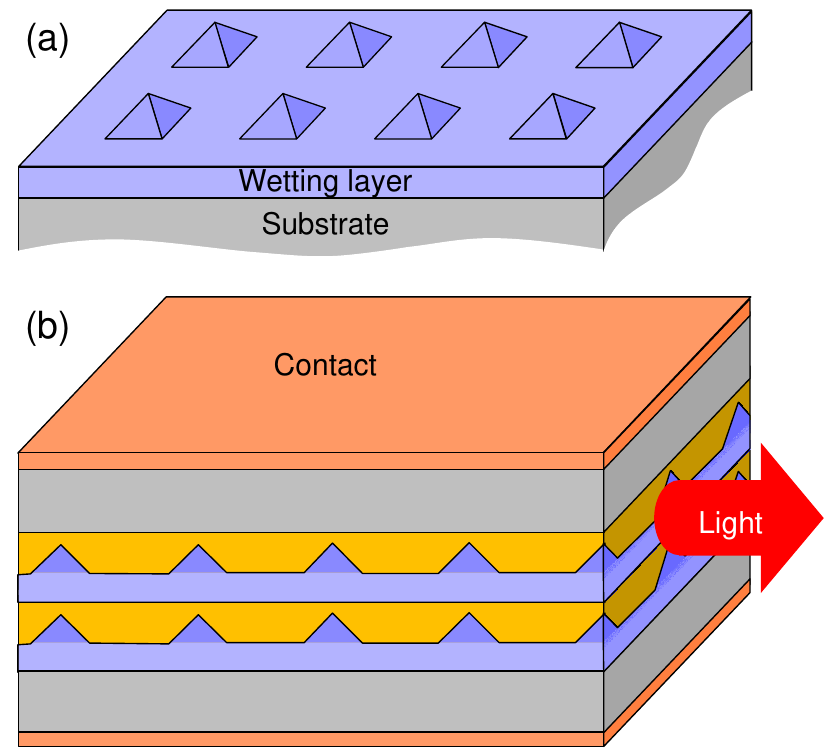}
\caption{Schematic illustration of (a) wetting layer with QDs and (b) QD
laser.}%
\label{fig:QD}%
\end{figure}

Optoelectronic applications employing large ensembles of QDs are often
fabricated utilizing self-assembly of the QDs on top of an initial
quasi-two-dimensional semiconductor layer, which is referred to as wetting
layer, as sketched in Fig.\thinspace\ref{fig:QD}(a). The resulting structure
is subsequently covered by another layer of suitable semiconductor material.
The wetting layer effectively forms a quantum well, which serves as a
reservoir for the carriers. The thus obtained QD layer forms the basis of
various devices such as QD lasers [see Fig.\thinspace\ref{fig:QD}(b)], where
commonly multiple layers are stacked on top of each other to increase the
optical gain.

\begin{figure}[ptb]
\includegraphics{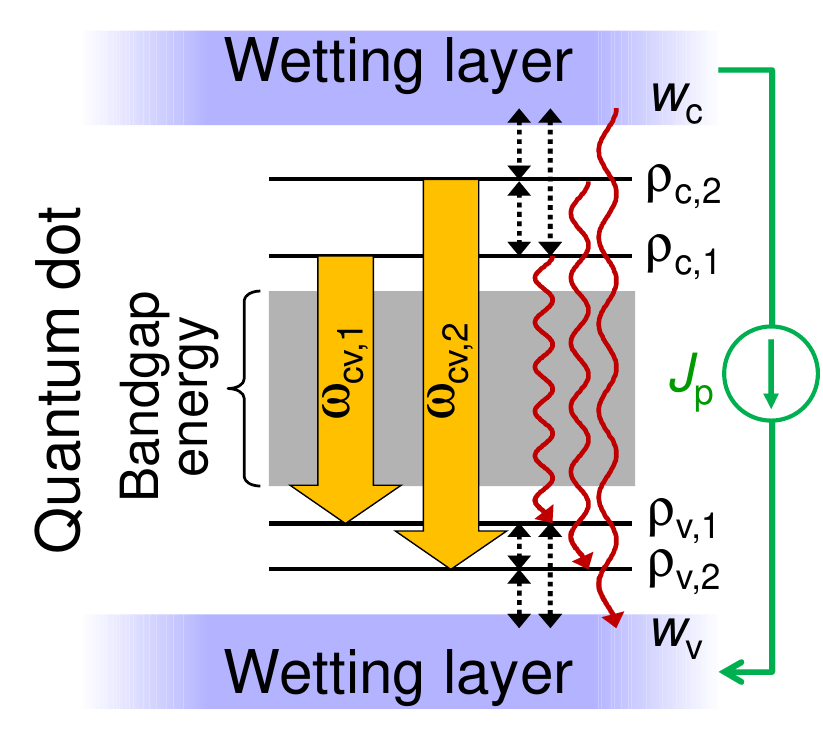}
\caption{Schematic energy diagram of QD and wetting layer. The big arrows
represent coherent light-matter interaction, the dotted arrows indicate
nonradiative intraband transitions, and the wavy arrows represent spontaneous
emission.}%
\label{fig:QDen}%
\end{figure}

In Fig.\thinspace\ref{fig:QDen}, a schematic energy diagram of the wetting
layer and a QD is shown. A description of the QD dynamics based on the
semiconductor Bloch equations, Eqs.\thinspace(\ref{eq:SB}) and
(\ref{eq:renorm}), features renormalized transition and Rabi frequencies due
to many-body Coulomb interactions \cite{chow2003theory,bidegaray2014nonlinear}%
. These renormalization effects are often neglected so that the conventional
Bloch equations, Eq.\thinspace(\ref{eq:MB2}), can be used as a starting point,
which are frequently supplemented by a detailed model for Coulomb scattering
and other scattering mechanisms in the dissipation term
\cite{majer2010cascading,kim2010maxwell}. In the following, the QD conduction
($\alpha=\mathrm{c}$) and valence ($\alpha=\mathrm{v}$) band states are
labeled by an index $i$. Furthermore, as described in Section \ref{sec:inh},
variations in QD\ size result in distributed resonance frequencies, and the
associated inhomogeneous broadening is included by dividing the QDs in
corresponding subensembles $s$ containing a fraction $p_{s}$ of QDs. The Bloch
equations, Eq.\thinspace(\ref{eq:MB2}), thus have to be adapted by replacing
the density matrix elements $\rho_{ij}$ with $\rho_{\alpha \alpha^{\prime},ij}^{s}$
for pairs of states $\left|  \alpha,i\right\rangle $ and $\left|
\alpha^{\prime},j\right\rangle $, where we write for compactness $\rho_{\alpha
\alpha,ij}^{s}=\rho_{\alpha,ij}^{s}$, $\rho_{\alpha \alpha^{\prime},ii}^{s}=\rho_{\alpha
\alpha^{\prime},i}^{s}$ and $\rho_{\alpha \alpha,ii}^{s}=\rho_{\alpha,i}^{s}$. For the
dipole matrix element vectors $\mathbf{d}_{ij}$, frequencies $\omega_{ij}$ and
dephasing rates $\gamma_{ij}$, we proceed analogously. Similarly as for
Eq.\thinspace(\ref{eq:SB}), the\ $\rho_{\mathrm{v},i}^{s}$ are taken as the
hole occupation probability of the $i$th QD valence band level, i.e., matrix
elements $\rho_{ii}$ in Eq.\thinspace(\ref{eq:MB2}) referring to the electron
occupation probabilities of QD valence band states have to be substituted
by $\left(  1-\rho_{\mathrm{v},i}^{s}\right)  $. Apart from the coherent
light-matter interaction, incoherent carrier transitions in QD systems mainly
occur due to carrier-carrier scattering which gives rise to Auger-type
processes, as well as carrier-phonon interactions and spontaneous photon
emission \cite{gehrig2002mesoscopic,majer2010cascading}. Rate equation terms
of the form Eq.\thinspace(\ref{eq:relax1}) with phenomenologically chosen
parameters are frequently used to model incoherent transitions in QD systems
\cite{gehrig2004dynamic,mukherjee2009spatial,protsenko2005dipole,cartar2017,arkhipov2016self,nielsen2007numerical}%
. For a more detailed modeling, it must be taken into account that important
dissipative processes in QDs depend on the occupations of two or more states,
and that Pauli blocking is not included in Eq.\thinspace(\ref{eq:relax1}).
This can be addressed by using an empirical nonlinear rate equation model
\cite{bardella2017self,capua2013finite}, or based on a microscopic treatment
\cite{gehrig2002mesoscopic,majer2010cascading,nielsen2004many,kim2010maxwell}.
For high pump currents, Auger processes, where two carriers scatter from their
respective initial to final levels, involving QD and wetting layer states,
constitute the dominant scattering process \cite{majer2010cascading}. This
includes scattering of two electrons or holes, as well as mixed processes
involving an electron transition in the conduction band and a hole transition
in the valence band. The associated change of the occupation $\rho_{\alpha
,i}^{s}$ is in the following represented by a generic intraband collision term
$\left[  \partial_{t}\rho_{\alpha,i}^{s}\right]  _{\mathrm{intra}}$
\cite{majer2010cascading}, which can be generalized to also include other
scattering-induced intraband carrier transitions, e.g., due to electron-phonon
interactions. Additionally, spontaneous electron-hole recombination is
typically taken into account as an important interband process, which depends
on the occupations of the initial and the final state. Within this model, the
dissipation terms in Eq.\thinspace(\ref{eq:MB2b}) are substituted by the more
general ansatz for incoherent processes \cite{majer2010cascading}%
\begin{equation}
\left[  \partial_{t}\rho_{\alpha,i}^{s}\right]  _{\mathrm{inc}}=\left[
\partial_{t}\rho_{\alpha,i}^{s}\right]  _{\mathrm{intra}}-\sum_{j}%
A_{\mathrm{cv},ij}^{s}\rho_{\alpha,i}^{s}\rho_{\beta\neq\alpha,j}^{s},
\label{eq:QDinc}%
\end{equation}
with the spontaneous recombination coefficient\ $A_{\mathrm{cv},ij}^{s}$. The
Lindblad dephasing rate approach of the form Eq.\thinspace(\ref{eq:relax2}),
as also used in Eq.\thinspace(\ref{eq:MB2a}), has been argued to be generally
well suited to model dephasing in QDs \cite{schneebeli2010zeno}. The dephasing
rates can be calculated based on microscopic models for carrier-carrier and
carrier-phonon scattering
\cite{nielsen2004many,nilsson2005homogeneous,kim2010maxwell,koprucki2011modeling}%
, or are phenomenologically chosen
\cite{majer2010cascading,wilkinson2013influence,nielsen2007numerical,schneebeli2010zeno}%
.

Considering that $\left[  \partial_{t}\rho_{\alpha,i}^{s}\right]
_{\mathrm{intra}}$ contains intra-QD transitions as well as carrier exchange
between the wetting layers and QD states, this term not only depends on the
occupations of the QD states involved, but also on the carrier densities in
the wetting layers. Thus, for a closed carrier transport model, the Bloch
equations have to be extended by equations for the wetting layers, which can
be modeled by \cite{majer2010cascading}%
\begin{equation}
\partial_{t}w_{\alpha}=\frac{J_{\mathrm{p}}}{e}-A_{\mathrm{cv}}w_{\mathrm{c}%
}w_{\mathrm{v}}-2n_{\mathrm{2D}}\sum_{i,s}p_{s}\left[  \partial_{t}%
\rho_{\alpha,i}^{s}\right]  _{\mathrm{intra}}. \label{eq:wet}%
\end{equation}
Here, $w_{\alpha}$ denotes the overall carrier sheet densities in the wetting layers,
i.e., $w_{\mathrm{c}}$ is the total number of conduction band electrons in all
wetting layers divided by the area $S$ of a wetting layer, and $w_{\mathrm{v}%
}$ is defined analogously for the valence band holes. $J_{\mathrm{p}}$ denotes
the electric pump current density. Furthermore, $n_{\mathrm{2D}}$ is the overall QD
sheet density, and the factor $2$ accounts for the spin degeneracy of the QD
states. $A_{\mathrm{cv}}$ in Eq.\thinspace(\ref{eq:wet}) is the rate
coefficient for spontaneous band-band recombination in the wetting layers.
Sometimes the carrier injection from the bulk to the quantum well wetting
layers is modeled by additional equations \cite{lingnau2012failure}. For
self-assembled quantum dash structures, the wetting layers can be considered in
an analogous manner \cite{hadass2005,capua2014}.

The extended Bloch equations, Eqs.\thinspace(\ref{eq:MB2}), (\ref{eq:QDinc})
and (\ref{eq:wet}), are then coupled to Maxwell's equations, Eq.\thinspace
(\ref{eq:maxw}), by the polarization current density for inhomogeneously
broadened media given in Eq.\thinspace(\ref{eq:dtP}),%
\begin{equation}
\mathbf{J}_{\mathrm{q}}=\frac{n_{\mathrm{2D}}}{d_{\mathrm{g}}}\sum
_{\alpha,\alpha^{\prime}}\sum_{i,j}\mathbf{d}_{\alpha^{\prime}\alpha,ji}\sum
_{s}p_{s}\left(  1-2\delta_{\alpha\mathrm{v}}\delta_{\alpha^{\prime}%
\mathrm{v}}\delta_{ij}\right)  \partial_{t}\rho_{\alpha \alpha^{\prime},ij}^{s},
\end{equation}
with the thickness of the gain medium $d_{\mathrm{g}}$. Here, the term
$\left(  1-...\right)  $ compensates for the fact that the electron occupation
probabilities of QD valence band states are in our density matrix convention
given by $\left(  1-\rho_{\mathrm{v},i}^{s}\right)  $. QD lasers and
amplifiers usually operate in TE mode, due to the character of the eigenstates
for the QD shapes and strains obtained with the widely employed
Stranski--Krastanov growth mode \cite{yasuoka2008demonstration}. A
corresponding one-dimensional MB model can be obtained by combining the
extended Bloch equations with Maxwell-type equations for TE operation,
Eq.\thinspace(\ref{eq:MaxwTE}), where the finite overlap of the QD active
region with the mode profile is considered by the field confinement factor,
Eq.\thinspace(\ref{eq:over}). The contribution of the spontaneous emission
processes in Eq.\thinspace(\ref{eq:QDinc}) to the optical field is in most
cases neglected, but can be considered as discussed in Section \ref{sec:noise}.

For interband QD devices, optical intersubband transitions can be neglected.
The QD interband dipole matrix elements are given by Eq.\thinspace
(\ref{eq:d_interb}). As discussed in Section \ref{sec:interdip}, for the
uppermost valence and lowest conduction band states, in good approximation
only optical interband transitions between states with equal quantum numbers
are allowed, and the corresponding envelope wavefunction overlap in
Eq.\thinspace(\ref{eq:d_interb}) is $\left\langle \varphi_{i}\right.  \left|
\varphi_{j}\right\rangle \approx1$
\cite{bimberg1999quantum,bimberg1999quantum2}. In fact, for these states close
to the band edge the index $i$ is typically associated with a single quantum
number \cite{majer2010cascading}, due to the typically small aspect ratio of
QDs. Under above assumptions, the Bloch equations simplify to%
\begin{align}
\partial_{t}\rho_{\mathrm{cv},i}^{s}  &  =-\mathrm{i}\omega_{\mathrm{cv}%
,i}^{s}\rho_{\mathrm{cv},i}^{s}+\mathrm{i}\hbar^{-1}\mathbf{d}_{\mathrm{cv}%
}\left(  1-\rho_{\mathrm{v},i}^{s}-\rho_{\mathrm{c},i}^{s}\right)
\mathbf{E} \nonumber\\
&\,\quad -\gamma_{\mathrm{cv},i}\rho_{\mathrm{cv},i}^{s},\nonumber\\
\partial_{t}\rho_{\alpha,i}^{s}  &  =2\hbar^{-1}\Im\left\{  \mathbf{d}%
_{\mathrm{vc}}\rho_{\mathrm{cv},i}^{s}\right\}  \mathbf{E}+\left[
\partial_{t}\rho_{\alpha,i}^{s}\right]  _{\mathrm{intra}} \nonumber\\
&\,\quad -A_{\mathrm{cv}%
,i}^{s}\rho_{\mathrm{c},i}^{s}\rho_{\mathrm{v},i}^{s}, \label{eq:MBQD}%
\end{align}
where $\alpha=\mathrm{c},\mathrm{v}$, and $\mathbf{d}_{\mathrm{cv},i}^{s}$ has
been approximated by an $s$ and $i$ independent value $\mathbf{d}%
_{\mathrm{cv}}$. For a closed description of the carrier dynamics,
Eq.\thinspace(\ref{eq:MBQD}) is again supplemented by Eq.\thinspace
(\ref{eq:wet}) \cite{majer2010cascading}. The radiative and nonradiative
transitions taken into account in the resulting model are illustrated in
Fig.\thinspace\ref{fig:QDen}. The RWA can be applied in the usual manner, as
described in Section \ref{sec:RWA}. The MB model has been demonstrated to
yield good agreement with experimental results for QD lasers and amplifiers,
and to be instrumental in interpreting the experimental findings. For example,
the ultrafast gain dynamics in a QD amplifier as well as the spatiotemporal
dynamics and emission characteristics of a QD laser were experimentally and
theoretically studied \cite{van2005ultrafast}. Furthermore, based on MB
simulations of ultrashort laser pulse propagation in a QD amplifier, it could
be confirmed that the experimentally observed reshaping was in part due to
coherent light--matter interaction \cite{kolarczik2013quantum,karni2013rabi}.

As discussed in Section \ref{sec:shb}, longitudinal spatial hole burning,
i.e., the formation of an inversion grating due to the standing wave pattern
in a Fabry-P\'{e}rot resonator, is automatically included in full-wave MB
simulations. Assuming that tunneling between adjacent QDs can be neglected,
the degradation of the inversion grating is governed by carrier diffusion in
the wetting layers, which can be modeled by adding to Eq.\thinspace(\ref{eq:wet})
a diffusion term of the form Eq.\thinspace(\ref{eq:diff})
\cite{asryan2000longitudinal,capua2013finite}.

\section{\label{sec:concl}Conclusion and Outlook}

The goal of this review has been to discuss in detail the underlying
theoretical framework of the MB model, its extension and adaption to certain
application areas and types of nanostructures, as well as special analytical
solutions and suitable numerical methods. Apart from the intuitive appeal of
the model and its adaptability, the relative compactness of the Bloch
equations make them highly suitable as an efficient quantum model for the
material polarization in computational electrodynamics. As shown in Section
\ref{sec:Num}, their representation as a system of ordinary differential
equations in time, where the position coordinates only enter as parameters,
allows an efficient coupling to numerical schemes for Maxwell's or related
propagation equations, such as the finite-difference time-domain method. This
compact form of the Bloch equations is enabled by a mostly phenomenological
treatment of dissipation based on the Lindblad formalism and restriction to
classical optical fields as well as discrete energy levels. Fully microscopic
descriptions of light-matter interaction in a semiconductor, such as the
semiconductor MB equations shortly discussed in Section \ref{sec:Bulk}
\cite{chow2012semiconductor,haug2009quantum}, illustrate the limitations of
semi-phenomenological Bloch equations, and can serve as a starting point to
develop improved compact Bloch equations. As an example, this strategy has
been used to model the carrier dynamics in a semiconductor structure with a
quasi-continuum of energy levels in the conduction and valence band by
macroscopic discrete-level Bloch equations
\cite{ning1997effective,yao1995semiconductor,balle1995effective}, as discussed
in Section \ref{sec:Bulk}.

The main requirement for computational models is generally to combine
numerical efficiency with accuracy, predictability and versatility. In this
context, detailed microscopic theories can quickly become very computationally
demanding, which renders them impractical for applications such as device
design \cite{valavanis2008theory}. Thus, a major goal is to further improve
the quantitative accuracy and adaptability of the macroscopic MB equations by
extending the model accordingly, however without substantially increasing its
numerical complexity. This implies that its general form as a system of a few
ordinary differential equations should not be compromised.

Probably the main limitation of the Bloch equations is the phenomenological
implementation of dissipation based on the Lindblad formalism. As shortly
discussed in Section \ref{sec:QD}, an empirical treatment of certain
processes, such as Pauli blocking or carrier-carrier scattering, requires a
generalization to nonlinear models. Here, special care must be taken to
preserve the properties of the density matrix guaranteeing its physical
character, which has for example been achieved by suitably extending the
Lindblad formalism \cite{rosati2014derivation}.

As mentioned in Section \ref{sec:val}, the Lindblad model is only realistic
from a microscopic point of view if the memory decay of the environment occurs
on a faster timescale than the coherent system dynamics and relaxation
processes \cite{breuer2002theory,le2011quantum}. Although the macroscopic MB
equations often work surprisingly well on the verge of, or even outside, this
microscopic validity range, advanced quantitative modeling requires going
beyond the Markovian approximation in such cases. An ad hoc extension of the
Lindblad approach is obtained by replacing $\mathcal{D}_{k}\left(  \hat{\rho
}\right)  \left(  t\right)  $ in Eq.\thinspace(\ref{eq:lindbl}) with $\int
_{0}^{t}K_{k}\left(  t-t^{\prime}\right)  \mathcal{D}_{k}\left(  \hat{\rho
}\right)  \left(  t^{\prime}\right)  \mathrm{d}t^{\prime}$, where $K_{k}$ is
the memory kernel \cite{barnett2001hazards}. In certain cases, it is
sufficient to treat the populations in the usual manner and include memory
effects only for dephasing, which requires substituting the dephasing terms
$\left[  \mathrm{d}_{t}\rho_{\alpha\beta}\right]  _{\mathrm{relax}}%
=-\gamma_{\alpha\beta}\rho_{\alpha\beta}$ in Eq.\thinspace(\ref{eq:relax2}%
)\ with $-\gamma_{\alpha\beta}\int_{0}^{t}K_{\alpha\beta}\left(  t-t^{\prime
}\right)  \rho_{\alpha\beta}\left(  t^{\prime}\right)  \mathrm{d}t^{\prime}$
\cite{adamashvili2008influence}. The characteristic memory time and functional
dependence of the memory kernel, such as Gaussian or exponential, depend on
the underlying scattering mechanism
\cite{butscher2005ultrafast,hu1999coherent}. Since the evaluation of
convolution integrals is numerically expensive, a representation based on
supplemental differential equations is preferential. For exponential memory
kernels $K_{\alpha\beta}=\tau_{\alpha\beta}^{-1}\exp\left(  -t/\tau
_{\alpha\beta}\right)  $, such a differential equation is easily derived, with
$\mathrm{d}_{t}s=-\tau_{\alpha\beta}^{-1}\left(  \gamma_{\alpha\beta}%
\rho_{\alpha\beta}+s\right)  $ where we have introduced $s=\left[
\mathrm{d}_{t}\rho_{\alpha\beta}\right]  _{\mathrm{relax}}$ for compactness.
However, since such modifications obviously do not preserve the Lindblad form
of the dissipation terms, a physical behavior of the density matrix is not
guaranteed, and in fact highly nonphysical behavior can emerge
\cite{barnett2001hazards}.

It should be pointed out that master equation models with memory effects do
not necessarily require convolution integrals \cite{breuer2004genuine}, and
that memory kernel master equations can even usually be cast into a time-local
form \cite{chruscinski2010non,laine2012local,kropf2016effective}. Thus, a
promising approach towards a more generalized treatment of dissipation is to
start with the Lindblad equation in the form Eq.\thinspace(\ref{eq:lindbl2}),
and to generalize the matrix $\mathcal{D}_{ijmn}$ given in Eq.\thinspace
(\ref{eq:Dijmn}) for an arbitrary set of Lindblad operators. As already
mentioned in Section \ref{sec:lindbl}, time dependent Lindblad operators
$\hat{L}_{k}\left(  t\right)  $, corresponding to time-varying dissipation
rates in Eqs.\thinspace(\ref{eq:Dijmn2}) and (\ref{eq:Dijmn3}), are
unproblematic \cite{breuer2004genuine,kropf2016effective}. Any further
generalization of $\mathcal{D}_{ijmn}$ comes at the price of potentially
unphysical results. One example is the occurrence of temporarily negative
rates in Eqs.\thinspace(\ref{eq:Dijmn2}) or (\ref{eq:Dijmn3}), which indeed
introduces memory effects into the Lindblad equation
\cite{laine2010measure,lu2010quantum,laine2012local,tang2012measuring,kropf2016effective,chruscinski2014degree}%
. In the construction of such a model, care should be taken to avoid
unphysical behavior, for example by adding certain constraints
\cite{laine2012local,kropf2016effective}. Furthermore, a widely used model of
the form Eq.\thinspace(\ref{eq:lindbl2}) is the Redfield equation, which is
derived from microscopic considerations, i.e., a perturbative treatment of a
quantum system weakly coupled to the environment
\cite{redfield1957theory,wangsness1953dynamical}. In this case $\mathcal{D}%
_{ijmn}$ corresponds to the generally time dependent \cite{whitney2008staying}
Redfield tensor, which is directly related to the system-environment coupling
and environment Hamiltonians \cite{egorova2003modeling}. The main advantages
of the Redfield model are its strong connection to microscopic physics, and to
some extent the inclusion of short-term memory effects
\cite{whitney2008staying,knezevic2013time}. However, in its commonly used
form, the Redfield equation does not guarantee positivity of the density
matrix which can lead to negative state occupations. In practice, the
emergence of this unphysical behavior appears to be a minor problem
\cite{2009PhRvB..79p5322W,vaz2008non,pan2017density}, and can also be cured
\cite{suarez1992memory,whitney2008staying}.

Finally, applying the Lindblad formalism to a suitably extended state space, a
non-Markovian evolution with arbitrarily long memory times and strong initial
correlations can be described \cite{breuer2007non}. This is achieved by
representing the reduced system density matrix $\rho_{\mathrm{s}}$ with
dimension $N$ as a sum of a certain number $M$ of positive matrices $\rho_{i}%
$, i.e., $\rho_{\mathrm{s}}=\sum_{i}\rho_{i}$ where the traces of the
$\rho_{i}$ must add up to one. Then, a big block diagonal density matrix
$\rho$ with dimension $MN$ is constructed from the $\rho_{i}$, and the
evolution of $\rho$ is modeled by a Lindblad equation for the extended system,
where the operators are required to preserve the block diagonal form of $\rho
$. This leads to $M$ coupled evolution equations for the $\rho_{i}$, where the
dynamics is now defined by $M$ arbitrary Hermitian operators $\hat{H}_{i}$ and
$M^{2}$ sets of arbitrary dissipation operators $\left\{  \hat{L}_{1}%
^{ij},\dots,\hat{L}_{K}^{ij}\right\}  $. In this way, although the dynamics of
$\rho$ is Markovian, the model can describe a highly non-Markovian evolution
of $\rho_{\mathrm{s}}$, while intrinsically preserving the physical properties
of $\rho_{\mathrm{s}}$. An interesting subcase is when the evolution equations
of the $\rho_{i}$ are decoupled, i.e., $\hat{L}_{k}^{ij}=\hat{L}_{k}%
^{ij}\delta_{ij}$ \cite{breuer2007non,budini2005random}. In this case, the
evolution of each $\rho_{i}$ is described by an equation of the Lindblad form
Eq.\thinspace(\ref{eq:lindbl}), but still a non-Markovian dynamics of
$\rho_{\mathrm{s}}$\ is obtained. Of course, for $M=1$, the standard Markovian
Lindblad dynamics is recovered.

\section*{Conflicts of Interest}

The authors declare no conflicts of interest.

\section*{Supporting Information}

Supporting information is available from the Wiley Online Library or from the author.

\section*{Acknowledgments}

We thank Gabriela Slavcheva for valuable comments and suggestions. The authors
acknowledge financial support by the European Union's Horizon 2020 research
and innovation programme under grant agreement No 820419 -- Qombs Project
''Quantum simulation and entanglement engineering in quantum cascade laser
frequency combs'' (FET Flagship on Quantum Technologies), and by the German
Research Foundation (DFG) within the Heisenberg program (JI 115/4-2).

\providecommand{\url}[1]{\texttt{#1}}
\providecommand{\urlprefix}{}
\providecommand{\foreignlanguage}[2]{#2}
\providecommand{\Capitalize}[1]{\uppercase{#1}}
\providecommand{\capitalize}[1]{\expandafter\Capitalize#1}
\providecommand{\bibliographycite}[1]{\cite{#1}}
\providecommand{\bbland}{and}
\providecommand{\bblchap}{chap.}
\providecommand{\bblchapter}{chapter}
\providecommand{\bbletal}{et~al.}
\providecommand{\bbleditors}{editors}
\providecommand{\bbleds}{eds.}
\providecommand{\bbleditor}{editor}
\providecommand{\bbled}{ed.}
\providecommand{\bbledition}{edition}
\providecommand{\bbledn}{ed.}
\providecommand{\bbleidp}{page}
\providecommand{\bbleidpp}{pages}
\providecommand{\bblerratum}{erratum}
\providecommand{\bblin}{in}
\providecommand{\bblmthesis}{Master's thesis}
\providecommand{\bblno}{no.}
\providecommand{\bblnumber}{number}
\providecommand{\bblof}{of}
\providecommand{\bblpage}{page}
\providecommand{\bblpages}{pages}
\providecommand{\bblp}{p}
\providecommand{\bblphdthesis}{Ph.D. thesis}
\providecommand{\bblpp}{pp}
\providecommand{\bbltechrep}{Tech. Rep.}
\providecommand{\bbltechreport}{Technical Report}
\providecommand{\bblvolume}{volume}
\providecommand{\bblvol}{Vol.}
\providecommand{\bbljan}{January}
\providecommand{\bblfeb}{February}
\providecommand{\bblmar}{March}
\providecommand{\bblapr}{April}
\providecommand{\bblmay}{May}
\providecommand{\bbljun}{June}
\providecommand{\bbljul}{July}
\providecommand{\bblaug}{August}
\providecommand{\bblsep}{September}
\providecommand{\bbloct}{October}
\providecommand{\bblnov}{November}
\providecommand{\bbldec}{December}
\providecommand{\bblfirst}{First}
\providecommand{\bblfirsto}{1st}
\providecommand{\bblsecond}{Second}
\providecommand{\bblsecondo}{2nd}
\providecommand{\bblthird}{Third}
\providecommand{\bblthirdo}{3rd}
\providecommand{\bblfourth}{Fourth}
\providecommand{\bblfourtho}{4th}
\providecommand{\bblfifth}{Fifth}
\providecommand{\bblfiftho}{5th}
\providecommand{\bblst}{st}
\providecommand{\bblnd}{nd}
\providecommand{\bblrd}{rd}
\providecommand{\bblth}{th}

\section*{Biographies}

\includegraphics[width=0.3\textwidth]{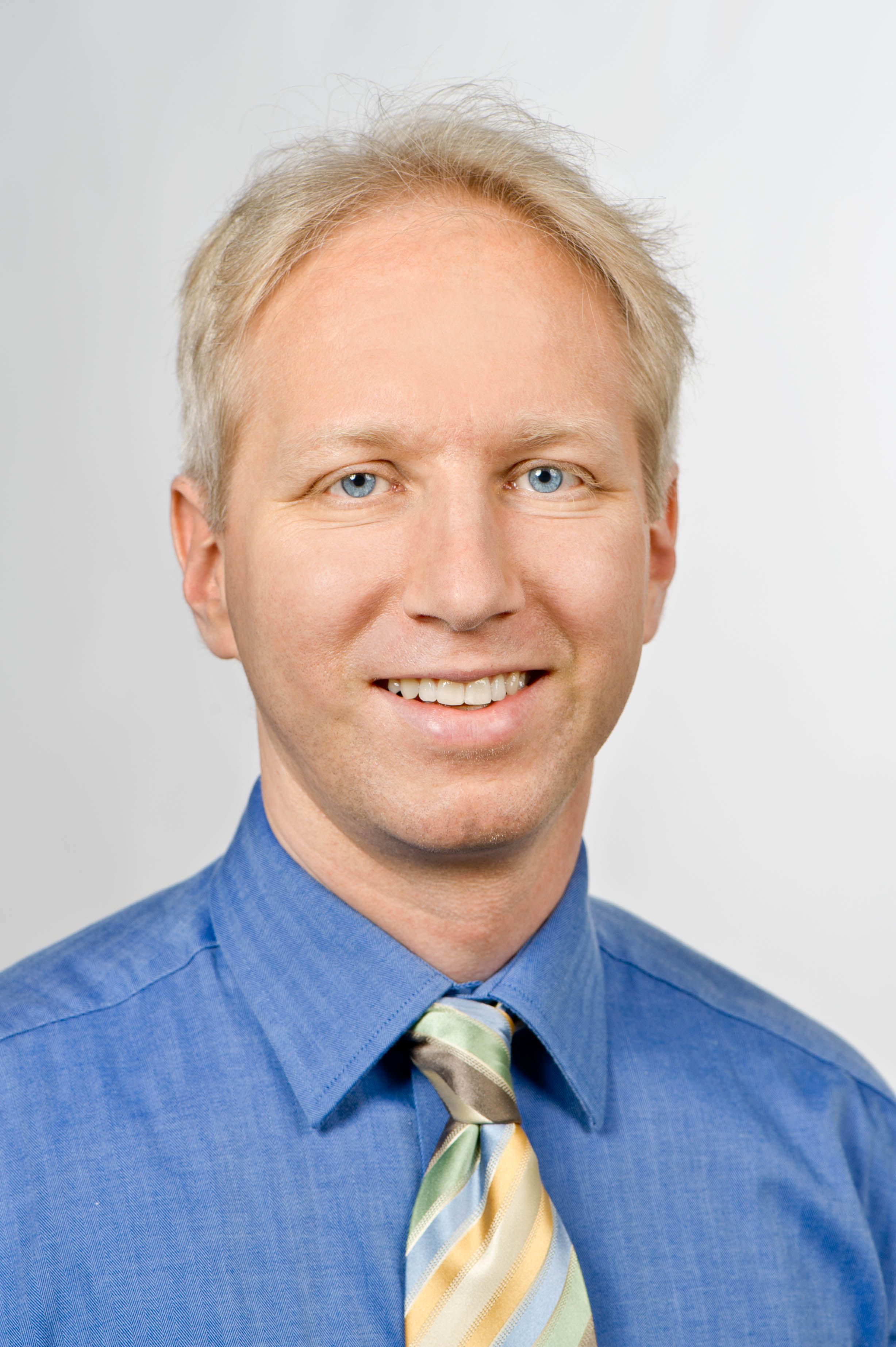}

Christian Jirauschek received the Dipl.-Ing. and Doctoral degrees in
Electrical Engineering from Universit\"{a}t Karlsruhe, Karlsruhe, Germany, in
2000 and 2004, respectively. From 2002 to 2005, he was with the Massachusetts
Institute of Technology (MIT). He then joined the Institute of
Nanoelectronics, Technical University of Munich (TUM), Munich, Germany, where,
starting from 2007 he headed an independent junior research group within the
Emmy Noether Program of the Deutsche Forschungsgemeinschaft (DFG). In 2015, he
was appointed Heisenberg professor of Computational Photonics at TUM. His
research interests include modeling in the areas of photonics and nanoelectronics.

\includegraphics[width=0.3\textwidth]{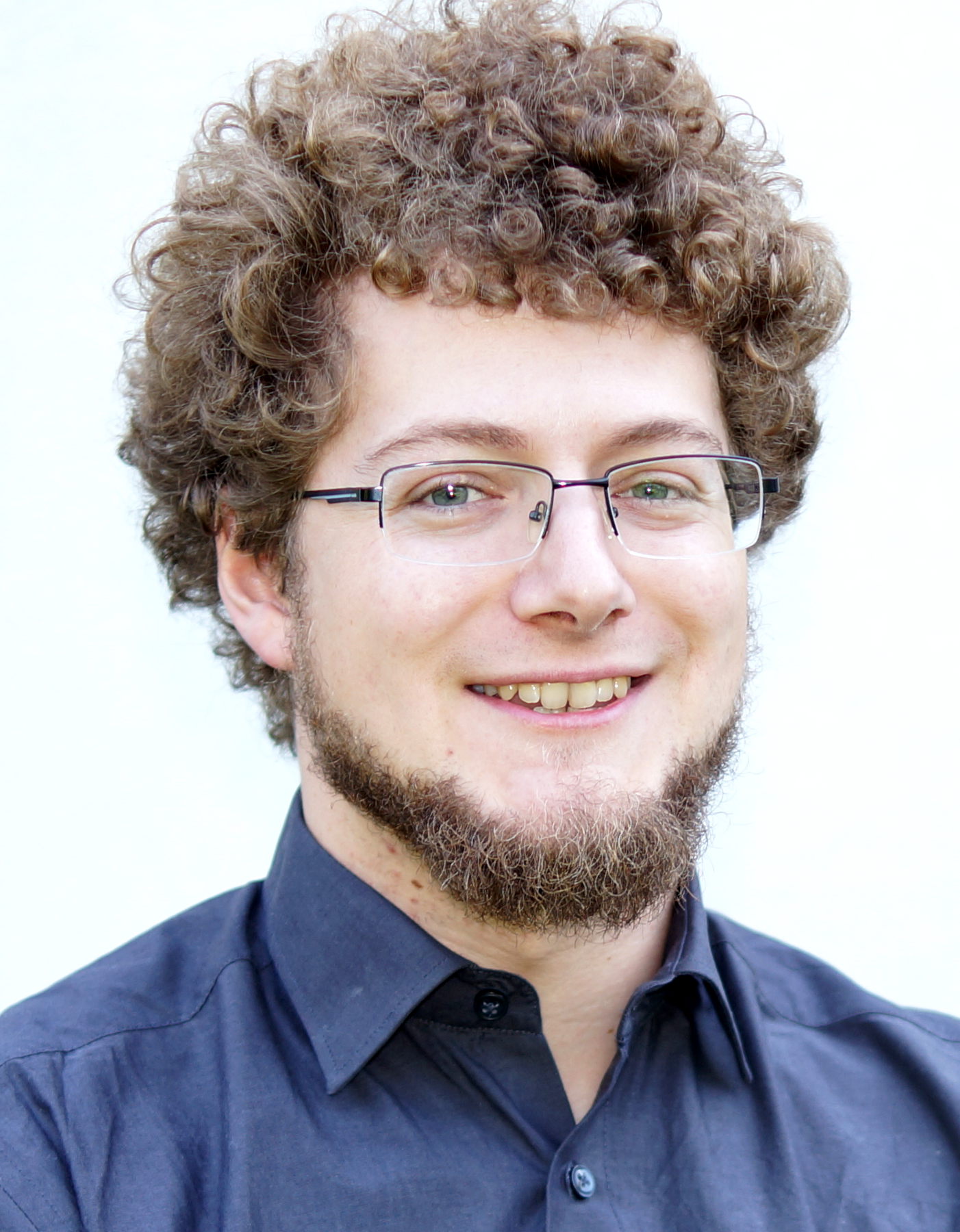}

Michael Riesch received the B.Sc. and M.Sc. degrees in Electrical and Computer
Engineering from the Technical University of Munich (TUM), Munich, Germany, in
2012 and 2015, respectively. He continued his studies at TUM and received the
M.Sc. degree in Computational Science and Engineering in 2016. He then joined
the Computational Photonics group at TUM as research assistant. His research
activities focus on the dynamical simulations of quantum cascade lasers and
include numerical methods and high performance computing.

\includegraphics[width=0.3\textwidth]{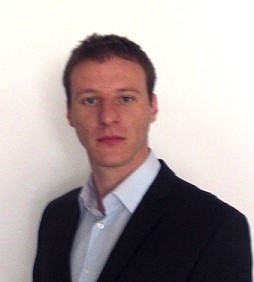}

Petar Tzenov received the B.S. degree in Applied Mathematics from Sofia
University (St. ''Kliment Ohrisdski''), Sofia, Bulgaria, in 2012. He received
the M.S. degree in Computational Science and Engineering from the Technical
University of Munich (TUM), Munich, Germany, in 2014, where between 2015 and
2018, as a member of the Computational Photonics group, he pursued a Ph.D.
degree in the areas of laser physics and nonlinear optics. His research
interests include classical and quantum optics, quantum electronics, and
computational physics, as well as high-performance computing.
\end{document}